\documentstyle[12pt,epsf,eqsection,subeqnarray,amsfonts,amssymb,amsbsy,indent]%
{article}

\newcommand{\CP}{ \hbox{\it CP\/} }
\newcommand{\QP}{ \hbox{\it QP\/} }

\begin{document}

\date{November 6, 2000}

\bibliographystyle{plain}

\title{Pathologies of the Large-$N$ Limit\break
       for $RP^{N-1}$, $\CP^{N-1}$, $\QP^{N-1}$ and\break
       Mixed Isovector/Isotensor $\sigma$-Models}
\author{
  \\
  {\small Alan D. Sokal}                  \\[-0.2cm]
  {\small Andrei O. Starinets}            \\[-0.2cm]
  {\small\it Department of Physics}       \\[-0.2cm]
  {\small\it New York University}         \\[-0.2cm]
  {\small\it 4 Washington Place}          \\[-0.2cm]
  {\small\it New York, NY 10003 USA}      \\[-0.2cm]
  {\small Internet: {\tt SOKAL@NYU.EDU}, {\tt AOS2839@SCIRES.NYU.EDU}}
                                          \\[-0.2cm]
  {\protect\makebox[5in]{\quad}}  
  \\
}
\vspace{0.5cm}

\maketitle
\thispagestyle{empty}   

\begin{abstract}
We compute the phase diagram in the $N \to\infty$ limit
for lattice $RP^{N-1}$, $\CP^{N-1}$ and $\QP^{N-1}$ $\sigma$-models
with the quartic action,
and more generally for mixed isovector/isotensor models.
We show that the $N=\infty$ limit exhibits phase transitions
that are forbidden for any finite $N$.
We clarify the origin of these pathologies by examining
the exact solution of the one-dimensional model:
we find that there are complex zeros of the partition function
that tend to the real axis as $N\to\infty$.
We conjecture the correct phase diagram for finite $N$
as a function of the spatial dimension $d$.
Along the way, we prove some new correlation inequalities for
a class of $N$-component $\sigma$-models,
and we obtain some new results concerning the complex zeros of
confluent hypergeometric functions.
\end{abstract}

\vspace{1cm}

\noindent
{\bf PACS CODES:}  11.15.Pg, 02.30.Gp, 05.50.+q, 11.10.Kk, 11.15.Ex,
64.60.Cn, 64.70.Md, 75.10.Hk.

\vspace{1cm}


\vspace*{-9in}
\hfill NYU-TH/00/11/01

\clearpage

\tableofcontents   
\clearpage

\newcommand{\be}{\begin{equation}}
\newcommand{\ee}{\end{equation}}
\newcommand{\<}{\langle}
\renewcommand{\>}{\rangle}
\newcommand{\para}{\|}
\renewcommand{\perp}{\bot}

\def\spose#1{\hbox to 0pt{#1\hss}}
\def\ltapprox{\mathrel{\spose{\lower 3pt\hbox{$\mathchar"218$}}
 \raise 2.0pt\hbox{$\mathchar"13C$}}}
\def\gtapprox{\mathrel{\spose{\lower 3pt\hbox{$\mathchar"218$}}
 \raise 2.0pt\hbox{$\mathchar"13E$}}}
\def\inapprox{\mathrel{\spose{\lower 3pt\hbox{$\mathchar"218$}}
 \raise 2.0pt\hbox{$\mathchar"232$}}}
\def\bv{\beta_V}
\def\bt{\beta_T}
\def\half{ {{1 \over 2 }}}
\def\smfrac#1#2{\textstyle{#1\over #2}}
\def\smhalf{ {\smfrac{1}{2}} }
\def\scra{{\cal A}}
\def\scrb{{\cal B}}
\def\scrc{{\cal C}}
\def\scrd{{\cal D}}
\def\scre{{\cal E}}
\def\scrf{{\cal F}}
\def\scrl{{\cal L}}
\def\scrm{{\cal M}}
\newcommand{\scrmvec}{\vec{\cal M}_V}
\def\scrmtens{{\stackrel{\leftrightarrow}{\cal M}_T}}
\def\scrp{{\cal P}}
\def\scrs{{\cal S}}
\def\ttens{{\stackrel{\leftrightarrow}{T}}}
\def\scrv{{\cal V}}
\def\scrw{{\cal W}}
\def\scry{{\cal Y}}
\def\scrz{{\cal Z}}
\def\tauss{\tau_{int,\,\scrm^2}}
\def\taux{\tau_{int,\,{\cal M}^2}}
\newcommand{\taum}{\tau_{int,\,\vec{\cal M}}}
\def\taue{\tau_{int,\,{\cal E}}}
\newcommand{\imag}{\mathop{\rm Im}\nolimits}
\newcommand{\real}{\mathop{\rm Re}\nolimits}
\newcommand{\tr}{\mathop{\rm tr}\nolimits}
\newcommand{\sgn}{\mathop{\rm sgn}\nolimits}
\newcommand{\diag}{\mathop{\rm diag}\nolimits}
\newcommand{\codim}{\mathop{\rm codim}\nolimits}
\newcommand{\rank}{\mathop{\rm rank}\nolimits}
\newcommand{\arccosh}{\mathop{\rm arccosh}\nolimits}
\def\textprime{{${}^\prime$ }}
\newcommand{\longto}{\longrightarrow}
\def\var{ \hbox{var} }
\newcommand{\btilde}{ {\widetilde{B}} }
\newcommand{\ctilde}{ {\widetilde{c}} }
\newcommand{\gtilde}{ {\widetilde{G}} }
\newcommand{\jtilde}{ {\widetilde{J}} }
\newcommand{\ktilde}{ {\widetilde{K}} }
\newcommand{\alphatilde}{ {\widetilde{\alpha}} }
\newcommand{\mutilde}{ {\widetilde{\mu}} }
\newcommand{\betatilde}{ {\widetilde{\beta}} }
\newcommand{\bvt}{ {\widetilde{\beta}_V} }
\newcommand{\btt}{ {\widetilde{\beta}_T} }
\def\deltann{ \delta^{\hbox{\scriptsize NN}}  }  
\newcommand{\USp}{ \hbox{\it USp} }
\def\hboxscript#1{ {\hbox{\scriptsize\em #1}} }
\def\ofo{ {{}_1 \! F_1} }
\newcommand{\s}{{s}}   

\newcommand{\plotdot}{\makebox(0,0){$\bullet$}}
\newcommand{\plotsmalldot}{\makebox(0,0){{\footnotesize $\bullet$}}}

\def\bsigma{\mbox{\protect\boldmath $\sigma$}}
\def\btau{\mbox{\protect\boldmath $\tau$}}
\def\br{{\bf r}}

\newcommand{\reff}[1]{(\ref{#1})}

%
\newcommand{\restrict}{\restriction}
\newcommand{\zed}{{\mathbb{Z}}}
\newcommand{\Z}{{\mathbb{Z}}}
\newcommand{\R}{{\mathbb{R}}}
\newcommand{\sR}{{\mathbb{R}}}
\newcommand{\C}{{\mathbb{C}}}
\newcommand{\Q}{{\mathbb{Q}}}
\newcommand{\K}{{\mathbb{K}}}




\newtheorem{theorem}{Theorem}[section]
\newtheorem{proposition}[theorem]{Proposition}
\newtheorem{corollary}[theorem]{Corollary}
\newtheorem{lemma}[theorem]{Lemma}
\def\proof{\bigskip\par\noindent{\sc Proof.\ }}
\newcommand{\proofof}[1]{\bigskip\par\noindent{\sc Proof of {#1}.}\quad}
\def\qed{\hbox{\hskip 6pt\vrule width6pt height7pt depth1pt \hskip1pt}\bigskip}

%
%
\newenvironment{sarray}{
          \textfont0=\scriptfont0
          \scriptfont0=\scriptscriptfont0
          \textfont1=\scriptfont1
          \scriptfont1=\scriptscriptfont1
          \textfont2=\scriptfont2
          \scriptfont2=\scriptscriptfont2
          \textfont3=\scriptfont3
          \scriptfont3=\scriptscriptfont3
        \renewcommand{\arraystretch}{0.7}
        \begin{array}{l}}{\end{array}}

\newenvironment{scarray}{
          \textfont0=\scriptfont0
          \scriptfont0=\scriptscriptfont0
          \textfont1=\scriptfont1
          \scriptfont1=\scriptscriptfont1
          \textfont2=\scriptfont2
          \scriptfont2=\scriptscriptfont2
          \textfont3=\scriptfont3
          \scriptfont3=\scriptscriptfont3
        \renewcommand{\arraystretch}{0.7}
        \begin{array}{c}}{\end{array}}

\section{Introduction}  \label{sec1}

Over the past decade there has been much interest in two-dimensional lattice
$\sigma$-models taking values in the real projective space $RP^{N-1}$
\cite{Hikami_82,Magnoli_87,Ohno_90,%
Kunz_89,Kunz_91,Kunz_92a,Kunz_92b,Butera_92,CEPS_LAT91,CEPS_LAT92,%
CEPS_PRL,Caracciolo_95,Hasenbusch_96,Niedermayer_96,Catterall_98,%
CEPS_rpn_static},
the complex projective space $\CP^{N-1}$
\cite{Hikami_82,DiVecchia_84,Kunz_89,Hasenbusch_92,Jansen_92,Irving_92,%
Wolff_92,Campostrini_92a,Campostrini_92b,Campostrini_93}
or the quaternionic projective space $\QP^{N-1}$
\cite{Gava_80}.\footnote{
   There is also, of course, an enormous literature on the continuum versions
   of these models, which we refrain from citing.
   Our point of view is that the continuum model can be given a {\em meaning}\/
   only by defining it properly as a continuum limit of ultraviolet-cutoff
   models (such as lattice models);  and in order to define such a continuum
   limit, one must first investigate the phase diagram of the cutoff models
   and search for critical points.
}
These models are good laboratories for investigating asymptotic freedom,
topological phenomena and universality;
they are intermediate in complexity between two-dimensional $N$-vector models
($\sigma$-models taking values in the sphere $S^{N-1}$)
and four-dimensional non-abelian gauge theories.\footnote{
   These models are also good laboratories
   for the development and application
   of new collective-mode Monte Carlo algorithms, such as
   multigrid \cite{MGMC_1,Hasenbusch_92,mgon} and
   cluster algorithms
   \cite{Wolff_89a,Wolff_90,CEPS_LAT90,CEPS_LAT91,CEPS_swwo4c2}.
}

In three dimensions, the $RP^2$ lattice theory has long been employed
as a model of nematic liquid crystals
\cite{Lasher_72,Lebwohl_72,Kohring_87}.
In addition, a mixed isovector/isotensor model generalizing the
$RP^2$ theory has been proposed as a model of
orientational transitions in crystals \cite{Krieger_54}
and of uniaxial liquid crystals exhibiting various kinds of head-tail asymmetry
\cite{Lin_87,Leung_87,Biscarini_91};
and a mixed isovector/isotensor model generalizing the
$RP^{N-1}$ theory arose in a study of $N$-component ferromagnets
with annealed random couplings \cite{Oku_82a,Oku_82b}.

The $N \to\infty$ limit of the $RP^{N-1}$ and $\CP^{N-1}$ $\sigma$-models
has been studied by several groups
\cite{Hikami_82,DiVecchia_84,Magnoli_87,Kunz_89,Ohno_90,Kunz_92b};
and Oku and Abe \cite{Oku_82a,Oku_82b},
Magnoli and Ravanini \cite{Magnoli_87}
and Ohno {\em et al.}\/ \cite{Ohno_90}
have studied the $N \to\infty$ limit of a mixed isovector/isotensor model
that generalizes the $RP^{N-1}$ model.\footnote{
   We have also profited from consulting an unpublished manuscript
   on this problem that Sergio Caracciolo and Andrea Pelissetto
   have kindly shared with us \cite{Caracciolo_95}.
}
The result (in lattice dimension $d=2$) is that
there is a first-order phase transition
in the $RP^{N-1}$ and $\CP^{N-1}$ models at
$\widetilde{\beta}_{c} \approx 0.956$.
The question is whether the phase transition observed {\em at}\/ $N=\infty$
occurs also at large but finite $N$.

The $N=\infty$ phase transition of the $RP^{N-1}$ and $\CP^{N-1}$ models
has two curious features
\cite{DiVecchia_84,Butera_private,Ohno_90}:
\begin{itemize}
   \item[(a)]  It occurs in all lattice dimensions, including $d=1$.
    Indeed, it occurs even for a lattice consisting of only two sites!
    (However, in dimension $d \le 3/2$ the transition
    is second-order rather than first-order \cite{Ohno_90}.)
   \item[(b)]  It is associated with the spontaneous breaking of the
    $Z_2$ or $U(1)$ local gauge invariance.
    That is, in the mixed isovector/isotensor model for
    $\widetilde{\beta}_T > \widetilde{\beta}_{c}$,
    the free energy has {\em unequal}\/ left and right derivatives
    with respect to $\widetilde{\beta}_V$ at $\widetilde{\beta}_V = 0$.
    (Equivalently, the isovector energy $E_V$ is a discontinuous function
    of $\widetilde{\beta}_V$ at $\widetilde{\beta}_V = 0$.)
\end{itemize}
Both of these features are, however, rigorously forbidden at finite $N$:
in any short-range one-dimensional model
(and of course on a finite lattice),
the free energy and all expectation values are real-analytic
functions of the parameters in the Hamiltonian\footnote{
   See \protect\cite{Dobrushin_73,Cassandro_81} and
   \protect\cite[Theorem II.5.3 and Remark 1 following it]{Simon_93}.
};
and spontaneous breaking of a local gauge symmetry cannot occur
under any conditions \cite{Elitzur_75}.
Thus, it seems unlikely that such a transition can survive to finite $N$.

In this paper we propose to clarify this behavior,
by comparing to the exact solution in the one-dimensional case.\footnote{
   The one-dimensional $RP^{N-1}$ and mixed isovector/isotensor models
   have been studied previously by
   Liu and Joseph \cite{Liu_72}, Thorpe and Blume \cite{Thorpe_72},
   Kohring and Shrock \cite{Kohring_87},
   Cucchieri {\em et al.}\/ \cite{Cucchieri_97},
   Seiler and Yildrim \cite{Seiler_97},
   and Hasenbusch and Horgan \cite{Hasenbusch_99}.
}
We shall see that, in $d=1$,
the phase transitions in the $RP^{N-1}$, $\CP^{N-1}$ and $\QP^{N-1}$ models
(and more generally in the mixed isovector/isotensor model)
are artifacts of the $N \to\infty$ limit,
arising from complex zeros of the partition function
that tend to the real axis as $N \to\infty$.
There is no phase transition for any finite $N$.\footnote{
   This phenomenon, in which a phase transition that is forbidden
   for all finite $N$ nevertheless occurs at $N=\infty$,
   was observed nearly two decades ago by
   Celmaster and Green \cite{Celmaster_83}.
   They noted that the one-dimensional $SU(N)$ and $U(N)$
   principal chiral models at $N=\infty$ show spontaneous breaking
   of a continuous global symmetry,
   and that the two-dimensional $SU(N)$ and $U(N)$
   lattice gauge theories at $N=\infty$ show spontaneous breaking
   of the local gauge symmetry
   --- transitions that are of course forbidden at finite $N$.
   They concluded that these examples ``illustrate the possible pitfalls
   of investigating symmetry breaking at $N\to\infty$.''
}
We conjecture that the same is true in all lattice dimensions $d \le 2$.

By contrast, in dimension $d>2$ there is a transition
from the isotropic phase to a phase with nematic long-range order
\cite{Angelescu_82,Tanaka_98,Campbell_99};
this transition is expected to be first-order for all $N>2$
\cite{Kohring_87,deGennes_93,Chaikin_95}.
In this paper we shall compute the location and properties
of this transition in the $N \to \infty$ limit.

The $RP^{N-1}$, $\CP^{N-1}$ and $\QP^{N-1}$ models to be considered here
are defined by the lattice Hamiltonian
\be
   H  \;=\;   - {\beta \over 2}
                \sum_{\<xy\>}  | \bsigma_x^* \cdot \bsigma_y |^2   \;,
\ee
where the sum runs over all nearest-neighbor pairs $\<xy\>$
(each pair counted once),
and each spin $\bsigma_x$ is a unit vector in
$\R^N$, $\C^N$ or $\Q^N$, respectively
(in the real case, the complex conjugation and absolute value
are of course redundant).
The {\em a~priori}\/ measure for each spin $\bsigma_x$
is normalized uniform measure on the appropriate unit sphere, which we shall
denote $d\Omega(\bsigma_x)$.
This model has a global symmetry group $G = O(N)$, $U(N)$ or $U(N,\Q)$
as well as a local gauge group $G_{loc} = Z_2$, $U(1)$ or
$U(1,\Q)$ [$\simeq SU(2)$].
In order to treat all three cases in parallel,
it is convenient to write $\K = \R$, $\C$ or $\Q$,
with corresponding real dimension $k = 1$, 2 or 4.

More generally, we shall consider the mixed isovector/isotensor model
defined by the Hamiltonian
\be
   H  \;=\;  - \sum_{\<xy\>}
     \left[ \beta_V \, \bsigma_x \cdot \bsigma_y   \,+\,
            {\beta_T\over 2}  (\bsigma_x \cdot \bsigma_y)^2
     \right]
   \;,
 \label{eq1.2}
\ee
where each spin $\bsigma_x$ is a unit vector in $\R^N$.
The pure $RP^{N-1}$ model corresponds to $\beta_V = 0$.
We shall restrict attention to the case $\beta_T \ge 0$.
For simplicity, we refrain from considering in this paper
the mixed models based on $\CP^{N-1}$ and $\QP^{N-1}$.
The central goal of this paper is to compute the phase diagram
of the model \reff{eq1.2} in the $N\to\infty$ limit ---
it turns out to be surprisingly intricate and dimension-dependent ---
and to discuss how much of this phase diagram should be expected
to survive to finite $N$.

The plan of this paper is as follows:
In Section~\ref{sec2} we show how to obtain the $N \to\infty$ limit
(and the $1/N$ expansion)
for the mixed isovector/isotensor model \reff{eq1.2},
paying particular attention to conceptual questions
(the order of integration,
 how to determine which saddle point is dominant, \ldots).
In Section~\ref{sec3} we specialize to the translation-invariant
nearest-neighbor case, and give two complementary methods for computing
the $N \to\infty$ limit:  a direct method and a parametric method.
In Sections~\ref{sec4} and \ref{sec5} we carry out these calculations
in detail for $\beta_V = 0$ and $\beta_V \neq 0$, respectively;
our main goal is to determine the phase diagram
in the $(\beta_V,\beta_T)$-plane as a function of the spatial dimension $d$
(which we treat as a continuous variable).
Some of the results in these sections are known
from previous work \cite{Magnoli_87,Ohno_90}, but most of them are new.
In Sections~\ref{sec6} and \ref{sec7} we solve the one-dimensional
model and extract the $N \to \infty$ limit;
we clarify the origin of the phase transition at $N=\infty$,
and explain why it does {\em not}\/ occur for any finite $N$.
In Section~\ref{sec8} we draw some conclusions,
and conjecture the phase diagram
in the $(\beta_V,\beta_T)$-plane for finite $N$.
In Appendix~\ref{appendix_b} we collect some formulae concerning
the lattice propagator and free energy in general dimension $d$.
In Appendix~\ref{appendix_c} we prove some asymptotic expansions
for confluent hypergeometric functions and their zeros;
these appear to be new.
In Appendix~\ref{app.correq} we prove some new correlation inequalities
for a class of $N$-component $\sigma$-models that includes
the mixed isovector/isotensor model.

\section{$N\to\infty$ Limit for Mixed Isovector/Isotensor Models:
         General Theory}   \label{sec2}

Let us begin by considering the mixed isovector/isotensor model
on an arbitrary {\em finite}\/ lattice $\scrl$ having $V$ sites.
The Hamiltonian is
\be
   H  \;=\;  - \sum_{\<xy\>}
     \left[ J_{xy} \bsigma_x \cdot \bsigma_y   \,+\,
            {K_{xy} \over 2}  (\bsigma_x \cdot \bsigma_y)^2
     \right]
   \;,
\ee
where the sum runs over all unordered pairs $\<xy\>$ of sites in $\scrl$
(each pair counted once).
We shall assume that $K_{xy} = K_{yx} \ge 0$,
while $J_{xy} = J_{yx}$ may have either sign.
Without loss of generality we assume $J_{xx} = K_{xx} = 0$.
The {\em a~priori}\/ measure for each spin $\bsigma_x$
is normalized uniform measure on the unit sphere in $\R^N$:
\be
   d\Omega(\bsigma_x)   \;=\;
      {\Gamma(N/2)  \over \pi^{N/2}} \,
      \delta(\bsigma_x^2 - 1) \, d^N\!\bsigma_x
   \;.
\ee
The partition function is therefore
\be
  Z  \;=\;  \left( {\Gamma(N/2)  \over  \pi^{N/2}} \right) ^{\! V}
       \int \exp\!\left\{  \sum_{\<xy\>}
                           \left[ J_{xy} \bsigma_x \cdot \bsigma_y   \,+\,
                                {K_{xy} \over 2}  (\bsigma_x \cdot \bsigma_y)^2
                           \right]
                  \right\}
    \,   \prod\limits_x \delta(\bsigma_x^2 - 1) \, d^N\!\bsigma_x
    \;.
  \label{Z_mixed}
\ee
Our main goal is to compute the leading term in the large-$N$ expansion
of the free energy $-\log Z$,
namely the limiting free energy per component
\be
   F  \;=\;   - \lim_{N\to\infty} {1 \over N} \log Z  \;.
\ee
We also want to compute the leading term in the large-$N$ expansion
of the $n$-point correlation functions
$\< \sigma_{x_1}^{(\alpha_1)} \cdots \sigma_{x_n}^{(\alpha_n)} \>$,
the most important of which are the
isovector and isotensor correlation functions
\begin{eqnarray}
   G_V(x,y)   & = &   \< \bsigma_x \cdot \bsigma_y \>             \\[1mm]
   G_T(x,y)   & = &   \< {\bf T}_x \cdot {\bf T}_y \>
      \;\equiv\;   \sum\limits_{\alpha,\beta = 1}^N
     \< T_x^{(\alpha\beta)} T_y^{(\alpha\beta)} \>             \nonumber \\
      & = &   \< (\bsigma_x \cdot \bsigma_y)^2 \> - {1 \over N}
\end{eqnarray}
where we have defined the symmetric traceless isotensor field
\be
   T_x^{(\alpha\beta)}   \;=\;   \sigma_x^{(\alpha)} \sigma_x^{(\beta)}
                                 \,-\,  {1 \over N} \delta^{\alpha\beta}   \;.
 \label{def_isotensor_field}
\ee


We begin by introducing auxiliary variables $\mu_{xy}$
so as to ``Gaussianize'' the isotensor part of the Boltzmann weight:
\be
   \exp\!\left[ {K_{xy} \over 2}  (\bsigma_x \cdot \bsigma_y)^2 \right]
     \;=\; (2\pi K_{xy})^{-1/2}
           \int\limits_{-\infty}^\infty
           \exp\!\left[ - {\mu_{xy}^2 \over 2 K_{xy}}
                        + \mu_{xy} \bsigma_x \cdot \bsigma_y
                 \right]
           \, d\mu_{xy}
           \;.
\ee
[When $K_{xy}=0$, the Gaussian $(2\pi K_{xy})^{-1/2} \exp[-\mu_{xy}^2/2K_{xy}]$
is of course interpreted as $\delta(\mu_{xy})$.]
%
%
It follows that the partition function \reff{Z_mixed}
of the mixed isovector/isotensor model
can be rewritten as that of a pure $N$-vector model with annealed random
couplings $J_{xy} + \mu_{xy}$,
where the $\mu_{xy}$ are Gaussian-distributed
with mean zero and variance $K_{xy}$.\footnote{
   This equivalence was noted almost two decades ago by
   Oku and Abe \cite{Oku_82a},
   who started from the $N$-vector model with annealed random couplings
   and showed its equivalence with the
   mixed isovector/isotensor model \reff{eq1.2}.
}

Next we introduce auxiliary variables $\{ \alpha_x \}$
in the standard way to ``Gaussianize'' the {\em a~priori}\/ measure:
\be
   \delta(\bsigma_x^2 - 1)
     \;=\;   {1 \over 2\pi i}
           \int\limits_{c_x - i\infty}^{c_x + i\infty}
           e^{- \alpha_x (\bsigma_x^2 - 1)}
           \, d\alpha_x
\ee
where $c_x = c_x(\{\mu\})$
is a real constant chosen large enough to make the
integral over $\{ \bsigma \}$ absolutely convergent.
We then get
\begin{eqnarray}
& &  Z  \;=\;  \left( {1 \over 2\pi i} \right) ^{\! V}
               \left( {\Gamma(N/2)  \over  \pi^{N/2}} \right) ^{\! V}
         \int\limits_{-\infty}^{\infty}  \prod\limits_{\<xy\>}
            (2\pi K_{xy})^{-1/2} \, d\mu_{xy}
    \,   \int\limits_{c_x - i\infty}^{c_x + i\infty}  \prod\limits_x d\alpha_x
    \,   \int\limits_{\sR^N}  \prod\limits_x d^N\!\bsigma_x
                                                             \nonumber \\
& & \qquad\qquad \exp\!\left[ \sum\limits_x \alpha_x
                         - \sum\limits_{\<xy\>} {\mu_{xy}^2 \over 2 K_{xy}}
                         - \half (\bsigma, B \bsigma)
                  \right]
\end{eqnarray}
where
\be
   B_{xy}   \;=\;   2\alpha_x \delta_{xy}  \,-\,
                    (J_{xy} + \mu_{xy})
   \;.
 \label{def_B}
\ee
Note that the order of integration here is crucial,
as the $\{c_x\}$ must be chosen in a $\{\mu\}$-dependent way
so that $\real B$ is positive definite.

We can now perform the Gaussian integral over $\{\bsigma\}$.
For convenience let us introduce new variables
$\jtilde_{xy} = J_{xy}/N$,
$\ktilde_{xy} = K_{xy}/N$,
$\mutilde_{xy} = \mu_{xy}/N$,
$\alphatilde_x = \alpha_x/N$ and hence
$\btilde_{xy} = B_{xy}/N$.
We get
\be
  Z  \;=\;  \left( {N \over 2\pi i} \right) ^{\! V}
            \left( {\Gamma(N/2)  \over  \pi^{N/2}} e^{N/2} \right) ^{\! V}
            \left( {2\pi \over N} \right) ^{\! NV/2}
         \int\limits_{-\infty}^{\infty}  \prod\limits_{\<xy\>}
            \left( {N \over 2\pi \ktilde_{xy}} \right) ^{\! 1/2} \,
            d\mutilde_{xy}
         \int\limits_{\ctilde_x - i\infty}^{\ctilde_x + i\infty}
            \prod\limits_x d\alphatilde_x
     \exp[ -N \scrf (\mutilde,\alphatilde) ]
 \label{Z_integral}
\ee
where
\be
   \scrf (\mutilde,\alphatilde)
     \;=\;
   -\sum\limits_x \alphatilde_x
     + \sum\limits_{\<xy\>} {\mutilde_{xy}^2 \over 2 \ktilde_{xy}}
     + \half \tr\log \btilde + {V \over 2}
   \;.
 \label{def_scrf}
\ee

Let us now perform the $\{\alphatilde\}$ integrals by the saddle-point method.
The saddle-point equations are
\be
   {\partial \scrf  \over  \partial \alphatilde_x}
      \;=\;  (\btilde ^{-1})_{xx} - 1  \;=\;  0 \qquad\hbox{for all } x  \;.
   \label{saddle_point.alpha}
\ee
It can be proven \cite{Sokal_n-vector_unpub}
that there exists a unique {\em real}\/ saddle point
$\alphatilde_\star = \alphatilde_\star (\mutilde)$
in the region where $\btilde$ is positive-definite;
we conjecture (but have not proven)
that this is the {\em only}\/ saddle point, real or complex,
in the region where $\real \btilde$ is positive-definite.
For the purposes of calculating the leading $N \to\infty$ contribution,
it suffices to replace $\alphatilde$ by its saddle-point value\footnote{
   In principle one should prove that the limit $N \to \infty$
   and the integration over $\{\mutilde\}$ can be interchanged.
   Unfortunately, we have no rigorous proof, though it seems eminently
   reasonable.
}:
\be
   \widehat{\scrf}(\mutilde)   \;\equiv\;
   \scrf(\mutilde, \alphatilde_\star (\mutilde))
   \;.
\ee
We then perform the integral over $\{\mutilde\}$ by Laplace's method:
it is dominated by the configuration(s) of $\{\mutilde\}$
that make $\widehat{\scrf}(\mutilde)$ an absolute minimum,
and so we have
\be
   F  \;=\;   \inf\limits_{\{\mutilde\}}  \widehat{\scrf}(\mutilde)
\ee
(the prefactors cancel to leading order).


The stationary points of $\widehat{\scrf}(\mutilde)$ are given by
\be
   {\partial \widehat{\scrf}  \over  \partial \mutilde_{xy}}   \;=\;
   \left. {\partial \scrf  \over  \partial \mutilde_{xy}} \right|
                                   _{\alphatilde = \alphatilde_\star (\mutilde)}
      \;=\;  {\mutilde_{xy} \over \ktilde_{xy}} \,-\, (\btilde ^{-1})_{xy}
         \;=\;  0   \qquad\hbox{for all } x,y  \;.
   \label{saddle_point.lambda}
\ee
In general, $\widehat{\scrf}$ has {\em many}\/ stationary points.
Indeed, for a pure $RP^{N-1}$ model ($J \equiv 0$),
any gauge transform $\mutilde_{xy} \to \eta_x \eta_y \mutilde_{xy}$
($\eta = \pm 1$) of a stationary point is also a stationary point with the
same value of $\scrf$, so in particular there are many degenerate
absolute minima.
If $J_{xy} \not\equiv 0$, there is no longer a gauge invariance,
but at least for small $J_{xy}$ all of the foregoing stationary points survive
(and move slightly).
We conjecture that if $J_{xy} \ge 0$ for all $x,y$ (``ferromagnetism''),
then one of the absolute minima has $\mutilde_{xy} \ge 0$ for all $x,y$,
and moreover there is a unique absolute minimum with this property;
but we have, as yet, no proof.
Let us also remark that if $J \equiv 0$,
then $\mutilde \equiv 0$ is always a stationary point
(with $\alphatilde_{\star x} = \half$ for all $x$),
but this is the dominant saddle point only for $K$ sufficiently small
(see Section~\ref{sec4.2}).

By expanding \reff{Z_integral}/\reff{def_scrf} around the saddle point,
one can obtain a systematic expansion in powers of $1/N$.
Examination of this expansion could potentially shed alternative light
on the reliability of the large-$N$ limit;
but we shall not pursue this approach here.

The saddle-point method also provides a $1/N$ expansion of the
correlation functions.  At leading order in $1/N$, the fields
$\bsigma = \{ \sigma_x^{(\alpha)} \} _{x \in \scrl ,\,  1 \le \alpha \le N}$
are Gaussian with mean zero and covariance matrix
\be
   \< \sigma_x^{(\alpha)} \sigma_y^{(\beta)} \>
   \;=\;
   {\delta^{\alpha\beta} \over N}
   \left[ (\widetilde{B}_\star^{-1})_{xy} \,+\,  O(1/N) \right]
\ee
where
\be
   \widetilde{B}_{\star xy}
   \;=\;
   2\widetilde{\alpha}_{\star x} \delta_{xy}  \,-\,
        (\widetilde{J}_{xy} + \mutilde_{xy})
   \;.
 \label{def_Bstar}
\ee
Thus,
\be
   G_V(x,y)  \;\equiv\;   \< \bsigma_x \cdot \bsigma_y \>
             \;=\;        (\widetilde{B}_\star^{-1})_{xy} \,+\,  O(1/N)
   \;,
 \label{eq2.GV}
\ee
where the saddle-point equation \reff{saddle_point.alpha}
ensures consistency when $x=y$ (since $\bsigma_x \cdot \bsigma_x = 1$).
Moreover, higher-point correlation functions
are given by sums of products of $\widetilde{B}_\star^{-1}$:
for example,
\be
   \< \sigma_{x_1}^{(\alpha_1)} \sigma_{x_2}^{(\alpha_2)}
      \sigma_{x_3}^{(\alpha_3)} \sigma_{x_4}^{(\alpha_4)} \>
   \;=\;
   {\delta^{\alpha_1\alpha_2} \delta^{\alpha_3\alpha_4} \over N^2}
   (\widetilde{B}_\star^{-1})_{x_1 x_2}
   (\widetilde{B}_\star^{-1})_{x_3 x_4}
   \,+\, \hbox{two permutations} \,+\,  O(1/N)
\ee
and likewise for all $2n$-point correlation functions.
One useful consequence is that, to leading order in $1/N$,
correlations involving products of $O(N)$-invariant combinations
{\em factorize}\/ \cite{Yaffe_82}:
\be
   \left\<  \prod\limits_{i=1}^n \bsigma_{x_i} \cdot \bsigma_{y_i} \right\>
   \;=\;
   \prod\limits_{i=1}^n \left\<  \bsigma_{x_i} \cdot \bsigma_{y_i} \right\>
   \,+\, O(1/N)
   \;.
 \label{eq2.factorization1}
\ee
In particular,
\be
   G_T(x,y)  \;\equiv\;  \< (\bsigma_x \cdot \bsigma_y)^2 \> - {1 \over N}
             \;=\;       (\widetilde{B}_\star^{-1})_{xy}^2 \,+\,  O(1/N)
   \;.
 \label{eq2.factorization1a}
\ee
For the pure $RP^{N-1}$ model, the factorization \reff{eq2.factorization1}
should be understood as applying
to the symmetry-broken measure $\< \,\cdot\, \> _{SB}$
obtained by letting $\widetilde{\beta}_V \downarrow 0$.
Alternatively, for $Z_2$-gauge-invariant correlations ---
that is, those for which each lattice site occurs an {\em even}\/
number of times in the sequence $x_1,\ldots,x_n,y_1,\ldots,y_n$ ---
we have (to leading order in $1/N$)
\be
   \left\<  \prod\limits_{i=1}^n \bsigma_{x_i} \cdot \bsigma_{y_i} \right\>
   \;=\;
   \prod\limits_{i=1}^n \left\< \bsigma_{x_i} \cdot \bsigma_{y_i} \right\>_{SB}
   \;=\;
   \prod\limits_{i=1}^n \left\< (\bsigma_{x_i} \cdot \bsigma_{y_i})^2 \right\>
        ^{\! 1/2}
   \;,
 \label{eq2.factorization2}
\ee
where $\< \,\cdot\, \>$ denotes expectation in the $Z_2$-gauge-invariant
Gibbs measure.

\section{$N\to\infty$ Limit for Mixed Isovector/Isotensor Models:
         Translation-Invariant Case}   \label{sec3}

\subsection{Basic Definitions}   \label{sec3.1}

Let us now restrict attention to the case where $\scrl$ is a $d$-dimensional
cube of side $L$ with periodic boundary conditions,
and the couplings are translation-invariant and nearest-neighbor:
$J_{xy} = \betatilde_V \deltann_{xy}$, $K_{xy} = \betatilde_T \deltann_{xy}$
with $\betatilde_T > 0$.
Let us also restrict attention to solutions
for $\{\alphatilde\}$ and $\{\mutilde\}$
that are translation-invariant and
for which $\mutilde$ is the same in all lattice directions.
(In principle we should {\em prove}\/ that these are
 the dominant saddle points;
 but since we are unable to do so, we shall simply {\em assume}\/ that
 they are, like all previous workers!)
We thus have
\be
   \btilde_{xy}  \;=\;  2\alphatilde \delta_{xy}  \,-\,
                        (\betatilde_V + \mutilde) \deltann_{xy}
   \;,
\ee
or in Fourier space
\be
   \btilde(p)   \;=\;
     2\alphatilde \left[ 1 \,-\, \gamma \sum\limits_{i=1}^d \cos p_i \right]
 \label{def_btilde_p}
\ee
where
\be
   \gamma   \;\equiv\;   { \betatilde_V + \mutilde
                           \over
                           \alphatilde
                         }
   \;.
  \label{def_gamma}
\ee
Let us introduce the fundamental quantities
\begin{eqnarray}
   f_{d,L}(\gamma)   & = &
      L^{-d} \sum\limits_p
     \left[ 1 \,-\, \gamma \sum\limits_{i=1}^d \cos p_i \right] ^{-1}
   \label{def_sum_fdL}      \\
   g_{d,L}(\gamma)   & = &
      L^{-d} \sum\limits_p
      \log\!\left[ 1 \,-\, \gamma \sum\limits_{i=1}^d \cos p_i \right]
   \label{def_sum_gdL}
\end{eqnarray}
where the sum runs over momenta with components
$p_i = 2\pi k_i/L$ ($k_i = 0, 1, \ldots, L-1$).
These quantities are well-defined for $|\gamma| < 1/d$;
they are strictly increasing functions of $|\gamma|$,
and diverge as $|\gamma| \uparrow 1/d$.
They satisfy
\be
   {d g_{d,L}(\gamma)  \over  d\gamma}   \;=\;
      -\, {f_{d,L}(\gamma) - 1  \over   \gamma}
   \;.
   \label{deriv_gdL}
\ee
We then have
\begin{subeqnarray}
   (\btilde ^{-1})_{xx}  & = &
       {1 \over 2\alphatilde}  f_{d,L}(\gamma)
   \slabel{binv.a}  \\[2mm]
   (\btilde ^{-1})_{xy}  & = &
       {1 \over 2d\alphatilde\gamma} \left[ f_{d,L}(\gamma) - 1 \right]
       \qquad\hbox{for } |x-y| = 1
   \slabel{binv.b}
   \label{binv}
\end{subeqnarray}
where of course the right-hand side of \reff{binv.b}
must be interpreted as zero when $\gamma=0$.
After passing to the infinite-volume limit,
the correlation function $(\btilde ^{-1})_{xy}$ exhibits exponential decay
\be
   (\btilde ^{-1})_{xy} \;\sim\; e^{-m|x-y|}
       \qquad\hbox{as } |x-y| \to\infty
\ee
with mass gap
\be
   m   \;=\;  \arccosh \!\left[ {1 \over \gamma} - (d-1) \right]
   \;.
 \label{mass_gap}
\ee

Our goal is to compute the free energy per unit volume
\be
   \bar{F}  \;\equiv\;  L^{-d} F
\ee
in the infinite-volume limit $L \to\infty$.
We shall do this in two ways:
a direct approach (Section \ref{sec3.2})
and a parametric approach (Section \ref{sec3.3}).
These two approaches have complementary advantages,
and together they give added insight into the phase diagram.
The parametric approach has been used previously by
Magnoli and Ravanini \cite{Magnoli_87},
while the direct approach appears to be new.

\subsection{Direct Approach}   \label{sec3.2}

We implement the method set forth in Section \ref{sec2}:
namely, we compute the function
$\widehat{\scrf}(\mutilde) \equiv \scrf(\mutilde, \alphatilde_\star (\mutilde))$
and then find its absolute minimum.
The first step is to fix $\mutilde$ and solve
the saddle-point equation \reff{saddle_point.alpha} for $\alphatilde$:
from \reff{binv.a} we have
\be
   2\alphatilde_\star  \;=\;   f_{d,L}(\gamma_\star)   \;\equiv\;
      f_{d,L} \!\left( { \betatilde_V + \mutilde
                         \over
                         \alphatilde_\star
                       }
                \right)   \;.
    \label{saddle.1}                                                  \\[2mm]
\ee
It is not hard to see that
$\gamma_\star$ goes monotonically from $-1/d$ to $1/d$
as $\mutilde$ goes from $-\infty$ to $\infty$.\footnote{
   This is easiest seen in reverse:  write
   $\mutilde = {1 \over 2} \gamma_\star f_{d,L}(\gamma_\star) - \betatilde_V$.
   It follows from the properties of $f_{d,L}$ that
   $\mutilde$ goes monotonically from $-\infty$ to $\infty$
   as $\gamma_\star$ goes from $-1/d$ to $1/d$.
}
So we can use $\gamma_\star$ (henceforth denoted simply by $\gamma$)
as an independent variable in place of $\mutilde$.
The free energy per unit volume
$\overline{\scrf}(\mutilde) \equiv L^{-d} \widehat{\scrf}(\mutilde)$,
written as a function of $\gamma$,
is
\be
  \overline{\scrf}(\gamma)  \;=\;
  {1 \over 2}   \left[1  \,-\, f_{d,L}(\gamma)
                \,+\, {d \over \btt} \left( {\gamma f_{d,L}(\gamma) \over 2}
                                            - \bvt   \right)^2
                \,+\, \log f_{d,L}(\gamma) \,+\, g_{d,L}(\gamma)
                \right]   \;.
  \label{free_energy_gamma}
\ee
Let us write it as
\be
   \overline{\scrf}(\gamma)  \;=\;
   \scrf_0(\gamma)  \,+\,  \bvt \scrf_1(\gamma)  \,+\,
                           {d \over 2\btt} \bvt^2   \;,
  \label{free_energy_gamma_a}
\ee
where $\scrf_0(\gamma)$ [resp.\ $\scrf_1(\gamma)$] is an even (resp.\ odd)
function of $\gamma$:
\begin{eqnarray}
   \scrf_0(\gamma)   & = &  {1 \over 2}   \left[1 \,-\, f_{d,L}(\gamma)
        \,+\, {d \over 4 \btt} \gamma^2 f_{d,L}(\gamma)^2
        \,+\, \log f_{d,L}(\gamma) \,+\, g_{d,L} (\gamma) \right]
     \label{free_energy_even}  \\[3mm]
   \scrf_1(\gamma)   & = &   -\, {d \over 2 \btt} \, \gamma f_{d,L}(\gamma)
     \label{free_energy_odd}
\end{eqnarray}
The free energy per unit volume is then
\be
   \bar{F}   \;=\;  \inf_\gamma  \overline{\scrf}(\gamma)   \;.
\ee
Since $\scrf_1(\gamma)/\gamma < 0$ for all $\gamma$,
it follows that when $\bvt > 0$ (resp.\ $\bvt < 0$)
the absolute minimum of $\overline{\scrf}$ lies at
$\gamma > 0$ (resp.\ $\gamma < 0$).
Of course, when $\bvt = 0$, $\overline{\scrf}$ is an even function
of $\gamma$, so the sign of $\gamma$ is irrelevant.

Note also that $f_{d,L}(\gamma)$ goes monotonically from
1 to $+\infty$ as $|\gamma|$ goes from 0 to $1/d$.
So we can use $f \equiv f_{d,L}(\gamma)$ and $\sgn\gamma$
as independent variables in place of $\gamma$, if we prefer.
As we shall see, this is sometimes convenient.

Note, finally, that we can read off from the $\bvt = 0$ solution
the qualitative behavior of the free energy for small but nonzero $\bvt$.
Indeed, suppose that the absolute minima of $\scrf_0(\gamma)$
lie at $\pm \gamma_1,\ldots,\pm \gamma_n$,
with common minimum value $\scrf_0(\pm \gamma_i) = \bar{F}_0$.
Then, for small $\bvt$, we have from \reff{free_energy_gamma_a}
\be
   \bar{F}  \;\equiv\;  \inf_\gamma  \overline{\scrf}(\gamma)
       \;=\;  \bar{F}_0 \,-\, |\bvt| \max_{1 \le i \le n} |\scrf_1(\gamma_i)|
                        \,+\, O(\bvt^2)   \;.
 \label{eq3.15}
\ee
Thus, if the unique absolute minimum of $\scrf_0(\gamma)$ lies at
$\gamma=0$, we have $\bar{F} = \bar{F}_0 + O(\bvt^2)$;
but if there is any absolute minimum of $\scrf_0(\gamma)$ at $\gamma \neq 0$,
we have $\bar{F} = \bar{F}_0 - O(|\bvt|)$
and hence a first-order phase transition as $\bvt$ passes through 0.
Surprisingly, this happens even in finite volume ($L < \infty$).
Note also that the isovector energy $E_V \sim \partial\bar{F}/\partial\bvt$
is discontinuous at such a first-order transition,
indicating that the $Z_2$ gauge invariance of the $\bvt=0$ theory
is spontaneously broken.
Both these features are manifestly artifacts of the $N\to\infty$ limit,
as they are rigorously forbidden at any finite $N$.

Let us now consider the infinite-volume limit $L\to\infty$.
{\em Formally}\/ we can pass to the infinite-volume limit
in \reff{def_sum_fdL} and \reff{def_sum_gdL}, yielding
\begin{eqnarray} 
   f_d(\gamma)  \;\equiv\;  \lim\limits_{L\to\infty}  f_{d,L}(\gamma)   & = &
      \int {d^d p  \over (2\pi)^d}  \,
     \left[ 1 \,-\, \gamma \sum\limits_{i=1}^d \cos p_i \right] ^{-1}
   \label{def0_fd}   \\[2mm]
   g_d(\gamma)  \;\equiv\;  \lim\limits_{L\to\infty}  g_{d,L}(\gamma)   & = &
      \int {d^d p  \over (2\pi)^d}  \,
      \log\!\left[ 1 \,-\, \gamma \sum\limits_{i=1}^d \cos p_i \right]
   \label{def0_gd}
\end{eqnarray}
where the integral runs over the Brillouin zone $[-\pi,\pi]^d$.
This substitution gives the correct answer in dimension $d \le 2$.
However, in dimension $d>2$ we must be a bit more careful,
because the infinite-volume quantity $f_d(\gamma)$
tends to a finite value $f_*$ as $\gamma \uparrow 1/d$
even though the finite-volume quantity $f_{d,L}(\gamma)$
diverges due to the contribution of the $p=0$ mode.\footnote{
   The same happens as $\gamma \downarrow -1/d$,
   due to the contribution of the $p=(\pi,\ldots,\pi)$ mode.
}
In particular, the equation $f_{d,L}(\gamma) = f$
has a solution for arbitrarily large $f$,
which satisfies $\gamma = 1/d - O(L^{-d})$.
In infinite volume, we can think of all these solutions
with $f \ge f_*$ as arising from $\gamma = 1/d$;
we shall refer to these solutions as the ``condensate phase'',
and use $f$ to parametrize them rather than $\gamma$.
It is important to note that when $f = 2\alphatilde_\star$ exceeds $f_*$,
the finite-lattice Green's function $\btilde_\star(p)^{-1}$
defined by \reff{def_btilde_p},
which consists of delta functions located at the momenta
with components $p_i = 2\pi k_i/L$,
converges in distribution as $L\to\infty$
{\em not}\/ to the corresponding infinite-lattice Green's function
$\btilde_\star(p)^{-1}$, but rather to
\be
   \btilde_\star(p)^{-1}  \,+\,
       (2\pi)^d \left( 1 - {f_* \over f} \right) \delta(p)
   \;.
\ee
Here the delta-function contribution at $p=0$ corresponds to
the long-range order
\be
   \lim_{|x-y| \to \infty} (\btilde_\star^{-1})_{xy}   \;=\;
   1 - {f_* \over f}
   \;.
\ee
Note also that the zero mode does {\em not}\/ contribute to $g_{d,L}$
in the infinite-volume limit at fixed $f \ge f_*$,
since $1/d - |\gamma| \gtapprox L^{-d}$
and $\lim_{L\to\infty} (\log L^d)/L^d = 0$;
therefore we should consider $g_{d,L}(\gamma)$
to be fixed at the value $g_* \equiv g_d(1/d)$.

In summary, in dimension $d>2$ we should consider the additional branches
\be
   \overline{\scrf}_\pm (f)  \;=\;
   \scrf_{0*}(f)  \,\pm\,  \bvt \scrf_{1*}(f)  \,+\,
                           {d \over 2\btt} \bvt^2   \;,
  \label{free_energy_gamma_f}
\ee
(corresponding to $\gamma = \pm 1/d$)
as a function of the variable $f \in [f_*,\infty)$, where
\begin{eqnarray}
   \scrf_{0*}(f)   & = &  {1 \over 2}   \left[1 \,-\, f
        \,+\, {f^2 \over 4d \btt}
        \,+\, \log f \,+\, g_* \right]
     \label{free_energy_even_f}  \\[3mm]
   \scrf_{1*}(f)   & = &   -\, {f \over 2 \btt}
     \label{free_energy_odd_f}
\end{eqnarray}
and we should then minimize over both \reff{free_energy_gamma_a}
and \reff{free_energy_gamma_f}.
Since $\scrf_{1*}(f) < 0$, the optimal choice of sign in
\reff{free_energy_gamma_f} is clearly $\sgn\gamma = \sgn \bvt$.
Whenever the absolute minimum of the free energy is found in the
condensate phase $f > f_*$ at $\gamma=1/d$,
the correlation functions at $N=\infty$
exhibit both ferromagnetic and nematic long-range order\footnote{
   In the antiferromagnetic case $\gamma = -1/d$, one has instead
   $\lim\limits_{|x-y| \to \infty}  (-1)^{|x-y|} G_V(x,y) = 1 - f_*/f$.
}:
\begin{eqnarray}
   \lim_{|x-y| \to \infty}  G_V(x,y)   & = &  1 - {f_* \over f}   \\[2mm]
   \lim_{|x-y| \to \infty}  G_T(x,y)   & = &
       \left( 1 - {f_* \over f} \right) ^{\! 2}
\end{eqnarray}

In Appendix \ref{appendix_b} we collect some results concerning
the behavior of the functions $f_d(\gamma)$ and $g_d(\gamma)$.
In particular, we show how these quantities have a natural
analytic continuation in the parameter $d$,
which allows us to discuss noninteger dimensions $d>0$;
the ability to treat $d$ as a continuous variable is very useful.
The small-$\gamma$ Taylor expansions of $f_d(\gamma)$ and $g_d(\gamma)$
are given in \reff{expansion_fd} and \reff{expansion_gd}, respectively.
It follows that $\scrf_0(\gamma)$ and $\scrf_1(\gamma)$
have the following small-$\gamma$ expansions:
\begin{eqnarray}
\scrf_0(\gamma)   &=&
   {d\over 8}\left( {1\over \btt } - 1\right) \gamma^2  \,+\,
{d\over 64}\left( {8d\over \btt } + (3-10d) \right)\gamma^4  
  \nonumber \\ 
 & & \quad +\,  \frac{d}{192}
   \left( {6d(7d-3)\over \btt} -  (62d^2 - 63d +20)\right)\gamma^6
\,+\, \cdots   
     \label{free_energy_even.expansion}   \\[3mm]
\scrf_1(\gamma ) & = &    -\frac{d}{2\btt}\gamma
 \,-\, \frac{d^2}{4\btt}\gamma^3  \,-\,
 \frac{3 d^2 (2d-1)}{16 \btt}\gamma^5  \,+\, \cdots
     \label{free_energy_odd.expansion}
\end{eqnarray}
The behavior of $f_d(\gamma)$ and $g_d(\gamma)$
as $\gamma \uparrow 1/d$ is discussed at the end of
Appendix \ref{appendix_b.1}.

\subsection{Parametric Approach}   \label{sec3.3}

Thus far we have been analyzing the phase diagram by searching directly
for the absolute minimum of $\overline{\scrf}(\mutilde)$,
without bothering to use the stationarity equation
${\overline{\scrf}\,}'(\mutilde) = 0$. 
On the other hand, in searching for the absolute minimum
we have the {\em right}\/ to restrict attention to such stationary points
if we so choose.  As we shall see, this alternate approach has some
advantages:  in particular, it will allow us to visualize the phase diagram
simultaneously for all $\btt > 0$ (at any fixed value of $\bvt$).

We proceed as follows:
The stationarity equation \reff{saddle_point.lambda} for $\mutilde$,
when combined with \reff{def_gamma}/\reff{binv.b}
and the saddle-point equation \reff{saddle.1} for $\alphatilde$,
yields\footnote{
   Alternatively, this same equation can be obtained by
   differentiating \reff{free_energy_gamma_a} with respect to $\gamma$,
   using \reff{deriv_gdL}, and setting the result equal to zero.
   Miraculously, the derivative of $f_{d,L}$ enters here
   only through the overall factor $\gamma f_{d,L} + f'_{d,L}$
   [which never vanishes, see e.g.\ \reff{eqnB.ineq_fprime}],
   so the solution does not involve $f'_{d,L}$ at all.
}
\be
     {1 \over 2} \gamma f_{d,L}(\gamma) - \betatilde_V
     \;\equiv\; \mutilde  \;=\;
     { f_{d,L}(\gamma) - 1
       \over
       d \gamma f_{d,L}(\gamma)
     }
     \, \betatilde_T
     \;.
  \label{saddle.2}
\ee
If we restrict attention to solutions of \reff{saddle.2},
then the free energy \reff{free_energy_gamma} can be rewritten
so as to eliminate reference to either $\bvt$ or $\btt$:
\begin{subeqnarray}
  \overline{\scrf}  & = &
  {1 \over 2}   \left[1  \,-\, f_{d,L}(\gamma)
                \,+\, \log f_{d,L}(\gamma) \,+\, g_{d,L}(\gamma)
                \,+\, {\btt \over d} \left( {f_{d,L}(\gamma) - 1
                                             \over
                                             \gamma f_{d,L}(\gamma)
                                            }
                                     \right)^2
                \right]  \qquad
   \slabel{free_energy_gamma.parametric1}   \\[3mm]
   & = &
   {1 \over 2}   \left[{1  - f_{d,L}(\gamma)
                        \over
                        2}
                \,+\, \log f_{d,L}(\gamma) \,+\, g_{d,L}(\gamma)
                \,-\, \bvt {f_{d,L}(\gamma) - 1
                            \over
                            \gamma f_{d,L}(\gamma)
                           }
                \right] \;.
   \slabel{free_energy_gamma.parametric2}
   \label{free_energy_gamma.parametric}
\end{subeqnarray}

The solutions to \reff{saddle.2} come in two types:

(a)  $\gamma = 0$ is a solution to \reff{saddle.2} if and only if $\bvt = 0$
(i.e.\ the pure $RP^{N-1}$ model);
the corresponding free energy is $\overline{\scrf} = 0$.\footnote{
   We remark that Magnoli and Ravanini \cite[equation (3)]{Magnoli_87}
   made the mistake of multiplying equation \reff{saddle.2} by $\gamma$,
   thereby introducing a spurious $\gamma=0$ solution also for
   $\betatilde_V \neq 0$.
   Fortunately, this error had no effect on their final results,
   because the purported $\gamma=0$ solution was never the dominant
   saddle point anyway.
}

(b)  If $\gamma \neq 0$, we can solve \reff{saddle.2} for $\btt$:
\be
   \btt  \;=\;  {d \gamma^2 f_{d,L}(\gamma)^2
                 \over
                 2 [f_{d,L}(\gamma) - 1]
                }
                \,-\,
                {d \gamma f_{d,L}(\gamma)
                 \over
                 f_{d,L}(\gamma) - 1
                }  \bvt
    \;.
 \label{beta.gamma}
\ee
For each fixed $\betatilde_V$,
equations \reff{beta.gamma} and \reff{free_energy_gamma.parametric2}
give $\betatilde_T$ and $\overline{\scrf}$ parametrically as a function of
$\gamma \in (-1/d,1/d)$.
We must trace out this curve in the $(\betatilde_T,\overline{\scrf})$-plane,
and then take the smallest value of $\overline{\scrf}$
for each $\betatilde_T$.\footnote{
   Alternatively we can fix $\betatilde_T$ and then use
   equations \reff{beta.gamma} and \reff{free_energy_gamma.parametric1}
   to give $\betatilde_V$ and $\overline{\scrf}$
   parametrically as a function of $\gamma$.
   Yet another variant was employed by Magnoli and Ravanini \cite{Magnoli_87},
   who fixed $\rho \equiv \betatilde_T/\betatilde_V$
   and computed $\betatilde \equiv \betatilde_V + \betatilde_T$
   and $\overline{\scrf}$
   parametrically as a function of $\gamma$.
}

In Sections~\ref{sec4} and \ref{sec5} we will carry out this procedure
(in the limit $L \to\infty$)
and discuss the qualitative properties of the solution as a function
of the lattice dimension $d$.
As preparation, let us compute, using \reff{deriv_gdL}/\reff{deriv_gd},
the derivatives with respect to $\gamma$ of
\reff{free_energy_gamma.parametric2} and \reff{beta.gamma}:
\begin{eqnarray}
{d\overline{\scrf}\over d\gamma}   &=&
   -\, {\gamma (f-2)f' + 2f(f-1) \over 4\gamma f} \,-\,
   \bvt {f - f^2 +\gamma f'\over 2 \gamma^2 f^2}
 \label{F_prime_gamma}  \\[3mm]
{d\btt\over d\gamma}  &=&
   {d \gamma f [\gamma (f-2)f' + 2f(f-1)] \over 2(f-1)^2} \,+\,
   \bvt {d(f - f^2 +\gamma f')\over (f-1)^2}
 \label{beta_prime_gamma}  \\[3mm]
{d^2\btt\over d\gamma^2}  &=&
{d\gamma ( 2\bvt - 2 \gamma f +\gamma f^2) \over 2(f-1)^2} f'' \,+\,
{\gamma d (\gamma - 2\bvt )\over (f-1)^3} f'\,^2 \nonumber \\
 & &  \qquad +\;
{2d(\bvt - 2\gamma f + \gamma f^2)\over (f-1)^2} f' \,+\, {d f^2\over f - 1}
\end{eqnarray}
where $f' \equiv df/d\gamma$;
these equations hold both in finite volume and in the infinite-volume limit.
In particular, we have
\be
{d\overline{\scrf}\over d\btt} \;=\; -\, {(f-1)^2\over 2d\gamma^2 f^2}
   \;\le\; 0   \;.
\label{F_prime_beta}
\ee
It is a curious fact that the derivative $d\overline{\scrf}/d\btt$ 
does not depend on $\bvt$.
Note also that $d\overline{\scrf}/d\btt$ becomes more negative
$|\gamma|$ grows (in all dimensions $d>0$):
this strict monotonicity follows from inequality \reff{eqnB.ineq_fprime}.

In the infinite-volume theory in dimension $d>2$,
we need to consider also the ``condensate phase''
parametrized by $f \in [f_*,\infty)$
with $\gamma \equiv (1/d) \sgn\bvt$ and $g \equiv g_*$.
Here \reff{free_energy_gamma.parametric} is replaced by
\begin{subeqnarray}
  \overline{\scrf}  & = &
  {1 \over 2}   \left[1  \,-\, f \,+\, \log f \,+\, g_*
                \,+\, d \btt \left( {f-1 \over f} \right)^2
                \right]  \qquad
   \slabel{free_energy_gamma.parametric_f1}   \\[3mm]
   & = &
   {1 \over 2}   \left[{1-f \over 2}
                \,+\, \log f \,+\, g_*
                \,-\, d |\bvt| {f-1 \over f}
                \right]
   \slabel{free_energy_gamma.parametric_f2}
   \label{free_energy_gamma.parametric_f}
\end{subeqnarray}
while \reff{beta.gamma} is replaced by
\be
   \btt  \;=\;  {f^2  \over  2d(f-1)}
                \,-\,
                {f \over f-1}  |\bvt|
    \;.
 \label{beta.gamma_f}
\ee
The derivatives of \reff{free_energy_gamma.parametric_f2}
and \reff{beta.gamma_f} with respect to $f$ are
\begin{eqnarray}
{d\overline{\scrf}\over df}  &=&
   -\, {f^2 - 2f + 2d |\bvt| \over 4 f^2}
  \label{F_prime_beta_f1}   \\[3mm]
{d\btt \over df}  &=&   {f^2 - 2f + 2d |\bvt| \over 2d(f-1)^2}
  \label{F_prime_beta_f2}
\end{eqnarray}
and hence
\be
{d\overline{\scrf} \over d\btt } \;=\; -\, {d (f-1)^2 \over 2 f^2 }  
   \;<\; 0 \;.
 \label{F_prime_beta_f3} 
\ee
Once again, $d\overline{\scrf}/d\btt$ is independent of $\bvt$.
Moreover, \reff{F_prime_beta} and \reff{F_prime_beta_f3}
coincide in the condensate phase $|\gamma| = 1/d$.\footnote{
   Both these properties follow alternatively from the fact that
   \reff{F_prime_beta} holds also in finite volume.
}
In particular, they coincide at their joining point $|\gamma| = 1/d$, $f=f_*$;
this means that any phase transition between these two branches
must be of second (or higher) order.
Note also that $d\overline{\scrf}/d\btt$ is a monotonic function of $f$:
it becomes more negative as $f$ grows.

For $d>2$ it is convenient to use $f$ as the independent variable in {\it both}
the ``normal phase'' ($|\gamma |<1/d$) and the condensate phase:
see e.g.\ Figure~\ref{fig_F0f} below.
In particular, equations \reff{F_prime_gamma}/\reff{beta_prime_gamma}
for the normal phase can be converted to yield
${d\overline{\scrf}/ df}$ and $d\btt / df$ by dividing them
by $f'(\gamma )$.
It is curious to note that in dimensions $2<d\leq4$, where
$f_*'(d)\equiv \lim\limits_{\gamma \uparrow 1/d} df_d(\gamma )/d\gamma$
is infinite, the limiting slopes ${d\overline{\scrf}/ df}$
and ${d\btt / df}$ for $\gamma \uparrow 1/d$
in the normal phase coincide with the values 
\reff{F_prime_beta_f1}/\reff{F_prime_beta_f2} 
obtained for $f\downarrow f_*$ in the condensate phase (this matching
of slopes is seen e.g. in Figure~\ref{fig_F0f}). On the other hand,
for $d>4$ this matching of slopes does not occur.

\section{Behavior of Solutions for $\bvt =0$}   \label{sec4}

\subsection{Overview}  \label{sec4.1}

Our conclusions concerning the phase diagram of the
$N=\infty$ $RP^{N-1}$ model,
as a function of the spatial dimension $d$ (treated as a real variable),
are summarized in Figure~\ref{phasediag_bv=0}.
There are three possible phases:
a white-noise phase ($\gamma = 0$)
   in which the isotensor correlation length is zero;
an isotropic phase ($0 < |\gamma| < 1/d$)
   in which the isotensor correlation length is finite but nonzero;
and a nematic ordered phase ($|\gamma| = 1/d$)
   in which the isotensor correlation function exhibits long-range order.
For $0 < d \le 3/2$, there is a second-order transition at $\btt = 1$
from the white-noise phase to the isotropic phase.
For $3/2 < d < d_* \approx 2.38403$, this transition is first-order
and takes place at a dimension-dependent $\widetilde{\beta}_{T,c} < 1$.
For $2 < d < d_*$, there is a subsequent second-order transition
from the isotropic phase to a nematic ordered phase
at a dimension-dependent
$\widetilde{\beta}_{T,c'} \equiv f_*^2 / [2d(f_* -1)] < \infty$.
At $d = d_*$ the two transitions collide at
$\widetilde{\beta}_{T,*} \approx 0.873264$,
and for $d > d_*$ there is only a first-order transition
from the white-noise phase directly into the nematic ordered phase,
at a dimension-dependent $\widetilde{\beta}_{T,c''} < 1$.\footnote{
   Ohno {\em et al.}\/ \cite{Ohno_90} were the first to note
   the existence of the special dimension $d_*$;
   but they did not attempt to determine it numerically,
   stating only that $2 < d_* < 3$.
   Caracciolo and Pelissetto, in an unpublished manuscript
   \cite{Caracciolo_95}, found essentially the same value of $d_*$
   as we report here.
}

Let us now show how these conclusions can be derived using
either the direct or the parametric approach.

\subsection{Direct Approach}   \label{sec4.2}

For $\bvt = 0$, the free energy is given simply by $\scrf_0(\gamma)$.
Since $\scrf_0$ is an even function of $\gamma$,
one of its stationary points is always $\gamma=0$.
To learn about other possible stationary points,
we begin by examining the Taylor expansion \reff{free_energy_even.expansion}
of $\scrf_0(\gamma)$ near $\gamma=0$.
The coefficient of $\gamma^2$ is positive for $\btt < 1$
and negative for $\btt > 1$.
The coefficient of $\gamma^4$ is positive for $\btt < 8d/\max(10d-3, 0)$
and negative for $\btt > 8d/\max(10d-3, 0)$.
It follows that the local behavior near $\gamma=0$ falls into three cases:
\begin{itemize}
   \item[(a)] $0 < d < 3/2$:  The coefficient of $\gamma^4$ is positive
      when $\btt = 1$.  Thus, as $\btt$ passes through 1,
      the local minimum at $\gamma=0$ turns into a local maximum,
      and a pair of degenerate minima at $\gamma = \pm \gamma_0(\btt)$
      bifurcates off it.  These minima satisfy the equation
\be
   \btt  \;=\;  {d \gamma_0^2 f_d(\gamma_0)^2  \over  2[f_d(\gamma_0) -1]}
   \;.
\ee
As $\btt \downarrow 1$ we have
\begin{eqnarray}
   \gamma_0(\btt)  & = &
       \left( {2 \over {3 \over 2} -d} \right)^{\!1/2} (\btt - 1)^{1/2}
       \,\times\,
     \nonumber \\
     & & \quad
       \left[ 1 \,+\, {20d^2 - 54d + 31 \over 4 ({3 \over 2}-d)^2}
                      (\btt - 1)
                \,+\, O\!\left((\btt - 1)^2\right)
       \right] \qquad
       \label{gammastar.d<32}   \\[3mm]
   \scrf_0(\gamma_0(\btt), \btt)   & = &
       - \, {d \over 8 ({3 \over 2} -d)} (\btt - 1)^2
       \,-\, {d (4d^2 - 18d +13) \over 48 ({3 \over 2}-d)^3} (\btt - 1)^3
     \nonumber \\
     & & \quad +\, O \!\left( (\btt - 1)^4\right)
       \label{F.d<32}
\end{eqnarray}
   \item[(b)] $d=3/2$:   The coefficient of $\gamma^4$ is zero
      when $\btt = 1$, but the coefficient of $\gamma^6$ is positive.
      Therefore, the qualitative behavior is the same as for $d < 3/2$,
      but the powers are changed:
\begin{eqnarray}
   \gamma_0(\btt)  & = &
       {2 \over 5^{1/4}} (\btt - 1)^{1/4}
       \,-\, {3 \over 10 \cdot 5^{3/4}}  (\btt - 1)^{3/4} 
           \,+\,  O\!\left((\btt - 1)^{5/4}\right)
       \nonumber \\ \label{gammastar.d=32}   \\
   \scrf_0(\gamma_0(\btt), \btt)   & = &
       - {1\over 2\sqrt{5}} (\btt - 1)^{3/2} \,-\,
        {9\over 10} (\btt - 1)^{2} \,+\,
        O \!\left( (\btt - 1)^{5/2} \right)
       \nonumber \\ \label{F.d=32}
\end{eqnarray}
   \item[(c)] $d > 3/2$:  The coefficient of $\gamma^4$ is negative
      when $\btt = 1$.  The behavior for $\btt \approx 1$ and
      $\gamma \approx 0$ will turn out to be irrelevant,
      as a first-order phase transition occurs
      {\em before}\/ $\btt = 1$, at a $\gamma$ far away from 0.
\end{itemize}

The local behavior near $\gamma = 0$ does not, of course,
necessarily determine the global behavior of $\scrf_0(\gamma)$.\footnote{
   This point was overlooked by Hikami and Maskawa \cite[Appendix]{Hikami_82},
   who asserted (incorrectly) that there is a phase transition at $\btt=1$
   in all dimensions $d$.
}
But our numerical computations
confirm ---
although we have been unable to prove it analytically ---
that $\scrf_0(\gamma)$ always has at most one pair of local minima
$\gamma = \pm \gamma_0(\btt) \neq 0$, and that:
\begin{itemize}
   \item[(a)]  $0 < d < 3/2$:
      For $\btt \le 1$, the unique absolute minimum of $\scrf_0(\gamma)$
        lies at $\gamma=0$.
      For $\btt > 1$, $\scrf_0(\gamma)$ has precisely two
        degenerate absolute minima,
        at $\gamma = \pm \gamma_0(\btt)$ satisfying \reff{gammastar.d<32}.
      The free energy $\bar{F}$ therefore equals 0 for $\btt \le 1$,
        and it satisfies \reff{F.d<32} for $\btt > 1$.
      There is thus a second-order phase transition at $\btt =1$,
        in which the specific heat
        $C_H \sim \partial^2 \bar{F} / \partial \btt^2$
        has a jump discontinuity (see Figure~\ref{fig_F0gamma}a).
   \item[(b)]  $d = 3/2$:
      The qualitative behavior is the same as for $d < 3/2$, but
        $\gamma_0(\btt)$ satisfies \reff{gammastar.d=32},
        and the free energy for $\btt > 1$ satisfies \reff{F.d=32}.
      There is thus a second-order phase transition at $\btt =1$,
        in which the specific heat has
        a $(\btt - 1)^{-1/2}$ singularity as $\btt \downarrow 1$
        (see Figure~\ref{fig_F0gamma}b).
   \item[(c)]  $3/2 < d \le 2$:
      For $\btt < \hbox{ a certain } \betatilde_{T,c}(d) < 1$,
        the unique absolute minimum of $\scrf_0(\gamma)$ lies at $\gamma=0$.
      For $\btt > \betatilde_{T,c}(d)$, $\scrf_0(\gamma)$ has precisely two
        degenerate absolute minima, at $\gamma = \pm \gamma_0(\btt)$,
        where
\be
   \gamma_0(\btt) \;\to\;
   \cases{ \gamma_c > 0   & as $\btt \downarrow \betatilde_{T,c}$  \cr
           \noalign{\vskip 2mm}
           1/d               & as $\btt \to +\infty$ \cr
         }
   \label{def_gammac}
\ee
      There is thus a first-order phase transition at a dimension-dependent
        $\betatilde_{T,c}(d) < 1$
        (see Figure \ref{fig_F0gamma}c for the case $d=2$).\footnote{
   For $d=2$, the transition at $\betatilde_{T,c} \approx 0.956$
   was found previously by Magnoli and Ravanini \cite{Magnoli_87},
   using the parametric approach (see Section~\ref{sec4.3} below).
}
      We have
\begin{eqnarray}
   \betatilde_{T,c}   & = &
       1 \,-\, \smfrac{3}{20} (d-\smfrac{3}{2})^2
         \,-\, {63 \over 100} (d-\smfrac{3}{2})^3
         \,+\, O\Bigl( (d-\smfrac{3}{2})^4 \Bigr)
    \label{betaTc_d_near_3/2}   \\[2mm]
   \gamma_c  & = &
   \sqrt{\smfrac{6}{5}} \left[ (d-\smfrac{3}{2})^{1/2}
                              \,+\, {51 \over 20} (d-\smfrac{3}{2})^{3/2}
                              \,+\, O\Bigl( (d-\smfrac{3}{2})^{5/2} \Bigr)
                    \right]
    \label{gammac_d_near_3/2}
\end{eqnarray}
as $d \downarrow 3/2$.
\end{itemize}
For $d > 2$ we need to consider also the ``condensate phase''
\reff{free_energy_even_f}
parametrized by $f \ge f_*$ at $\gamma = 1/d$ and $g = g_*$.
For convenience we use $f$ as the independent variable in the plots
also in the ``ordinary phase'' $0 < \gamma < 1/d$.
We have:
\begin{itemize}
   \item[(d)]  $2 < d < d_* \approx 2.38403$:
      The qualitative behavior near the first-order transition point
        $\betatilde_{T,c}$ is the same as for $d \le 2$.
        However, beyond this point the behavior is different:
        $\gamma_0(\btt)$ now reaches $1/d$ when $\btt$ reaches
        the {\em finite}\/ value
\be
   \betatilde_{T,c'}(d) \;\equiv\;  {f_*^2  \over  2d(f_* - 1)}
   \;.
  \label{def_betatilde_T_c'}
\ee
      For $\btt > \betatilde_{T,c'}$, the function $\scrf_0(\gamma)$
        is strictly decreasing on the interval $[0, 1/d]$,
        and the absolute minimum of the free energy
        is found in the condensate phase at
\begin{eqnarray}
   f & = &  d \btt \left[ 1 \,+\, \left( 1 - {2 \over d\btt} \right) ^{\! 1/2}
                   \right]
      \;>\; f_*
      \label{f_condensate_min}  \\[2mm]
   \scrf_0 & = &  -\, {f-1 \over 4} \,+\, {\log f + g_* \over 2}  \;<\; 0
      \label{F_condensate_min} 
\end{eqnarray}
      There is a second-order isotropic-to-nematic phase transition
      at $\btt = \betatilde_{T,c'}$ (see Figure \ref{fig_F0f}a).
      The two transitions $\betatilde_{T,c} < \betatilde_{T,c'}$
      come closer together as $d$ grows, and they collide at the
      dimension $d_*$ where $\gamma_c$ reaches $1/d$
      (see Figure~\ref{fig.dgammac_versus_d}).
      Equivalently, $d_*$ is defined by the condition
\be
   \scrf_* \;\equiv\; -\, {f_* - 1 \over 4} \,+\, {\log f_* + g_* \over 2}
           \; \cases{ < 0     & for $2 < d < d_*$   \cr
                      = 0     & for $d = d_*$       \cr
                      > 0     & for $d > d_*$       \cr
                    }
\ee
       (see Figure \ref{fig.Fstar_versus_d}).
   \item[(e)]  $d = d_*$:  The two transitions coincide:
       $\betatilde_{T,c} = \betatilde_{T,c'} =
        \widetilde{\beta}_{T,*} \approx 0.873264$.
       Thus, for $\btt < \widetilde{\beta}_{T,*}$,
          the unique minimum lies at $\gamma = 0$;
       for $\btt = \widetilde{\beta}_{T,*}$,
          it lies exactly at $\gamma = 1/d$ with $f = f_*$;
       for $\btt > \widetilde{\beta}_{T,*}$,
          it lies in the condensate phase $\gamma = 1/d$ at
          the value \reff{f_condensate_min}.
       There is a first-order phase transition as
          $\btt$ passes through $\widetilde{\beta}_{T,*}$
          (see Figure \ref{fig_F0f}b).
   \item[(f)] $d > d_*$:
   For $\btt \le f_*^2 / [2d(f_* - 1)]$, $\scrf_0(\gamma)$ is a strictly
       increasing function of $\gamma$ on $[0,1/d]$;
   for $f_*^2 / [2d(f_* - 1)] < \btt < 1$,
       it first increases and then decreases;
   for $\btt \ge 1$, it is strictly decreasing.
   In all three cases, $\scrf_0(\gamma)$ never has a local minimum
       in the interval $0 < \gamma \le 1/d$.
   Therefore, the absolute minimum is either at $\gamma = 0$
       or else in the condensate phase at
       \reff{f_condensate_min}/\reff{F_condensate_min}.
   There is a $\betatilde_{T,c''}(d)$ such that the
       former behavior occurs for $\btt < \betatilde_{T,c''}(d)$
       and the latter for $\btt > \betatilde_{T,c''}(d)$
       (see Figure \ref{fig_F0f}c):
       it is given by
\be
       \betatilde_{T,c''}(d)   \;\equiv\;  {f_{c''}^2  \over 2d(f_{c''} -1)}
  \label{def_betatilde_T_c''}
\ee
       where $f_{c''}$ is the unique solution of
       $\log f + g_* = (f-1)/2$ satisfying $f > f_*$.\footnote{
   Since $-(f-1)/4 + (\log f + g_*)/2$ is a concave function of $f$
   that is $>0$ at $f=f_*$ and tends to $-\infty$ as $f \to +\infty$,
   it has a unique zero in the interval $(f_*,\infty)$.
}
   As $d\to\infty$, we have $f_{c''} = f_\infty + O(d^{-1})$ and
   $\betatilde_{T,c''} = f_\infty^2/[2(f_\infty -1)] d^{-1} + O(d^{-2})$,
   where $f_\infty \approx 3.512863$ is the unique solution of
   $\log f = (f-1)/2$ on $f > 1$.
\end{itemize}

\subsection{Parametric Approach}   \label{sec4.3}

Since $\overline{\scrf}$ and $\btt$ are {\em even}\/ functions of $\gamma$\
when $\bvt = 0$, it suffices to consider $\gamma \ge 0$.
{}From \reff{beta.gamma} we have
\be
\btt \;=\; {d \gamma^2 f^2  \over 2\left( f - 1 \right)} \;=\;
1 \,+\, {3 - 2d\over 4}\gamma^2 \, - \, {20d^2 - 54d +31\over 16}\gamma^4
\,+\, \ldots \;.
\label{beta_series}
\ee
In Figure~\ref{fig_btt_versus_gamma}  we plot $\btt$ versus $\gamma$
for dimensions $d= 1, 3/2, 2, 9/4, d_*, 3$.
Moreover, for $d>2$ we need to consider the ``condensate phase''
\be
   \btt  \;=\;  {f^2 \over 2d(f-1)}
\ee
parametrized by $f \in [f_*,\infty)$.

The following assertions appear to be true:
\begin{itemize}
   \item[(a)]  For $0 < d \le 3/2$,
     $\btt$ is a strictly increasing function of $\gamma$ on $[0,1/d)$.
     It tends to $+\infty$ as $\gamma \uparrow 1/d$.
   \item[(b)]  For $3/2 < d < \bar{d} \approx 2.55391$, 
     there exists $\gamma_{cusp}(d) \in (0,1/d)$ such that
     $\btt$ is a strictly decreasing function of $\gamma$
     on $[0,\gamma_{cusp}]$
     and a strictly increasing function of $\gamma$ on $[\gamma_{cusp},1/d)$.
     The value of $\gamma_{cusp}$ is determined by
     the condition $\gamma = 2f(f-1)/[(2-f)f']$.
     Moreover, $d\gamma_{cusp}$ is an increasing function of $d$:
     it moves from $\gamma_{cusp} = 0$ at $d=3/2$
     to $\gamma_{cusp} = 1/d$ at $d=\bar{d}$.
     The value of $\bar{d}$ is determined by the condition $f_*(\bar{d}) = 2$.
   \begin{itemize}
       \item[(b${}_1$)] For $3/2 < d \le 2$,
              $\btt$ tends to $+\infty$ as $\gamma \uparrow 1/d$.
       \item[(b${}_2$)]  For $2 < d < \bar{d}$, $\btt$ tends to a {\em finite}\/
              value $f_*^2 / [2d (f_* -1)]$
              as $\gamma \uparrow 1/d$.
              However, the slope $d\btt/d\gamma$ at this point is $+\infty$.
              In the condensate phase $f \ge f_*$,
              $\btt$ is a strictly increasing function of $f$.
   \end{itemize}
   \item[(c)] For $d \ge \bar{d}$, $\btt$ is a strictly decreasing function
      of $\gamma$ on $[0,1/d)$.  It tends to a {\em finite}\/ value 
      $f_*^2 / [2d (f_* -1)] < 1$ as $\gamma \uparrow 1/d$.
   \begin{itemize}
       \item[(c${}_1$)] For $\bar{d} \le d \le 4$, the slope $d\btt/d\gamma$
              is $-\infty$ at $\gamma = 1/d$.
       \item[(c${}_2$)] For $d > 4$, the slope is a finite negative number.
   \end{itemize}
   In either case, in the condensate phase $f \ge f_*$,
   $\btt$ at first decreases and then increases;
   it reaches its absolute minimum at $f=2$ and $\btt=2/d$.
\end{itemize}
The assertions about the behavior as $\gamma \uparrow 1/d$
are simple rigorous consequences of known facts about $f_d(\gamma)$
[see Appendix \ref{appendix_b}],
and the assertions about the condensate phase are trivial calculus.
But we have been unable to demonstrate analytically
the observed monotonicity properties of $\btt(\gamma)$.

Let us next look at the free energy $\overline{\scrf}(\gamma)$
defined by \reff{free_energy_gamma.parametric2} with $\bvt=0$:
\be
\overline{\scrf} \;=\; -\, {f-1 \over 4} \,+\, {\log f + g \over 2}
   \;=\;   {d(2d - 3)\over 64} \gamma^4 \,+\,
    {d(7d^2 - 18d +10)\over 48}\gamma^6 \,+\, \ldots
  \;\,.
\label{F_series}
\ee
{}From \reff{F_prime_beta} it follows that
$\overline{\scrf}$ is decreasing wherever $\btt$ is increasing,
and conversely.
As $\gamma \uparrow 1/d$,
$\overline{\scrf}$ tends to $-\infty$
[resp.\ to a finite number $\scrf_* = -(f_*-1)/4 + (\log f_* + g_*)/2$]
in dimension $d \le 2$ (resp.\ $d>2$).
In dimension $d>2$, we need to consider also the condensate phase
$f \ge f_*$, in which $\overline{\scrf}$ is given by
\reff{free_energy_gamma.parametric_f2}.
Again we have to distinguish $2 < d \le \bar{d}$ (corresponding to $f_* \ge 2$)
from $d > \bar{d}$ (corresponding to $f_*  < 2$).
In the former case,
$\overline{\scrf}$ decreases monotonically from $\scrf_*$ to $-\infty$
as $f$ rises from $f_*$ to $+\infty$.
In the latter case,
$\overline{\scrf}$ first rises from $\scrf_*$ to $(2\log 2 - 1)/4 + g_*/2$
as $f$ rises from $f_*$ to 2,
then decreases monotonically to $-\infty$
as $f$ rises from 2 to $+\infty$.
We refrain from showing the plots of $\overline{\scrf}$ versus $\gamma$.

We are now ready to plot $\overline{\scrf}$ parametrically versus $\btt$.
As in Section~\ref{sec4.2}, six cases need to be distinguished
(see Figure~\ref{fig_F_versus_btt}):

{\bf Case (a), $\boldsymbol{0 < d < 3/2}$:}
As $\gamma$ increases from 0,
$\btt$ increases smoothly from 1,
with an initial correction of order $\gamma^2$,
while $\overline{\scrf}$ decreases smoothly from 0,
with an initial correction of order $\gamma^4$.
It follows that $\overline{\scrf}$ is a smooth and decreasing function of $\btt$
in the regime $\btt \ge 1$;
from \reff{beta_series} and \reff{F_series} we reobtain
the expansion \reff{F.d<32} as $\btt \downarrow 1$.
This of course beats the $\gamma=0$ solution
$\overline{\scrf} = 0$ whenever $\btt > 1$.
On the other hand, for $\btt \le 1$ the only solution is
given by $\gamma=0$ and $\overline{\scrf}=0$.
We therefore have a second-order phase transition at $\btt = 1$\
(see Figure~\ref{fig_F_versus_btt}a).
[In Section~\ref{sec4.4} we give explicit formulae for the case $d=1$.]

{\bf Case (b), $\boldsymbol{d = 3/2}$:}
The parametric plot $\overline{\scrf}(\btt)$ is qualitatively
like that for $d < 3/2$, but since $\btt -1 \sim \gamma^4$
while $\overline{\scrf} \sim \gamma^6$,
the expansion as $\btt \downarrow 1$ has fractional powers
$\overline{\scrf} \sim (\btt -1)^{3/2}$,
as already seen in \reff{F.d=32}.
There is again a second-order
phase transition at $\btt =1$ (see Figure~\ref{fig_F_versus_btt}b).

{\bf Case (c), $\boldsymbol{3/2 < d \le 2}$:}
A typical parametric plot $\overline{\scrf}(\btt)$ is shown in
Figure~\ref{fig_F_versus_btt}c.
As $\gamma$ increases from 0,
$\btt$ initially {\em decreases}\/ from 1,
while $\overline{\scrf}$ initially {\em increases}\/ from 0;
the dependence $\overline{\scrf}(\btt )$ is thus initially backward-bending.
At $\gamma = \gamma_{cusp}$,
both $d\btt/d\gamma$ and $d\overline{\scrf}/d\gamma$ pass through zero,
so the parametric plot exhibits a cusp.
After this, $\btt$ increases and $\overline{\scrf}$ decreases,
with the curve staying below the $\gamma < \gamma_{cusp}$ curve
[the latter property follows from the monotonicity
 of the slope \reff{F_prime_beta}];
$\btt$ tends to $+\infty$ and $\overline{\scrf}$ tends to $-\infty$
as $\gamma \uparrow 1/d$.
Of course, the $\gamma=0$ solution $\overline{\scrf}=0$
is also present for all $\btt$.
Since the stable phase is the one with the
minimal $\overline{\scrf}$, the point where the ``cusped branch''
crosses the $\overline{\scrf}=0$ solution corresponds to a
first-order phase transition with a $d$-dependent
critical inverse temperature $\widetilde{\beta}_{T,c} <1$.
This value can be located numerically
by solving $\overline{\scrf}(\gamma) = 0$ for $\gamma$
using \reff{free_energy_gamma.parametric2},
and then substituting into \reff{beta.gamma} to find $\btt$.
For $d$ slightly greater than 3/2 it is given by
\reff{betaTc_d_near_3/2}/\reff{gammac_d_near_3/2}.

{\bf Cases (d--f), $\boldsymbol{d>2}$:}
These cases are qualitatively similar to case (c).
The only difference is that for $d>2$,
$\btt$ reaches a finite value at $\gamma = 1/d$;
beyond this point we must consider the condensate phase
parametrized by $f \ge f_*$ at $\gamma = 1/d$ and $g=g_*$.
{}From \reff{beta.gamma_f} and \reff{free_energy_gamma.parametric_f2}
this branch is given by
\begin{eqnarray}
   \btt             & = & {f^2 \over 2 d ( f - 1 ) }
      \label{beta.gamma_f_bvt=0}   \\[2mm]
  \overline{\scrf}  & = & -\, {f-1  \over  4}  \,+\,  {\log f + g_*  \over 2}
      \label{free_energy_gamma.parametric_f2_bvt=0}
\end{eqnarray}
(shown as dashed lines in Figures~\ref{fig_F_versus_btt}d--f).
The condensate phase may begin after the cusp
($d < \bar{d} \approx 2.55391$), at the cusp ($d = \bar{d}$)
or before the cusp ($d > \bar{d}$);
this distinction is, however, of little interest.
What {\em does}\/ matter is where the cusped branch crosses
the $\overline{\scrf} = 0$ solution:
for $d < d_* \approx 2.38403$ this crossing occurs within the
``ordinary branch'' $0 < \gamma < 1/d$ (Figure \ref{fig_F_versus_btt}d),
while for $d > d_*$ it occurs within the ``condensate branch''
$f > f_*$ at $\gamma = 1/d$ (Figure \ref{fig_F_versus_btt}f);
for $d=d_*$ it occurs precisely at the joining point $f=f_*$
(Figure \ref{fig_F_versus_btt}e).
For $d < d_*$, this crossing corresponds to the
first-order white-noise--to--isotropic transition;
we solve numerically for the corresponding $\widetilde{\beta}_{T,c}$
in the same manner as in case (c).
For these dimensions there is also a subsequent
second-order isotropic-to-nematic transition
where the ordinary and condensate branches join:
this occurs at
$\widetilde{\beta}_{T,c'} \equiv f_*^2/[2d(f_* - 1)]$.
For $d > d_*$, the crossing corresponds to the
first-order white-noise--to--nematic transition;
we solve numerically for $\widetilde{\beta}_{T,c''}$
by solving $\overline{\scrf}(f) = 0$ for $f$
using \reff{free_energy_gamma.parametric_f2_bvt=0}
[taking only the solution with $f > f_*$]
and then substituting into \reff{beta.gamma_f_bvt=0}.
We have, for example,
$\betatilde_{T,c''} (3) \approx 0.734865$,
$\betatilde_{T,c''} (4) \approx 0.572931$,
$\betatilde_{T,c''} (5) \approx 0.466744$ and
$\betatilde_{T,c''} (6) \approx 0.393066$.
As $d\rightarrow \infty$,
we have
$\betatilde_{T,c''} = f_\infty^2/[2(f_\infty -1)] d^{-1} + O(d^{-2})$,
where $f_\infty \approx 3.512863$ is the unique solution of
$\log f = (f-1)/2$ on $f > 1$.

\subsection{One-Dimensional Case}   \label{sec4.4}

For the special case $d=1$, we have available the exact formulae
\reff{exact_fd}/\reff{exact_gd} for $f(\gamma)$ and $g(\gamma)$:
\begin{eqnarray}
   f(\gamma)   & = &  (1-\gamma^2)^{-1/2}        \\
   g(\gamma)   & = &  \log\!\left( {1 + (1-\gamma^2)^{1/2} \over 2} \right)
\end{eqnarray}
It is convenient to  introduce a new variable $\s \in [0,1)$ via
\begin{eqnarray}
   \gamma &=& {2\s \over 1+\s^2}
     \label{var_change_gamma}        \\[2mm]
   f      &=& {1+\s^2 \over 1-\s^2}
     \label{var_change_f}
\end{eqnarray}
[From \reff{mass_gap} we see that $\s = e^{-m}$
 where $m$ is the isovector mass gap.]
Using the direct approach \reff{free_energy_gamma},
we find
\be
   \overline{\scrf}(\s)  \;=\;  \scrf_0(\s)   \;=\;
      {1 \over 2} \left[ {-2\s^2 \over 1-\s^2} + {\s^2  \over \btt (1-\s^2)^2}
                         - \log(1-\s^2)\right]\,.
\label{free_energy_d=1_bvt=0} 
\ee
The derivative of \reff{free_energy_d=1_bvt=0} with respect to $\s^2$ is
\be
{d \scrf_0 \over d (\s^2)} \;=\; {1+\s^2 \over 2  (1-\s^2)}\left[
{1\over \btt}  - 1 + \s^2 \right]\,.
\ee
It follows then that for $\btt \le 1$,
$\scrf_0$ is a strictly increasing function of $\s^2$ on $[0,1)$,
so the absolute minimum lies at $\s_\star=0$ ($\gamma=0$) and has $\scrf_0 = 0$.
For $\btt \ge 1$,
$\scrf_0$  has a unique local minimum at $\s_\star^2 = 1 - 1/\btt$,
which is of course its absolute minimum.
We therefore have
\be
   \bar{F}  \;\equiv \inf\limits_{0 \le \s < 1 } \scrf_0( \s )  \;=\;
   \cases{ 0                                    & for $0 \le \btt \le 1$   \cr
           \noalign{\vskip 2mm}
           -{1 \over 2} (\btt - 1 - \log\btt)   & for $\btt \ge 1$  \cr
         }
  \label{solution_d=1_bvt=0}
\ee
exhibiting a second-order phase transition at $\btt = 1$.

Finally, let us compute the correlation functions.
Recalling that in $d=1$ the free propagator is a pure exponential
\be
   (\btilde^{-1})_{xy}  \;=\;  e^{-m|x-y|}  \;=\;  \s^{|x-y|}
   \;,
\ee
we find from \reff{eq2.GV} and \reff{eq2.factorization1a}
the isovector and isotensor correlation functions
\begin{eqnarray}
   G_V(x,y)  & = &   \s_\star^{|x-y|}   \\[2mm]
   G_T(x,y)  & = &   \s_\star^{2|x-y|}
      \label{eq4.28}
\end{eqnarray}
Of course, the result for $G_V$ reflects the (spurious) spontaneous breaking
of $Z_2$ gauge invariance at $\btt > 1$, when $N=\infty$.

In Section \ref{sec6} we will confirm these results by
computing the $N\to\infty$ limit of the exact finite-$N$ solution.

\section{Behavior of Solutions for $\bvt \neq 0$}   \label{sec5}

\subsection{Overview}  \label{sec5.1}

We begin by summarizing our conclusions concerning the
phase diagram of the $N=\infty$ mixed isovector/isotensor model.
The complete phase diagram is a two-dimensional surface in the space of
parameters $\bvt,\btt,d$.
In Figures~\ref{phase_diag_d<=32}--\ref{phase_diag_d=3}
we show the qualitative phase diagrams in the $(\bvt,\btt)$-plane
for different ranges of spatial dimension $d$:
first-order transitions are shown in solid lines,
and second-order transitions in dashed lines.

There is always a line of first-order transitions at $\bvt = 0$,
$\btt > \betatilde_{T,c}(d)$.
The remaining features of the phase diagram depend on the spatial dimension:
\begin{itemize}
   \item[(a)]  $0 < d \leq 3/2$:
      There are no phase transitions other than
      the second-order transition at $(\bvt,\btt) = (0,1)$
      and the first-order line at $\bvt = 0$, $\btt > 1$
      (Figure~\ref{phase_diag_d<=32}).
   \item[(b)]  $3/2 < d \le 2$:
      Three curves of first-order transitions meet at
      the point $(\bvt,\btt) = (0,\betatilde_{T,c})$:
      the first-order line at $\bvt = 0$, $\btt > \betatilde_{T,c}$;
      and a pair of symmetrical curves $\betatilde_{T,c}(\bvt )$.
      Second-order phase transitions occur at the endpoints
      $(\bvt,\btt) = (\pm \betatilde_{V,e}, \betatilde_{T,e})$
      of the latter curves (Figure~\ref{phase_diag_d=2}).
   \item[(c)]  $2 < d < d_* \approx 2.38403$:
      In addition to the transitions present for $3/2 < d \le 2$,
      there is an additional pair of second-order transition lines
      located at
\be
   \betatilde_{T,c'}  \;=\;  {f_*^2 \over 2d(f_* - 1)}
 \,-\,   {f_* \over f_* - 1} |\bvt|
  \label{nematic_lines}
\ee
      (Figure~\ref{phase_diag_2<d<dstar}).\footnote{
   Equation (10) of \cite{Oku_82b} contains an error:
   where Oku writes $\widetilde{K}$ (= our $\btt$)
   he should have written $\widetilde{K}/\bar{K}^2$ (= our $\btt/\bvt^2$);
   with this change, the equation agrees with our \reff{nematic_lines}.
   This error can be traced back to a notational confusion in
   an earlier paper of the same group \cite[p.~125]{Oku_82a},
   where the authors say ``here we have scaled $a$ as $a \to \bar{K} a$''
   but without explicitly defining a new symbol;
   they then presumably confused the scaled and unscaled meanings of $a$.
   We have been unable to understand where equation (11) of \cite{Oku_82b}
   comes from.
}
      The intersection of these lines with the $\bvt=0$ axis at
      $\betatilde_{T,c'} = f_*^2/[2d(f_*-1)]$ 
      corresponds to the
      isotropic-to-nematic transition of the $RP^{N-1}$ model,
      while their intersections with the $\btt=0$ axis at $\bvt =\pm f_*/2d$
      correspond to the
      paramagnetic-to-ferromagnetic ($\bvt > 0$)
      and paramagnetic-to-antiferromagnetic ($\bvt < 0$)
      transitions of the $N$-vector model.
   \item[(d)]  $d = d_*$:
      The two transitions of the $RP^{N-1}$ model
      (at $\betatilde_{T,c}$ and $\betatilde_{T,c'}$) coincide.
      For $\bvt \neq 0$, the second-order transition lines
      \reff{nematic_lines} remain above the first-order curves
      (Figure~\ref{phase_diag_d=dstar}).
   \item[(e)]  $d_* < d < d_{**}=3$:
      Second-order transitions occur on that part of
      \reff{nematic_lines} where they lie above
      the first-order curves (Figure~\ref{phase_diag_dstar<d<dstarstar}).
   \item[(f)]  $d \ge 3$:
      The second-order transition lines \reff{nematic_lines}
      join onto the endpoints $(\pm \betatilde_{V,e}, \betatilde_{T,e})$
      of the first-order transition curves
      (Figure~\ref{phase_diag_d=3}).
\end{itemize}
The dependence of $\betatilde_{T,e}$ and $\betatilde_{V,e}$ on $d$
is shown in Figures~\ref{beta_T_endpoint_d} and \ref{beta_V_endpoint_d}.
Together with Figure~\ref{phasediag_bv=0},
which illustrates the dependence of $\betatilde_{T,c}$ on $d$ at $\bvt =0$,
these plots describe the principal features of the phase diagram. 

Let us now show how these conclusions can be derived using
either the direct or the parametric approach.

\subsection{Direct Approach}   \label{sec5.2}

We now have to look at the entire free energy
$\overline{\scrf}(\gamma) =
 \scrf_0(\gamma) + \bvt \scrf_1(\gamma) + {\rm const}$.
Since $\scrf_1(\gamma)/\gamma < 0$ for all $\gamma$,
it follows immediately that the absolute minimum of $\overline{\scrf}(\gamma)$
lies in the region $\sgn\gamma = \sgn\bvt$.
In what follows, we restrict attention without loss of generality
to $\bvt > 0$ and $\gamma > 0$.

The plots for $\bvt = 0$ (Figures~\ref{fig_F0gamma} and \ref{fig_F0f})
get smoothly perturbed as we turn on $\bvt > 0$.
In particular, for $d \le 3/2$ we observe empirically
(though we have been unable to prove it analytically) that
$\overline{\scrf}(\gamma)$ has a {\em unique}\/ local minimum
in the region $\gamma > 0$;  it is of course the absolute minimum.
Moreover, it varies smoothly with $\btt > 0$ and $\bvt > 0$.
Therefore, there are no phase transitions at $\bt \neq 0$.

For $d > 3/2$ the story is more complicated.
The local minima lying for $\bvt = 0$ at $\gamma = 0$
and at $\gamma = \gamma_0(\btt)$ both move smoothly as
a small $\bvt > 0$ is turned on.
Let us call the resulting minima $\gamma_a(\bvt,\btt)$ and
$\gamma_b(\bvt,\btt)$, respectively.
Three nontrivial things can then happen:
\begin{itemize}
   \item[(a)]  A first-order phase transition can occur when the two minima
 become degenerate [$\overline{\scrf}(\gamma_a) = \overline{\scrf}(\gamma_b)$
 with $\gamma_a \neq \gamma_b$].
   \item[(b)]  A second-order phase transition can occur when the two minima
 collide ($\gamma_a = \gamma_b$).
   \item[(c)]  A second-order phase transition to a phase with long-range order
 can occur when $\gamma_b$ reaches $1/d$, provided that it is the absolute
 minimum [$\gamma_b = 1/d$ with
 $\overline{\scrf}(\gamma_b) \le \overline{\scrf}(\gamma_a)$].
\end{itemize}
These correspond, respectively, to the first-order curve
$\betatilde_{T,c}(\bvt)$,
the critical endpoints $(\pm \betatilde_{V,e}, \betatilde_{T,e})$,
and the second-order lines $\betatilde_{T,c'}(\bvt )$ given by
\reff{nematic_lines}.

One could now construct plots of $\overline{\scrf}$ versus $\gamma$
(or versus $f$ for the condensate phase) 
for various values of $\bvt$ and  $\btt$. For example, 
one could look at ``slices'' at fixed $\bvt$ and show how the shape of 
 $\overline{\scrf}(\gamma )$ curves is deformed as $\btt$ increases
(analogous to Figures~\ref{fig_F0gamma} and \ref{fig_F0f} for $\bvt =0$).
One would then see the behavior (a), (b) and (c) corresponding
to the first-order transition curve, the critical endpoint and the
second-order transition line.
Rather than pursuing this approach in detail, however, we turn to the
parametric approach, in which the same features can be observed more
efficiently because one studies {\em all}\/ $\btt$ in a single plot,
for each fixed $\bvt$.

\subsection{Parametric Approach}  \label{sec5.3}

We know that the absolute minimum of the free energy
lies in the region $\sgn\gamma = \sgn\bvt$.
We therefore restrict attention to $\bvt > 0$
and consider only $\gamma > 0$.
For each fixed $\bvt$,
we plot $\btt$ and $\overline{\scrf}$
parametrically as a function of $\gamma$,
using \reff{beta.gamma} and \reff{free_energy_gamma.parametric2}.
Since $f_d(\gamma)$ is a monotonically increasing function of $\gamma$
on $[0,1/d]$, we can (and do) alternatively use
$f$ as an independent parameter in place of $\gamma$;
this has the advantage of allowing us to treat normal and condensate phases
using the same independent parameter $f$ for both of them.
We use \reff{beta.gamma}/\reff{free_energy_gamma.parametric2}
for the normal phase $1 \le f < f_*$,
and \reff{beta.gamma_f}/\reff{free_energy_gamma.parametric_f2}
for the condensate phase $f \ge f_*$.

Figure~\ref{new_triangle_guide} contains a ``notations guide''
illustrating various features of the parametric plots
of $\overline{\scrf}$ versus $\btt$,
to be discussed presently.
In particular, points c and c\textprime correspond, respectively,
to the first-order curve $\betatilde_{T,c}(\bvt)$
and the second-order lines $\betatilde_{T,c'}(\bvt)$ given by
\reff{nematic_lines}.
The condensate phase $f \ge f_*$ is shown
(both here and in subsequent figures) as a dashed line.

In Figures~\ref{fig_param_F_versus_btt_d=2}--\ref{fig_param_F_versus_btt_d=5}
we plot $\btt$ versus $f$, and $\overline{\scrf}$ versus $\btt$,
at different values of $\bvt$ in various ranges of dimension $d$.
Two qualitatively different cases need to be distinguished:

\bigskip

{\bf Case (a), $\boldsymbol{0 < d \leq 3/2}$:}
Here $\btt$ is a strictly increasing function of $f$ on
$[1,\infty)$ for any value of $\bvt$;
it tends to $+\infty$ as $f\rightarrow +\infty$ ($\gamma \uparrow 1/d$).
Then $\overline{\scrf}$ is a smooth function of $\btt$ for any $\bvt \neq 0$.
In particular, no phase transition occurs for nonzero $\bvt$.
The complete phase diagram is schematically shown in
Figure~\ref{phase_diag_d<=32}.
(We refrain from showing here the parametric plots, as they are so boring.)

\bigskip

{\bf Cases (b--f), $\boldsymbol{d > 3/2}$:}
For $d>3/2$, the generic behavior $\btt (f)$ is the following:
For $0 < \bvt < \betatilde_{V,e}$ there
 exist $f_1$, $f_2$ $\in [1,\infty )$ such that
$\btt$ is a strictly increasing function of $f$ for $1 \le f \le f_1$,
a strictly decreasing function for $f_1 \le f \le f_2$,
and a strictly increasing function for $f_2 \le f < \infty$.
As $\bvt$ increases, the maximum at $f_1$ and the minimum at $f_2$
come closer together, and they finally collide at $f = f_e$
(or equivalently $\gamma = \gamma_e$) when $\bvt = \betatilde_{V,e}$.
This point is given by the solution of the simultaneous equations
$d\btt /d\gamma =0$ and $d^2\btt /d\gamma^2 =0$
(see Figure~\ref{fig_d_gamma_e_versus_d} for a plot of $\gamma_e$ versus $d$).
Beyond this point, $\btt$ is a strictly increasing function of $f$
on the entire domain $1 \le f < \infty$.

This behavior of $\btt(f)$ is reflected in the 
parametric plot $\overline{\scrf}(\btt)$,
where the values $f_1$ and $f_2$ correspond to the 
cusp singularities ``1'' and ``2'' of the triangle-shaped structure
shown schematically in Figure~\ref{new_triangle_guide}.
This is exemplified in Figure~\ref{fig_param_F_versus_btt_d=2},
which shows the $d=2$ case.
The ``triangle'' becomes smaller as $\bvt$ increases
 (Figure \ref{fig_param_F_versus_btt_d=2}a,b,c)
and disappears as $\bvt \uparrow \betatilde_{V,e}$ 
 (Figure \ref{fig_param_F_versus_btt_d=2}d).

\begin{figure}
\begin{center}
\epsffile{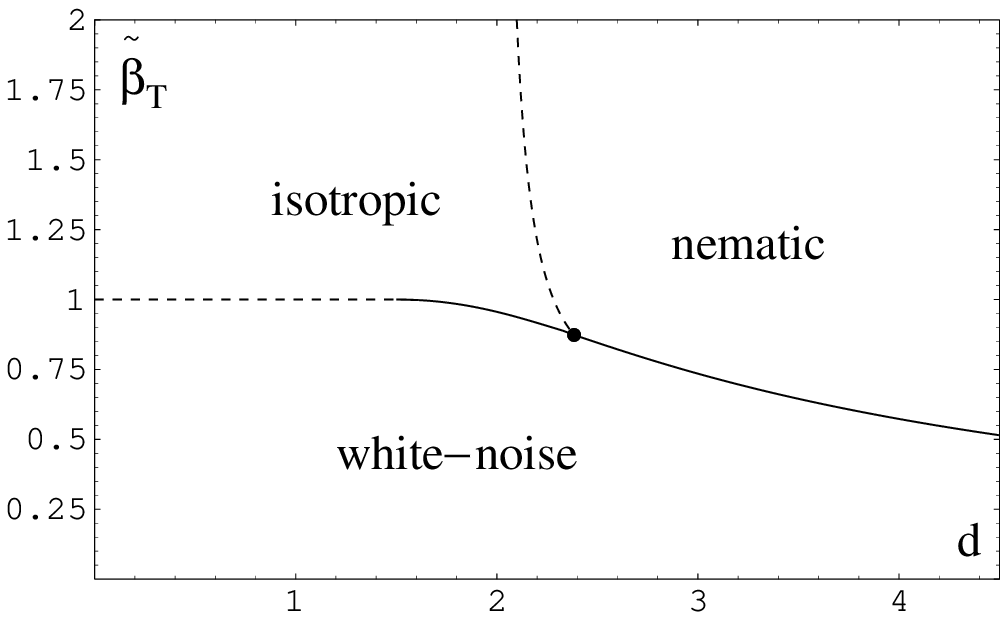}
\end{center}
\vspace*{5mm}
\caption{
     Phase diagram of the $RP^{N-1}$ model at $N=\infty$
     in the $(\btt,d)$-plane.
     Transition from white-noise phase to isotropic phase
     is second-order for $0 < d \le 3/2$
     and first-order for $3/2 < d < d_* \approx 2.38403$.
     Transition from isotropic phase to nematic ordered phase
     is second-order for $2 < d < d_*$.
     Transition from white-noise phase to nematic ordered phase
     is first-order for $d \ge d_*$.
   }
\label{phasediag_bv=0}
\end{figure}

\begin{figure}[p]
\vspace*{-2cm} \hspace*{-0cm}
\begin{center}
\leavevmode\epsffile{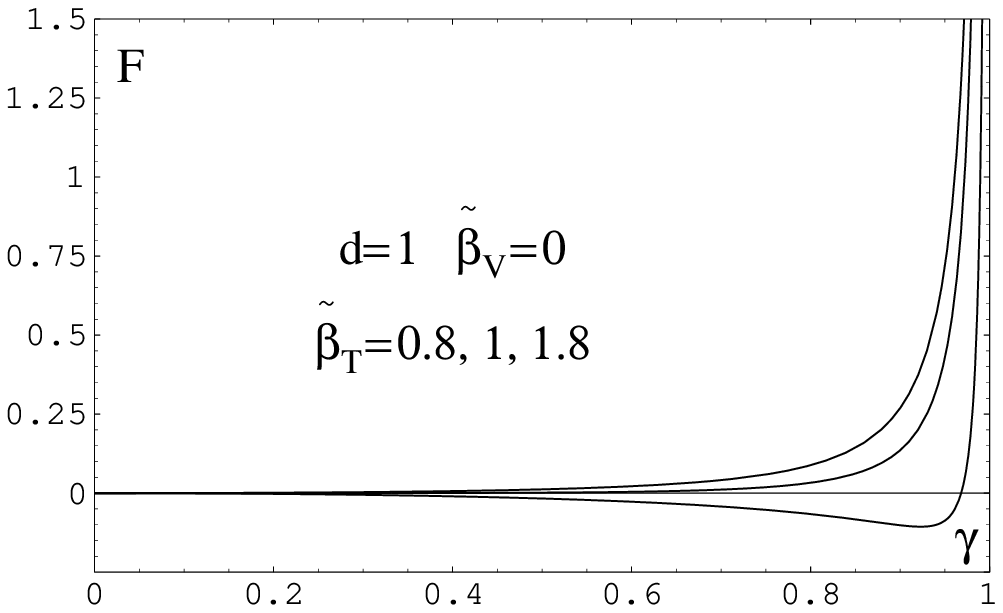} \\
\vspace{1cm}
\leavevmode\epsffile{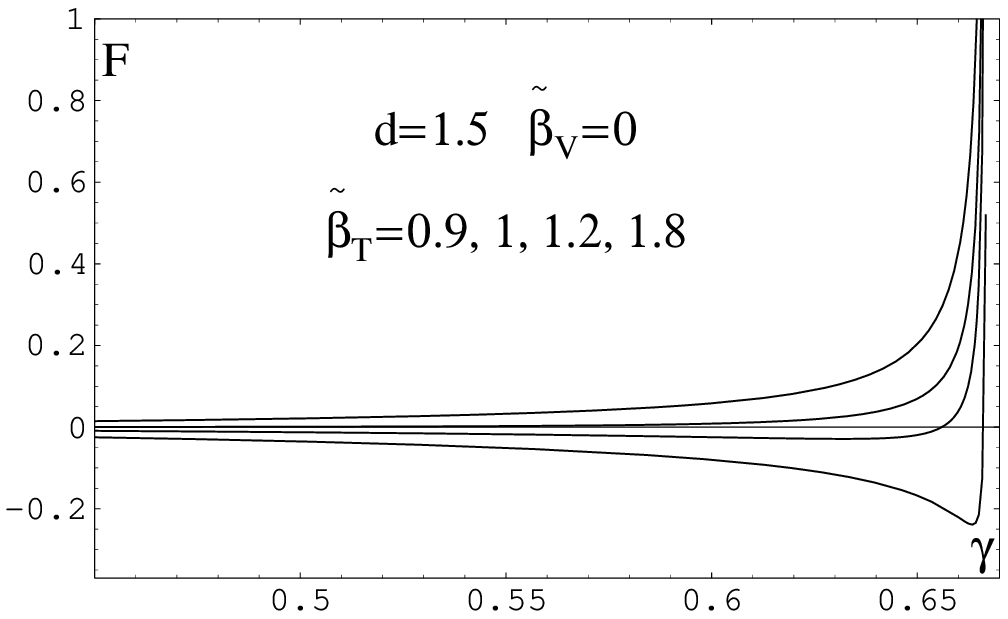} \\
\vspace{1cm}
\leavevmode\epsffile{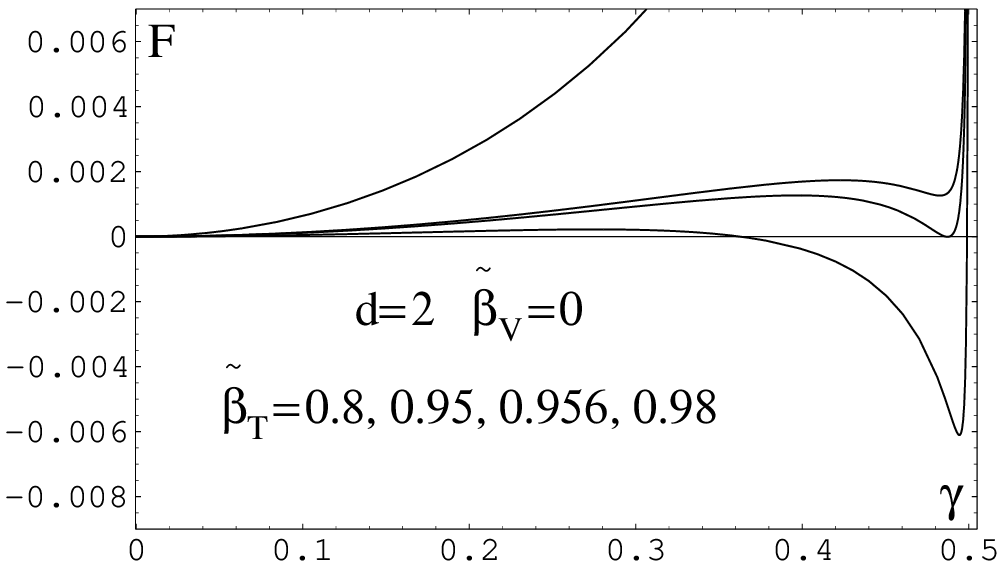}
\end{center}
\vspace{5mm}
\caption{
   $\scrf_0(\gamma)$ versus $\gamma$ for various values of $\btt$,
   for (a) $d=1$, (b) $d=3/2$, (c) $d=2$.
   Since $\scrf_0(\gamma) = \scrf_0(-\gamma)$,
   we show only $\gamma \ge 0$.
}
\label{fig_F0gamma}
\end{figure}

\begin{figure}[p]
\vspace*{-2cm} \hspace*{-0cm}
\begin{center}
\leavevmode\epsffile{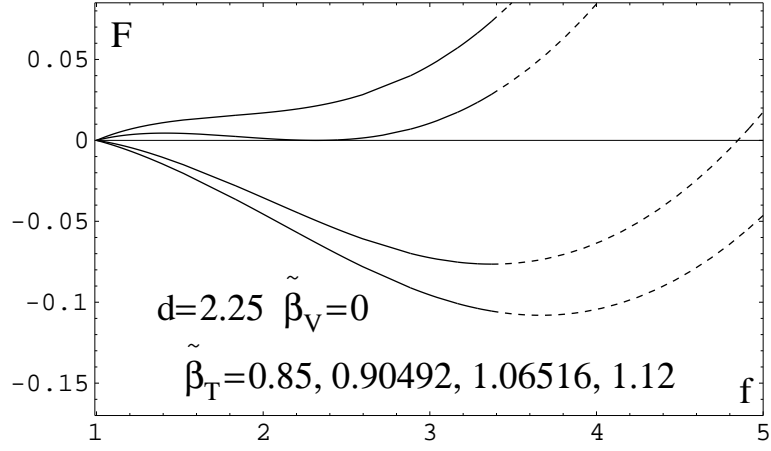} \\
\vspace{1cm}
\leavevmode\epsffile{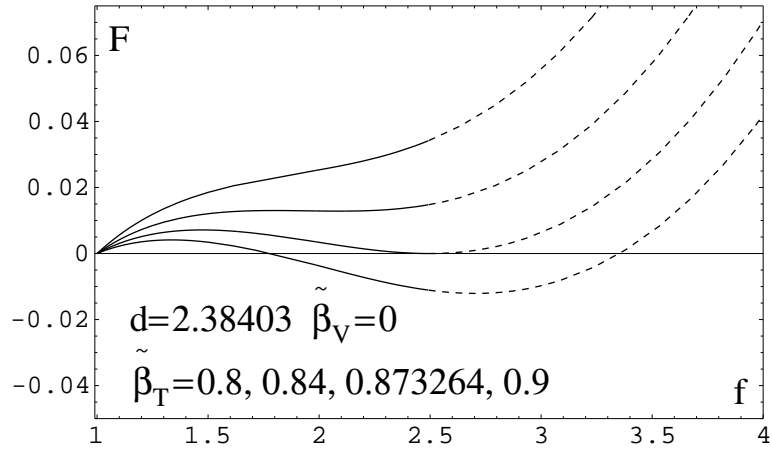} \\
\vspace{1cm}
\leavevmode\epsffile{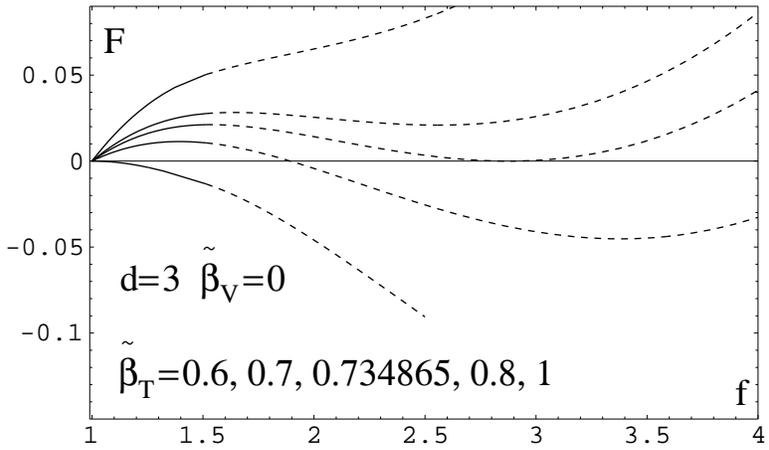}
\vspace{5mm}
\end{center}
\caption{
   $\scrf_{0}(\gamma)$ (solid curve) and $\scrf_{0*}(f)$ (dashed curve)
   plotted versus $f$ for various values of $\btt$,
   for (a) $d=9/4$, (b) $d=d_* \approx 2.38403$, (c) $d=3$.
   The junction between the solid and dashed curves
   corresponds to $\gamma = 1/d$, $f=f_*$.
}
\label{fig_F0f}
\end{figure}

\begin{figure}[p]
\begin{center}
\epsffile{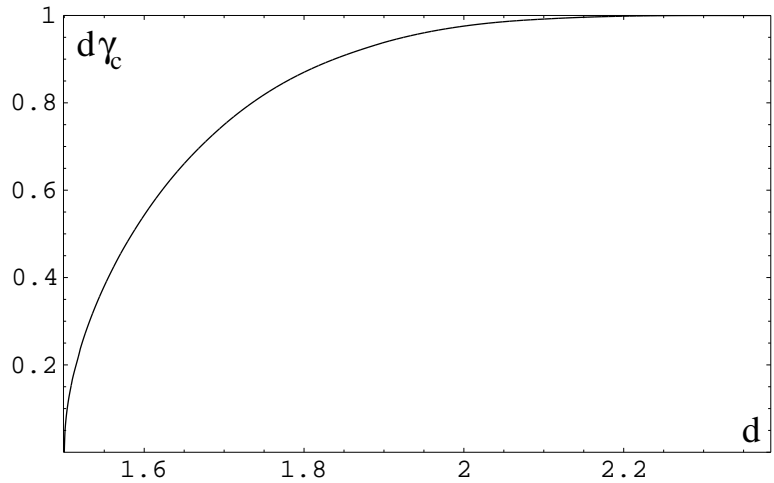}
\end{center}
\caption{
   $d\gamma_c$ plotted versus $d$ for $3/2 \le d \le d_* \approx 2.38403$.
   Here $\gamma_c$ is the $\gamma$ value at the first-order phase transition
   of the $RP^{\infty}$ model [see \reff{def_gammac}];
   it tends to zero as $d \downarrow 3/2$ and to $1/d$ as $d \uparrow d_*$.
}
\label{fig.dgammac_versus_d}
\end{figure}

\begin{figure}[p]
\begin{center}
\epsffile{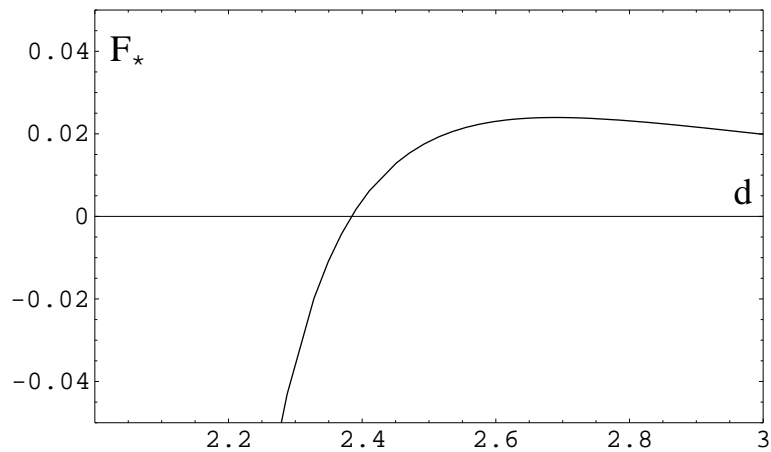}
\end{center}
\caption{
   $\scrf_* \equiv -(f_* -1)/4 + (\log f_* + g_*)/2$
   plotted versus $d$.
   We have $\scrf_* = 0$ at the dimension $d_* \approx 2.38403$.
}
\label{fig.Fstar_versus_d}
\end{figure}

\begin{figure}[p]
\vspace*{-2cm} \hspace*{-0cm}
\begin{center}
\vspace*{0cm} \hspace*{-0cm}
\epsfxsize=0.45\textwidth
\leavevmode\epsffile{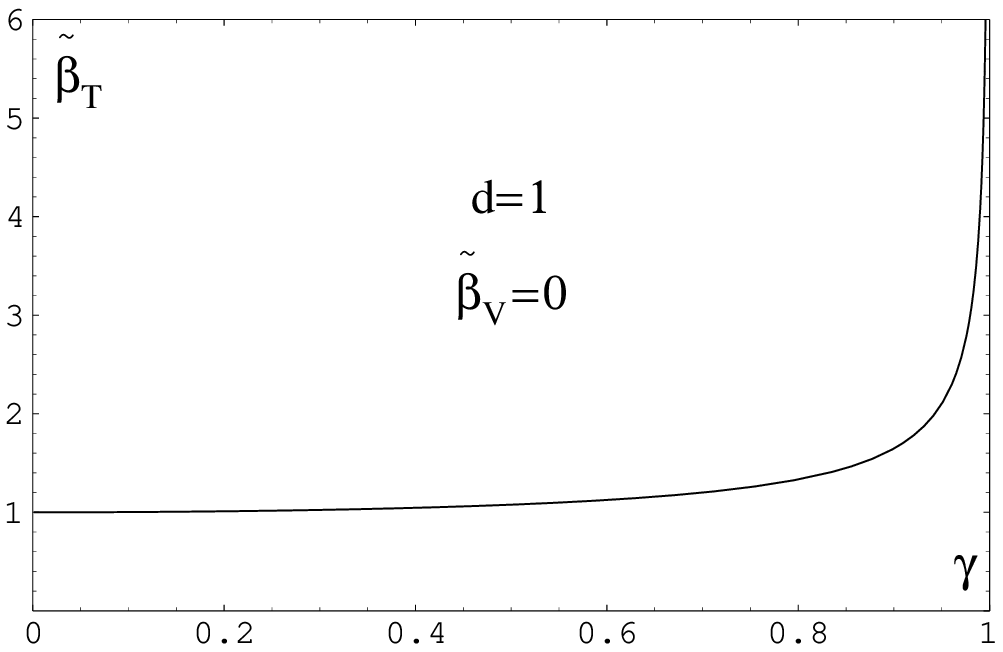}
\hspace{0.5cm}
\epsfxsize=0.45\textwidth
\epsffile{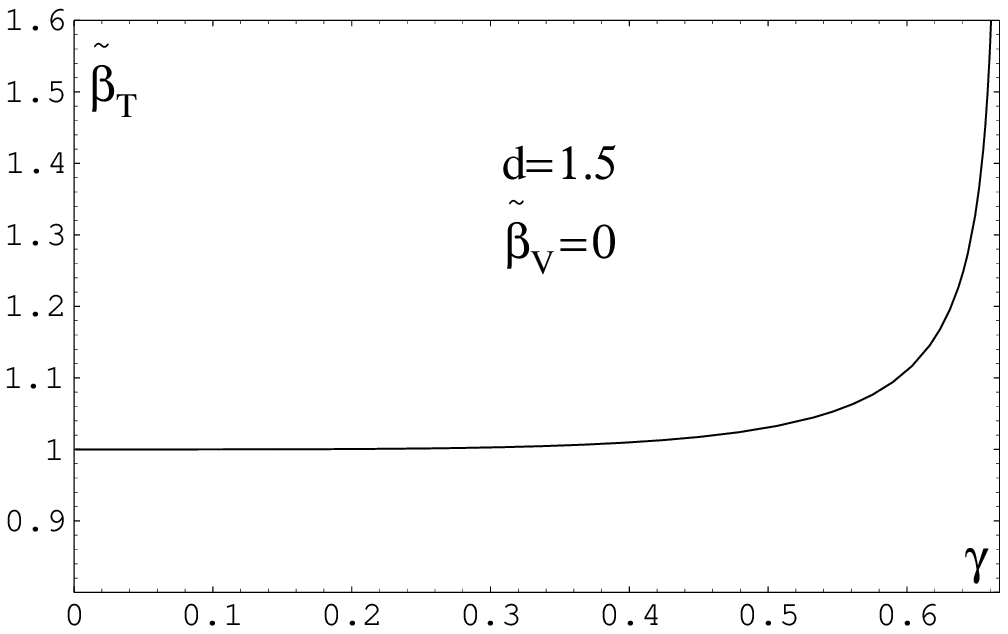}  \\
\vspace*{1cm}
\epsfxsize=0.45\textwidth
\leavevmode\epsffile{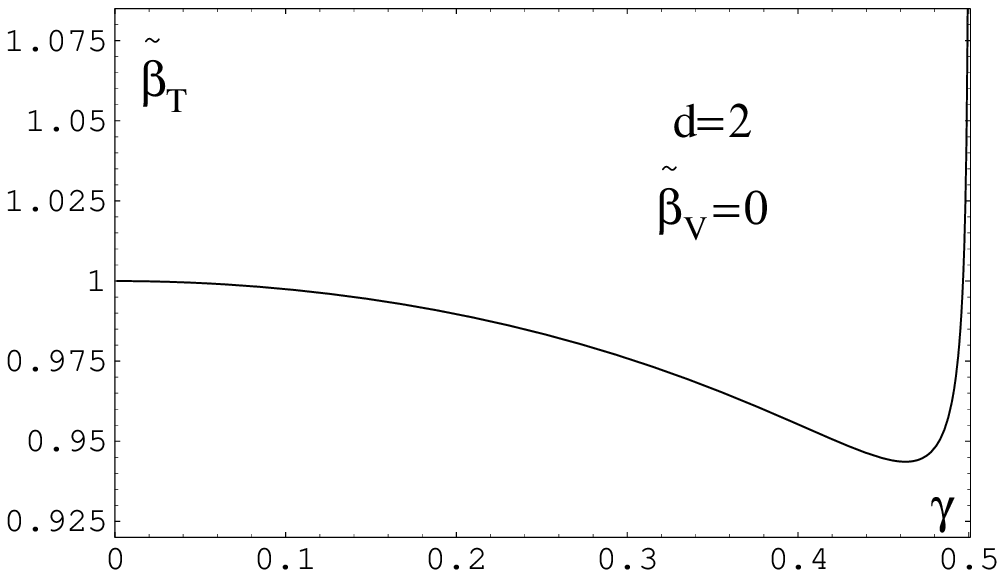}
\hspace{0.5cm}
\epsfxsize=0.45\textwidth
\epsffile{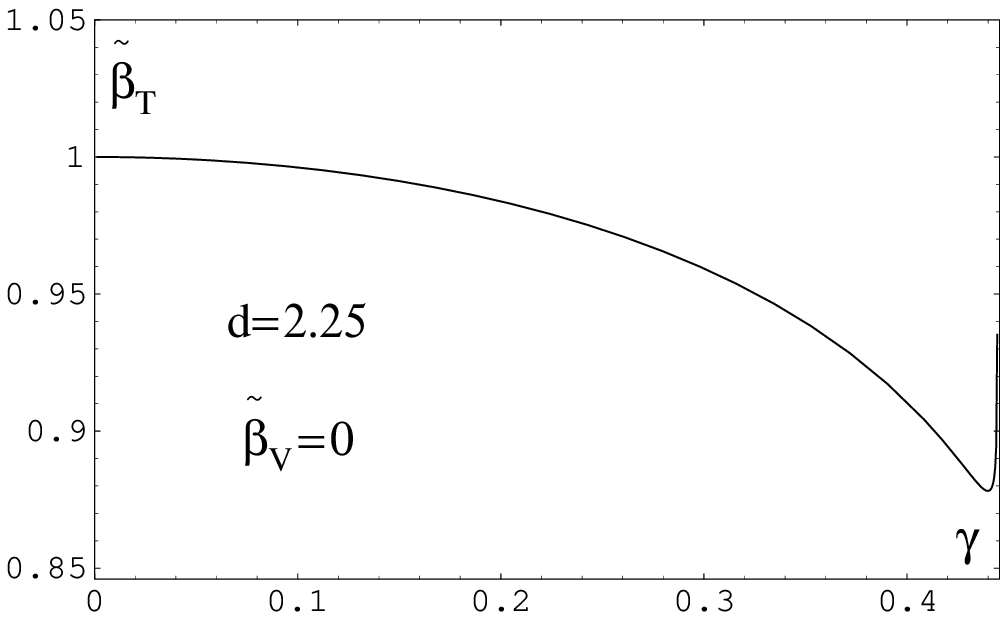}  \\
\vspace*{1cm}
\epsfxsize=0.45\textwidth
\leavevmode\epsffile{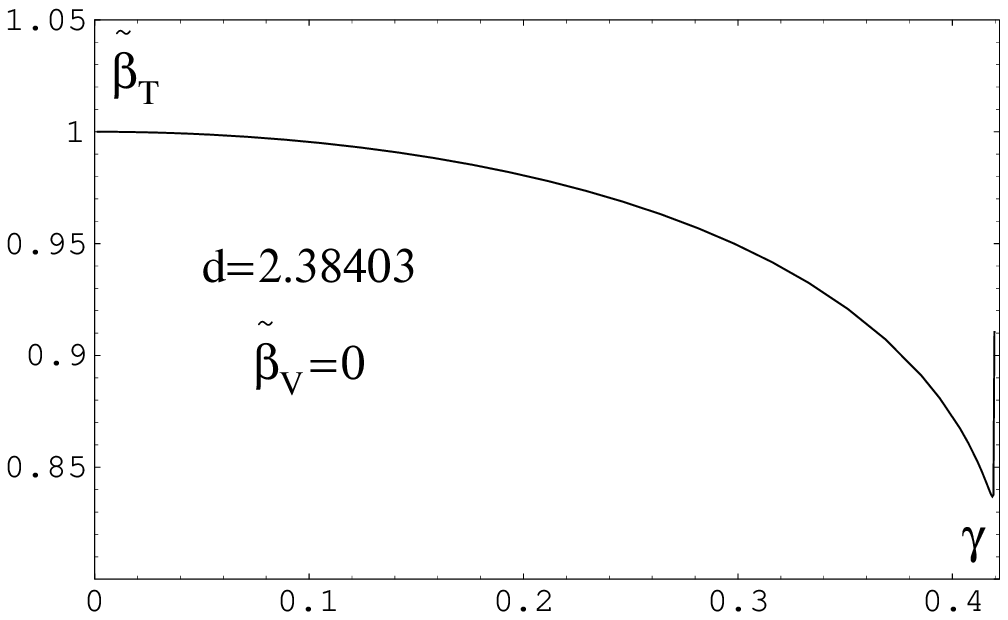}
\hspace{0.5cm}
\epsfxsize=0.45\textwidth
\epsffile{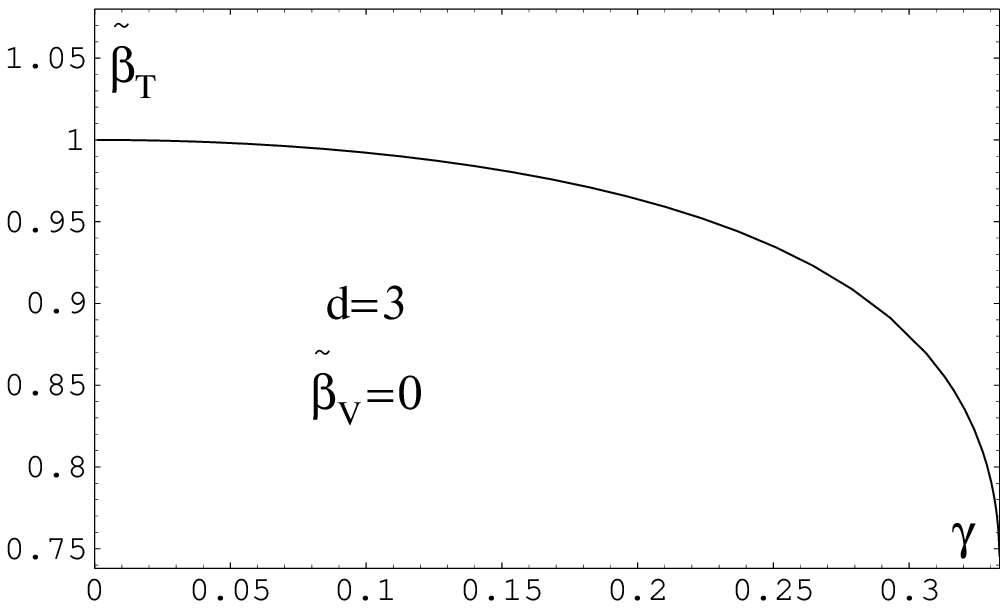}  \\
\end{center}
\vspace*{5mm}
\caption{
    $\btt$ versus $\gamma$ at $\bvt=0$,
    for (a) $d=1$, (b) $d=3/2$, (c) $d=2$, (d) $d=9/4$,
    (e) $d=d_* \approx 2.38403$, (f) $d=3$.
    For $3/2 < d < \bar{d} \approx 2.55031$,
    $\btt$ reaches a minimum at $\gamma = \gamma_{cusp}(d)$.
}
\label{fig_btt_versus_gamma}
\end{figure}

\begin{figure}[p]
\vspace*{-2cm} \hspace*{-0cm}
\begin{center}
\epsfxsize=0.45\textwidth
\leavevmode\epsffile{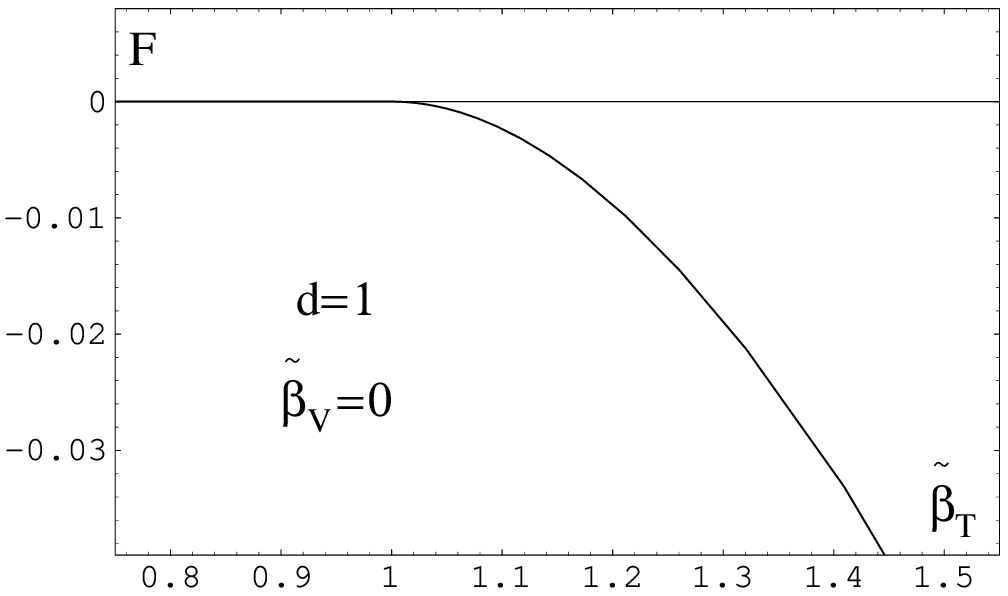}
\hspace{0.5cm}
\epsfxsize=0.45\textwidth
\epsffile{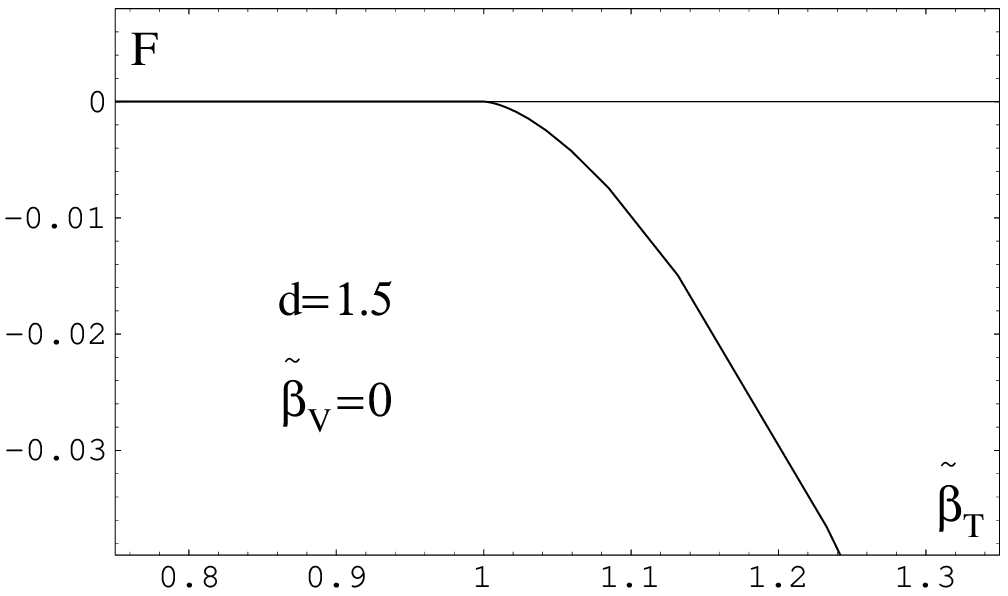}  \\
\vspace*{1cm}
\epsfxsize=0.45\textwidth
\leavevmode\epsffile{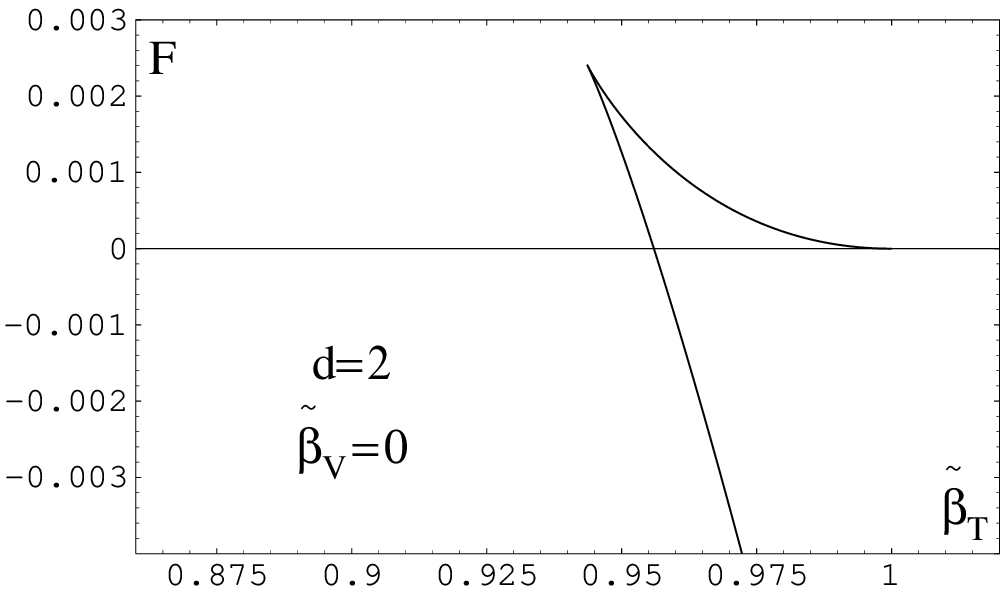}
\hspace{0.5cm}
\epsfxsize=0.45\textwidth
\epsffile{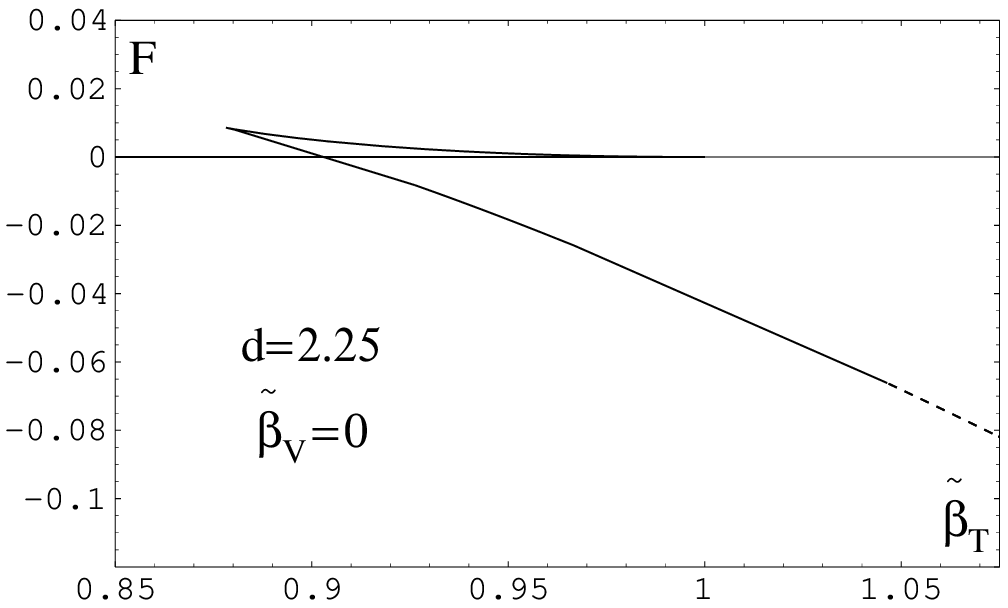}  \\
\vspace*{1cm}
\epsfxsize=0.45\textwidth
\leavevmode\epsffile{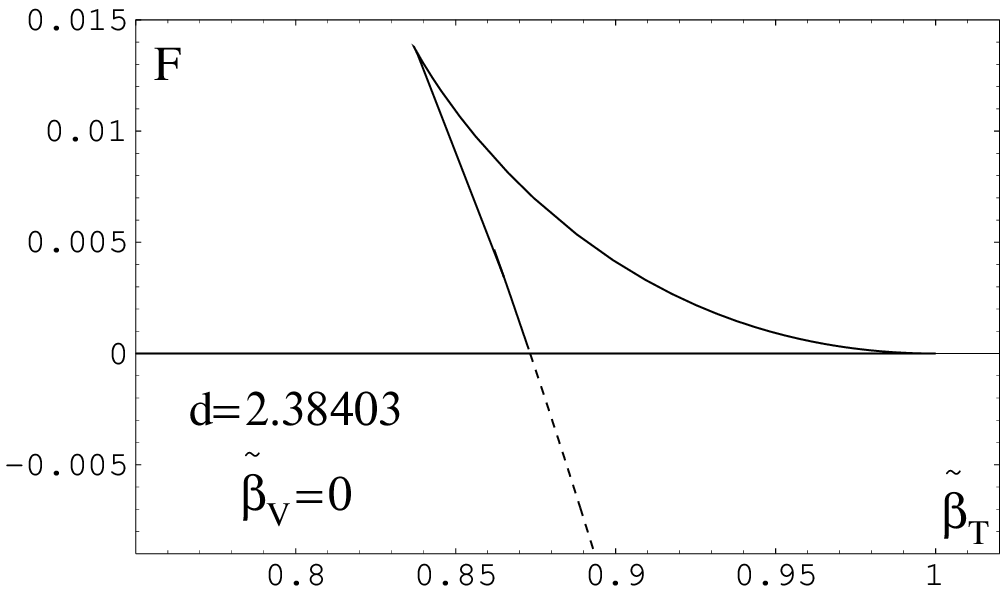}
\hspace{0.5cm}
\epsfxsize=0.45\textwidth
\epsffile{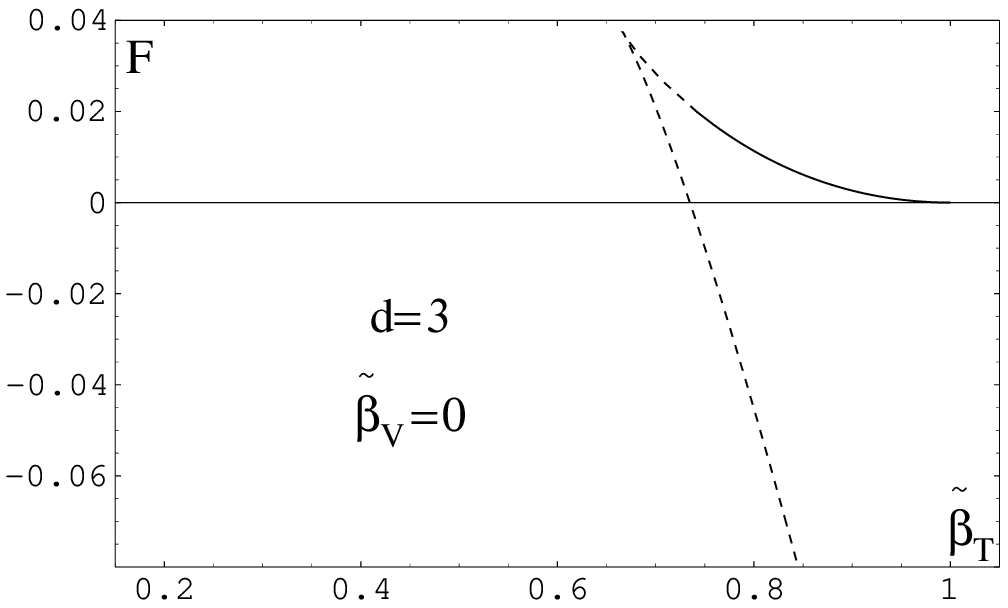}  \\
\vspace*{5mm}
\end{center}
\caption{
    Free energy $\overline{\scrf}$ versus $\btt$ at $\bvt = 0$,
    for (a) $d=1$, (b) $d=3/2$, (c) $d=2$, (d) $d=9/4$,
    (e) $d=d_* \approx 2.38403$, (f) $d=3$.
    In (d)--(f), the dashed curve corresponds to the condensate phase.
}
\label{fig_F_versus_btt}
\end{figure}

\begin{figure}[p]
\begin{center}
\epsffile{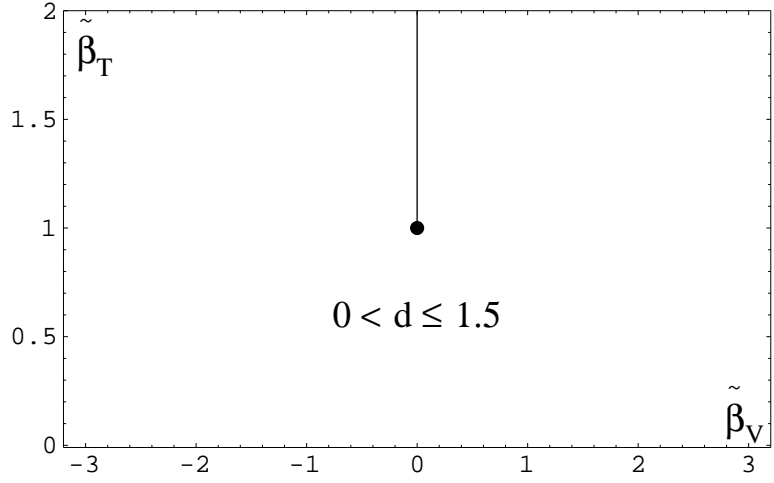}
\end{center}
\caption{
   Phase diagram in the $(\bvt,\btt)$-plane for dimensions $0 < d \leq 3/2$.
   There is a second-order transition at $(\bvt,\btt) = (0,1)$
   and a line of first-order transitions at $\bvt = 0$, $\btt > 1$.
}
\label{phase_diag_d<=32}
\end{figure}

\begin{figure}[p]
\begin{center}
\epsffile{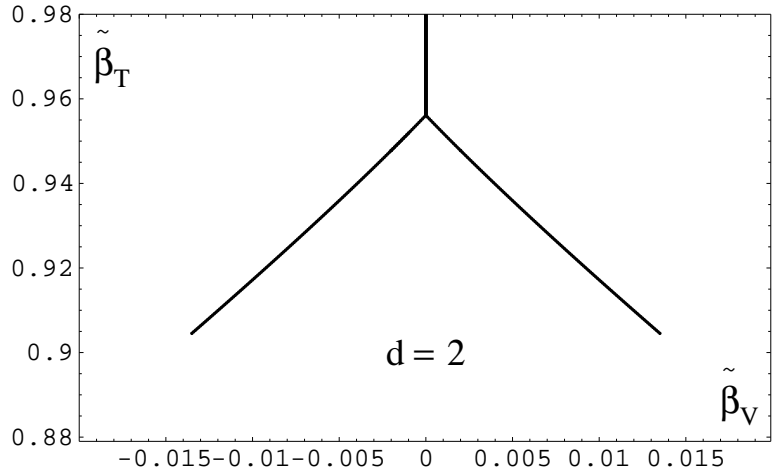}
\end{center}
\caption{
   Phase diagram in $(\bvt,\btt)$-plane for dimensions $3/2 < d \le 2$,
   shown here for the case $d=2$.
   The first-order transition curves $\betatilde_{T,c}(\bvt)$
   are almost straight lines but are in fact slightly convex.
   In $d=2$, the triple point lies at
      $(\bvt,\btt) \approx (0, 0.956)$ and $\gamma_c \approx 0.487$,
      while the endpoints of the first-order transition curves lie at
      $(\pm \betatilde_{V,e},\betatilde_{T,e} ) 
      \approx (\pm 0.0134, 0.905)$ and $\gamma_e \approx 0.411$.
}
\label{phase_diag_d=2}
\end{figure}

For  $3/2 < d \leq 2$, a first-order phase transition occurs
at the point of self-intersection
(point c in  Figure~\ref{new_triangle_guide})
for all $0 < \bvt < \betatilde_{V,e}$.
The coordinates $\betatilde_{T,c}(\bvt)$ of this transition
are found by solving the system
$\btt (\gamma_1,d,\bvt)=\btt (\gamma_2,d,\bvt)$ and
$\overline{\scrf}(\gamma_1,d,\bvt)=\overline{\scrf}(\gamma_2,d,\bvt)$.
They form two symmetric slightly convex branches on the phase diagram,
as shown in Figure \ref{phase_diag_d=2} for the case $d=2$.\footnote{
   Magnoli and Ravanini \cite[Figure 2]{Magnoli_87}
   correctly found the endpoint
   $(\betatilde_{V,e},\betatilde_{T,e}) \approx (0.0134, 0.905)$
   with $\gamma_e \approx 0.411$,
   but incorrectly drew the transition curve as
   concave and as smooth at $\bvt = 0$.
   They also failed to notice the first-order transition curve at
   $\bvt=0$, $\btt > \betatilde_{T,c}$.
}
The triple point $\betatilde_{T,c}(0)$ is a slightly decreasing
function of $d$, as shown in Figure~\ref{phasediag_bv=0}.

\begin{figure}[p]
\begin{center}
\epsffile{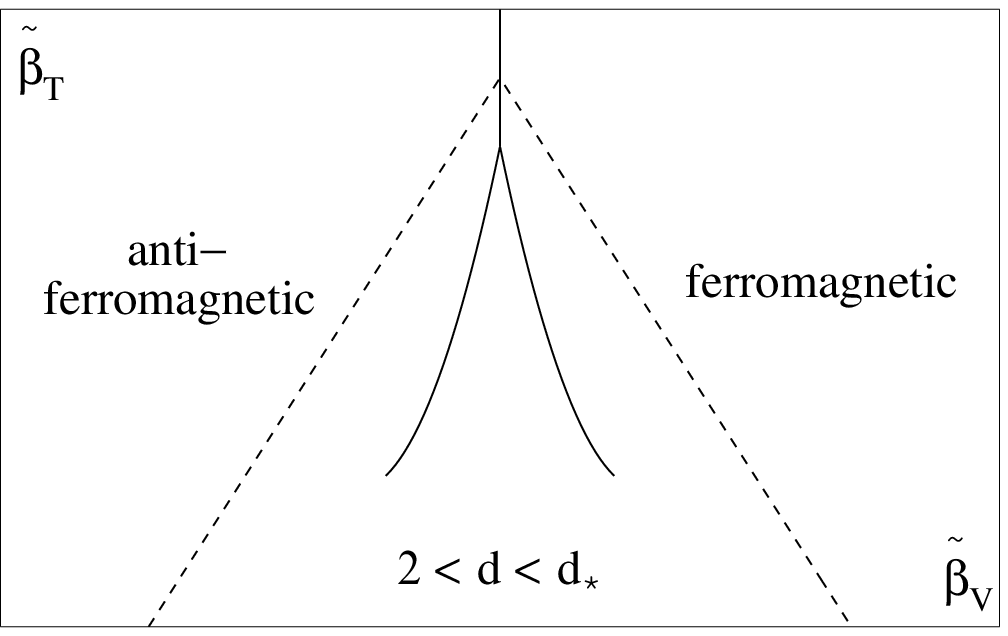}
\end{center}
\caption{
   Phase diagram in the $(\bvt,\btt)$-plane for dimensions
   $2 < d < d_* \approx 2.38403$.
   In addition to the first-order transitions present for $3/2 < d \le 2$
   at $\betatilde_{T,c}(\bvt)$,
   there is an additional pair of second-order transition lines (dashed)
   at $\betatilde_{T,c'}(\bvt)$ given by \reff{nematic_lines}.
   The phase at the upper right (resp.\ upper left) of the diagram
   exhibits ferromagnetic (resp.\ antiferromagnetic) long-range order.
}
\label{phase_diag_2<d<dstar}
\end{figure}

\begin{figure}[p]
\begin{center}
\epsffile{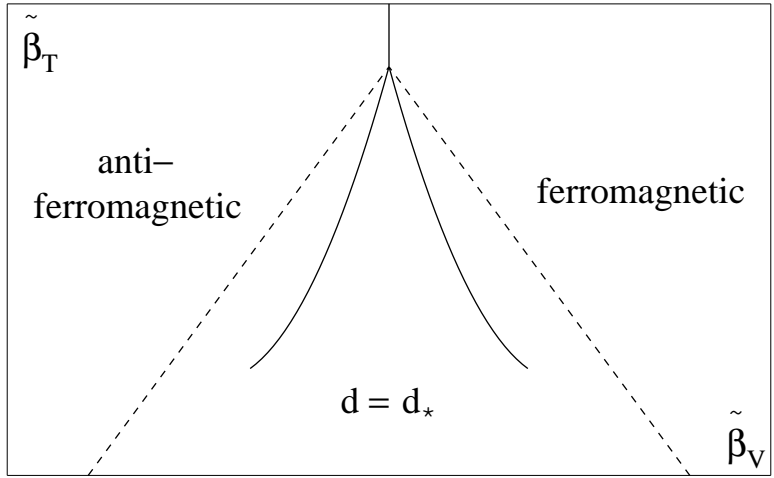}
\end{center}
\caption{
   Phase diagram in the $(\bvt,\btt)$-plane for dimension
   $d = d_* \approx 2.38403$.
   The two transitions coincide at $\bvt=0$.
   The phase at the upper right (resp.\ upper left) of the diagram
   exhibits ferromagnetic (resp.\ antiferromagnetic) long-range order.
}
\label{phase_diag_d=dstar}
\end{figure}

\begin{figure}[p]
\begin{center}
\epsffile{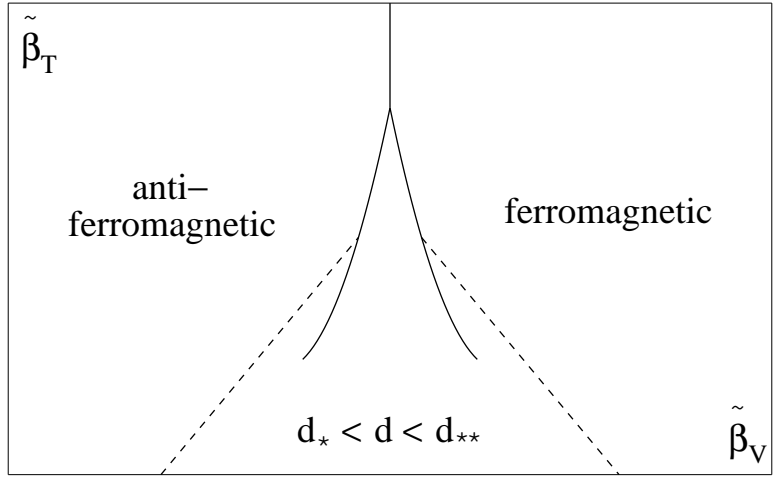}
\end{center}
\caption{
   Phase diagram in the $(\bvt,\btt)$-plane for dimensions
   $d_* \approx 2.38403 < d < d_{**}=3$.
   Second-order transitions (dashed lines) occur only where
   they lie above the first-order transition curves;
   this occurs for $|\bvt |>\betatilde_{V,cc'}$.
   The phase at the upper right (resp.\ upper left) of the diagram
   exhibits ferromagnetic (resp.\ antiferromagnetic) long-range order.
}
\label{phase_diag_dstar<d<dstarstar}
\end{figure}

\begin{figure}[p]
\begin{center}
\epsffile{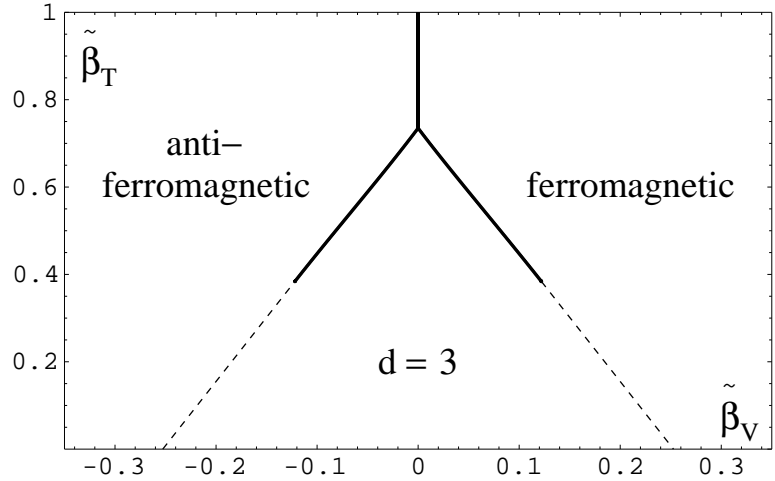}
\end{center}
\caption{
   Phase diagram in $(\bvt,\btt)$-plane for $d \ge d_{**}=3$,
   shown here for the case $d=3$.
   The first-order transition curves $\betatilde_{T,c}(\bvt )$
   are almost straight lines but are in fact slightly convex.
   In $d=3$, the triple point lies at
   $(\bvt,\btt) \approx (0, 0.73487)$,
   while the endpoints of the first-order transition curves lie at
   $(\betatilde_{V,e},\betatilde_{T,e}) 
    = (\pm f_*(3)[2-f_*(3)]/6, f_*(3)^2/6)\approx (\pm 0.122224, 0.383238)$.
   The phase at the upper right (resp.\ upper left) of the diagram
   exhibits ferromagnetic (resp.\ antiferromagnetic) long-range order.
}
\label{phase_diag_d=3}
\end{figure}

\begin{figure}[p]
\begin{center}
\epsffile{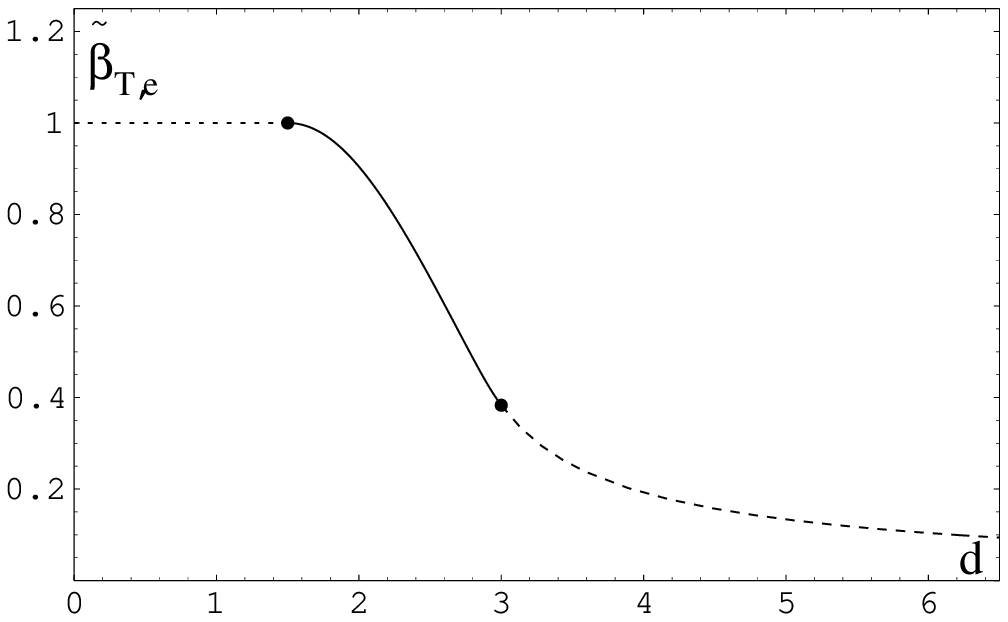}
\end{center}
\caption{
   Coordinate $\betatilde_{T,e}$ of the endpoint of the
   first-order transition curve,
   as function of dimension $d$.
   For $d \ge d_{**}=3$ it is given by $\betatilde_{T,e} =f_*^2/2d$
   (dashed line) and tends to zero as $d\rightarrow \infty$. 
}
\label{beta_T_endpoint_d}
\end{figure}

\begin{figure}[p]
\begin{center}
\epsffile{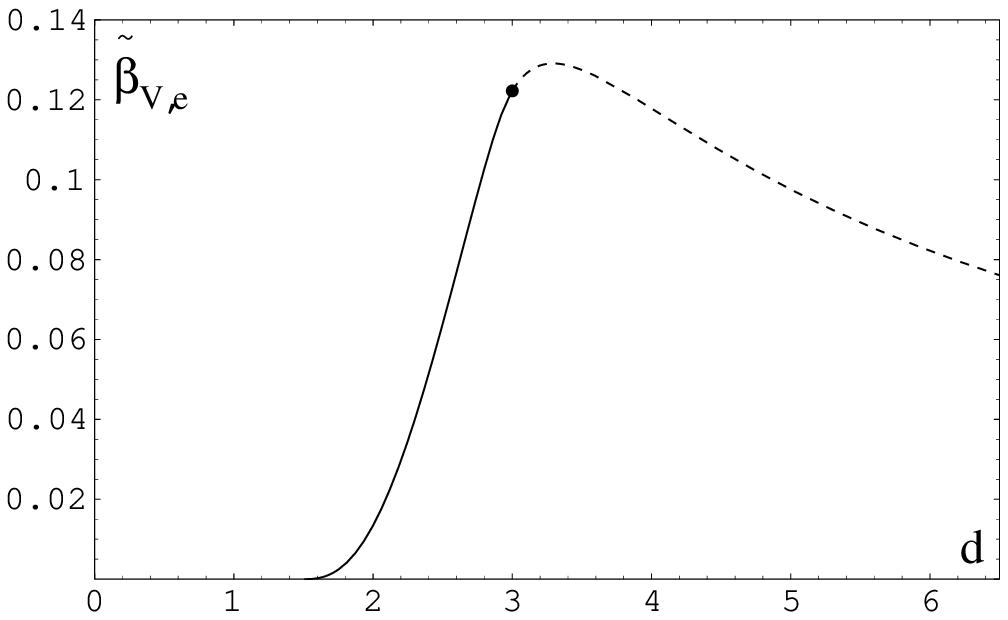}
\end{center}
\caption{
   Coordinate $\betatilde_{V,e}$ of the endpoint of the
   first-order transition curve,
   as function of dimension $d$.
   For $d \ge d_{**}=3$ it is given by
   $\betatilde_{V,e} = \betatilde_{V,2}=f_* (2-f_*)/2d$
   (dashed line) and tends to zero as $d\rightarrow \infty$. 
}
\label{beta_V_endpoint_d}
\end{figure}

\begin{figure}[p]
\begin{center}
\epsfxsize=0.55\textwidth
\leavevmode\epsffile{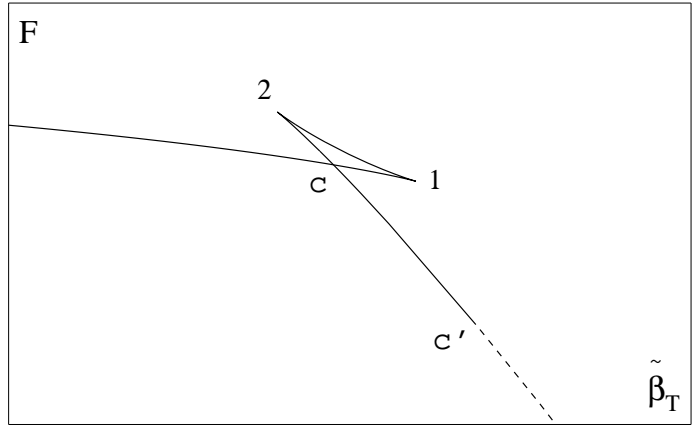}
\end{center}
\caption{
   ``Notations guide'' for the parametric plots of
   $\overline{\scrf}$ versus $\btt$.
   Points 1 and 2 are, respectively,
   the local maximum and minimum of $\btt$ versus $f$.
   Point c is the first-order transition at $\betatilde_{T,c}(\bvt)$.
   Point c\textprime is the second-order normal-to-condensate transition
   at $f=f_*$.  The condensate phase $f \ge f_*$ is shown as a dashed line.
}
\label{new_triangle_guide}
\end{figure}

\begin{figure}[p]
\vspace*{-2cm} \hspace*{-0cm}
\begin{center}
\vspace*{0cm} \hspace*{-0cm}
\epsfxsize=0.3\textwidth
\leavevmode\epsffile{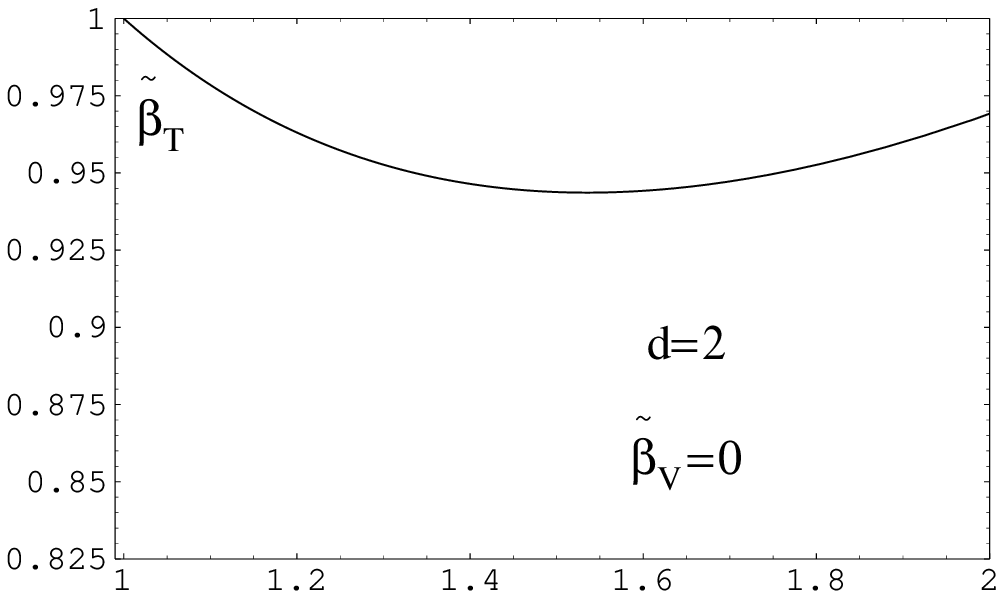}
\hspace{1.5cm}
\epsfxsize=0.3\textwidth
\epsffile{new_param_F_beta_d=2_v=0.eps}  \\
\vspace*{0.7cm}
\epsfxsize=0.3\textwidth
\leavevmode\epsffile{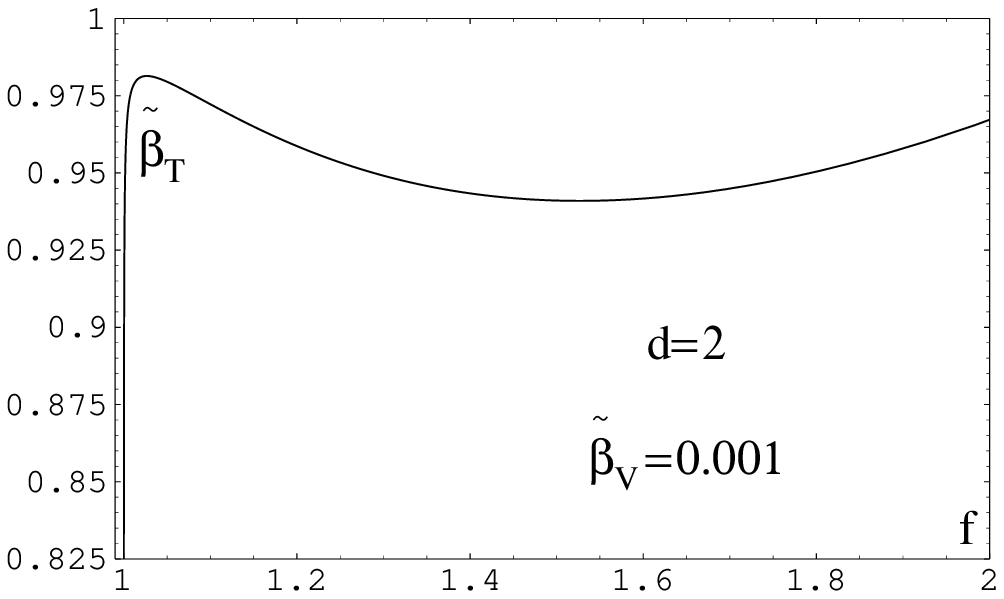}
\hspace{1.5cm}
\epsfxsize=0.3\textwidth
\epsffile{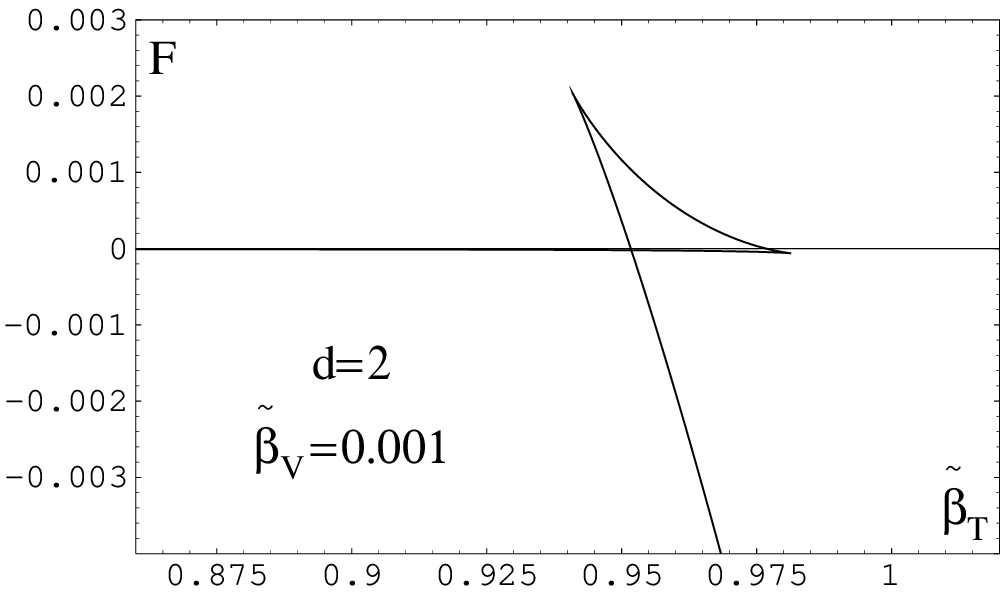}  \\
\vspace*{0.7cm}
\epsfxsize=0.3\textwidth
\leavevmode\epsffile{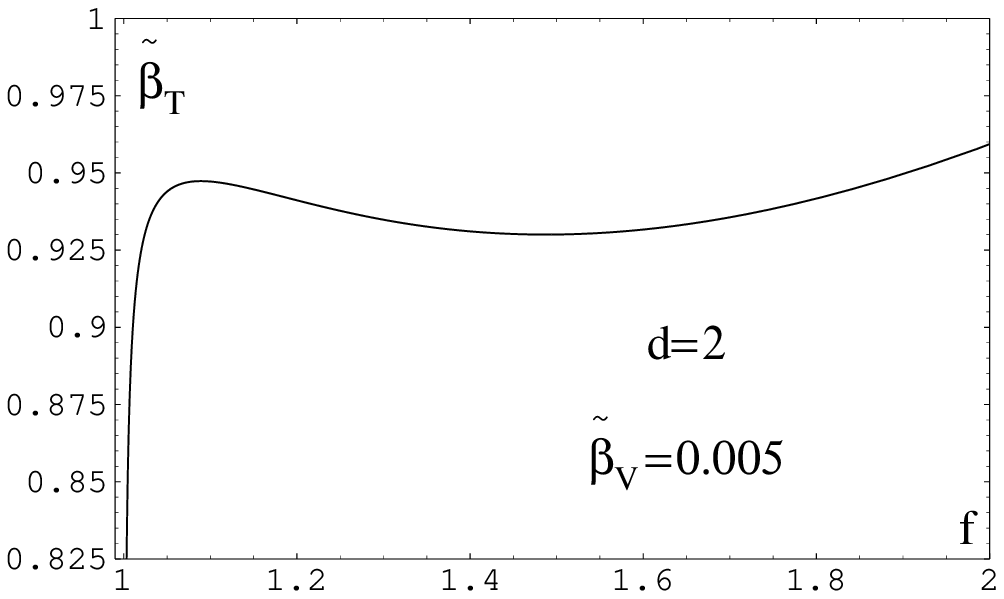}
\hspace{1.5cm}
\epsfxsize=0.3\textwidth
\epsffile{ 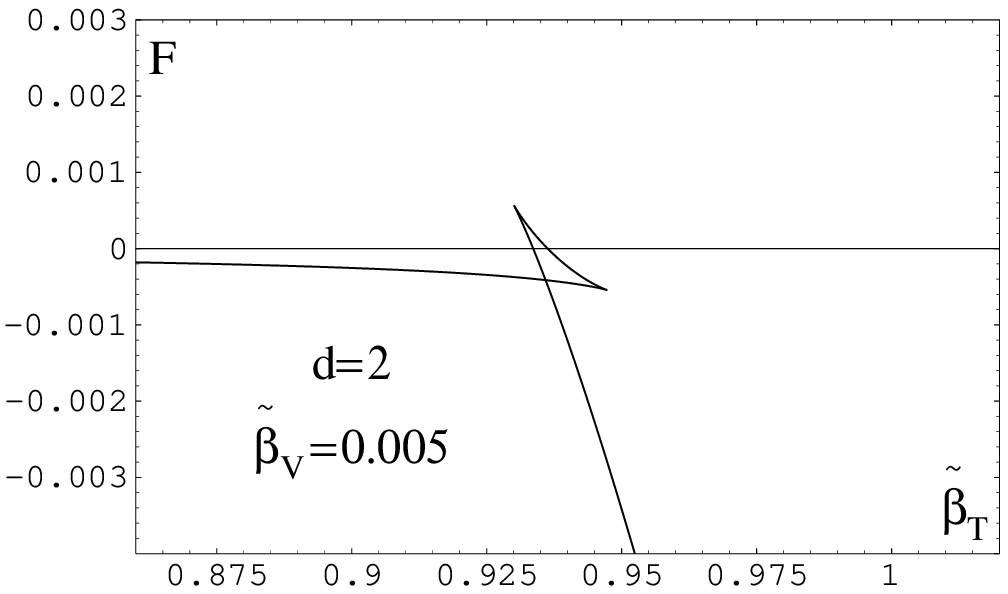}  \\
\vspace*{0.7cm}
\epsfxsize=0.3\textwidth
\leavevmode\epsffile{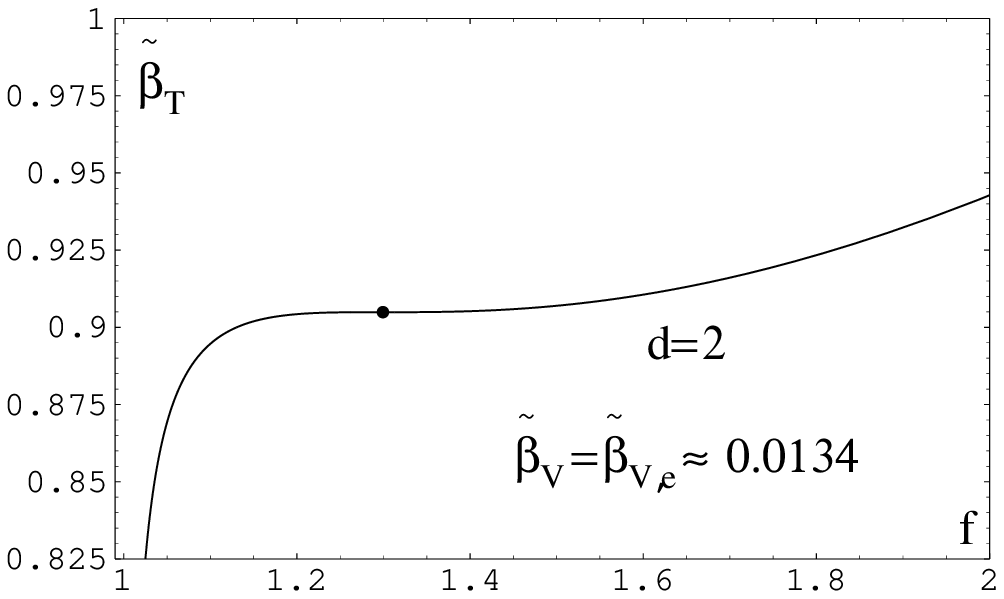}
\hspace{1.5cm}
\epsfxsize=0.3\textwidth
\epsffile{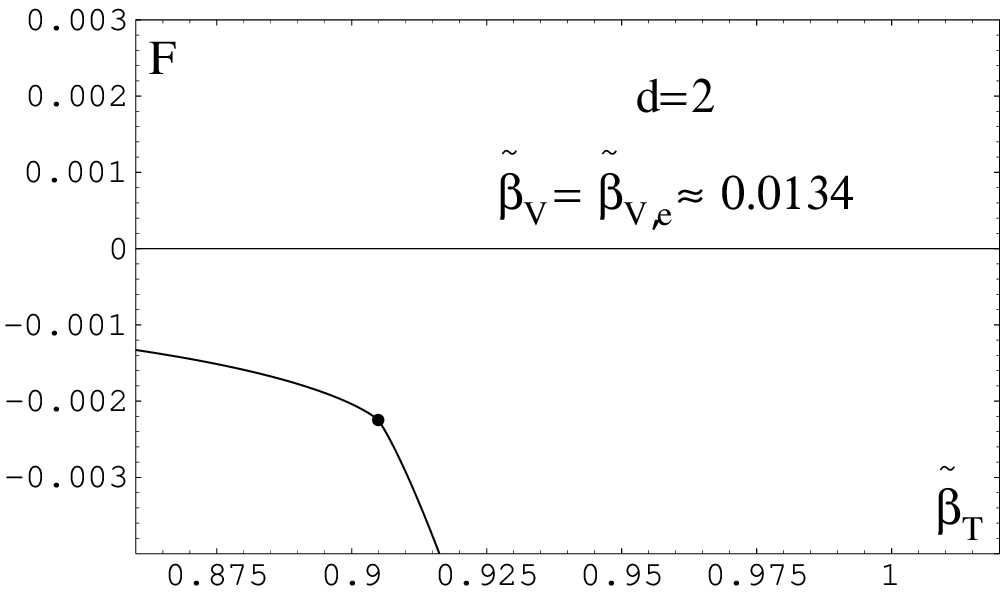}  \\
\vspace*{0.7cm}
\epsfxsize=0.3\textwidth
\leavevmode\epsffile{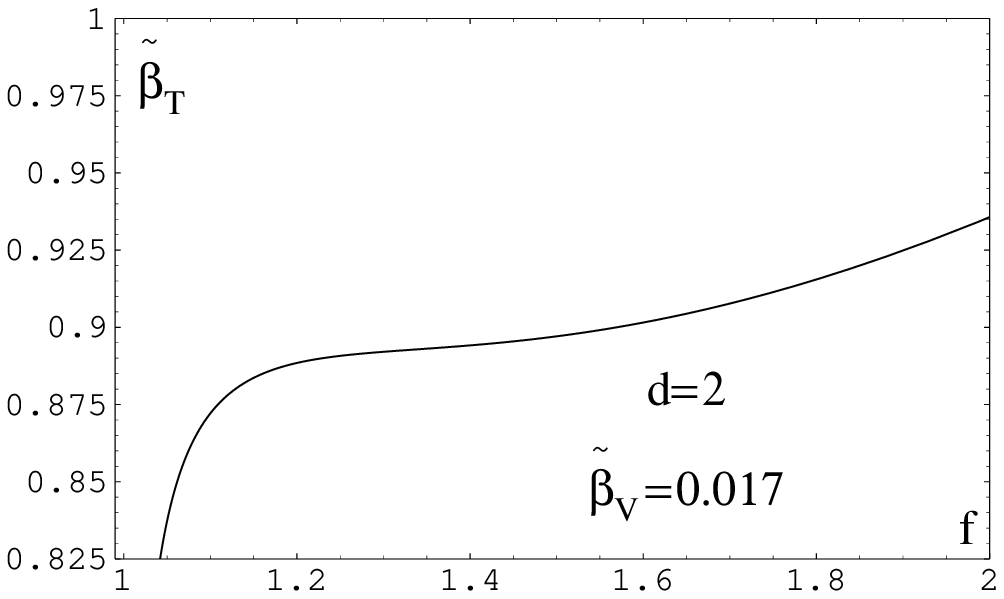}
\hspace{1.5cm}
\epsfxsize=0.3\textwidth
\epsffile{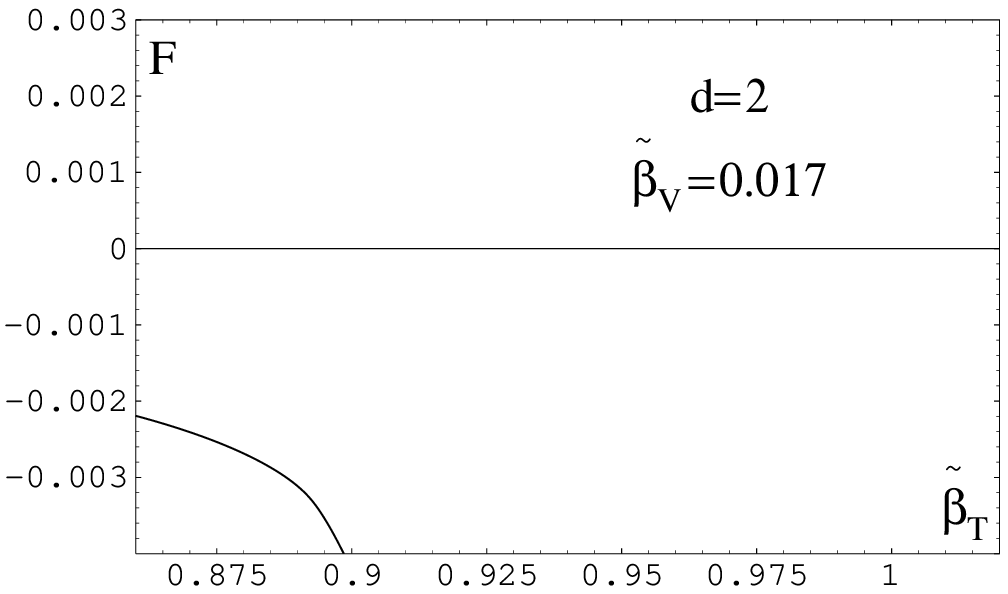}  \\
\vspace*{0.7cm}
\epsfxsize=0.3\textwidth
\leavevmode\epsffile{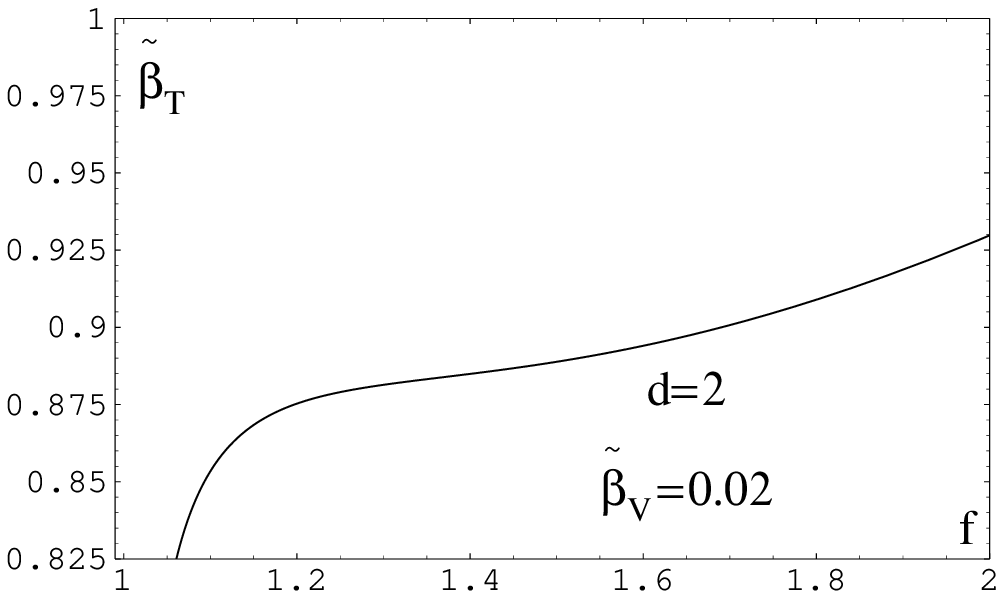}
\hspace{1.5cm}
\epsfxsize=0.3\textwidth
\epsffile{ 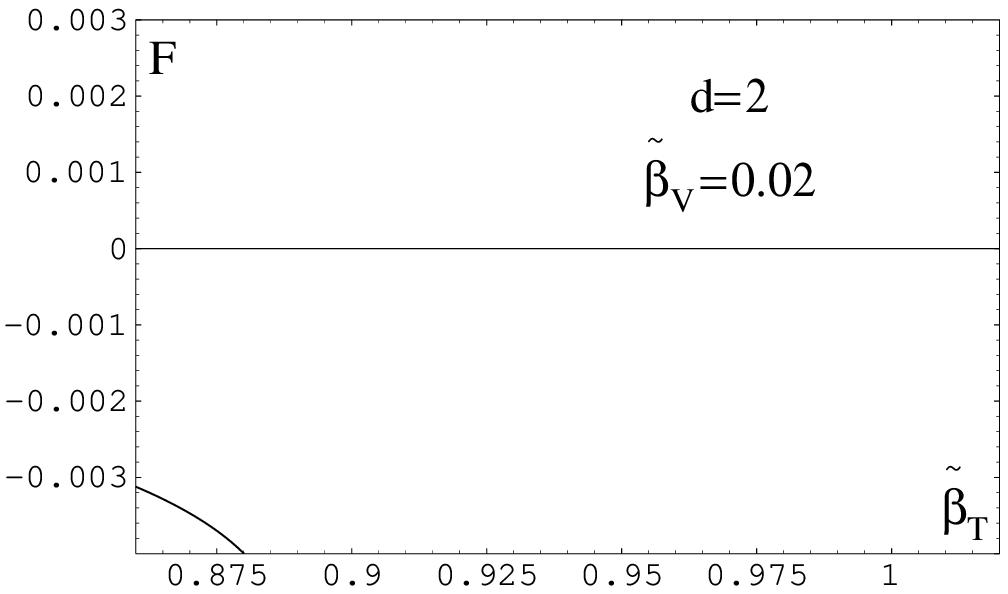}  \\
\end{center}
\vspace*{5mm}
\caption{$\btt$ versus $f$, and 
    $\overline{\scrf}$ versus $\btt$, for $d=2$ at
 (a) $\bvt =0$, (b) $\bvt = 0.001$, (c) $\bvt = 0.005$,
(d) $\bvt = \betatilde_{V,e}\approx 0.0134$ [the dot indicates the endpoint
$(\betatilde_{V,e},\betatilde_{T,e})$], (e) $\bvt = 0.017$,
    (f) $\bvt =0.02$.
}
\label{fig_param_F_versus_btt_d=2}
\end{figure}

\begin{figure}[p]
\vspace*{-2cm} \hspace*{-0cm}
\begin{center}
\vspace*{0cm} \hspace*{-0cm}
\epsfxsize=0.3\textwidth
\leavevmode\epsffile{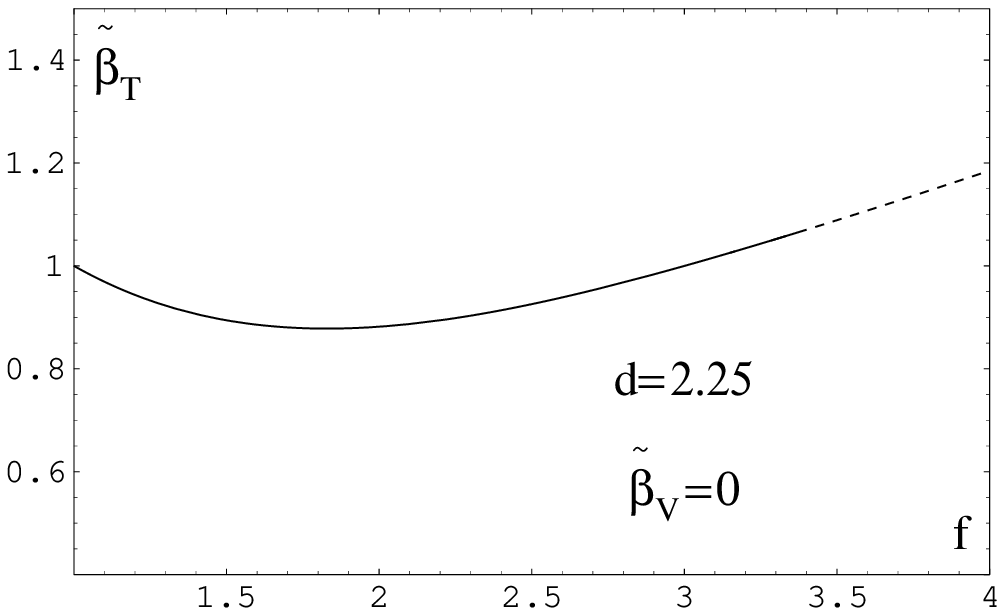}
\hspace{1.5cm}
\epsfxsize=0.3\textwidth
\epsffile{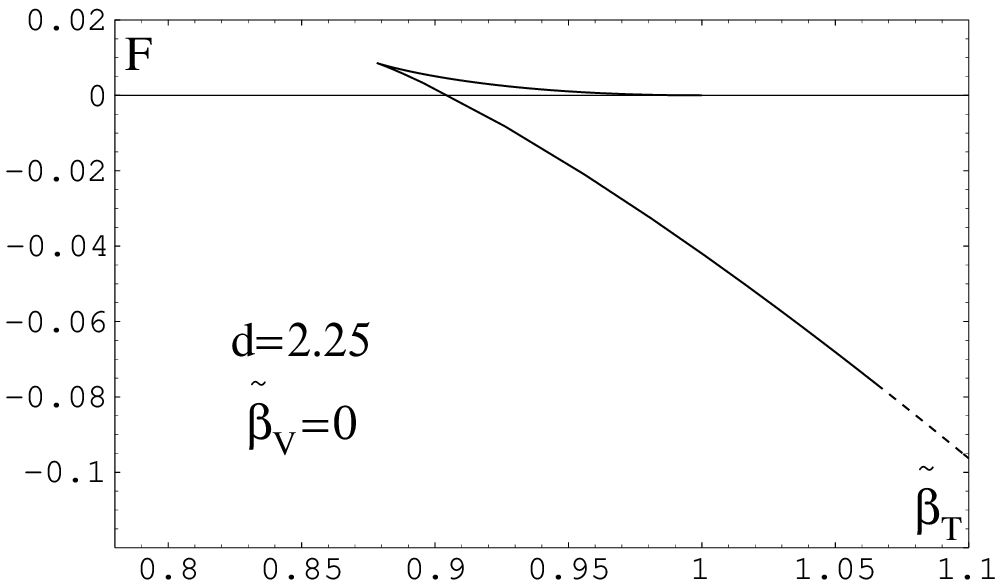}  \\
\vspace*{0.7cm}
\epsfxsize=0.3\textwidth
\leavevmode\epsffile{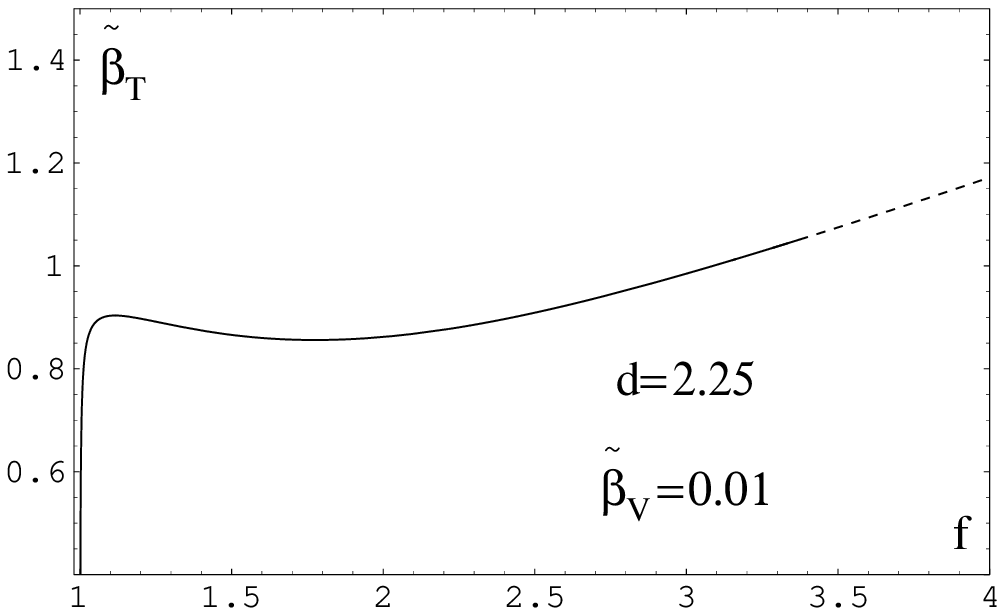}
\hspace{1.5cm}
\epsfxsize=0.3\textwidth
\epsffile{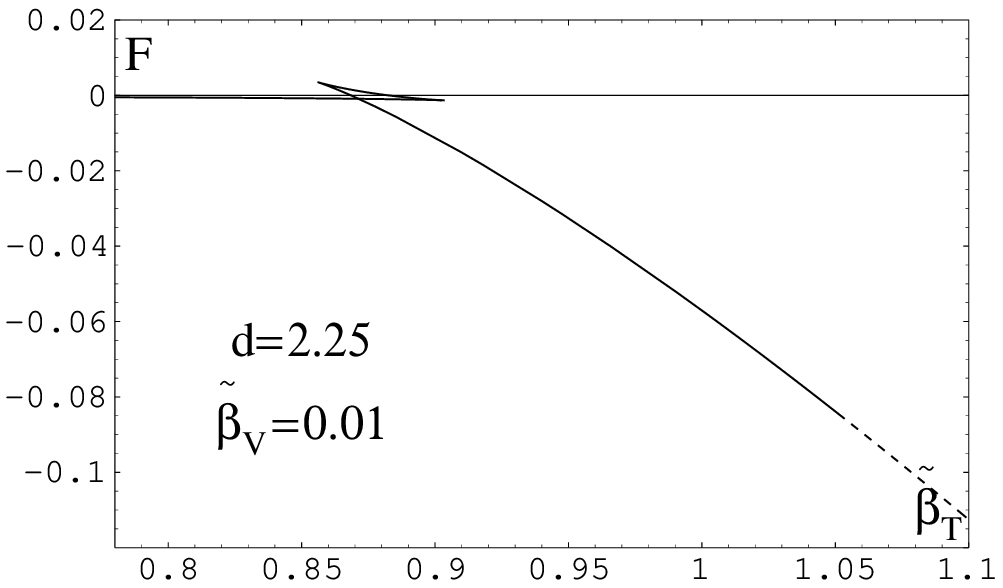}  \\
\vspace*{0.7cm}
\epsfxsize=0.3\textwidth
\leavevmode\epsffile{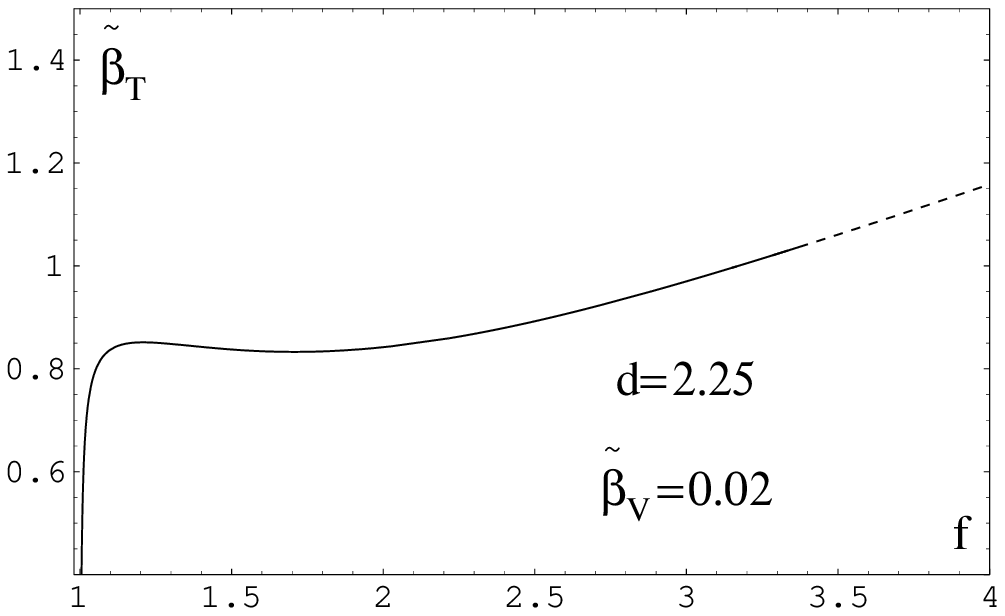}
\hspace{1.5cm}
\epsfxsize=0.3\textwidth
\epsffile{ 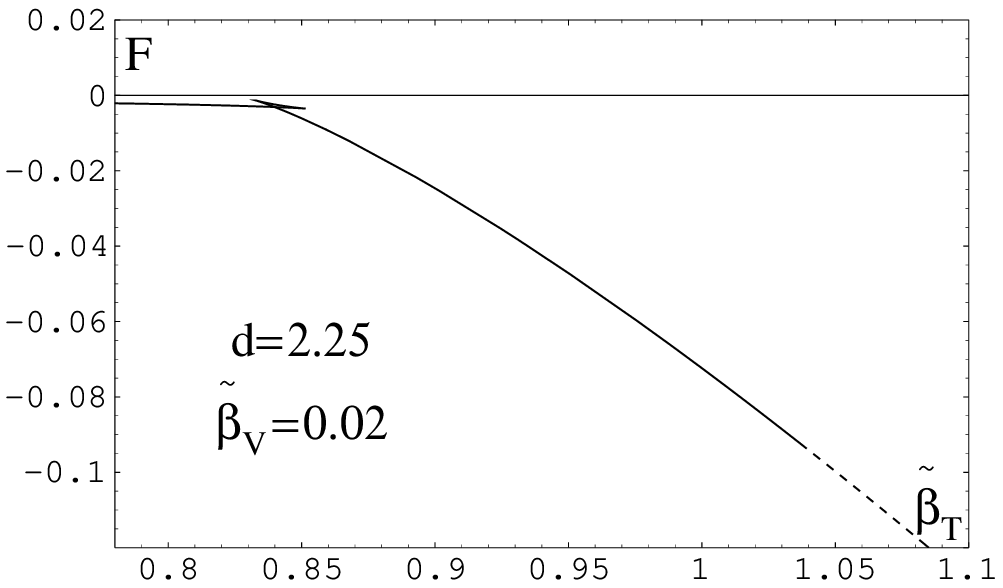}  \\
\vspace*{0.7cm}
\epsfxsize=0.3\textwidth
\leavevmode\epsffile{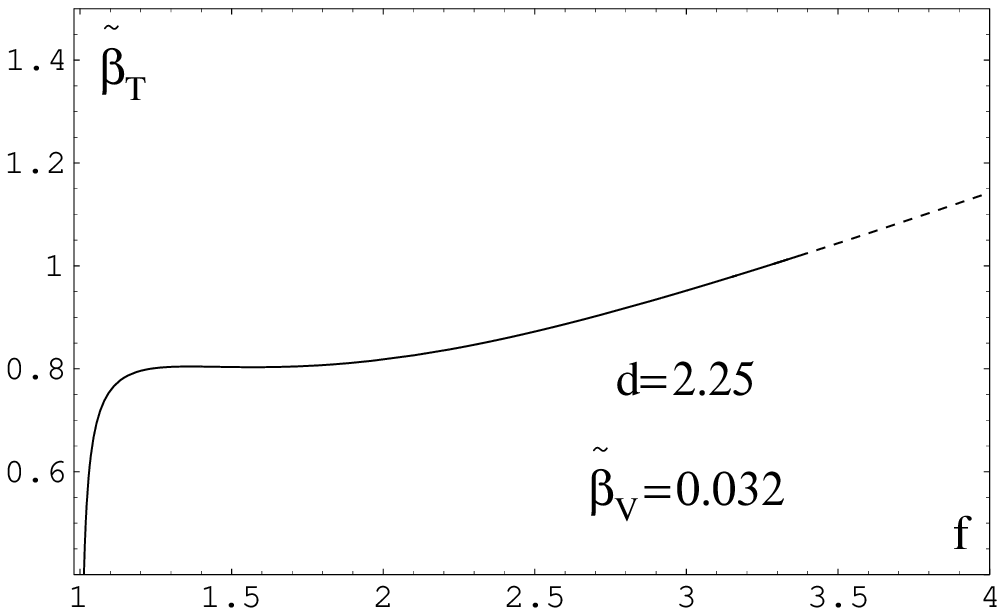}
\hspace{1.5cm}
\epsfxsize=0.3\textwidth
\epsffile{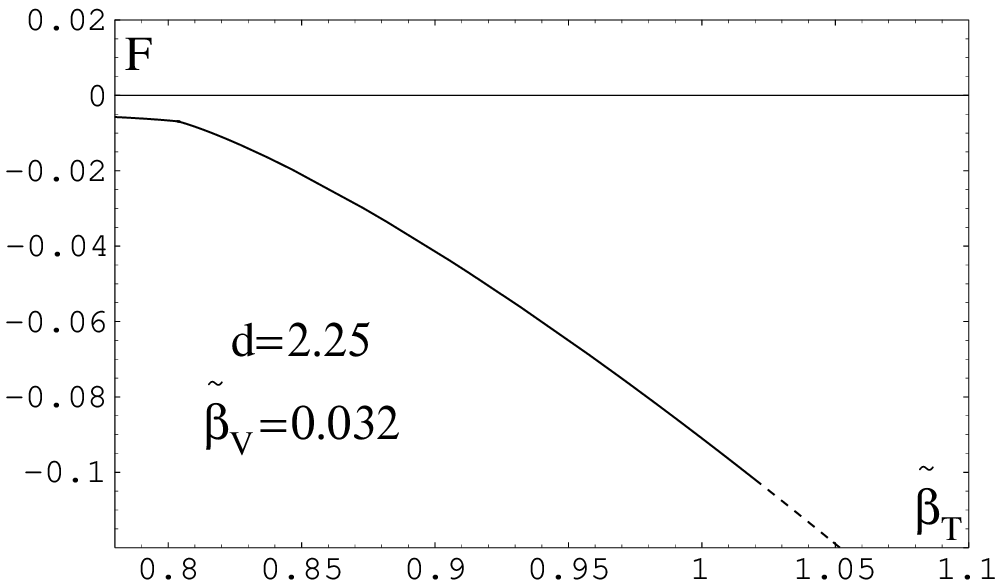}  \\
\vspace*{0.7cm}
\epsfxsize=0.3\textwidth
\leavevmode\epsffile{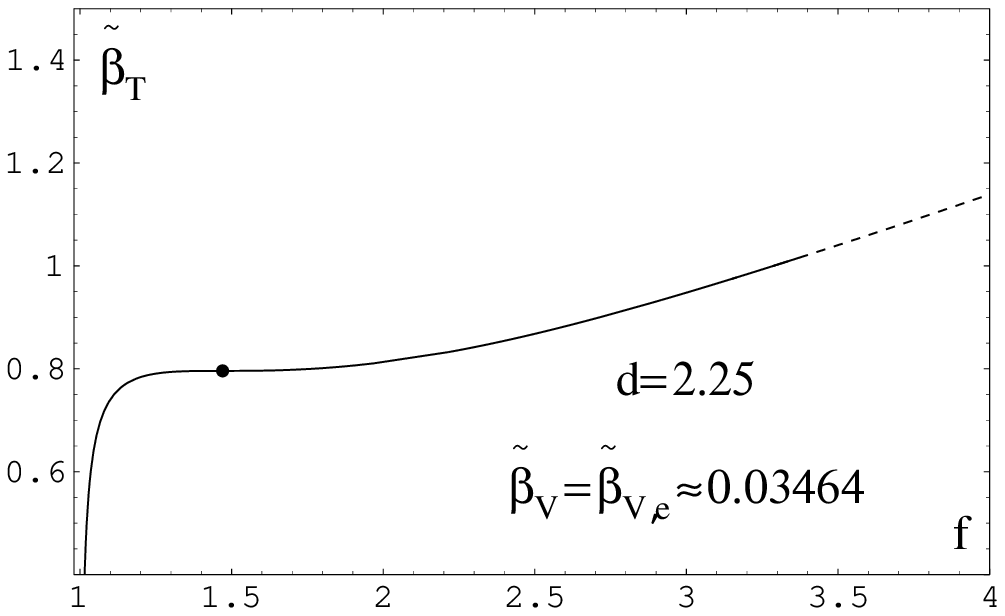}
\hspace{1.5cm}
\epsfxsize=0.3\textwidth
\epsffile{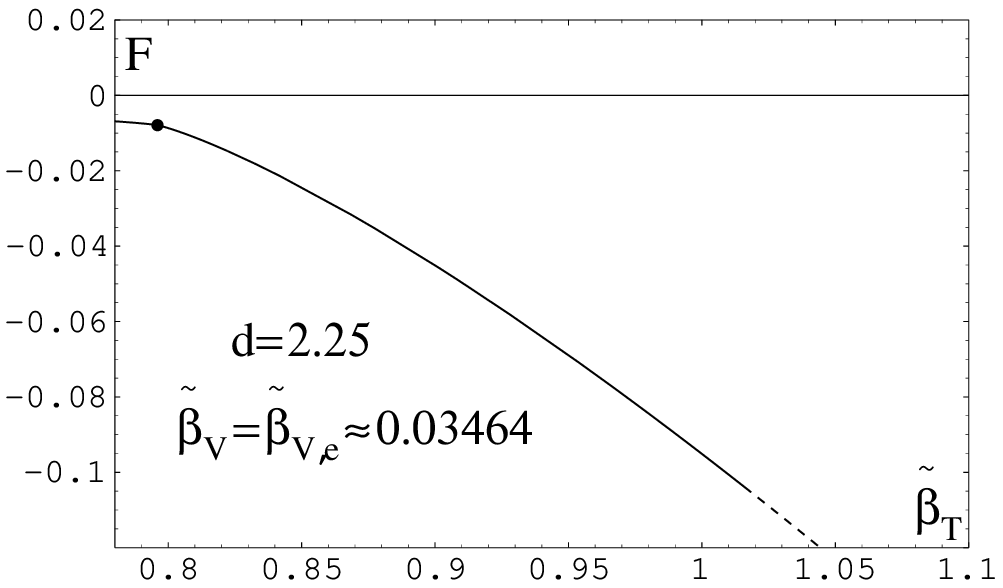}  \\
\vspace*{0.7cm}
\epsfxsize=0.3\textwidth
\leavevmode\epsffile{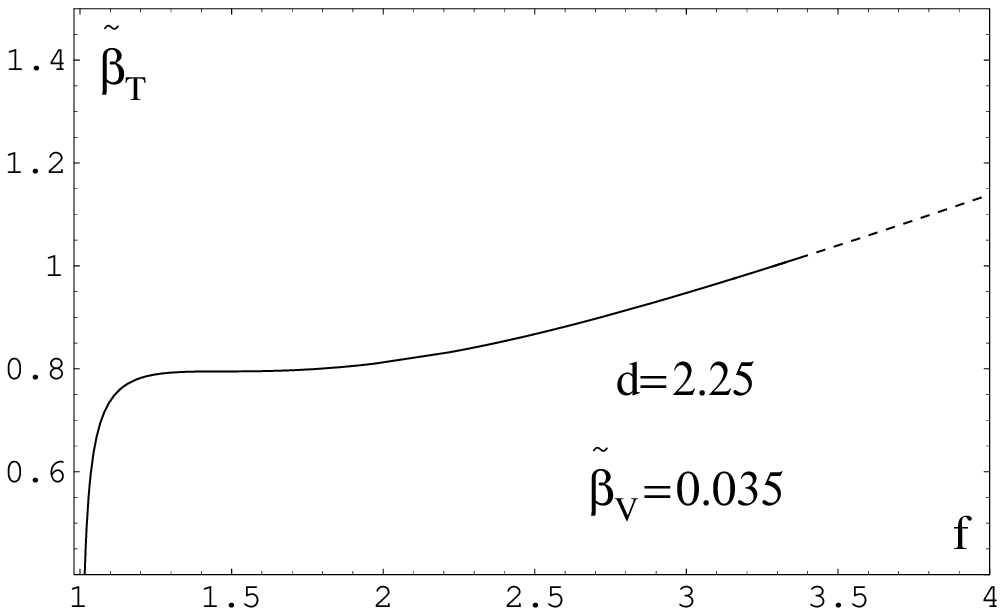}
\hspace{1.5cm}
\epsfxsize=0.3\textwidth
\epsffile{ 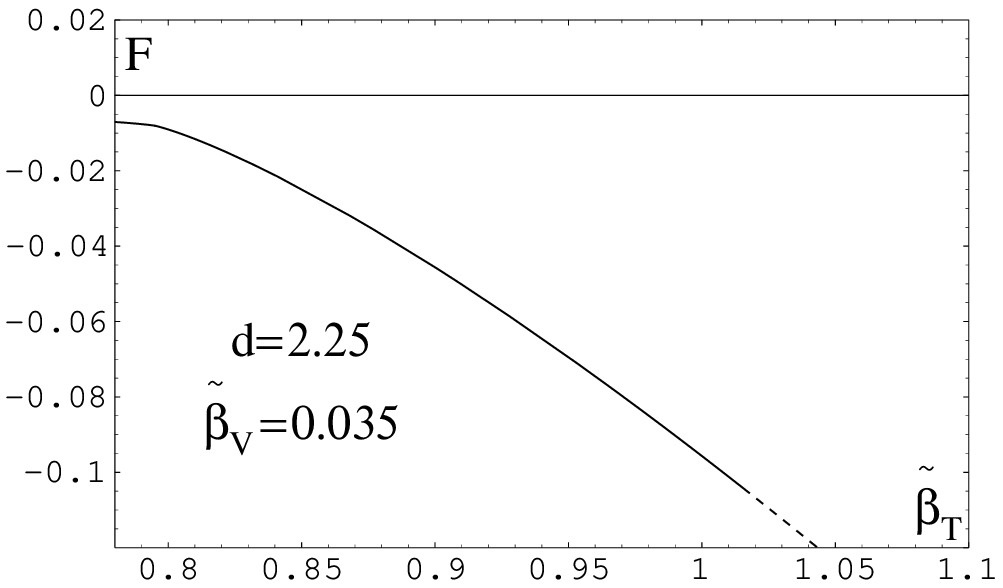}  \\
\end{center}
\vspace*{5mm}
\caption{$\btt$ versus $f$, and 
    $\overline{\scrf}$ versus $\btt$, for $d=2.25$ at
     (a) $\bvt =0$, (b) $\bvt = 0.01$, (c) $\bvt = 0.02$,
(d) $\bvt =0.032$, (e) $\bvt = \betatilde_{V,e}
        \approx 0.03464$ (note that $\betatilde_{T,e}\approx 0.795978$
        and $\gamma_e \approx 0.416596$), (f) $\bvt =0.035$.
}
\label{fig_param_F_versus_btt_d=2.25}
\end{figure}

\begin{figure}[p]
\vspace*{-2cm} \hspace*{-0cm}
\begin{center}
\vspace*{0cm} \hspace*{-0cm}
\epsfxsize=0.3\textwidth
\leavevmode\epsffile{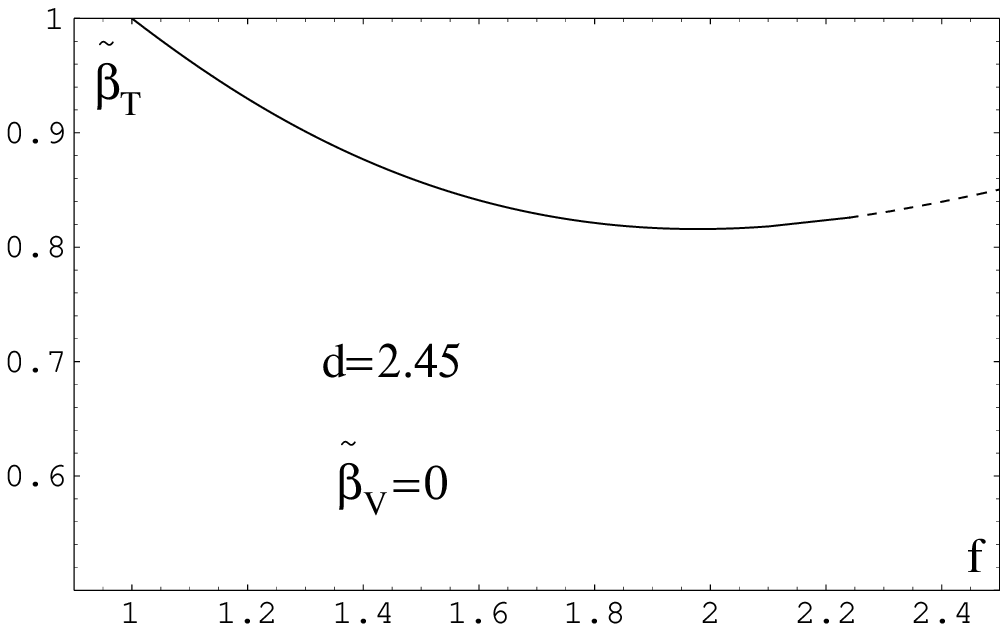}
\hspace{1.5cm}
\epsfxsize=0.3\textwidth
\epsffile{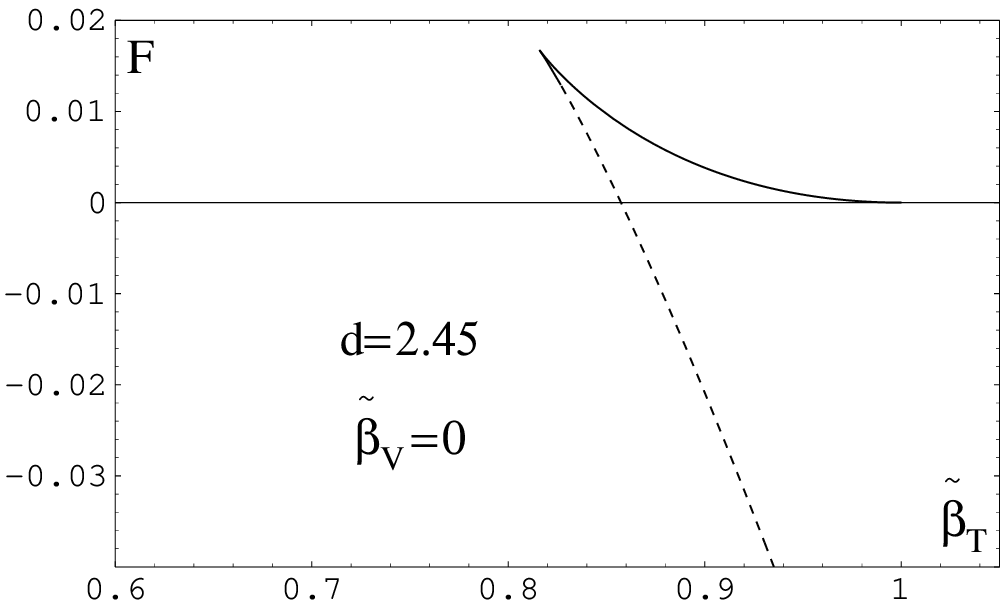}  \\
\vspace*{0.7cm}
\epsfxsize=0.3\textwidth
\leavevmode\epsffile{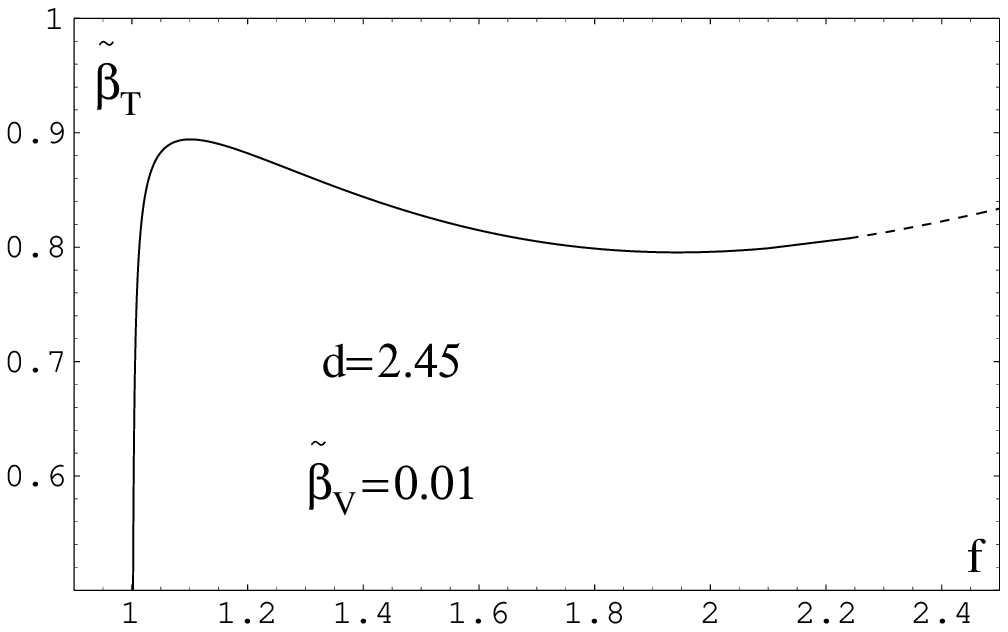}
\hspace{1.5cm}
\epsfxsize=0.3\textwidth
\epsffile{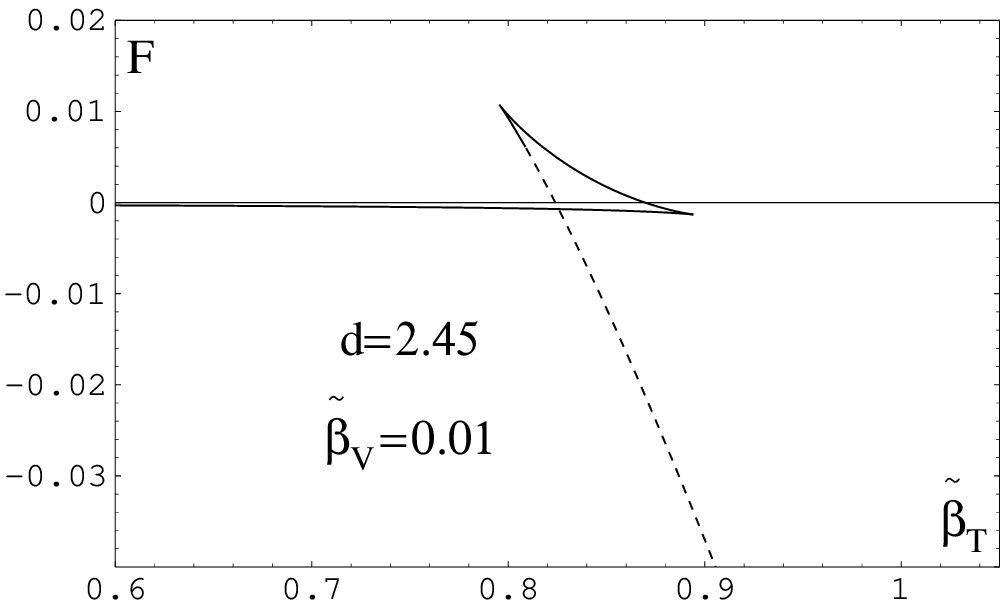}  \\
\vspace*{0.7cm}
\epsfxsize=0.3\textwidth
\leavevmode\epsffile{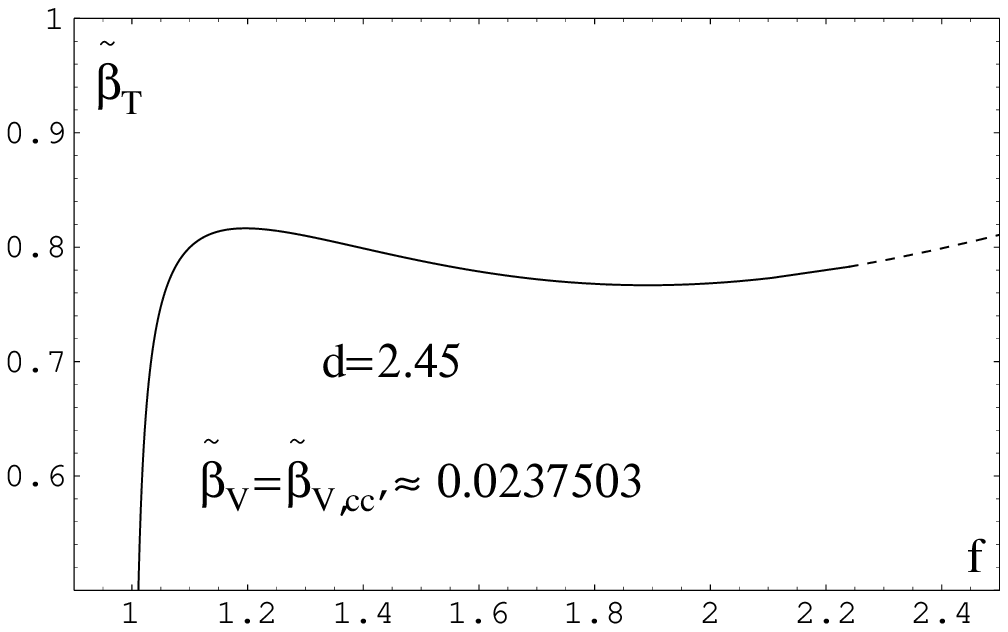}
\hspace{1.5cm}
\epsfxsize=0.3\textwidth
\epsffile{ 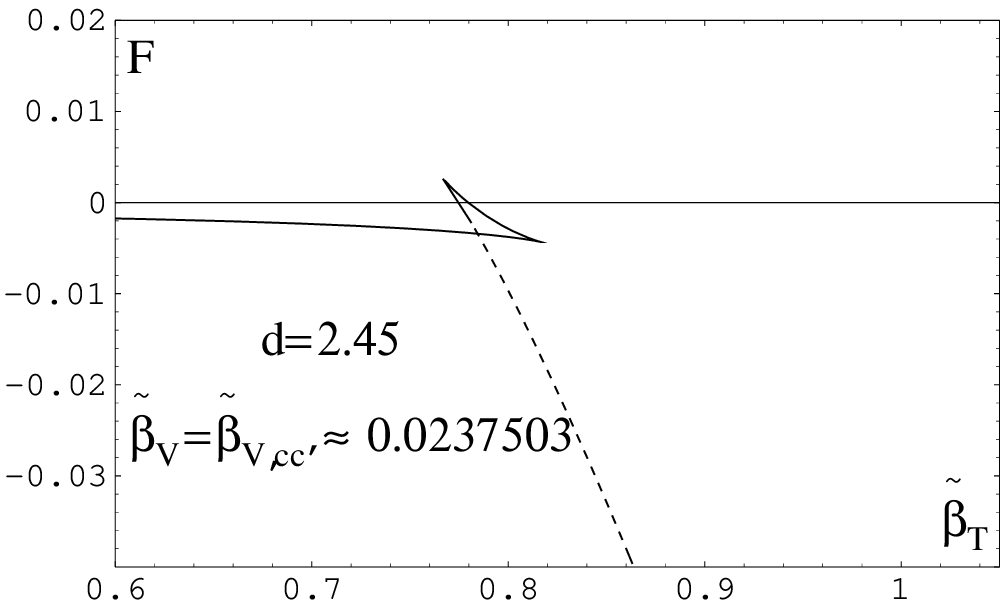}  \\
\vspace*{0.7cm}
\epsfxsize=0.3\textwidth
\leavevmode\epsffile{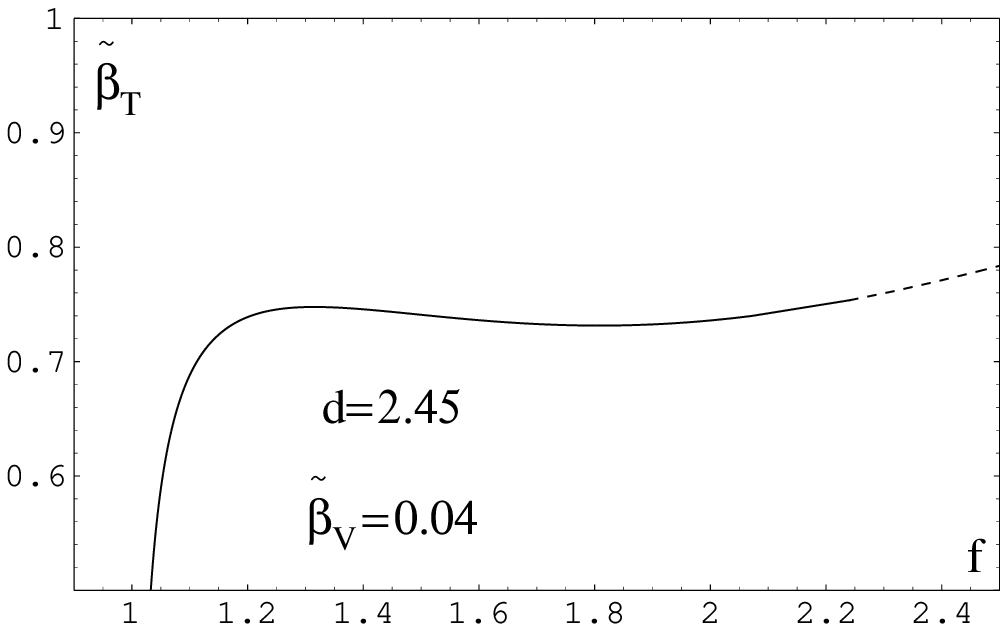}
\hspace{1.5cm}
\epsfxsize=0.3\textwidth
\epsffile{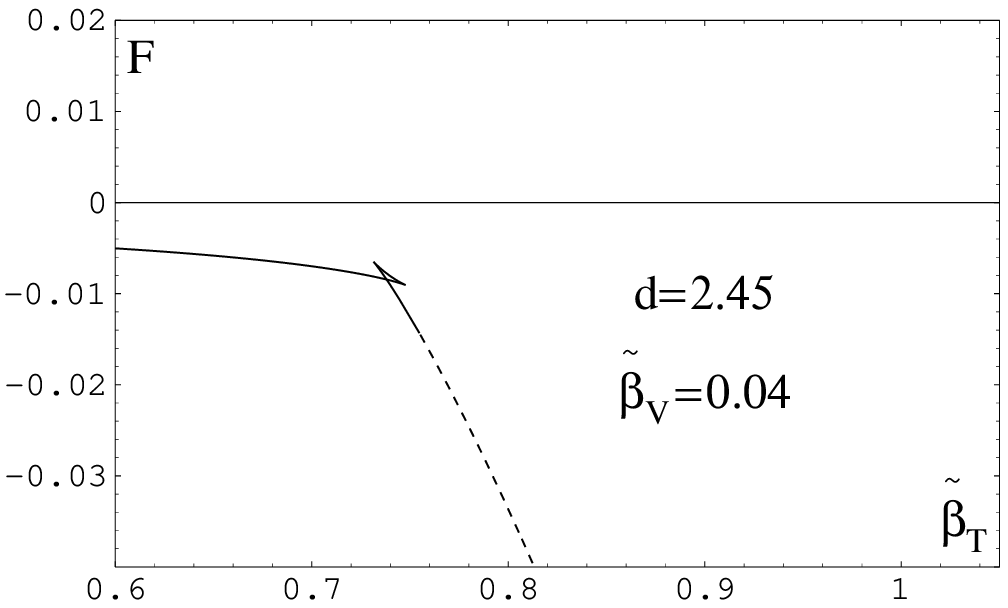}  \\
\vspace*{0.7cm}
\epsfxsize=0.3\textwidth
\leavevmode\epsffile{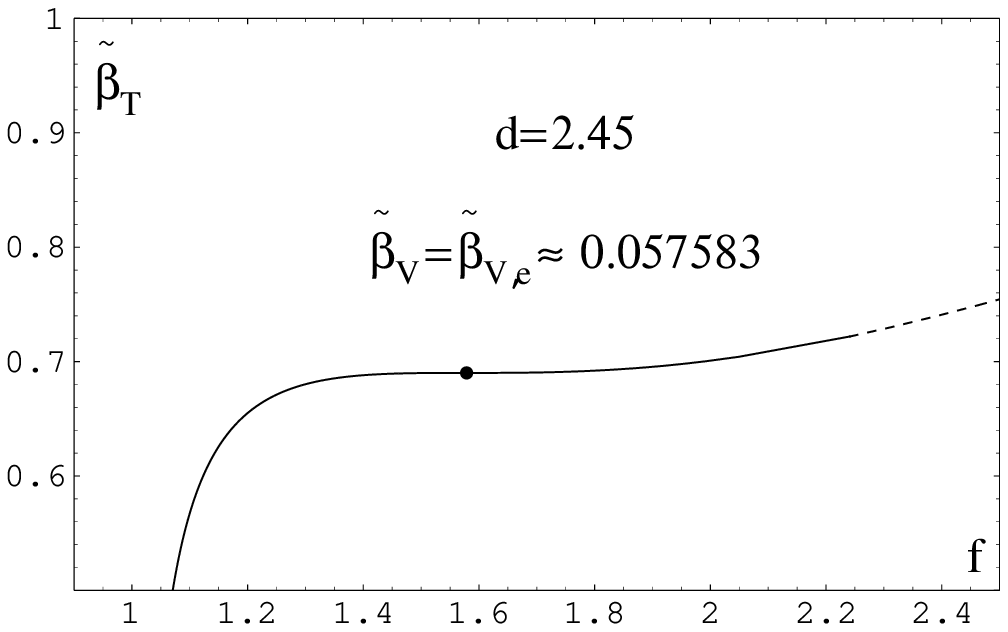}
\hspace{1.5cm}
\epsfxsize=0.3\textwidth
\epsffile{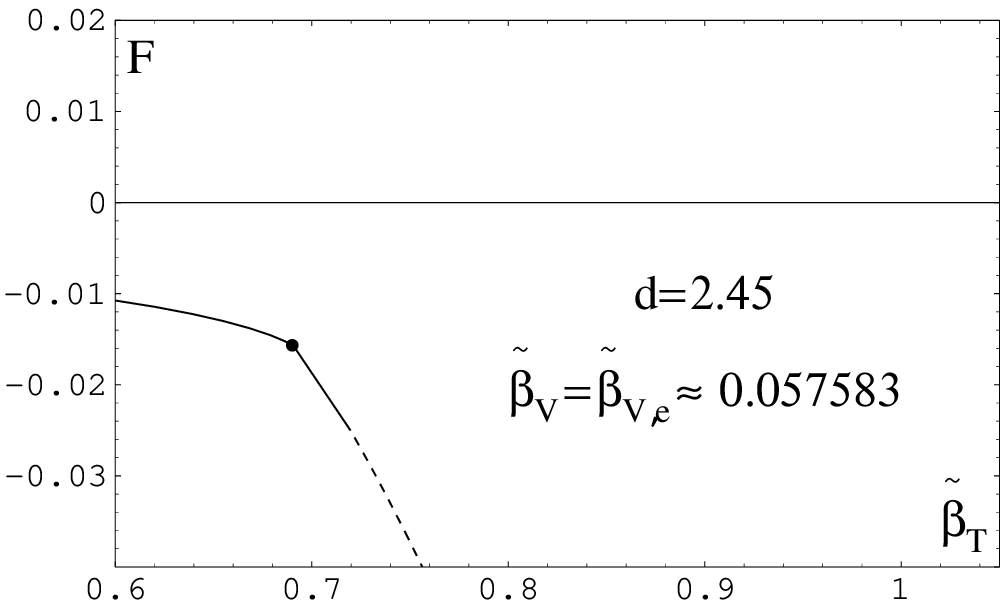}  \\
\vspace*{0.7cm}
\epsfxsize=0.3\textwidth
\leavevmode\epsffile{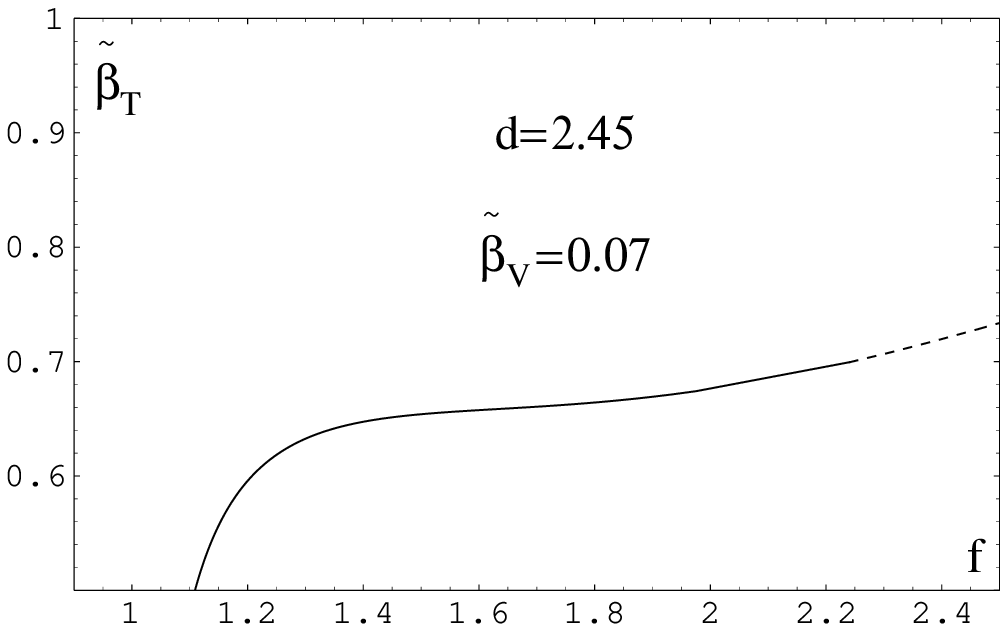}
\hspace{1.5cm}
\epsfxsize=0.3\textwidth
\epsffile{ 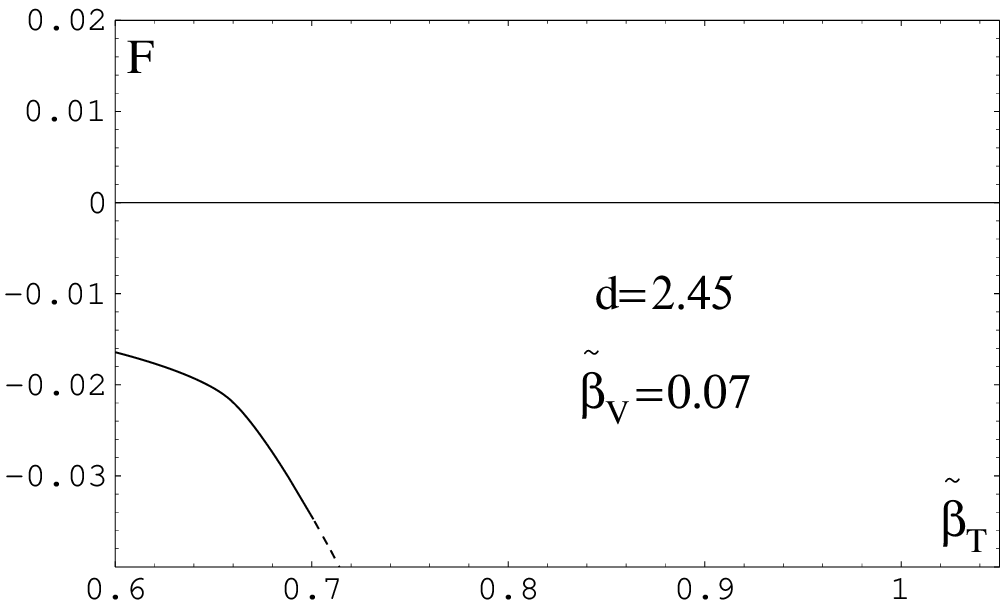}  \\
\end{center}
\vspace*{5mm}
\caption{$\btt$ versus $f$, and 
    $\overline{\scrf}$ versus $\btt$, for $d=2.45$ at
    (a) $\bvt =0$, (b) $\bvt = 0.01$, (c) $\bvt = \betatilde_{V,cc'}
        \approx 0.0237503$,
    (d) $\bvt =0.04$, (e) $\bvt = \betatilde_{V,e} \approx 0.057583$
        (note that $\betatilde_{T,e}\approx 0.69010393$
         and $\gamma_e\approx 0.399979$),
        (f) $\bvt =0.07$.
}
\label{fig_param_F_versus_btt_d=2.45}
\end{figure}

\begin{figure}[p]
\vspace*{-2cm} \hspace*{-0cm}
\begin{center}
\vspace*{0cm} \hspace*{-0cm}
\epsfxsize=0.3\textwidth
\leavevmode\epsffile{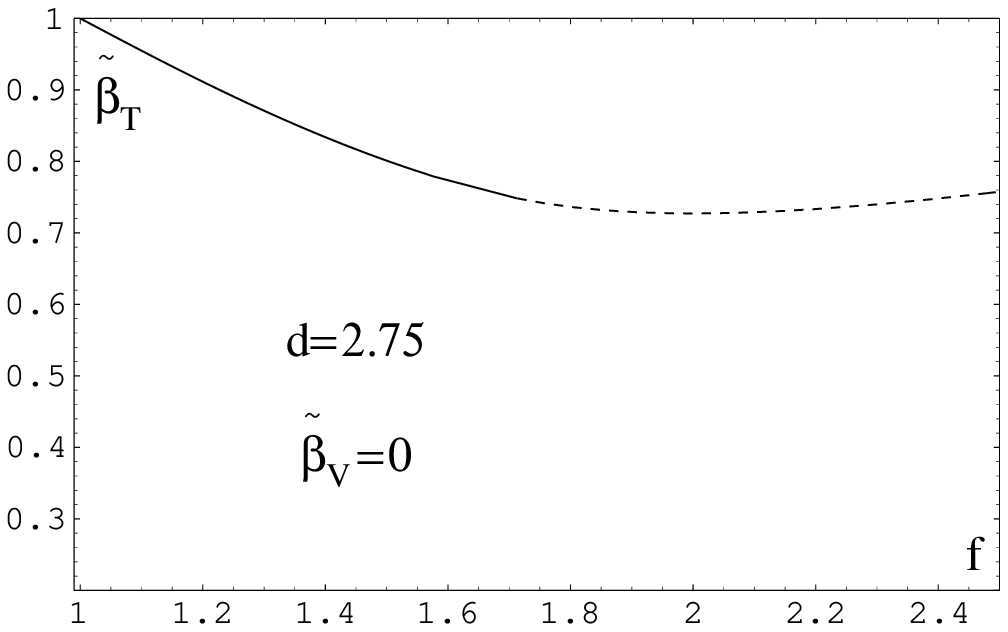}
\hspace{1.5cm}
\epsfxsize=0.3\textwidth
\epsffile{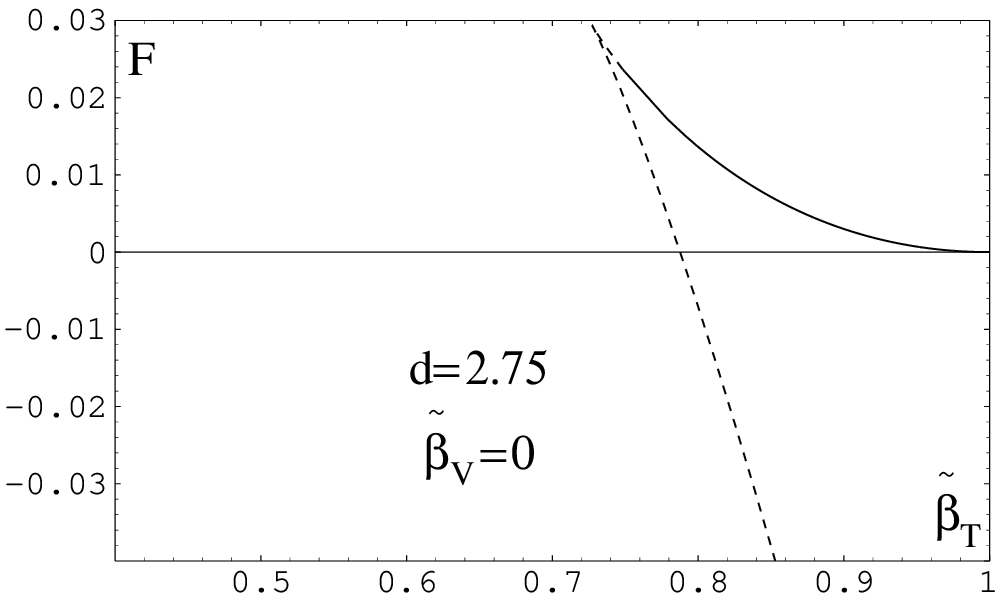}  \\
\vspace*{0.7cm}
\epsfxsize=0.3\textwidth
\leavevmode\epsffile{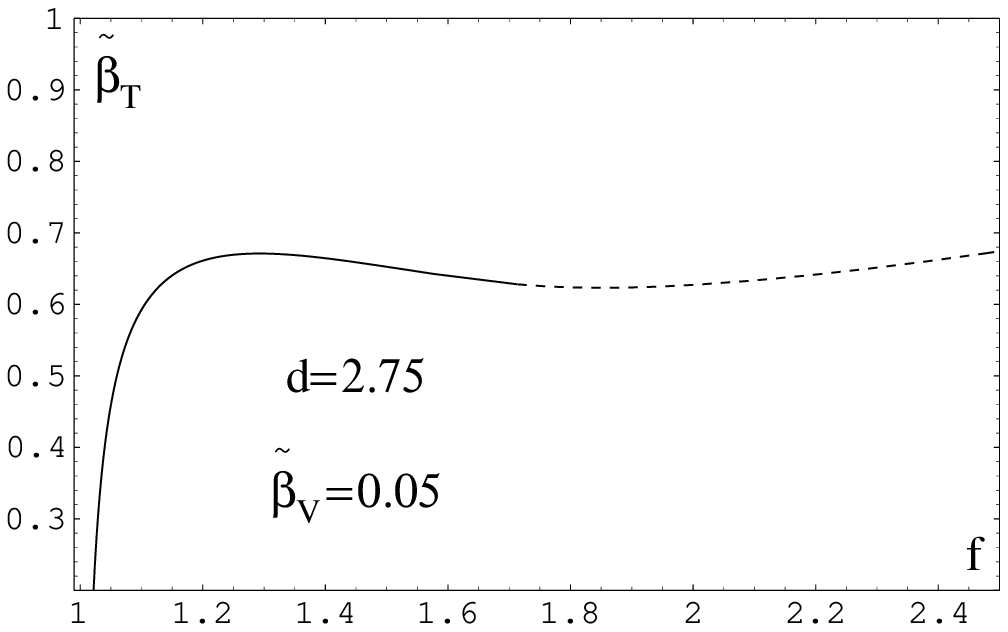}
\hspace{1.5cm}
\epsfxsize=0.3\textwidth
\epsffile{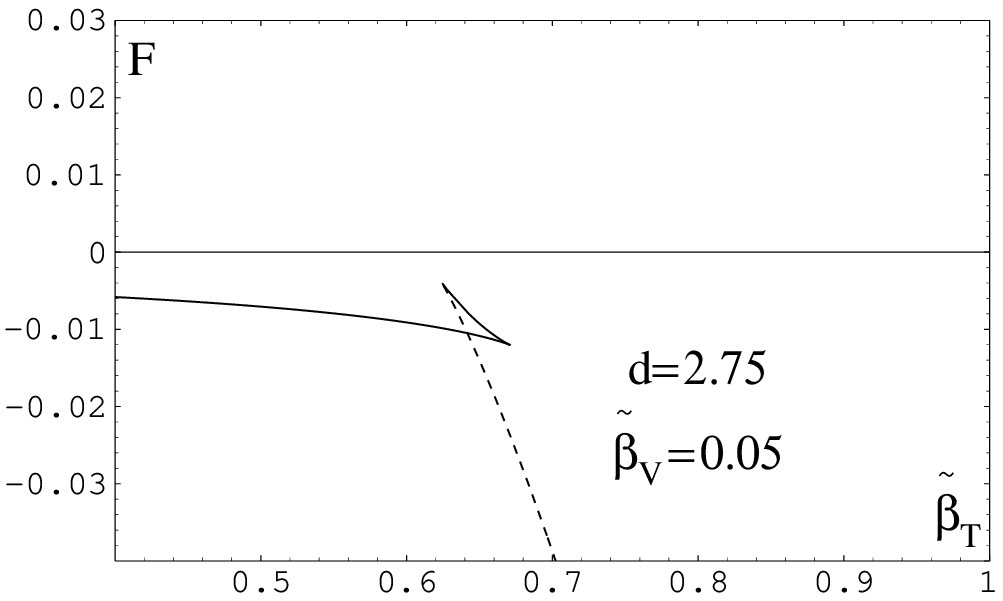}  \\
\vspace*{0.7cm}
\epsfxsize=0.3\textwidth
\leavevmode\epsffile{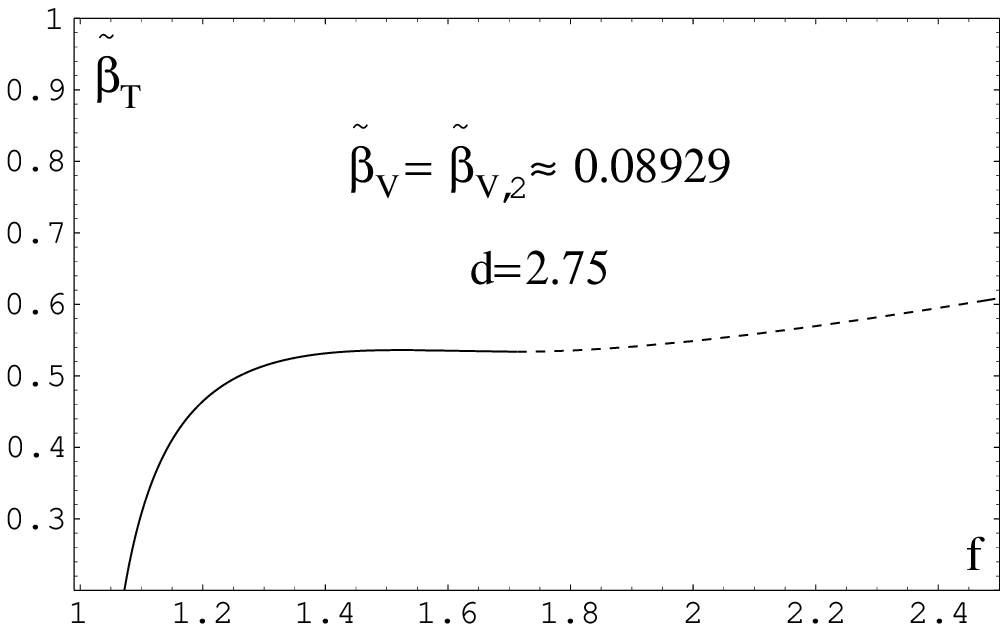}
\hspace{1.5cm}
\epsfxsize=0.3\textwidth
\epsffile{ 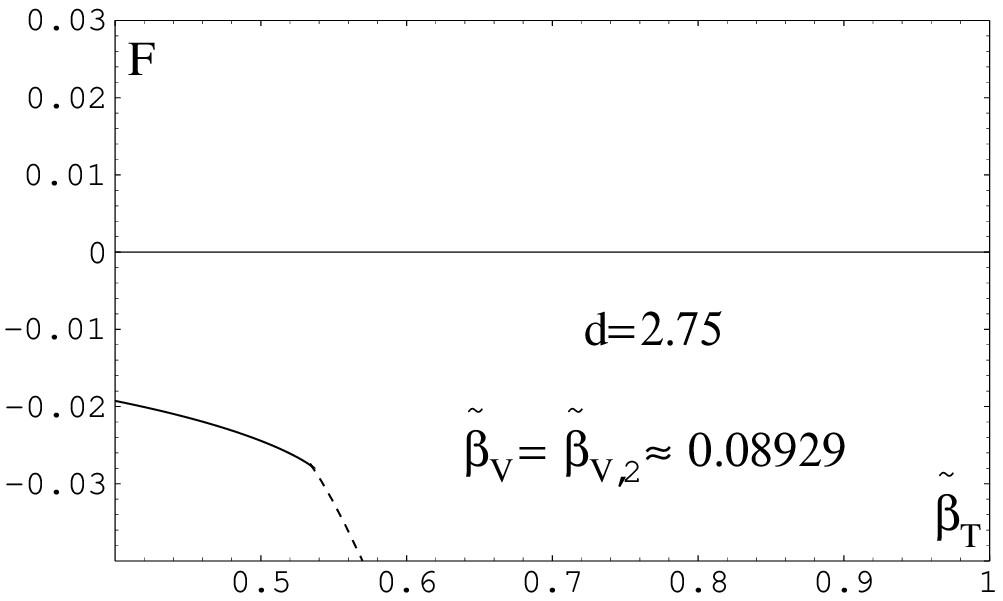}  \\
\vspace*{0.7cm}
\epsfxsize=0.3\textwidth
\leavevmode\epsffile{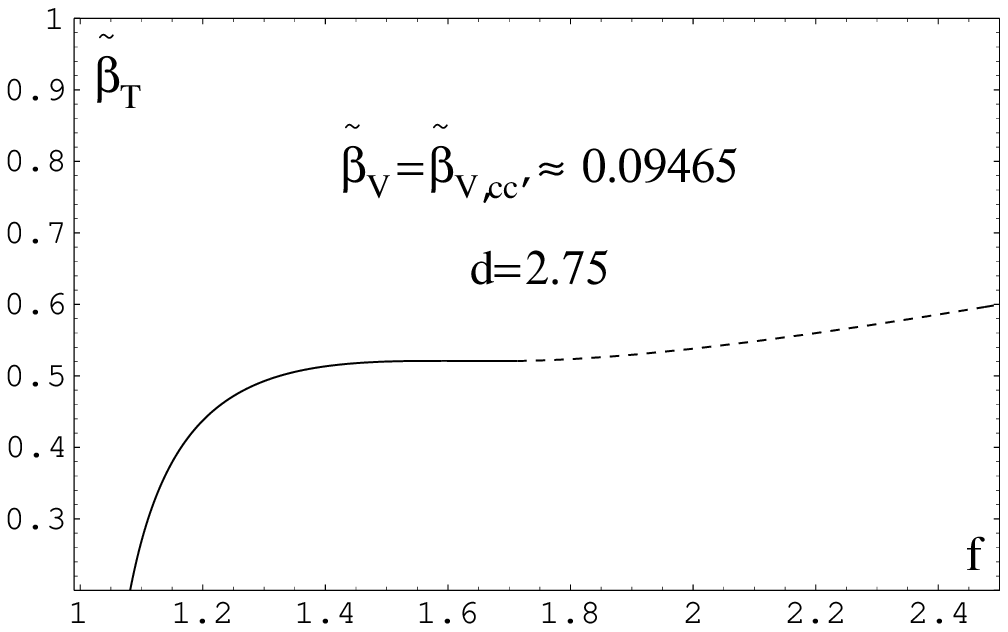}
\hspace{1.5cm}
\epsfxsize=0.3\textwidth
\epsffile{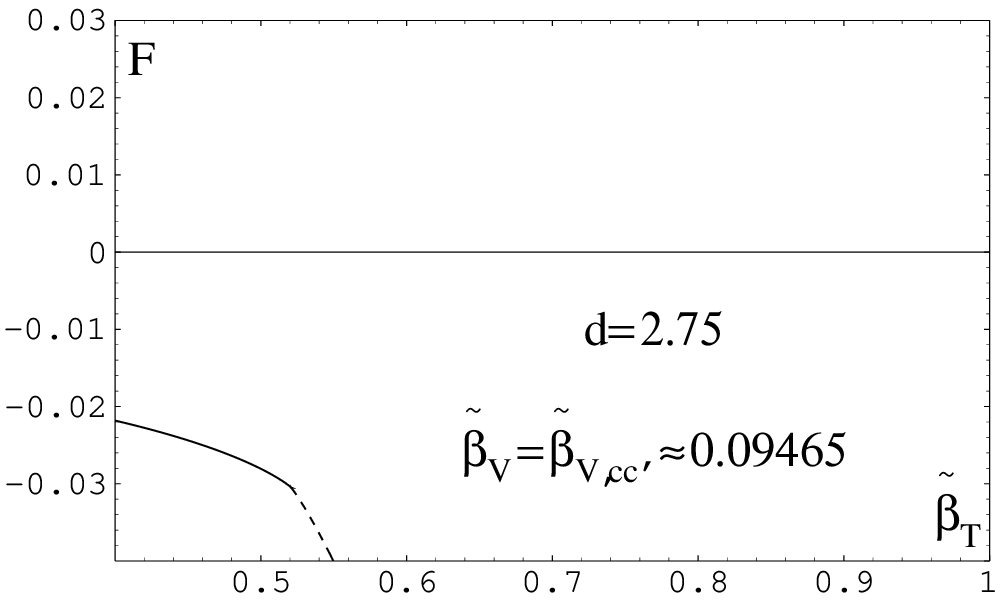}  \\
\vspace*{0.7cm}
\epsfxsize=0.3\textwidth
\leavevmode\epsffile{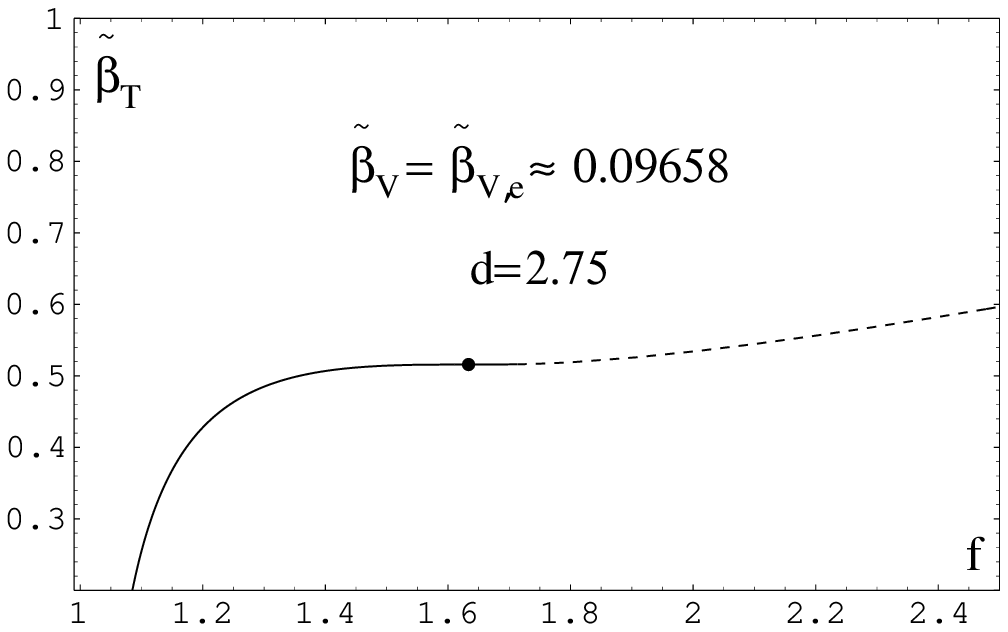}
\hspace{1.5cm}
\epsfxsize=0.3\textwidth
\epsffile{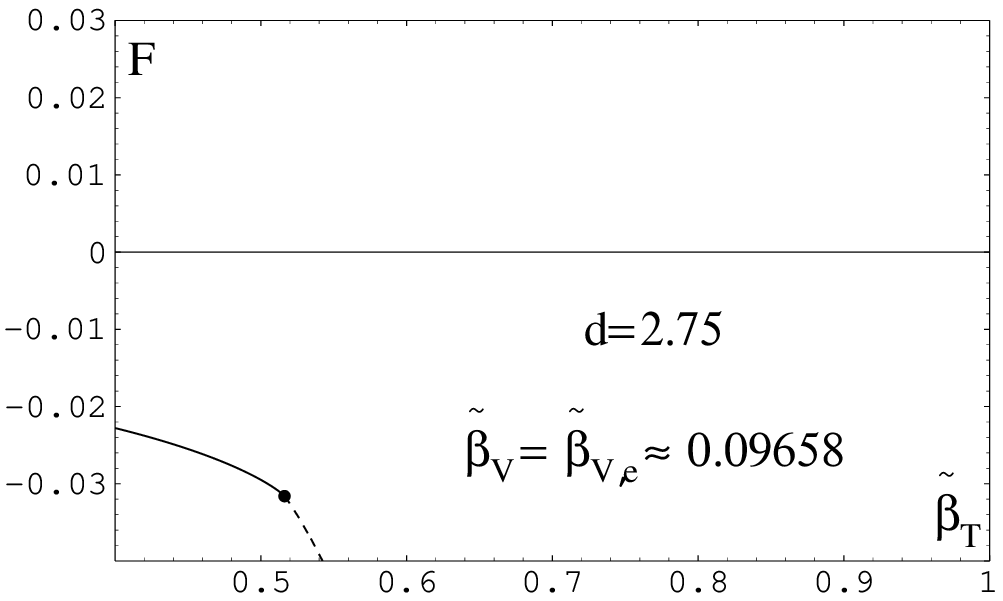}  \\
\vspace*{0.7cm}
\epsfxsize=0.3\textwidth
\leavevmode\epsffile{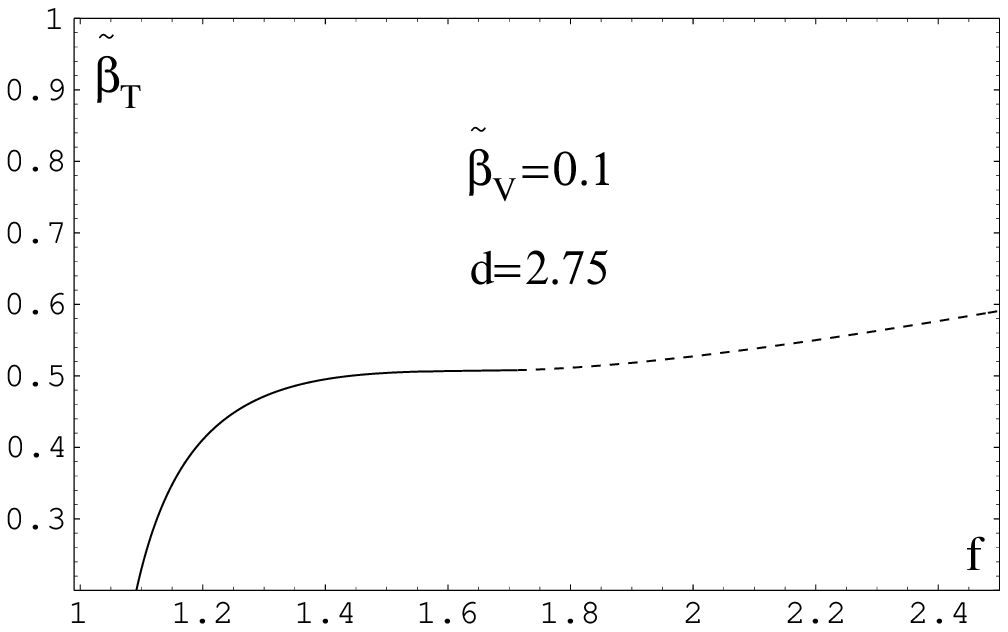}
\hspace{1.5cm}
\epsfxsize=0.3\textwidth
\epsffile{ 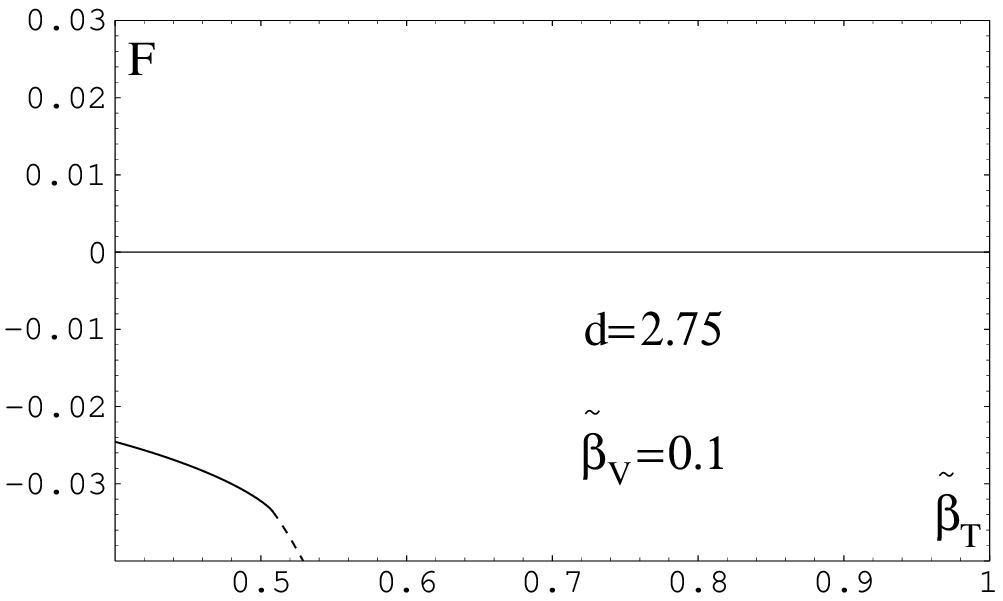}  \\
\end{center}
\caption{$\btt$ versus $f$, and 
    $\overline{\scrf}$ versus $\btt$, for $d=2.75$ at
    (a) $\bvt =0$, (b) $\bvt = 0.05$, (c) $\bvt = \betatilde_{V,2}
        \approx 0.0892914$
    (d) $\bvt = \betatilde_{V,cc'}\approx 0.0946534$,
        (e) $\bvt = \betatilde_{V,e} \approx 0.0965845$
            (note that $\betatilde_{T,e}\approx 0.5158930$
             and $\gamma_e\approx 0.363383$),
        (f) $\bvt =0.1$.
}
\label{fig_param_F_versus_btt_d=2.75}
\end{figure}

\begin{figure}[p]
\vspace*{-2cm} \hspace*{-0cm}
\begin{center}
\leavevmode\epsffile{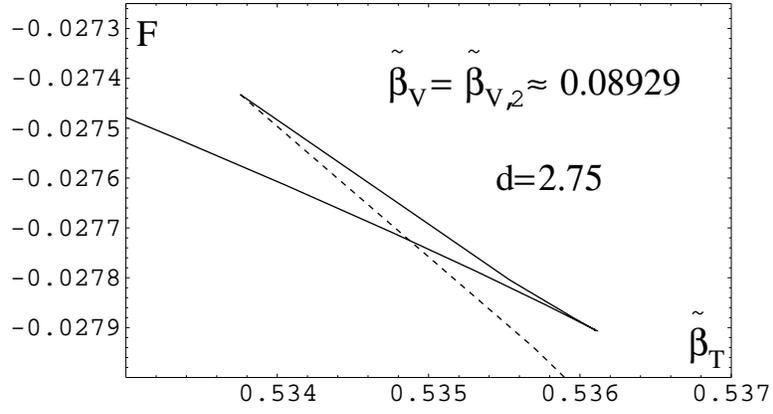} \\
\vspace{1cm}
\leavevmode\epsffile{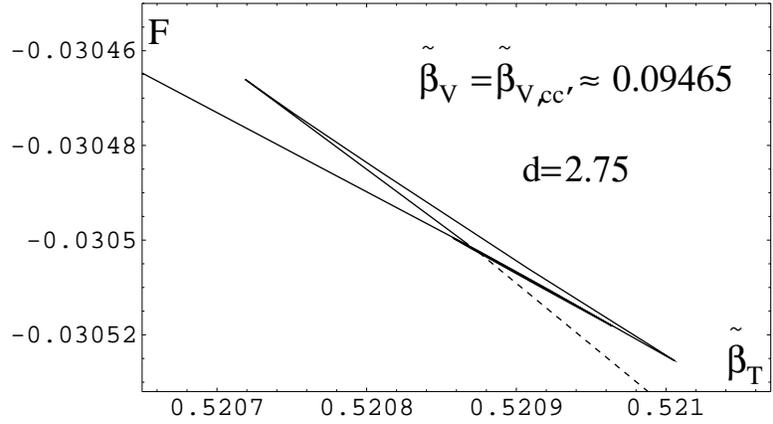} \\
\vspace{1cm}
\leavevmode\epsffile{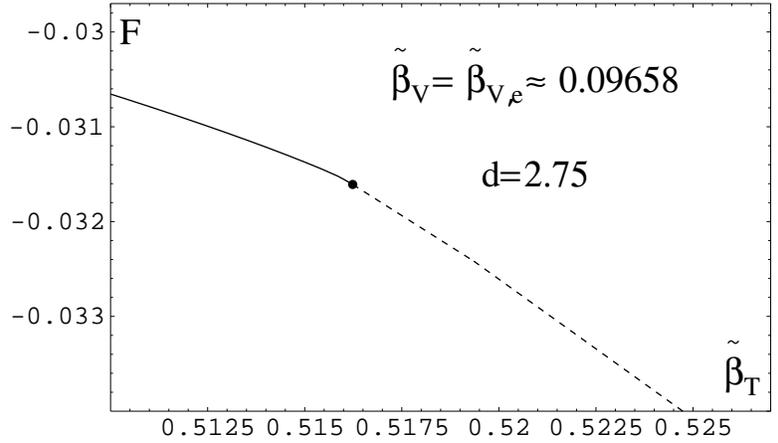}
\vspace{5mm}
\end{center}
\caption{
   Blow-ups of the parametric plots for $d=2.75$.
   (a) At $\bvt = \betatilde_{V,2} \approx 0.08929$,
       points c\textprime and 2 coincide.
   (b) At $\bvt = \betatilde_{V,cc'} \approx 0.09465$,
       points c\textprime and c coincide.
   (c) At $\bvt = \betatilde_{V,e} \approx 0.09658$,
       the ``triangle'' disappears;  this occurs within the normal phase.
}
\label{blow_2.75}
\end{figure}

\begin{figure}[p]
\vspace*{-2cm} \hspace*{-0cm}
\begin{center}
\vspace*{0cm} \hspace*{-0cm}
\epsfxsize=0.35\textwidth
\leavevmode\epsffile{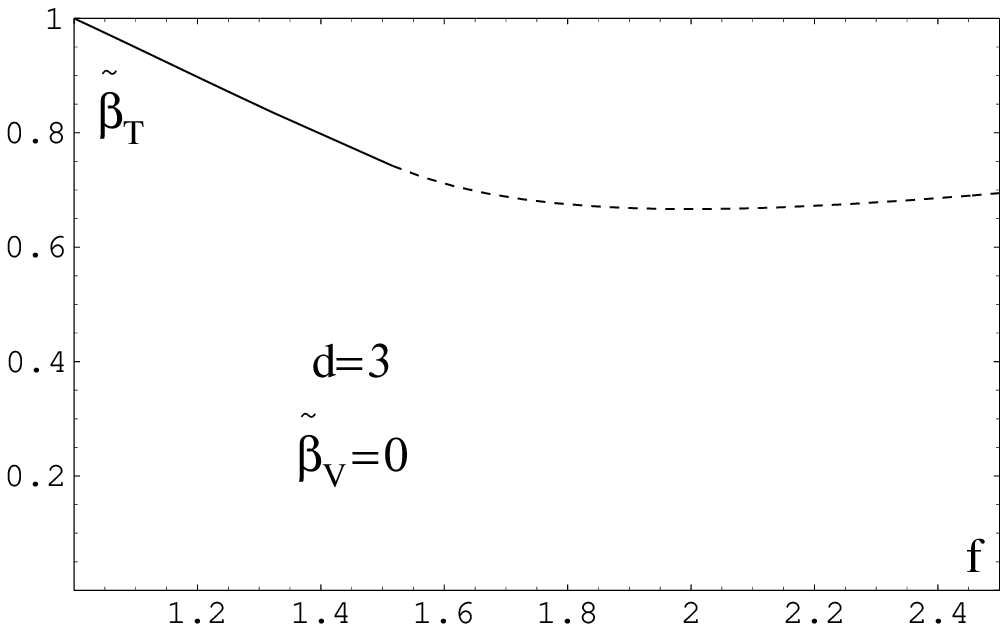}
\hspace{0.5cm}
\epsfxsize=0.35\textwidth
\epsffile{new_param_F_beta_d=3_v=0.eps    }  \\
\vspace*{0.5cm}
\epsfxsize=0.35\textwidth
\leavevmode\epsffile{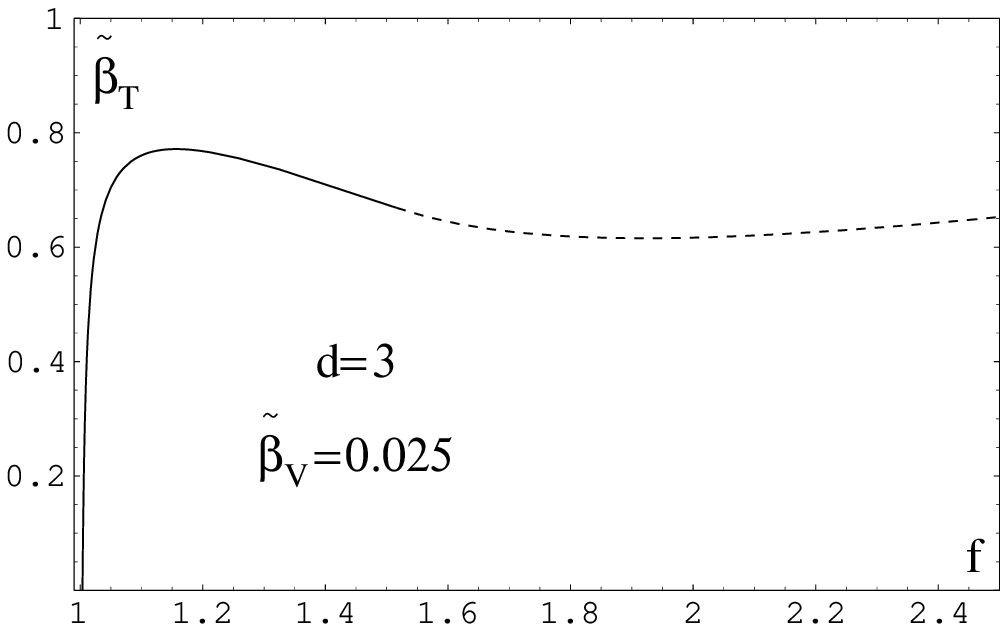}
\hspace{0.5cm}
\epsfxsize=0.35\textwidth
\epsffile{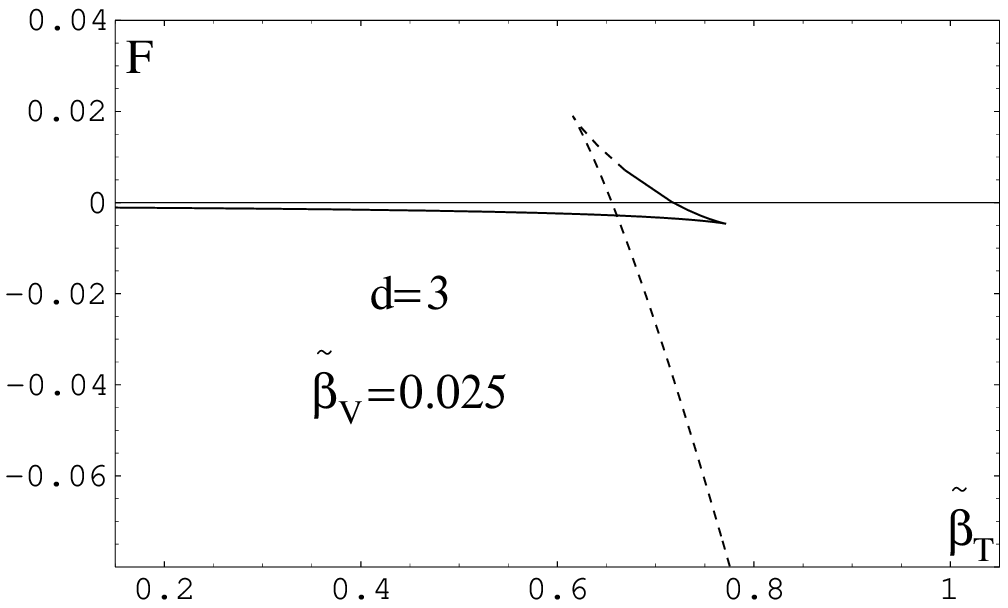    }  \\
\vspace*{0.5cm}
\epsfxsize=0.35\textwidth
\leavevmode\epsffile{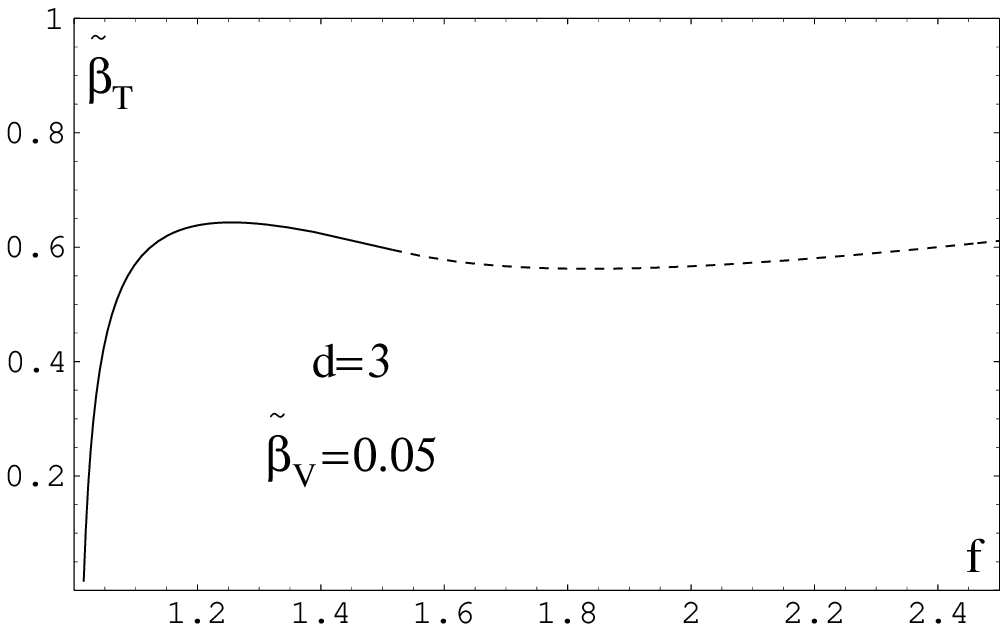}
\hspace{0.5cm}
\epsfxsize=0.35\textwidth
\epsffile{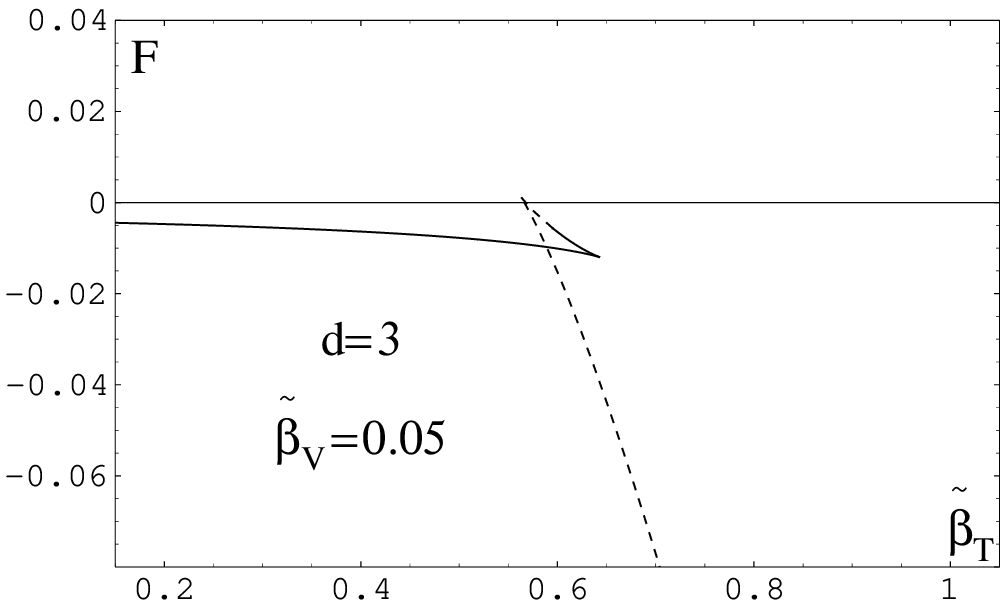    }  \\
\vspace*{0.5cm}
\epsfxsize=0.35\textwidth
\leavevmode\epsffile{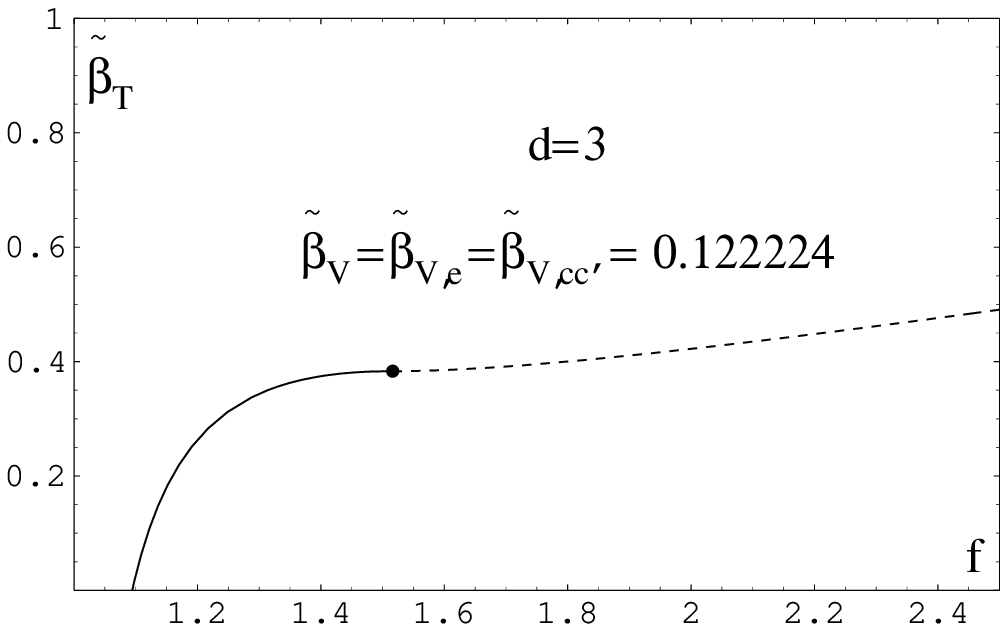}
\hspace{0.5cm}
\epsfxsize=0.35\textwidth
\epsffile{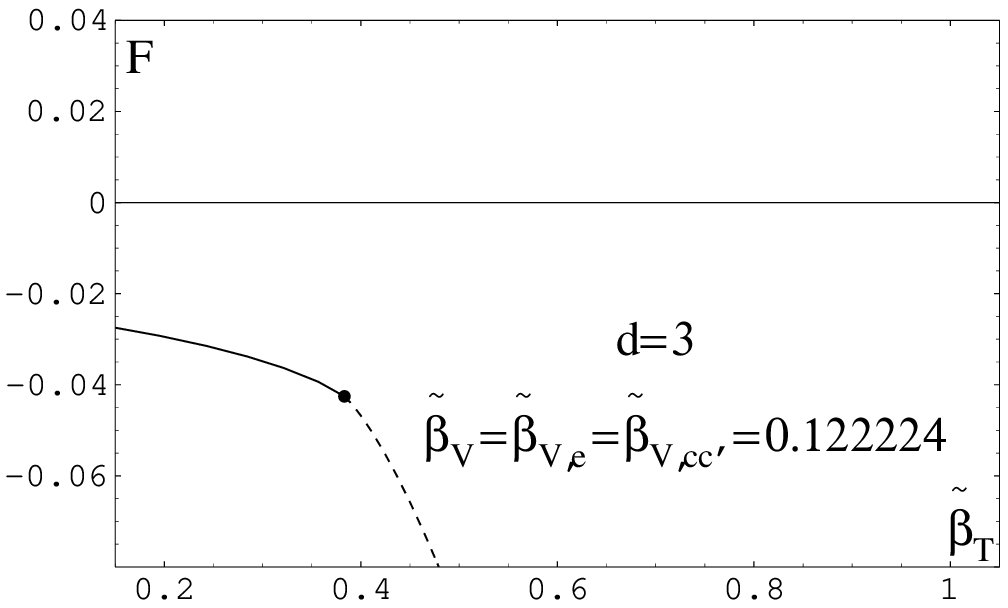    }  \\
\vspace*{0.5cm}
\epsfxsize=0.35\textwidth
\leavevmode\epsffile{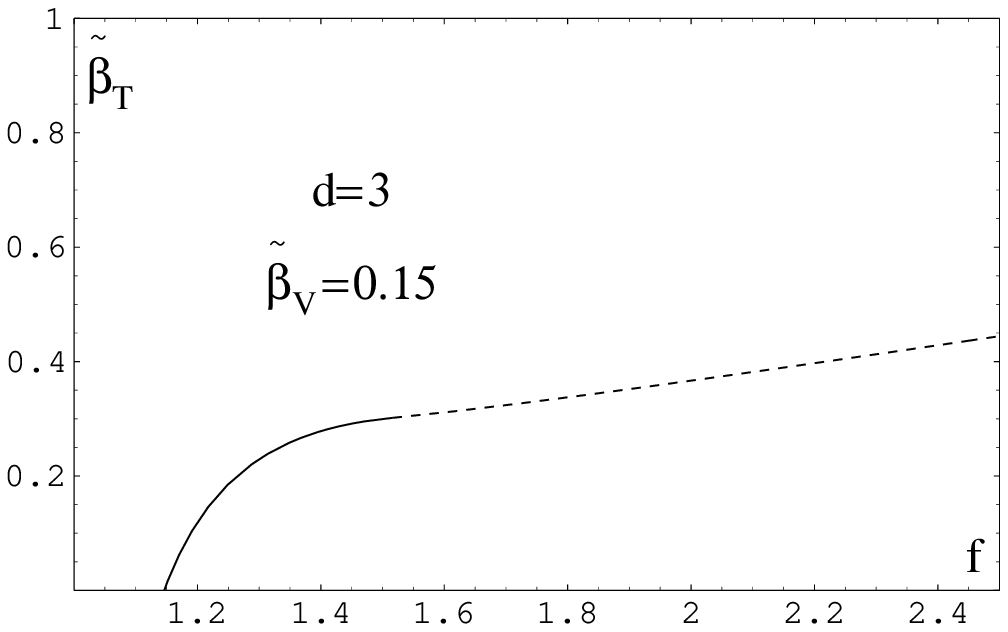}
\hspace{0.5cm}
\epsfxsize=0.35\textwidth
\epsffile{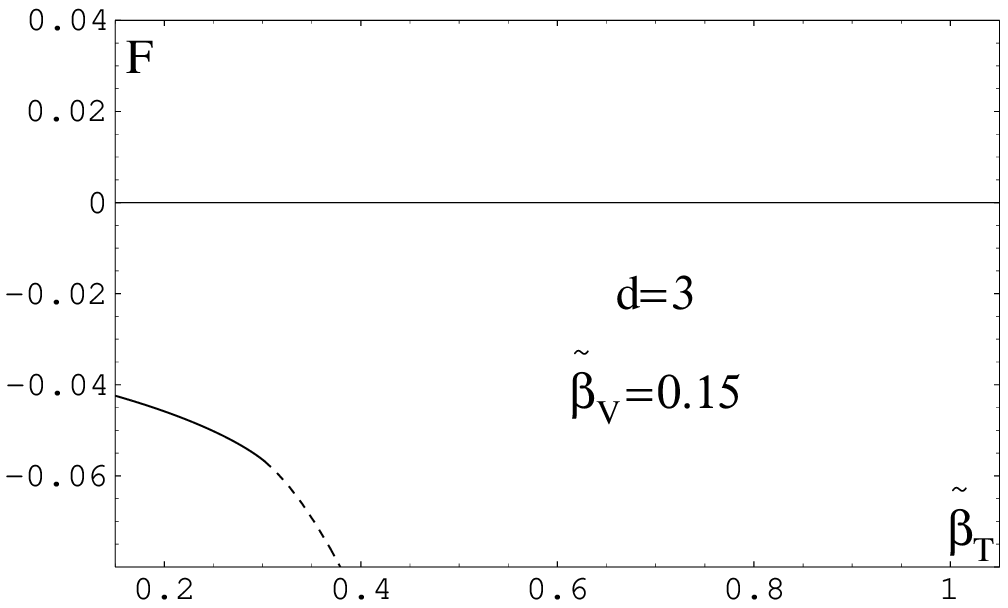    }  \\
\end{center}
\vspace*{5mm}
\caption{ $\btt$ versus $f$, and $\overline{\scrf}$ versus $\btt$,
    for $d=3$ at
    (a) $\bvt =0$, (b) $\bvt = 0.025$, (c) $\bvt = 0.05$,
    (d) $\bvt = \betatilde_{V,2} = \betatilde_{V,cc'} = \betatilde_{V,e}
            \approx 0.122224$,
    (e) $\bvt = 0.15$.
}
\label{fig_param_F_versus_btt_d=3}
\end{figure}

\begin{figure}[p]
\vspace*{-2cm} \hspace*{-0cm}
\begin{center}
\vspace*{0cm} \hspace*{-0cm}
\epsfxsize=0.3\textwidth
\leavevmode\epsffile{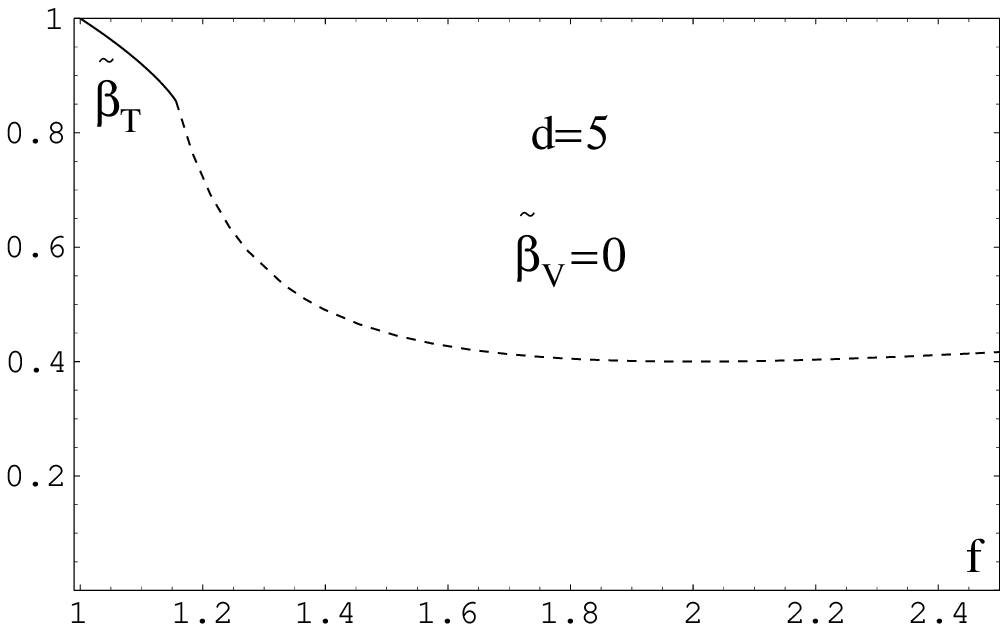}
\hspace{1.5cm}
\epsfxsize=0.3\textwidth
\epsffile{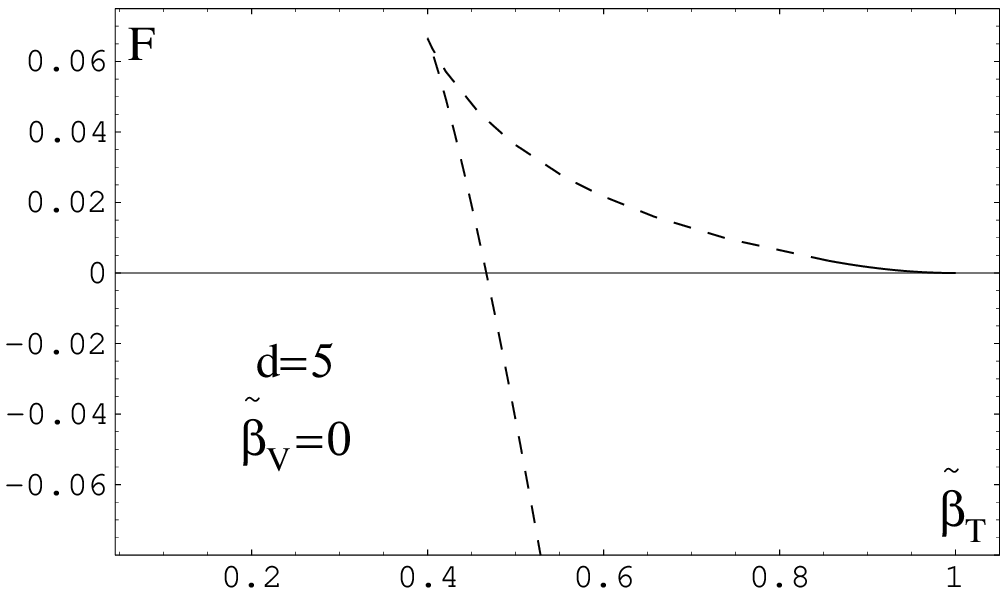}  \\
\vspace*{0.7cm}
\epsfxsize=0.3\textwidth
\leavevmode\epsffile{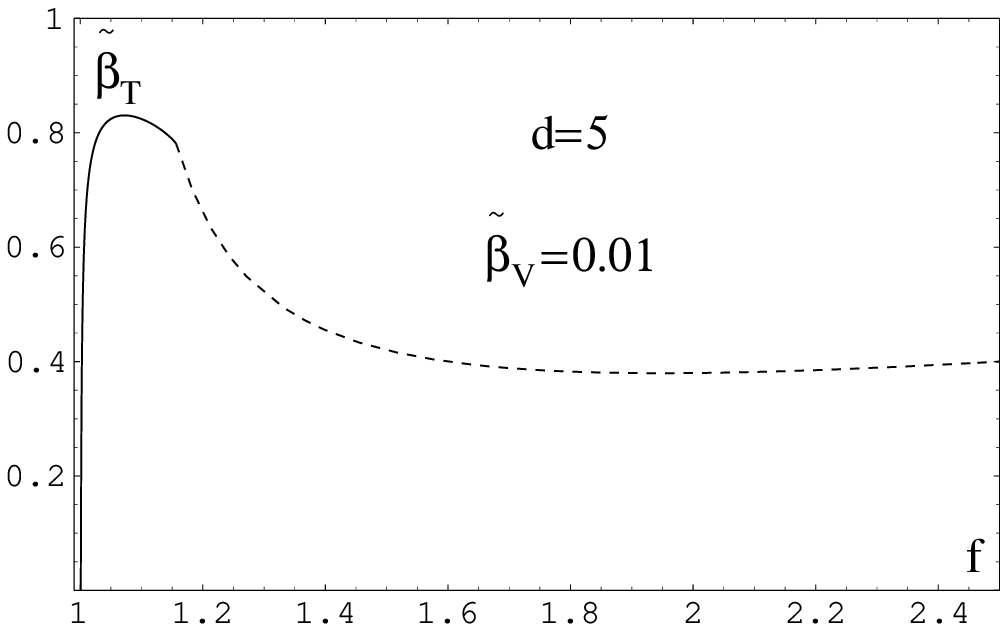}
\hspace{1.5cm}
\epsfxsize=0.3\textwidth
\epsffile{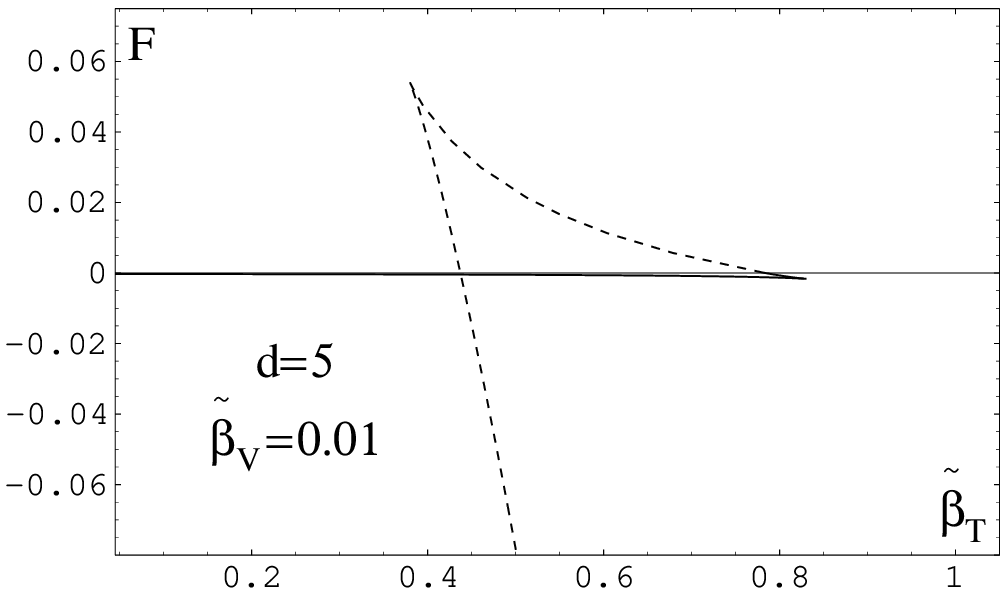}  \\
\vspace*{0.7cm}
\epsfxsize=0.3\textwidth
\leavevmode\epsffile{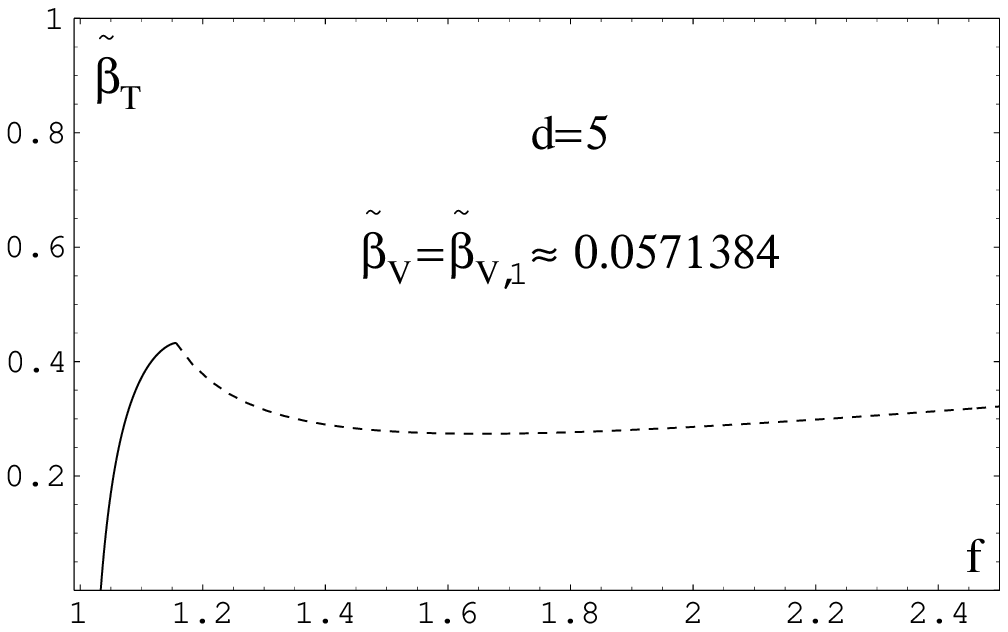}
\hspace{1.5cm}
\epsfxsize=0.3\textwidth
\epsffile{ 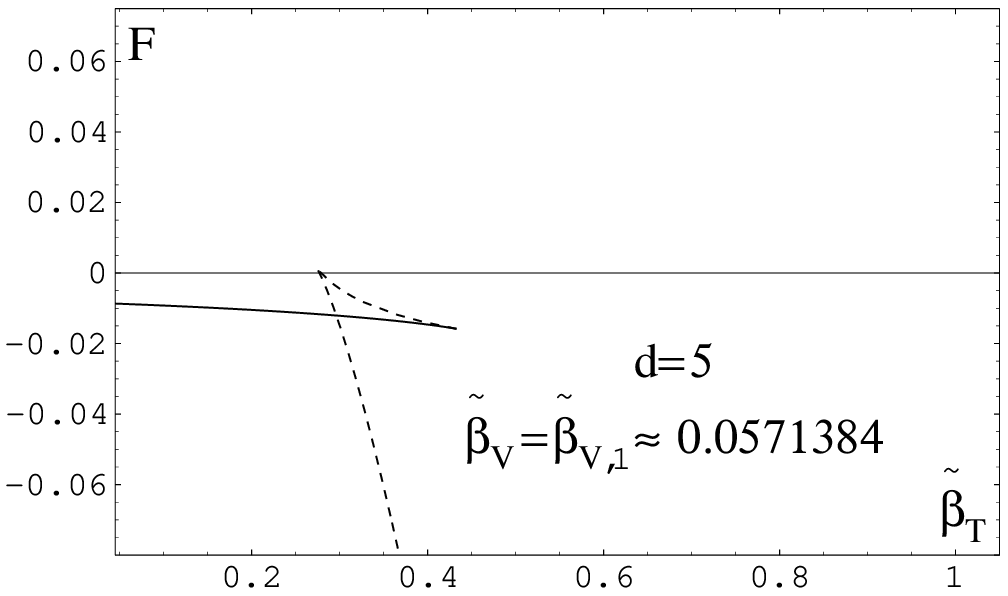}  \\
\vspace*{0.7cm}
\epsfxsize=0.3\textwidth
\leavevmode\epsffile{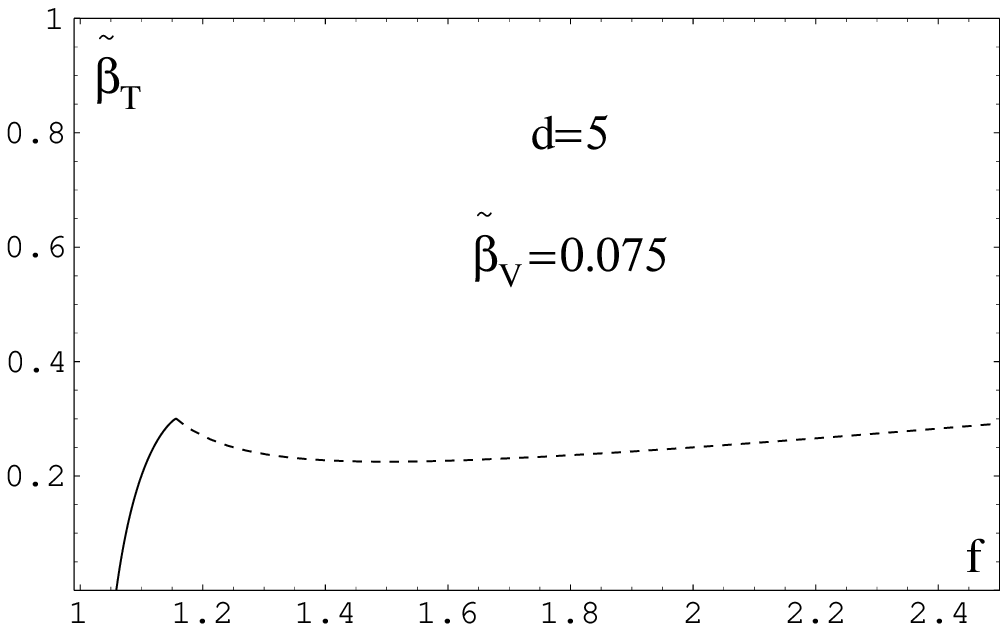}
\hspace{1.5cm}
\epsfxsize=0.3\textwidth
\epsffile{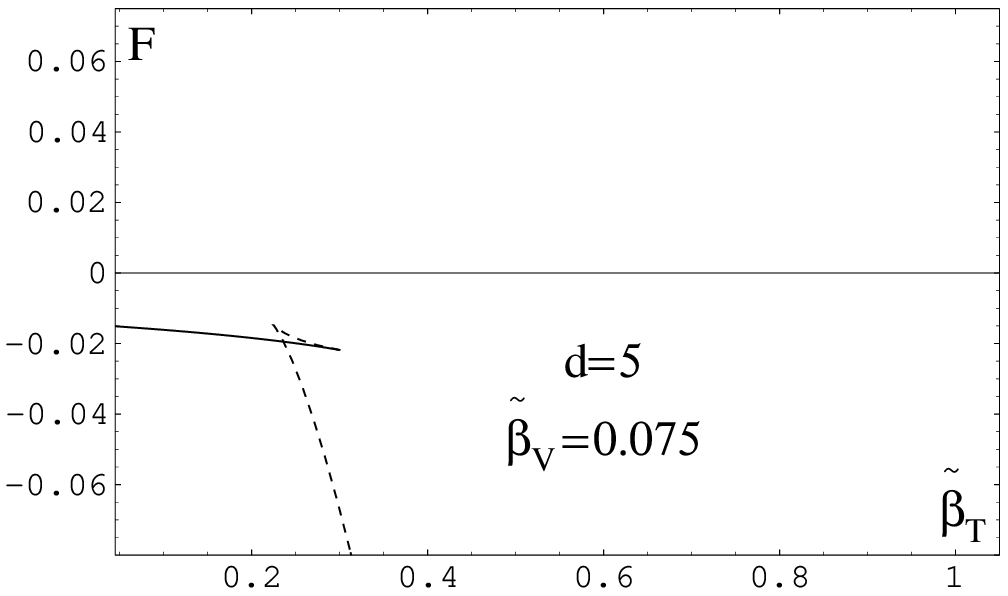}  \\
\vspace*{0.7cm}
\epsfxsize=0.3\textwidth
\leavevmode\epsffile{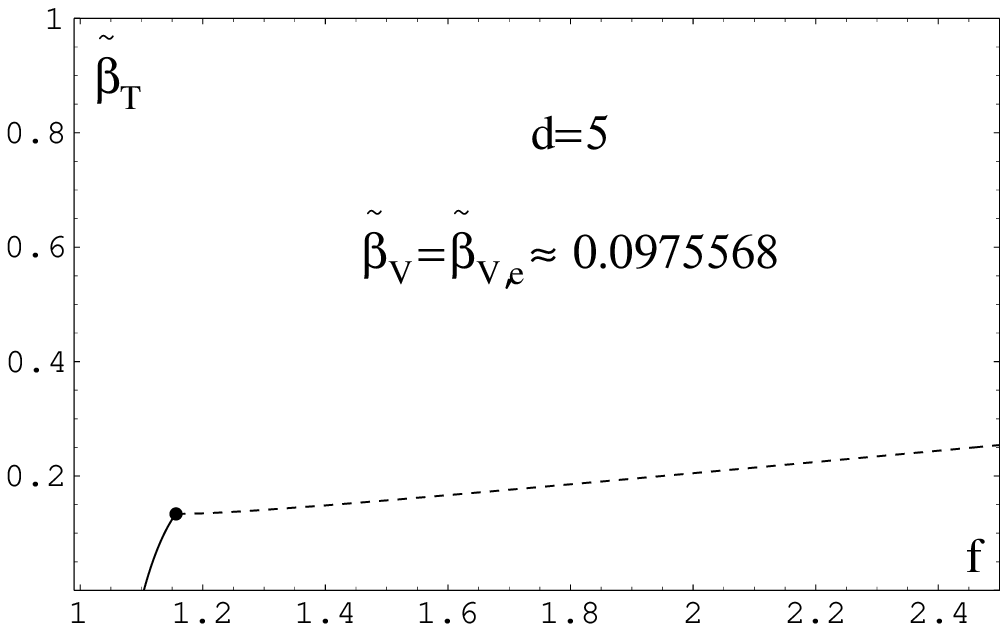}
\hspace{1.5cm}
\epsfxsize=0.3\textwidth
\epsffile{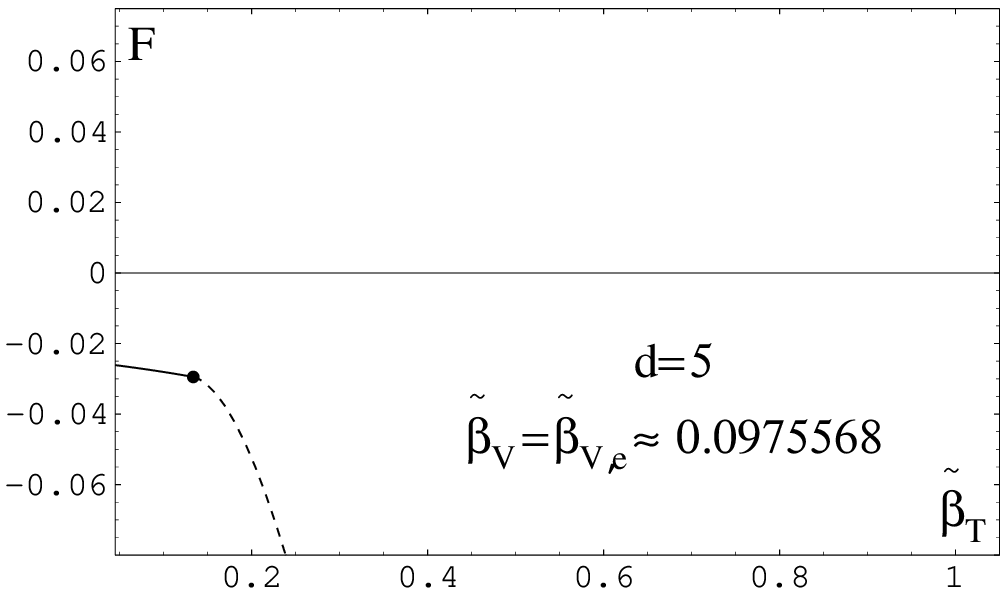}  \\
\vspace*{0.7cm}
\epsfxsize=0.3\textwidth
\leavevmode\epsffile{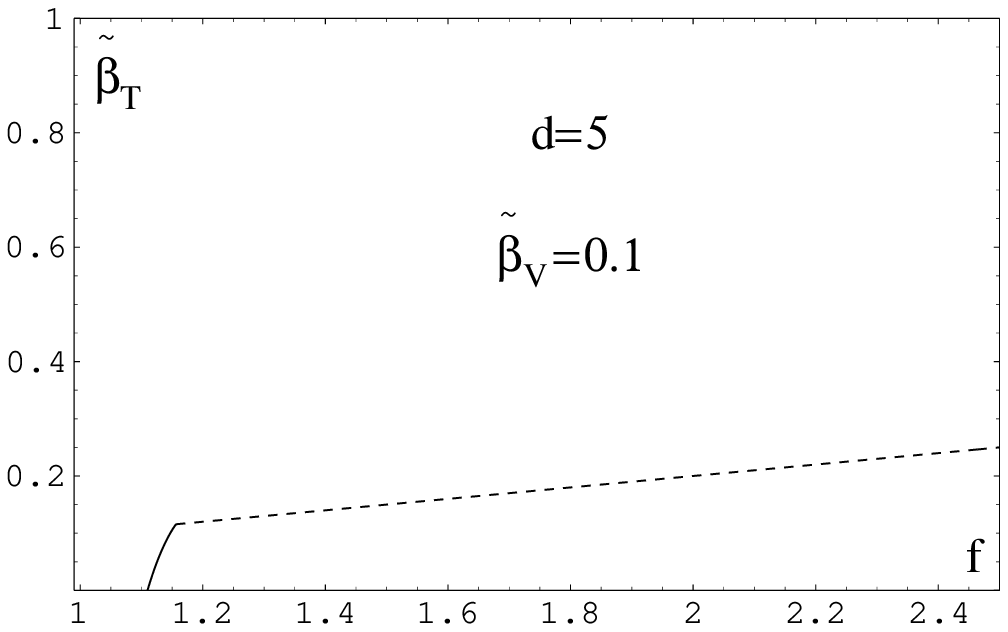}
\hspace{1.5cm}
\epsfxsize=0.3\textwidth
\epsffile{ 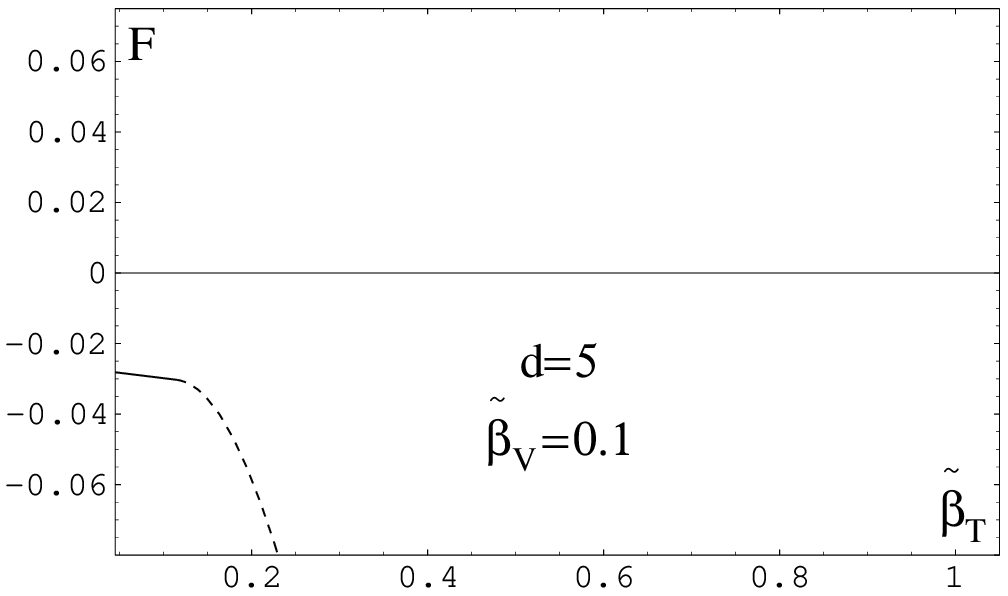}  \\
\end{center}
\vspace*{5mm}
\caption{$\btt$ versus $f$, and 
    $\overline{\scrf}$ versus $\btt$, for $d=5$ at
    (a) $\bvt =0$, (b) $\bvt = 0.01$, (c) $\bvt = \betatilde_{V,1}
        \approx 0.0571384$,
       (d) $\bvt = 0.075 $,
       (e) $\bvt = \betatilde_{V,2} = \betatilde_{V,cc'}
        =\betatilde_{V,e}\approx 0.0975568$, (f) $\bvt =0.1$.
}
\label{fig_param_F_versus_btt_d=5}
\end{figure}

\begin{figure}[p]
\begin{center}
\epsfxsize=0.55\textwidth
\leavevmode\epsffile{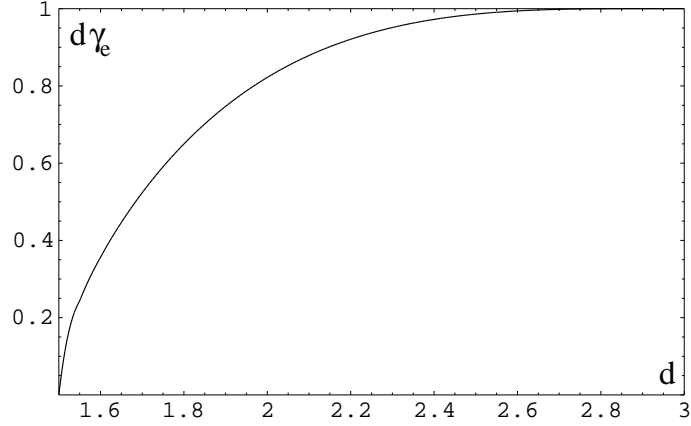}
\end{center}
\caption{$d\gamma_e$ plotted versus $d$ for $3/2 \le d \le d_{**}=3$.}
\label{fig_d_gamma_e_versus_d}
\end{figure}

For $d > 2$, we need to consider also the branch coming from
the condensate phase $f \ge f_*$ with $\gamma=1/d$
(shown by a dashed line in
Figures~\ref{new_triangle_guide}--\ref{fig_param_F_versus_btt_d=5}).
The beginning of this branch at point c\textprime
(Figure~\ref{new_triangle_guide})
corresponds to $f=f_*$.
A key role is played by the position of c\textprime relative to
the points c, 2 and 1
(and therefore to the internal structure of the ``triangle'');
this depends on $d$ and on $\bvt$.
We have to distinguish several cases:

\bigskip

{\bf Case (c), $\boldsymbol{2 < d < d_* \approx 2.38403}$:}
The evolution of $\btt(f)$ follows the generic pattern
described earlier for $d > 3/2$.
For small values of $\bvt$
(Figure~\ref{fig_param_F_versus_btt_d=2.25}a,b,c,d)
there are two phase transitions:
a first-order transition at point c of the ``triangle'',
and a normal-to-condensate second-order transition at point c\textprime,
which lies {\em outside}\/ the ``triangle''.
These correspond on the phase diagram (Figure~\ref{phase_diag_2<d<dstar})
to the first-order curve $\betatilde_{T,c}(\bvt)$
and the second-order lines $\betatilde_{T,c'}(\bvt)$, respectively.
As we increase $\bvt$, the ``triangle'' shrinks and finally
disappears at the point $(\betatilde_{V,e},\betatilde_{T,e})$,
shown as a dot in Figure~\ref{fig_param_F_versus_btt_d=2.25}e.
This point corresponds to the endpoint of the solid curve in
Figure~\ref{phase_diag_2<d<dstar},
and is a second-order phase transition.
Its evolution as a function of $d$ is
is shown in Figures~\ref{beta_T_endpoint_d} and \ref{beta_V_endpoint_d}.
For $\bvt > \betatilde_{V,e}$,
only the transition at point c\textprime remains
(dashed line in Figure~\ref{phase_diag_2<d<dstar}).

\bigskip

{\bf Case (d), $\boldsymbol{d = d_* \approx 2.38403}$:}
As $d$ increases, the first-order transition curve $\betatilde_{T,c}(\bvt)$
and the second-order transition lines $\betatilde_{T,c'}(\bvt)$
come closer together;
at $d=d_*$ they collide at $\bvt=0$,
$\betatilde_{T,c} = \betatilde_{T,c'} = \betatilde_{T,*} \approx 0.873264$
(Figures~\ref{phasediag_bv=0} and \ref{phase_diag_d=dstar}).
In terms of the parametric plot, points c and c\textprime collide when $\bvt=0$
(Figure~\ref{fig_F_versus_btt}e).
For all $\bvt > 0$, point c\textprime continues to lie
{\em outside}\/ the ``triangle'',
so that the rest of the scenario has the same features as the 
case $2 < d < d_*$ considered above.

\bigskip

\begin{figure}[p]
\vspace*{-2cm} \hspace*{-0cm}
\begin{center}
\epsfxsize=0.65\textwidth
\leavevmode\epsffile{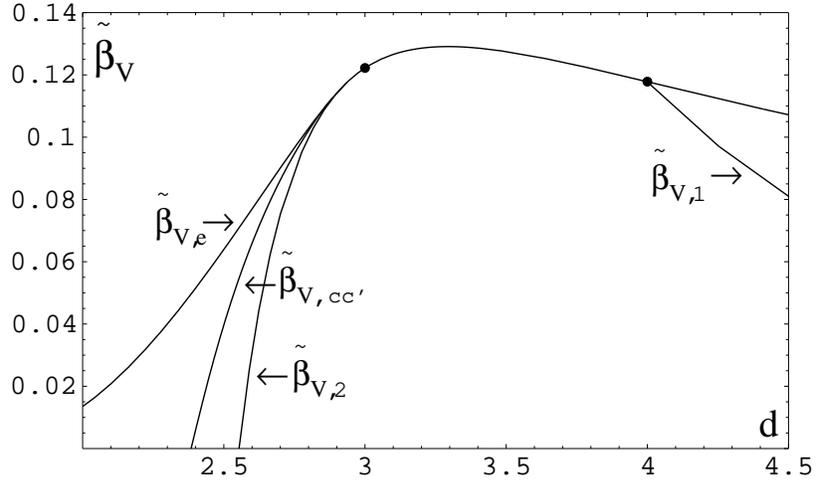}
\end{center}
\caption{
   $\betatilde_{V,e}$, $\betatilde_{V,cc'}$, $\betatilde_{V,2}$ and
   $\betatilde_{V,1}$ as functions of dimension $d$.
   The curve $\betatilde_{V,e}$ smoothly reaches $0$ at $d=3/2$;
   the curve $\betatilde_{V,cc'}$ hits 0 at $d=d_*\approx 2.38403$;
   the curve $\betatilde_{V,2} = f_* (2-f_*)/(2d)$
       hits 0 at $d=\bar{d} \approx 2.55391$;
   the curves $\betatilde_{V,e}$, $\betatilde_{V,cc'}$ and $\betatilde_{V,2}$
      all merge at $d=d_{**}=3$;
   the curve $\betatilde_{V,1}$ branches off the
      $\betatilde_{V,e} = \betatilde_{V,cc'} = \betatilde_{V,2}$ curve
      at $d=4$, and then approaches zero as $d \to\infty$.
} 
\label{phase_diag_beta_v_vs_dimension}
\end{figure}

{\bf Case (e${}_{\boldsymbol 1}$),
     $\boldsymbol{d_* < d < \bar{d} \approx 2.55391}$:}
At small $\bvt$, the point c\textprime lies {\em inside}\/ the ``triangle'',
between points 2 and c;
the first-order phase transition at point c
thus corresponds to a passage directly into the condensate phase
(Figure~\ref{fig_param_F_versus_btt_d=2.45}a,b).
As $\bvt$ increases, the condensate branch gradually gets ``expelled''
from the ``triangle'';
we define $\betatilde_{V,cc'}$ to be the value of $\bvt$
at which points c and c\textprime coincide
(Figure~\ref{fig_param_F_versus_btt_d=2.45}c).
In Figure~\ref{phase_diag_beta_v_vs_dimension}
we plot $\betatilde_{V,cc'}$ as a function of $d$.
For $\bvt >  \betatilde_{V,cc'}$,
the point c\textprime lies outside the ``triangle''
(Figure~\ref{fig_param_F_versus_btt_d=2.45}d),
so that the first-order transition at point c 
is a transition within the normal phase
(just as in the case $2 < d < d_*$);
there is a subsequent second-order transition at point c\textprime
to the condensate phase.
Finally, at $\bvt =\betatilde_{V,e}$,
the ``triangle'' disappears (Figure~\ref{fig_param_F_versus_btt_d=2.45}e);
this occurs within the normal phase.
For $\bvt >\betatilde_{V,e}$ the only phase transition is the
second-order transition to the condensate phase at point c\textprime.  
The complete phase diagram is shown in
Figure~\ref{phase_diag_dstar<d<dstarstar}.

\bigskip

{\bf Case (e${}_{\boldsymbol 2}$),
     $\boldsymbol{d = \bar{d} \approx 2.55391}$:}
The dimension $\bar{d}$ is, by definition, the one in which
point c\textprime coincides with point 2 of the ``triangle'' when $\bvt =0$.
For any $\bvt >0$, the description is identical to the one given above for
the case $d_* < d < \bar{d}$.

\bigskip

{\bf Case (e${}_{\boldsymbol 3}$),
     $\boldsymbol{\bar{d} < d < d_{**}=3}$:}
At $\bvt =0$, the start of the condensate branch (point c${}'$)
lies between points between $2$ and $1$ of the ``triangle''
(Figure~\ref{fig_param_F_versus_btt_d=2.75}a,b).
As $\bvt$ increases, point c\textprime is gradually expelled
from the ``triangle'':
first it reaches point 2 when $\bvt$ equals
\be
   \betatilde_{V,2} \;=\; {f_* (2-f_*) \over 2d}
   \;,
\ee
where \reff{F_prime_beta_f2} vanishes at $f=f_*$;
then it reaches point c at $\bvt = \betatilde_{V,cc'}$.
Finally, the ``triangle'' disappears at $\bvt = \betatilde_{V,e}$;
it does so in the normal phase
(since point c\textprime has already been expelled from the ``triangle''),
so that $\gamma_e < 1/d$ (see Figure~\ref{fig_d_gamma_e_versus_d}).
This evolution is shown in detail in Figure~\ref{blow_2.75}a--c
for the case $d=2.75$.
For $\bvt > \betatilde_{V,e}$, the only remaining transition is 
the normal-to-condensate transition at point c\textprime
(Figure~\ref{fig_param_F_versus_btt_d=2.75}f).
The phase diagram is identical to that for $d_* < d < \bar{d}$.
In Figure~\ref{phase_diag_beta_v_vs_dimension}
we plot $\betatilde_{V,2} < \betatilde_{V,cc'} < \betatilde_{V,e}$
as a function of $d$.

\bigskip

{\bf Case (f), $\boldsymbol{d \ge d_{**}=3}$:}
The three points $\betatilde_{V,2}$, $\betatilde_{V,cc'}$ and
$\betatilde_{V,e}$ all merge at $d=d_{**}=3$
(we demonstrate this analytically below).
For $d \ge 3$, point c\textprime stays inside the ``triangle''
until the ``triangle''  disappears at $\bvt = \betatilde_{V,e}$
(see Figure~\ref{fig_param_F_versus_btt_d=3} for $d=3$
 and Figure~\ref{fig_param_F_versus_btt_d=5} for $d=5$).
The phase diagram for $d \ge 3$ is shown in Figure~\ref{phase_diag_d=3}.
The dimension $d_{**} = 3$ is equivalently characterized
by noting that $\gamma_e$ reaches $1/d$ as $d \uparrow 3$
(see Figure~\ref{fig_d_gamma_e_versus_d}).\footnote{
   The existence of the special dimension $d_{**}$,
   as well as the fact that $d_{**} = 3$,
   were found independently by Caracciolo and Pelissetto \cite{Caracciolo_95}.
}

A curious thing happens in dimensions $d>4$,
where $f'_* \equiv f'_d(1/d)$ is finite.
Then $d\btt/d\gamma$ for the normal phase,
given by \reff{beta_prime_gamma}, can vanish at $\gamma = 1/d$:
this happens when $\bvt$ equals
\be
   \betatilde_{V,1} \;=\;
   {f_* \over 2d}
   \left[ 2 \,-\, {f_* \over 1 \,-\, d f_* (f_*-1)/f'_*} \right]
   \;,
\ee
which lies strictly below $\betatilde_{V,2}$
(since $f_* > 1$ and $f'_* < \infty$)
and corresponds to point c\textprime coinciding with point 1
of the ``triangle''.
Indeed, this coincidence occurs for all $\bvt$ in the range
$\betatilde_{V,1} \le \bvt \le \betatilde_{V,2} = \betatilde_{V,e}$;
it ends only when the ``triangle'' disappears.
See Figures~\ref{fig_param_F_versus_btt_d=5}c,d,e
for an illustration of this phenomenon in $d=5$.

Let us now prove that $d_{**}=3$.
Using the asymptotic expansion \reff{f_d_asymptotic} for $f_d(\gamma)$
as $\gamma \uparrow 1/d$, we obtain
\be
{d\btt \over df} \;=\; p_1(d) \, \epsilon^{{d\over 2}-1} \,+\, 
 p_2(d,\bvt) \, \epsilon^{2 - {d\over 2}} \,+\, O(\epsilon)   \;,
\label{as_1}
\ee
where $\epsilon = 1 - \gamma d > 0$ and
\begin{eqnarray}
p_1(d)  & = &  \left( {d\over 2\pi} \right)^{\! d/2} \,
               {\Gamma(1 - d/2) \over d (f_* - 1)}        \\[1mm]
p_2(d,\bvt )  & = &  \left( { 2\pi\over d} \right)^{\!d/2} \,
               {f_* (f_* - d \bvt) \over
                (f_* - 1) \, \Gamma(2 - d/2)}
\end{eqnarray}
Alternatively, we can write \reff{as_1} as a function of $f_*-f$
as $f\uparrow f_*$:
\be
{d\btt \over df} \;=\;
- {f_*-f\over d(f_*-1)} 
\,+\, 
 p_2 (d,\bvt ) \left( {- 2\pi \over d \, \Gamma(1-d/2)}\right)^{{4-d\over d-2}}
(f_*-f)^{{4-d\over d-2}}
 \,+\, O\!\left( (f_*-f)^{{2\over d-2}}\right)   \;.
\label{as_1s}
\ee

Let us now fix $\bvt$ to the value $\betatilde_{V,2} =f_*(2-f_*)/(2d)$
where point c\textprime coincides with point 2 of the ``triangle''.
At $\bvt = \betatilde_{V,2}$ the coefficient $p_2$ becomes
\be
p_2(d,\betatilde_{V,2}) \;=\; \left( { 2\pi\over d} \right)^{\! d/2} \,
     {f_*^3\over 2 (f_* - 1) \, \Gamma(2 - d/2)}   \;.
\ee
As $\epsilon \downarrow 0$ (i.e.\ $f\uparrow f_*$), 
the first term in \reff{as_1}/\reff{as_1s} [which is negative for $2<d<4$]
dominates for $d<3$,
while the second term (which is positive for $0\leq d<4$) dominates for $d>3$;
at $d=3$ both terms are of the order $\epsilon^{1/2}$,
with net coefficient $p_1(3) + p_2(3,\betatilde_{V,2}(3)) > 0$.
Therefore, for any $2<d<3$,
we have $d\btt / df < 0$ for sufficiently small $\epsilon > 0$
(i.e.\ for $f$ slightly less than $f_*$).
The condition $d\btt / df < 0$ corresponds to the existence
of the back-bending branch of the ``triangle'' (between points 1 and 2).
It means that the point $\betatilde_{V,e}$ where the ``triangle'' disappears
is strictly larger than $\betatilde_{V,2}$.
By contrast, for $d\geq 3$, we have $d\btt / df > 0$
for sufficiently small $\epsilon > 0$.
This means that there is no ``triangle'' structure for any
$\bvt >\betatilde_{V,2}$; rather, the ``triangle'' disappears exactly at
$\bvt =\betatilde_{V,2}$.
That is, the points $\betatilde_{V,2}$, $\betatilde_{V,cc'}$
and $\betatilde_{V,e}$ all coincide for $d \geq d_{**}=3$.

Finally, as 
 $d\rightarrow\infty$, we have the following asymptotic expressions for
the endpoints $\betatilde_{T,e}$ and $\betatilde_{V,e}$:
\begin{eqnarray}
   \betatilde_{T,e}  & = & {1\over 2d} \,+\,{1\over 2d^2} \,+\,{7\over 8 d^3}
 \,+\,{15\over 8 d^4} \,+\,{153\over  32 d^5} \,+\, {451\over 32 d^6} \,+\,
   \cdots   \\[2mm]
   \betatilde_{V,e}  & = & {1\over 2d} \,-\,{1\over 8d^3} \,-\,{3\over 8 d^4}
 \,-\,{33\over 32 d^5} \,-\,{3\over  d^6} \,-\, \cdots
\end{eqnarray}

\subsection{One-Dimensional Case}   \label{sec5.4}

The free energy in the direct approach is given by \reff{free_energy_gamma}.
We use the explicit expressions \reff{exact_fd}/\reff{exact_gd}
to eliminate $\gamma$, $f(\gamma)$ and $g(\gamma )$ in favor
of the variable $\s$ defined in \reff{var_change_gamma}/\reff{var_change_f}:
\be
   \overline{\scrf}(\s)   \;=\;
      {1 \over 2} \left[ {-2\s^2 \over 1-\s^2} + {\s^2  \over \btt (1-\s^2)^2}
                         - \log(1-\s^2)
                         - {\bvt \over \btt} {2\s \over 1-\s^2}
                         + {\bvt^2 \over \btt}  \right]
      \label{free_energy_d=1_a}  \\
\ee
The derivative of \reff{free_energy_d=1_a} with respect to $\s$ is
\be
{d  \overline{\scrf}\over d \s}  \;=\;  {(1 + \s^2)\over \btt (1 - \s^2)^3}
\left[ \s +\btt \left( \s + {\bvt \over \btt} \right) (\s^2-1)\right]
\ee
and therefore the extremum $\s_\star$  satisfies the cubic equation
\be
   \s^3 \,+\, v\s^2 \,-\, (1 - r)\s \,-\, v \;=\;  0   \;,
  \label{eq_x_plus}
\ee
where $v=\bvt/\btt$ and $r=1/\btt$.
For $\bvt > 0$ this cubic is strictly convex on $\s>0$,
negative at $\s=0$ and positive at $\s=1$;
it therefore has exactly one root $\s_\star > 0$,
which is simple and lies in $0 < \s_\star < 1$;
moreover, $\s_\star$ is a real-analytic function of $\bvt, \btt > 0$.
For $\bvt < 0$, the relevant solution is obtained by noting that
\reff{eq_x_plus} is invariant under the combined substitution
$\bvt \to -\bvt$, $\s \to -\s$.
There is thus no phase transition when $\bvt \neq 0$.
Note also that when $0 < \btt < 1$, the root $\s=0$ when $v=0$ is simple,
so that $\s_\star$ is a real-analytic function of $\bvt$ also {\em at}\/
$\bvt = 0$.
So there is no phase transition when $\bvt = 0$ and $0 < \btt < 1$.
(The same obviously holds also for $\btt=0$ and in fact also
for $\btt < 0$.)

For small $\bvt > 0$ we have:

1) For $\btt < 1$:
\begin{eqnarray}
\s_{\star}  &=&  \frac{\bvt}{1 - \btt} \,+\, O( \bvt^{3} )   \\[2mm]
\bar{F}(\bvt ,\btt ) &=& - \, \frac{\bvt^2}{2(1 - \btt)} \,+\, O( \bvt^4 )
\end{eqnarray}

2) For $\btt = 1$:
\begin{eqnarray}
\s_{\star} & = &  |\bvt |^{1/3} - \frac{|\bvt |}{3}
     + \frac{|\bvt|^{5/3}}{9} \,+\, O( |\bvt |^{7/3} )      \\[2mm]
\bar{F}(\bvt ,\btt ) &=& - \, \frac{3}{4}|\bvt |^{4/3} +
\frac{|\bvt |^2}{6} \,+\, O( |\bvt |^{8/3} )
\end{eqnarray}

3) For $\btt > 1$:
\begin{eqnarray}
\s_{\star} &=& \left(1 - {1 \over \btt}\right)^{\! 1/2}
    \,+\, \frac{\bvt}{2\btt (\btt - 1)} \,+\,  O( \bvt^2)   \\[2mm]
\bar{F}(\bvt ,\btt ) &=&
    - \smfrac{1}{2} (\btt - 1 -\log \btt)
    \,-\,  |\bvt| \left(1 - {1 \over \btt}\right)^{\! 1/2}  \,+\, O(\bvt^2)
\end{eqnarray}
We conclude that for $\btt > 1$ there is a first-order phase transition
as $\bvt$ passes through 0:
the free energy has an absolute-value cusp,
and the isovector energy $E_V$ has a jump discontinuity.
In particular, at $\bvt=0$ the $Z_2$ gauge symmetry is spontaneously broken.
Of course, this is an artifact of the $N\to\infty$ limit,
and does not occur for any finite $N$.

{\bf Remark.}
 To obtain the free energy in the parametric approach,
 we plug this $\s_\star$ into either
\reff{free_energy_gamma.parametric1} or
\reff{free_energy_gamma.parametric2},
using the explicit expressions \reff{exact_fd}/\reff{exact_gd}
to eliminate both $\gamma$ and $g$ in favor of $f$ and hence $\s$:
\begin{subeqnarray}
   \overline{\scrf}  
      & = &
      {1 \over 2} \left[ {-2\s^2 \over 1-\s^2}
                         - \log(1-\s^2)
                         + \btt \s^2 \right]
      \slabel{free_energy_d=1_b}  \\[2mm]
      & = &
      {1 \over 2} \left[ {-\s^2 \over 1-\s^2}
                         - \log(1-\s^2)
                         - \bvt \s \right]
      \slabel{free_energy_d=1_c}
\end{subeqnarray}

Finally, by the same reasoning as in Section~\ref{sec4.4}
the correlation functions are
\begin{eqnarray}
   G_V(x,y)  & = &   \s_\star^{|x-y|}     \label{eq5.21}  \\[2mm]
   G_T(x,y)  & = &   \s_\star^{2|x-y|}    \label{eq5.22}
\end{eqnarray}

\section{One-Dimensional Case:
         Pure $RP^{N-1}$, $\CP^{N-1}$ and $\QP^{N-1}$ Models}
\label{sec6}

In this section we compute the exact finite-$N$ solution
for the one-dimensional $RP^{N-1}$, $\CP^{N-1}$ and $\QP^{N-1}$ models
and discuss the behavior as $N \to\infty$.
We reproduce the $N=\infty$ phase transition at $\btt=1$
found in Section~\ref{sec4.4}, and we explain why it is an artifact
of the $N \to\infty$ limit.

\subsection{General $N$}  \label{sec6.1}

Note first that the one-link integral
\be
  \scrz_N(\beta)  \;=\;
  \int \exp\left[ {\beta \over 2}  |\bsigma_1^* \cdot \bsigma_2|^2 \right] \,
   d\Omega(\bsigma_2)
\ee
is independent of the unit vector $\bsigma_1$
[since by $G$-invariance $\bsigma_1$ can be rotated to $(1,0,\ldots,0)$].
It follows that for a one-dimensional {\em open}\/ chain
--- or more generally, for any lattice which has no closed cycles
(i.e.\ is a {\em forest}\/) ---
the partition function factorizes as
\be
   Z   \;=\;   \scrz_N(\beta)^\ell
\ee
where $\ell$ is the number of links in the lattice.
In particular, the infinite-volume free energy in one dimension
(which is independent of boundary conditions and can thus be computed
using an open chain) is
\be
   \bar{F}_N(\beta)   \;=\;
   \lim_{L\to\infty}  \left[ - {1 \over L} \log \scrz_N(\beta)^{L-1} \right]
   \;=\;  -\log \scrz_N(\beta)   \;.
\ee
So it suffices to study the one-link integral $\scrz_N(\beta)$.

Let us study the more general integral
\be
 \label{eq3.1}
   \widetilde{\scrz}_n(M)   \;=\;
   \int e^{{\bf x} \cdot M {\bf x}}  \, d\Omega({\bf x})
\ee
where $M$ is an $n \times n$ symmetric real matrix
and $d\Omega$ is normalized uniform measure on the unit sphere in $\R^n$.
It is easy to see that $\scrz_N(\beta)$ equals $\widetilde{\scrz}_{kN}(M)$
where $k=1,2,4$ for $\K = \R,\C,\Q$
provided we take $M$ to have $k$ eigenvalues equal to $\beta/2$
and $k(N-1)$ eigenvalues equal to 0.

Diagonalizing $M$ and then making the change of variables
$\lambda_j = x_j^2$ in \reff{eq3.1}, we obtain
\be
   \widetilde{\scrz}_n(M)   \;=\;
   c_n \int\limits_0^\infty
   \left( \prod\limits_{j=1}^n {d\lambda_j \over \lambda_j^{1/2}} \right)
   \, \delta\!\left( \sum\limits_{j=1}^n \lambda_j  - 1 \right)
   \, \exp\!\left[ \sum\limits_{j=1}^n \mu_j \lambda_j \right]
   \;,
\ee
where $\mu_1,\ldots,\mu_n$ are the eigenvalues of $M$;
here $c_n = \Gamma (n/2)/2\pi^{n/2}$ since $\widetilde{\scrz}_n(0) = 1$.
Specializing now to the case
$\mu_1 = \ldots = \mu_k = \beta/2$ and
$\mu_{k+1} = \ldots = \mu_n = 0$, we change variables to
$t = \lambda_1 + \ldots + \lambda_k$ and obtain
\begin{subeqnarray}
  \slabel{eq3.3a}
   \scrz_N(\beta)    & = &
      {\Gamma({n \over 2})  \over \Gamma({k \over 2}) \Gamma({n-k \over 2})}
      \int\limits_0^1  e^{\beta t/2} \, t^{{k \over 2} - 1} \,
          (1-t)^{{n-k \over 2} - 1} \, dt                           \\[2mm]
   & = &  \ofo \!\left( {k \over 2} ; {n \over 2} ; {\beta \over 2} \right)
\end{subeqnarray}
where $n=kN$ and
\begin{subeqnarray}
   \ofo(a;c;z)   & = &
   1 \,+\, {a \over c} z  
     \,+\, {a(a+1) \over c(c+1)} {z^2 \over 2!}
     \,+\, \ldots                                                   \\[2mm]
   & = &  {\Gamma(c)  \over  \Gamma(a) \Gamma(c-a)}
          \int\limits_0^1  e^{zt} \, t^{a-1} \, (1-t)^{c-a-1} \, dt
     \qquad[\hbox{for }  \real c > \real a > 0]                 \nonumber \\
  \slabel{eq3.4b}
\end{subeqnarray}
is the confluent hypergeometric function
\cite{Buchholz,Tricomi,Slater}.
Thus,
\be
   \scrz_N(\beta)   \;=\;
   \cases{ \ofo(\half; {N \over 2}; {\beta \over 2})     & for $RP^{N-1}$   \cr
           \noalign{\vskip 2pt}
           \ofo(  1  ;     N      ; {\beta \over 2})     & for $\CP^{N-1}$  \cr
           \noalign{\vskip 2pt}
           \ofo(  2  ;    2N      ; {\beta \over 2})     & for $\QP^{N-1}$  \cr
         }
\ee
We remark that the $\CP^{N-1}$ case can be written in terms of
elementary functions:
\be
   \ofo \!\left(1; N; {\beta \over 2} \right)   \;=\;
      {(N-1)! \over (\beta/2)^{N-1}}
      \left[ e^{\beta/2} -
             \sum\limits_{j=0}^{N-2} {(\beta/2)^j \over j!} \right]
   \;.
\ee
(A slightly more complicated derivation of the $\CP^{N-1}$ result
has been given by DiVecchia {\em et al.}\/ \cite[Appendix B.2]{DiVecchia_84}.
The $RP^{N-1}$ result is also well known \cite{Liu_72,Kohring_87,Ohno_90}.)

Finally, let us discuss the correlation functions,
restricting attention for simplicity to the $RP^{N-1}$ model.
For the isovector correlation function we have
\be
   G_V(x,y) \;\equiv\; \< \bsigma_x \cdot \bsigma_y \>  \;=\;  \delta_{xy}
   \;;
\ee
it vanishes for $x \neq y$ as a consequence of $Z_2$ gauge invariance.
The isotensor correlation function can be computed using
standard results on one-dimensional $O(N)$-invariant $\sigma$-models\footnote{
   See e.g.\ \cite[equations (2.33), (3.16) and (3.19)]{Cucchieri_97}.
};
the result is
\be
   G_T(x,y) \;\equiv\; \< {\bf T}_x \cdot {\bf T}_y \>
            \;\equiv\; \< (\bsigma_x \cdot \bsigma_y)^2 \> \,-\, {1 \over N}
            \;=\; \left( 1 - {1 \over N} \right)  v_{N,2}^{|x-y|}
 \label{GT_d=1_RPN}
\ee
where
\begin{subeqnarray}
   v_{N,2}  & = &  {\beta \over N(N+2)} \,
                   {\ofo({3 \over 2}; {N \over 2}+2; {\beta \over 2})
                    \over
                    \ofo({1 \over 2}; {N \over 2}; {\beta \over 2})
                   }
   \\[2mm]
   & = &  {1 \over N-1}
          \left( {\ofo({3 \over 2}; {N \over 2}+1; {\beta \over 2})
                  \over
                  \ofo({1 \over 2}; {N \over 2}; {\beta \over 2})
                 }
                 \,-\, 1
          \right)
   \\[2mm]
   & = &  {N \over (N-1)\beta}
          {\ofo({3 \over 2}; {N \over 2}; {\beta \over 2})
           \over
           \ofo({1 \over 2}; {N \over 2}; {\beta \over 2})
          }
          \,+\,
          {1 \over N-1} \left( {N \over \beta} + 1 \right)
   \slabel{vN2_d=1_RPNc}
   \label{vN2_d=1_RPN}
\end{subeqnarray}
(the equalities being proven by standard recurrence formulae for $\ofo$).

\subsection{$N \to\infty$ Limit}  \label{sec6.2}

The large-$N$ limit is obtained by taking $N \to\infty$
with $\betatilde \equiv \beta/kN$ fixed.\footnote{
   Some authors define $\betatilde = \beta/N$.
   But the present definition is more suited to treating the
   $RP^{N-1}$, $\CP^{N-1}$ and $\QP^{N-1}$ cases in a unified manner.
}
For the free energy, this limit can be read off
by applying Laplace's method to the integral \reff{eq3.3a}:
\begin{subeqnarray}
   \bar{F}(\betatilde)   \;=\;
   - \lim\limits_{N \to\infty} {1 \over kN} \log \scrz_N(kN\betatilde)
   & = &
   - \sup\limits_{0 \le t \le 1}
             \smhalf [ \betatilde t  \,+\,  \log(1-t)]     \\[2mm]
   & = &
   \cases{ 0                              & for $\betatilde \le 1$  \cr
           \noalign{\vskip 4pt}
           - {1 \over 2} (\betatilde - 1 -\log\betatilde)
                                          & for $\betatilde \ge 1$  \cr
                }
   \qquad
 \label{eq6.10}
\end{subeqnarray}
in agreement with \reff{solution_d=1_bvt=0}.
Thus, in the $N = \infty$ limit, the free energy exhibits a second-order
phase transition at $\betatilde = 1$:
the specific heat $C_H \sim \partial^2 \bar{F}/\partial \betatilde^2$
has a jump discontinuity.

\begin{figure}[p]
\vspace*{0cm} \hspace*{-0cm}
\begin{center}
\epsfxsize=0.7\textwidth
\leavevmode\epsffile{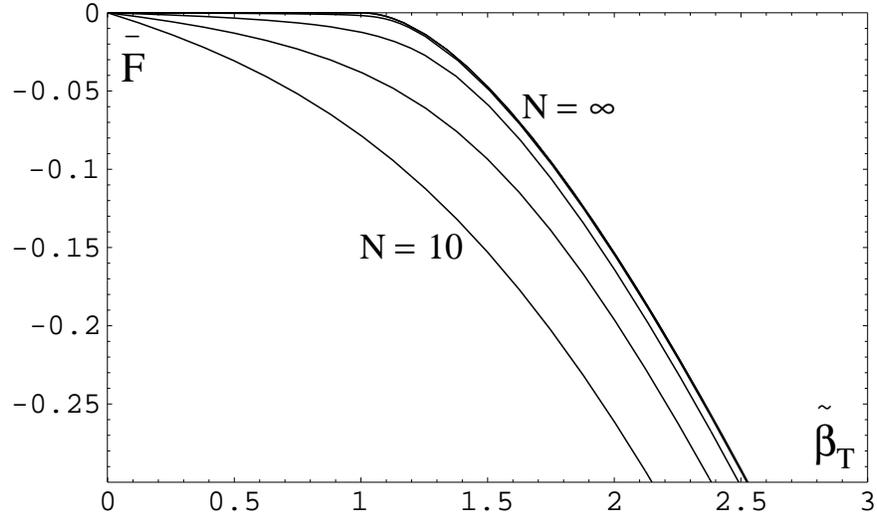} \\
\vspace{2cm}
\epsfxsize=0.7\textwidth
\leavevmode\epsffile{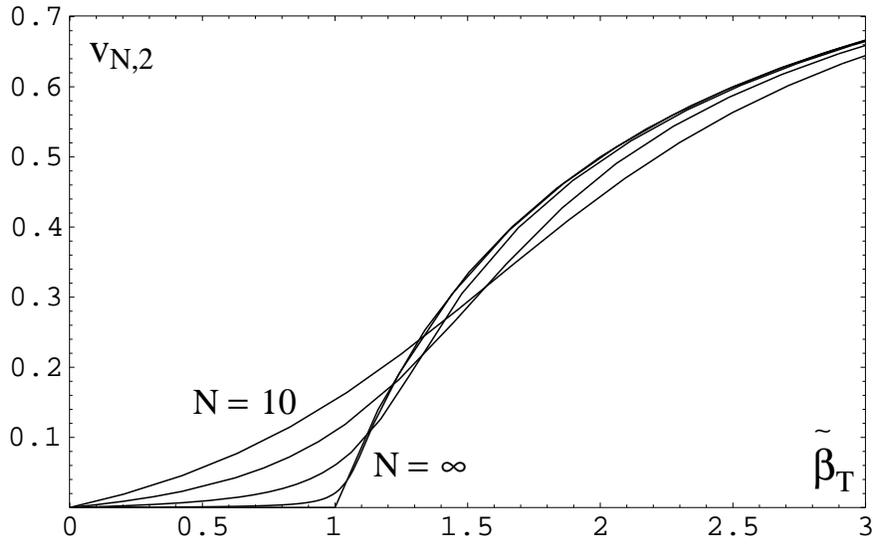}
\vspace{5mm}
\end{center}
\caption{
   Approach to the $N=\infty$ limit in the one-dimensional $RP^{N-1}$ model:
   curves shown are $N=10,25,100,1000,\infty$.
   (a) Free energy $\bar{F}(\betatilde) = - (1/N) \log \scrz_N(N\betatilde)$.
   (b) Correlation decay rate $v_{N,2} = e^{-m_T}$
       where $m_T$ is the isotensor mass gap.
}
\label{d=1_plots}
\end{figure}

On the other hand, we know that such a phase transition is impossible
for any finite $N$.
Indeed, from the integral \reff{eq3.3a}, it is clear that $\scrz_N(\beta)$ is
an entire analytic function of $\beta$;
and since $\scrz_N(\beta) > 0$ for $\beta$ real,
it follows that $\log \scrz_N(\beta)$ is a real-analytic function of $\beta$
(see Figure~\ref{d=1_plots}a).
How, then, can the nonanalyticity at $\betatilde = 1$ arise in the
limit $N \to\infty$?
The answer is that a limit of real-analytic functions need not be
real-analytic:  singularities can come in from the complex plane
and pinch the real axis.
Since $\scrz_N(\beta)$ is analytic,
the only singularities of $\log \scrz_N(\beta)$
are at the (complex) zeros of $\scrz_N(\beta)$.
So there must be zeros of $\scrz_N(\beta)$ very close to $\betatilde = 1$:

\begin{theorem}  \label{thm6.1}
Fix $a > 0$.  Then there exist, for all sufficiently large $c > 0$,
complex numbers $z^*_c$ satisfying $\ofo(a;c;z^*_c) = 0$
and $\lim\limits_{c\to\infty} (z^*_c/c) = 1$.
\end{theorem}

\noindent
In Appendix \ref{appendix_c} we give two proofs of this theorem:
a ``soft'' proof based on a direct transcription of the intuition
just sketched (Appendix \ref{appendix_c.2}),
and a ``hard'' proof that provides an explicit asymptotic expansion
for $z^*_c$ in powers of $c^{-1/2}$ (Appendix \ref{appendix_c.3}).
Surprisingly, we have been unable to find this result anywhere
in the extensive mathematical literature on confluent hypergeometric functions.

For the isotensor correlation function \reff{GT_d=1_RPN}/\reff{vN2_d=1_RPN},
we can apply Theorem~\ref{thmC.1} to \reff{vN2_d=1_RPNc}
to obtain
\be
   \lim_{N\to\infty} v_{N,2}(N\betatilde)   \;=\;
   \cases{ 0                        & for $\betatilde \le 1$  \cr
           \noalign{\vskip 4pt}
           1 - 1/\betatilde         & for $\betatilde \ge 1$  \cr
         }
 \label{vN2_d=1_RPN_N=infty}
\ee
in agreement with \reff{eq4.28}.
The approach to the $N\to\infty$ limit is shown in Figure~\ref{d=1_plots}b.

\section{One-Dimensional Case: Mixed Isovector/Isotensor Models}
\label{sec7}

Let us now generalize these one-dimensional results to the
mixed isovector/isotensor model \reff{eq1.2}.

\subsection{General $N$}  \label{sec7.1}

Consider the mixed isovector/isotensor Hamiltonian
\be
   H  \;=\;  - \sum_{\<xy\>}
     \left[ \beta_V \,\bsigma_x \cdot \bsigma_y   \,+\,
            {\beta_T\over 2}  (\bsigma_x \cdot \bsigma_y)^2
     \right]
\ee
with $\bsigma_x \in S^{N-1}$, on a one-dimensional lattice of $L$ sites
with free boundary conditions.
We obtain immediately the partition function
\be
  Z  \;=\; \scrz_N (\beta_V,\beta_T)^\ell \;,
\ee
where $\ell = L-1$ is the number of links and
\be
\scrz_N (\beta_V,\beta_T) \;=\;
 \int \exp   \left[ \beta_V \bsigma_1 \cdot \bsigma_2   \,+\,
            {\beta_T\over 2}  (\bsigma_1 \cdot \bsigma_2)^2
     \right] d\Omega ( \bsigma_2 )
  \label{eq5.3}
\ee
is the one-link integral.
The free energy is
\be
   \bar{F}_N(\beta_V,\beta_T)   \;=\;  \lim_{L\to\infty}
       \left[ - {1 \over L} \log \scrz_N(\beta_V,\beta_T)^{L-1} \right]
   \;=\;  -\log \scrz_N(\beta_V,\beta_T)   \;.
\ee

To evaluate the one-link integral \reff{eq5.3},
let us fix $\bsigma_1 = (1,0,\ldots ,0)$
and write $\bsigma_2 \;=\;(\s_1,\ldots, \s_N )$, so that
\begin{subeqnarray}
\scrz_N(\beta_V,\beta_T)  & = &
  {\Gamma (N/2) \over 2\pi^{N/2}}
  \int \delta(\s_1^2+\cdots + \s_N^2 -1) \,
   e^{\beta_V \s_1 +(\beta_T / 2) \s_1^2} \, d\s_1 \,\cdots\, d\s_N  \qquad \\[2mm]
  & = &
  {\Gamma ({N\over 2})\over \Gamma ({1\over 2}) \Gamma ({N-1\over 2})}
  \int_{-1}^{1} ( 1-\s^2)^{(N-3)/2}
   \exp\left[ \bv \s + {\bt\over 2} \s^2 \right]  \, d\s  \;.
  \slabel{eq_mixed}
\end{subeqnarray}
%
%
In the two limiting cases $\bt =0$ and $\bv = 0$, we have
\begin{eqnarray}
\scrz_N(\bv ,0) & = & \Gamma(N/2) \,
   \left( {2\over \bv}\right)^{\! (N-2)/2} I_{(N-2)/2}(\bv)
  \label{ZN_Bessel}  \\[2mm]
\scrz_N(0,\bt ) & = & \ofo\!\left({1\over 2}; {N\over 2};
                                  {\beta_T\over 2}\right)
\end{eqnarray}
where $I_\nu$ is a modified Bessel function.
Expanding \reff{eq_mixed} in powers of either $\bt$ or $\bv$,
we get the series representations
\be
\scrz_N(\bv,\bt) \;=\; \sum_{k=0}^{\infty}{(\bt /2)^k\over k!} \,
{(1/2)_k\over (N/2)_k}
   \;{{}_1 \! F_2} \!\left(k+\smhalf ;\smhalf ;{N\over 2}+k ;
{\bv^2\over 4}\right)
  \label{eq5.11}
\ee
and
\begin{subeqnarray}
\scrz_N(\bv,\bt)  & = &
   \sum_{k=0}^{\infty}  {\bv^{2k} \over (2k)!}  \,
                        {(1/2)_k\over (N/2)_k} \;
    {{}_1 \! F_1}\!\left(k+\smhalf ; {N\over 2}+k ; {\bt\over 2} \right)
  \\[1mm]
   & = & \sum_{k=0}^{\infty} {(\bv/2)^{2k} \over k!} \,
         {1 \over (N/2)_k} \;
         {{}_1 \! F_1}\!\left(k+\smhalf ; {N\over 2}+k ; {\bt\over 2} \right)
  \label{eq5.12}
\end{subeqnarray}
where $(a)_n = \Gamma (a+n)/\Gamma (a)$ is Pochhammer's symbol.
Or we can have the double series
\be
   \scrz_N(\bv,\bt)  \;=\;
   \sum_{k,l=0}^{\infty}  { (1/2)_{k+l}  \over  (N/2)_{k+l} \, (1/2)_l }  \,
         {(\bt/2)^k (\bv^2/4)^l  \over  k! \, l!}
   \;,
\ee
which is a special case of the Kamp\'e de Feriet
double hypergeometric series
\cite[p.~150, equation 29]{Appell_26}  
\cite[Section 1.5]{Exton_76}
\cite[pp.~26--27]{Srivastava_85}.

It is worth remarking that $\scrz_N(\bv,\bt)$
has a kind of ``Lee-Yang property'':
for all $N \ge 1$ (not necessarily integer) and all $\bt \ge 0$,
the zeros of $\scrz_N(\bv,\bt)$ in the complex $\bv$-plane
all lie on the imaginary axis.
Indeed, for $\bt = 0$ this is a well-known property
of the Bessel functions \reff{ZN_Bessel}\footnote{
   The zeros of $I_\nu$ are pure imaginary for all $\nu > -1$
   \cite[section 15.27]{Watson_44},
   so the right-hand side of \reff{ZN_Bessel}
   has the Lee-Yang property for all $N > 0$.
   But the integral \reff{eq_mixed} is convergent only for $N > 1$
   (or, by a limiting process, also for $N=1$).
};
while for $\bt > 0$ it follows from the formula
\be
   \scrz_N(\bv,\bt)  \;=\;
   \exp\!\left( {\beta_T\over 2} \, {\partial^2 \over \partial \bv^2} \right)
   \, \scrz_N(\bv,0)
\ee
combined with the fact that the operator
$\exp[(\beta_T/2) \, (\partial^2 / \partial \bv^2)]$
preserves the Lee-Yang property \cite{Lieb-Sokal}.
This Lee-Yang property fails for sufficiently negative $\bt$,
except when $N=1$ \cite{Newman_76}.

Let us now compute the isovector and isotensor correlation functions.
Standard results on one-dimensional $O(N)$-invariant $\sigma$-models
\cite{Cucchieri_97} yield
\begin{subeqnarray}
   G_V(x,y) &\equiv& \< \bsigma_x \cdot \bsigma_y \>
            \;=\;  v_{N,1}^{|x-y|}
   \slabel{GV_d=1_mixed}  \\[2mm]
   G_T(x,y) &\equiv& \< {\bf T}_x \cdot {\bf T}_y \>
            \;\equiv\; \< (\bsigma_x \cdot \bsigma_y)^2 \> \,-\, {1 \over N}
            \;=\; \left( 1 - {1 \over N} \right)  v_{N,2}^{|x-y|}
   \slabel{GT_d=1_mixed}
   \label{GV+T_d=1_mixed}
\end{subeqnarray}
where
\begin{subeqnarray}
   v_{N,1}  & = &
   { \int\limits_{-1}^{1}
          \s \, (1-\s^2)^{(N-3)/2} \, e^{\bv \s + (\bt/2) \s^2} \, d\s
     \over
     \int\limits_{-1}^{1}
                (1-\s^2)^{(N-3)/2} \, e^{\bv \s + (\bt/2) \s^2} \, d\s
   }
   \slabel{vN1_d=1_mixed}  \\[4mm]
   v_{N,2}  & = &
   { \int\limits_{-1}^{1}
          {\s^2 - 1/N \over 1-1/N} \,
          (1-\s^2)^{(N-3)/2} \, e^{\bv \s + (\bt/2) \s^2} \, d\s
     \over
     \int\limits_{-1}^{1}
          (1-\s^2)^{(N-3)/2} \, e^{\bv \s + (\bt/2) \s^2} \, d\s
   }
   \slabel{vN2_d=1_mixed}
   \label{vN1+2_d=1_mixed}
\end{subeqnarray}
For the $N$-vector model ($\beta_T = 0$) we recover the well-known result
\be
  v_{N,k}  \;=\;  { I_{(N/2) + k-1}(\beta_V)
                    \over
                    I_{(N/2)-1}(\beta_V)
                  }
   \;,
\ee
while for the pure $RP^{N-1}$ model ($\beta_V = 0$)
we recover \reff{vN2_d=1_RPN}.

\subsection{$N\rightarrow\infty$ Limit in Mixed Models with Various Scalings}
\label{sec7.2}

Let us now consider the limit $N\to\infty$
with $\bv$ and $\bt$ scaled as
\begin{subeqnarray}
\bv & = & N^{\varepsilon} \, \bvt \\
\bt & = & N \, \btt
\end{subeqnarray}
where $0 \le \varepsilon \le 1$.
We apply Laplace's method to the integral \reff{eq_mixed},
which can be written in the form
\be
\scrz_N(\bv,\bt) \;=\; 
   {\Gamma ({N\over 2})\over \Gamma ({1\over 2}) \Gamma ({N-1\over 2})}
   \int_{-1}^{1} \left( 1-\s^2\right)^{-3/2}  e^{-N \scrf(\s)} \, d\s
\label{eq5.20}
\ee
where
\be
\scrf(\s) \;=\; - {1\over 2} \log (1-\s^2) - {\btt \over 2} \s^2
                    - N^{\varepsilon-1} \bvt \s  \;.
   \label{eq5.20a}
\ee

If $0\leq \varepsilon < 1$,
the behavior at $N=\infty$ is exactly as in the pure $RP^{N-1}$ model
($\bvt = 0$):
for $\btt \le 1$, $\scrf(\s)$ has an absolute minimum at $\s=0$;
while for $\btt > 1$, $\scrf(\s)$ has a local maximum at $\s=0$
and absolute minima at $\s = \pm \s_0$, where
\be
   \s_0 \;=\; \left( 1 - {1\over \btt} \right) ^{\! 1/2}  \;.
\ee
There is therefore a second-order phase transition for $\btt =1$,
and the value of $\bvt$ plays no role.

However, for $\varepsilon =1$ the situation is very different.  
The stationary points of the function $\scrf(s)$ are the solutions of
the cubic equation
\be
   \s^3 + v  \s^2  - (1 - r) \s - v  \;=\; 0 \;,
\label{eq5.23}
\ee
where $v = \bvt/\btt$ and $r = 1/ \btt$.
The absolute minimum of $\scrf$ is given by the unique root $\s_{\star}$
of \reff{eq5.23} lying in the interval $0 \le (\sgn \bvt) \s \le 1$.\footnote{
   Using elementary monotonicity arguments one can show that for any $\btt$
   and nonzero $\bvt$, equation \reff{eq5.23} always has a unique real root
   in the interval $0 \le (\sgn \bvt) \s \le 1$.
}  
This is exactly the same cubic equation \reff{eq_x_plus}
found in Section~\ref{sec5.4},
even though the free energy is different
[compare \reff{free_energy_d=1_a} with \reff{eq5.20a}].\footnote{
   We remark that
   \reff{eq5.20a} = 2\reff{free_energy_d=1_c} $-$ \reff{free_energy_d=1_b}.
}
We therefore draw the same conclusions:
there is no phase transition when $\bvt \neq 0$ or $0 \le \btt < 1$;
but for $\btt > 1$ (resp.\ $\btt = 1$)
there is a first-order (resp.\ second-order)
phase transition as $\bvt$ passes through 0.

\begin{figure}[p]
\begin{center}
\epsfxsize=0.7\textwidth
\epsffile{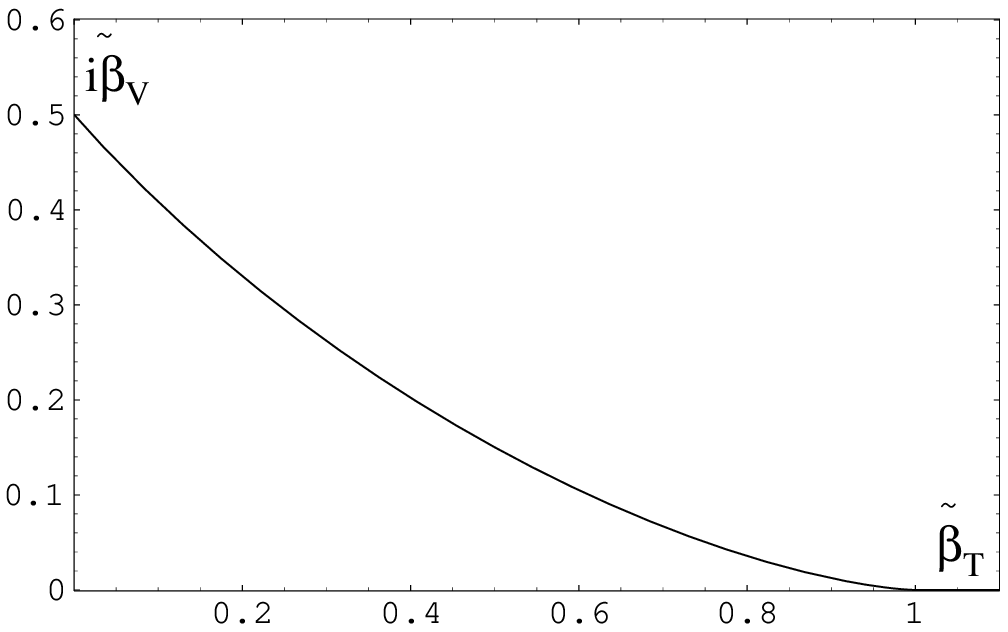}
\end{center}
\caption{
   Magnitude of the first imaginary zero of $\scrz_N(N\bvt,N\btt)$
   in the limit $N\to\infty$, as a function of $\btt$.
}
\label{fig_bvt_zero}
\end{figure}

\begin{figure}[p]
\begin{center}
\epsfxsize=0.7\textwidth
\epsffile{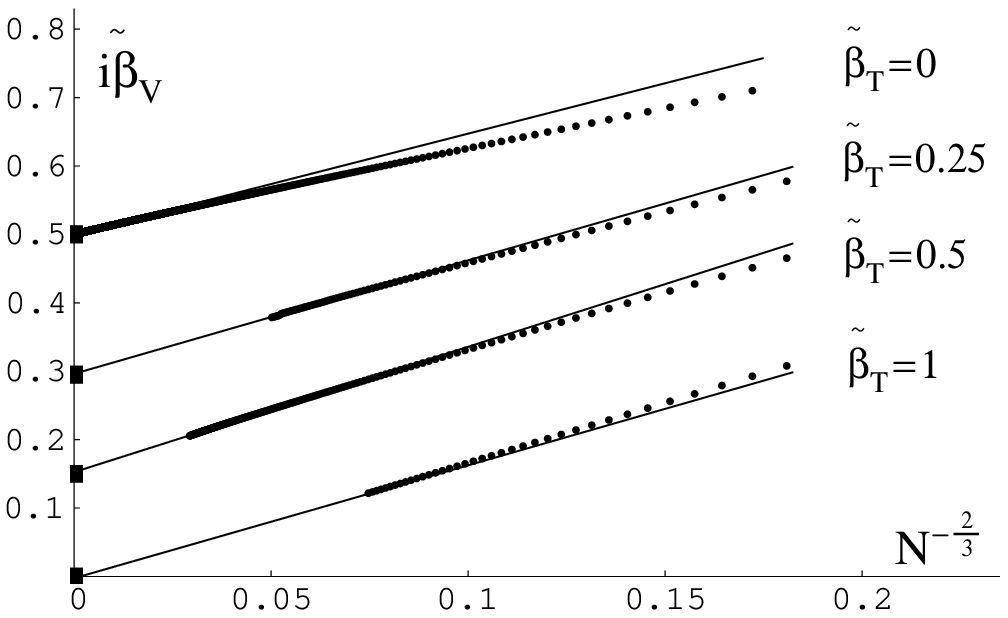}
\end{center}
\caption{
   Magnitude of the first imaginary zero of $\scrz_N(N\bvt,N\btt)$
   plotted versus $N^{-2/3}$,
   for $\btt = 0, 1/4, 1/2, 1$.
   Solid box indicates the exact limiting value \reff{bvt_zero_scaling}
   for $N=\infty$.
   Slope of straight line is given by a best fit to the data for
   $N \gtapprox 50$, except for $\btt=0$, when it is given by the
   exact asymptotic formula (see footnote \ref{footnote_jnu1}).
}
\label{fig_infeld_zeros}
\end{figure}

The phase transition observed as $\bvt$ passes through 0
when $\btt \ge 1$ can be understood as arising from pure-imaginary
zeros of $\scrz_N(\bv,\bt)$ in the complex $\bvt$-plane that approach
the origin as $N\to\infty$.
Indeed, Laplace's method applied to \reff{eq_mixed} immediately yields
\be
   \lim\limits_{N\to\infty}  {\scrz_N(\bv, N\btt)  \over \scrz_N(0, N\btt)}
   \;=\;
   \cosh(\bv \s_0)
 \label{eq7.20}
\ee
where
\be
   \s_0  \;=\;  \cases{ 0     & for $\btt \le 1$  \cr
                       \noalign{\vskip 2mm}
                       \left( 1 - {1\over \btt} \right) ^{\! 1/2}
                             & for $\btt \ge 1$  \cr
                     }
 \label{eq7.21}
\ee
It follows that if $\btt > 1$,
$\scrz_N(\bv, N\btt)$ has (for all sufficiently large $N$)
zeros in the complex $\bv$-plane
that tend to $\pm (2k+1) \pi i/2\s_0$ as $N \to\infty$.
(As mentioned previously, we know that these zeros lie on the
 imaginary axis for all $N$.)
In the variable $\bvt \equiv \bv/N$, these zeros tend to the origin
 at a rate $1/N$, producing a phase transition in the limit.
For $0 \le \btt < 1$, by contrast,
the zeros in the $\bv$-plane scale with $N$
and thus stay a finite distance away from the origin in the $\bvt$-plane.
Indeed, the closest zeros to the origin lie on the imaginary axis at
the $\bvt$ value where two saddle points given by \reff{eq5.23} collide, namely
\be
   \bvt  \;=\; \pm \,
     {(r-4 + \sqrt{r^2 +8r}) \, \sqrt{r+2 - \sqrt{r^2 +8r}}
      \over
      4 \sqrt{2} r
     }
 \label{bvt_zero_scaling}
\ee
where $r = 1/ \btt > 1$;
this goes monotonically from $\pm i/2$ at $\btt=0$
to zero at $\btt=1$ (see Figure~\ref{fig_bvt_zero}).\footnote{
   With a bit more work one can presumably show that the free energy
   $\bar{F} =$\break $- \lim_{N\to\infty} (1/N) \log \scrz_N(N\bvt,N\btt)$
   has a branch cut on the imaginary axis starting at the points
   \reff{bvt_zero_scaling}.
   From this it would follow, by the methods of Appendix~\ref{appendix_c.2},
   that zeros of $\scrz_N(N\bvt,N\btt)$ accumulate densely
   on this cut as $N\to\infty$.
}
The corrections to \reff{bvt_zero_scaling} are presumably of order $N^{-2/3}$,
as they are at $\btt = 0$.\footnote{
 \label{footnote_jnu1}
   At $\btt = 0$ we have $\scrz_N(\beta_V,0) \sim I_{(N-2)/2}(\beta_V)$,
   and the first root of $I_\nu(z)$ is given by $z = \pm i j_{\nu,1}$
   where
$$
   j_{\nu,1}   \;\simeq\;  \nu \,+\, 1.85575 \nu^{1/3}
                               \,+\, 1.03315 \nu^{-1/3}
                               \,-\, 0.00397 \nu^{-1}
                               \,+\, \ldots
$$
   \cite[section 15.8]{Watson_44}
   \cite[equation 9.5.14]{Abramowitz_70}.
   We have no proof that this same scaling of the correction terms
   persists for $0 < \btt < 1$, but it seems plausible
   and is consistent with our numerical calculations.
}
For $\btt = 1$, the zeros in the $\bv$-plane presumably scale
as a fractional power of $N$;
empirically we find that they scale like $N^{1/3}$,
exactly like the leading correction term for $0 \le \btt < 1$
(see Figure~\ref{fig_infeld_zeros}).

Let us conclude by considering the correlation functions
\reff{GV+T_d=1_mixed}/\reff{vN1+2_d=1_mixed}
in the limit $N \to\infty$ with $\bvt,\btt$ fixed.
In this limit, the integrals in both numerator and denominator of
\reff{vN1+2_d=1_mixed} are dominated by $s_\star$,
so we obtain
\be
   \lim_{N\to\infty} v_{N,k}(N\betatilde)   \;=\;  s_\star^k
\ee
for $k=1,2$,
in complete agreement with \reff{eq5.21}/\reff{eq5.22}.
We thus verify by explicit calculation the
factorization \reff{eq2.factorization1a} in the $N\to\infty$ limit.
Of course, this solution exhibits the (spurious) spontaneous breaking
of $Z_2$ gauge invariance at $\bvt = 0$ and $\btt > 1$, when $N=\infty$.

\section{Conclusions}   \label{sec8}

What is the phase diagram of the mixed isovector/isotensor model \reff{eq1.2}
for large but finite $N$?  We know some things for sure:
\begin{itemize}
   \item[(a)]  In dimension $d \le 1$, no phase transitions of any kind
       are possible in the finite $(\beta_V,\beta_T)$-plane
       \cite{Dobrushin_73,Cassandro_81,Simon_93}.
   \item[(b)]  In dimension $d \le 2$, no spontaneous breaking of the
       $SO(N)$ global symmetry is possible
       in the finite $(\beta_V,\beta_T)$-plane
       \cite{Mermin_67,Dobrushin_75,Pfister_81}.
       In particular, magnetic and nematic long-range order
       are impossible, and all correlation functions decay
       at least as an inverse power of the distance
       \cite{McBryan_77,Tanaka_98}.
   \item[(c)]  No spontaneous breaking of the $Z_2$ gauge invariance
       at $\beta_V = 0$ is possible in any dimension \cite{Elitzur_75}.
\end{itemize}
In addition, some properties of the phase diagram in dimension $d>2$
are known rigorously:
\begin{itemize}
   \item[(d)]  All correlation functions
      $\< (\bsigma_{x_1} \!\cdot \bsigma_{y_1}) \cdots
           (\bsigma_{x_n} \!\cdot \bsigma_{y_n})    \>$
      in the mixed isovector/isotensor model at $(\beta_V,\beta_T)$
      are bounded above by the corresponding Ising-model correlation functions
      $\< \varepsilon_{x_1} \varepsilon_{y_1} \cdots
          \varepsilon_{x_n} \varepsilon_{y_n}        \>$
      at $\beta = |\beta_V|$:
      we prove this in Appendix~\ref{app.correq}
      (Theorem~\ref{thmA.5} and Corollary~\ref{corA.5a}).
      In particular, for $|\beta_V| \le \beta_{c,Ising}$
      there is no ferromagnetic or antiferromagnetic long-range order.
   \item[(e)]  For $\beta_V > N f_*/(2d)$
      there is ferromagnetic long-range order \cite{Tanaka_98,Campbell_99}.
      Likewise, for $\beta_V < -N f_*/(2d)$
      there is antiferromagnetic long-range order.
   \item[(f)]  For $\beta_T > {N^2 (N+1) \over 2(N-1)} {f_* \over 2d}$
      there is nematic long-range order
      \cite{Angelescu_82,Tanaka_98,Campbell_99}.
\end{itemize}
Facts (e) and (f) are proven by the method of infrared bounds
\cite{FSS_76,Frohlich-Spencer_77,FILS_78}.
It is noteworthy that (e) is asymptotically sharp as $N\to\infty$
when $\beta_T=0$:
it captures the leading large-$N$ behavior \reff{nematic_lines}
of the pure $N$-vector model.
By contrast, (f) is a very poor bound for large $N$:
it behaves as $N^2$ while the true critical point is presumably of order $N$.

One of the most striking aspects of the $N=\infty$ solution
is that it violates fact (c):
spontaneous breaking of the $Z_2$ gauge invariance occurs at $N=\infty$
whenever the absolute minimum of the free energy in the pure $RP^{N-1}$ model
is located at $\gamma \neq 0$ [cf.\ the discussion surrounding \reff{eq3.15}].
But the distinction at $N=\infty$ between the white-noise phase
($\gamma=0$) and the isotropic phase ($0 < \gamma < 1/d$)
does not survive to finite $N$.
Indeed, at finite $N$ the correlation functions {\em never}\/
have the white-noise form (except of course at $\bt=0$).
Rather, the white-noise ``phase'' $0 < \btt < \betatilde_{T,c}$
(or $0 < \btt < \betatilde_{T,c''}$ in dimensions $d > d_*$)
simply corresponds to the region where $\gamma$ is of order $1/N$
[and hence the mass gap \reff{mass_gap} is of order $\log N$]
when $N\to\infty$ at fixed $\btt$,
while the isotropic ``phase'' corresponds to the region where $\gamma$
and the mass gap are of order 1 when $N\to\infty$ at fixed $\btt$.
There is no sharp distinction between these two regions at any finite $N$;
rather, for each finite $N$, the white-noise and isotropic ``phases''
both belong to a single paramagnetic phase in which the isotensor mass gap
$m_T$ is finite and nonzero.
This behavior is seen explicitly in the $d=1$ solution
\reff{GT_d=1_RPN}/\reff{vN2_d=1_RPN}/\reff{vN2_d=1_RPN_N=infty}
[see Figure~\ref{d=1_plots}b];
we conjecture that it constitutes the correct interpretation
of the ``white-noise-to-isotropic (pseudo)transition'' in all dimensions.
Likewise, the spontaneous breaking of the $Z_2$ gauge invariance
in the isotropic ``phase'' at $N=\infty$,
arising from the absolute-value cusp in the free energy
$\bar{F} = \bar{F}_0 - O(|\bvt|)$ as $\bvt$ passes through 0,
cannot survive to finite $N$;
the behavior at small $\bv$ {\em must}\/ be smooth \cite{Elitzur_75}.
Rather, we can understand the white-noise ``phase''
$0 < \btt < \betatilde_{T,c}$ (or $0 < \btt < \betatilde_{T,c''}$)
as corresponding to the region where this rounding takes place
over a $\bv$ interval of order $N$,
while the isotropic ``phase'' $\btt > \betatilde_{T,c}$
(or $\btt > \betatilde_{T,c''}$) corresponds to the region where
this rounding takes place over a $\bv$ interval of order $o(N)$.
Once again, this behavior is seen explicitly in the $d=1$ solution,
where the rounding in the isotropic ``phase''
takes place over a $\bv$ interval of order 1
[cf.\ the discussion surrounding \reff{eq7.20}/\reff{eq7.21}];
we conjecture that this constitutes the correct interpretation
of the ``spontaneous breaking of $Z_2$ gauge invariance'' in all dimensions.

\begin{figure}[p]
\begin{center}
\epsfxsize=0.8\textwidth
\epsffile{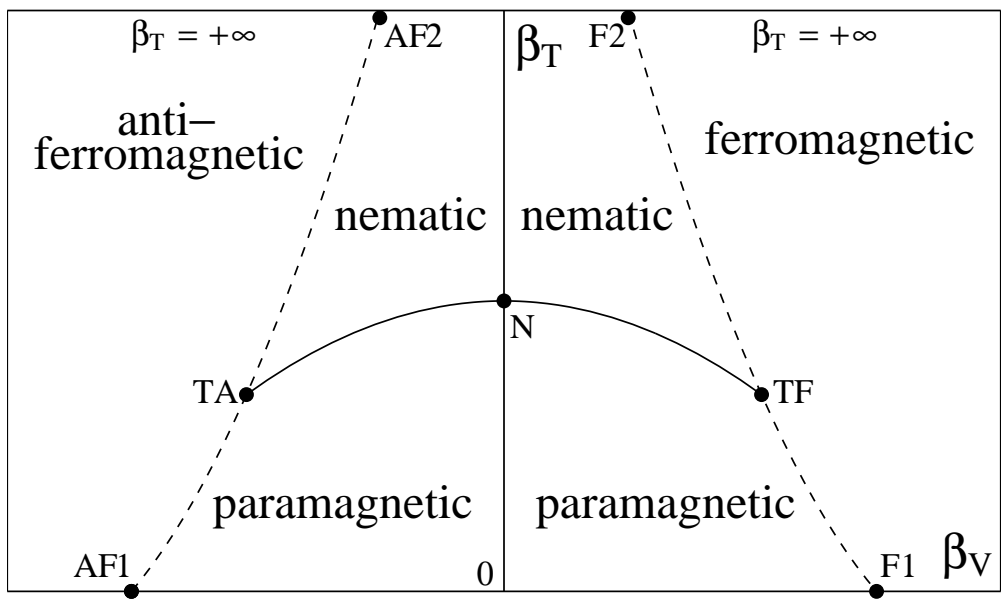}
\end{center}
\caption{
   Conjectured phase diagram in the $(\beta_V,\beta_T)$-plane
   for the mixed isovector/isotensor model with $2 < N < \infty$
   in dimension $d>2$.
   First-order transitions are shown as solid lines,
   and second-order transitions as dashed lines.
}
\label{conjectured_phase_diagram}
\end{figure}

Putting facts (a)--(f) together with our analysis of the large-$N$ limit
--- and its pathologies in $d=1$ --- we are led to conjecture the
phase diagram shown in Figure~\ref{conjectured_phase_diagram}
for the finite-$N$ mixed isovector/isotensor model in dimensions $d>2$.
When both $\beta_T$ and $|\beta_V|$ are small,
there is a paramagnetic phase in which
both the isovector and isotensor correlation functions decay exponentially;
in particular, there is no long-range order of any kind.
Along the $\beta_T=0$ axis (pure $N$-vector model),
there is a second-order paramagnetic-to-ferromagnetic transition
at $\beta_V = \beta_{c,N-vector}$,
and a second-order paramagnetic-to-antiferromagnetic transition
at $\beta_V = -\beta_{c,N-vector}$;
for large $N$ we have $\beta_{c,N-vector} = N f_*/(2d) + O(1)$.
At the opposite extreme $\beta_T = +\infty$,
the spins are forced to point up or down along a single axis,
so the $O(N)$-invariant correlations are those of
an Ising model with coupling $\beta_V$;
therefore, there is a second-order paramagnetic-to-ferromagnetic transition
at $\beta_V = \beta_{c,Ising}$
and a second-order paramagnetic-to-antiferromagnetic transition
at $\beta_V = -\beta_{c,Ising}$ (independent of $N$).
It is reasonable to conjecture that the paramagnetic-to-ferromagnetic
transitions at $\beta_T=0$ and $\beta_T=+\infty$
are the endpoints of a continuous curve of paramagnetic-to-ferromagnetic
transitions;
moreover, this curve $\beta_{V,c}(\beta_T)$ should be monotone decreasing
(since $\beta_V$ and $\beta_T$ both promote ferromagnetic ordering).
In the entire region to the right of this curve,
there is ferromagnetic long-range order.
It is also reasonable to conjecture,
though this is much less certain,
that the transition remains second-order everywhere along this curve.
There is, of course, a symmetrical paramagnetic-to-antiferromagnetic
transition curve in the half-plane $\beta_V < 0$.

Finally, along the $\beta_V = 0$ axis (pure $RP^{N-1}$ model),
there is a paramagnetic-to-nematic transition
at $\beta_T = \beta_{c,RP^{N-1}}$.
On the basis of mean-field theory
\cite{Kohring_87,Ohno_90,deGennes_93,Chaikin_95}
and Monte Carlo simulations (for $d=3$, $N=3,4,5$) \cite{Kohring_87},
this transition is expected to be first-order for all $N>2$,
at least in dimensions $d \ge 3$.\footnote{
   By contrast, for $N=2$ the $RP^{N-1}$ model is equivalent,
   via the change of variables $\theta' = 2\theta$,
   to the standard $XY$ model, so the transition is second-order.
}
(Perhaps the transition is second-order for $2 < d < d_*$.\footnote{
   Of course, it is far from clear what noninteger dimensions $d$
   should be taken to {\em mean}\/ nonperturbatively.
   But one can also study the 2-dimensional model with
   power-law-decaying long-range interaction
   $K_{xy} \sim |x-y|^{-(d+2-\sigma)}$ ($\sigma > 0$ small),
   which should behave qualitatively like the putative short-range model
   in dimension $d=2+\sigma$.
})
We expect that $\beta_{c,RP^{N-1}}$ scales for large $N$ proportional to $N$;
we predict (but with less confidence) that the
limiting proportionality constant is
$\betatilde_{T,c'}$ [defined by \reff{def_betatilde_T_c'}]
when $2 < d < d_*$,
and $\betatilde_{T,c''}$ [defined by \reff{def_betatilde_T_c''}]
when $d \ge d_*$.

What about the first-order transition curve $\betatilde_{T,c}(\bvt)$
found at $N=\infty$ in dimensions $d > 3/2$
(Figures~\ref{phase_diag_d=2}--\ref{phase_diag_d=3}) ---
does it survive to finite $N$?
We doubt it:  this transition is the extension to $\bvt \neq 0$
of the ``white-noise-to-isotropic transition'' of the $RP^{N-1}$ model,
which we have argued is an artifact of the $N=\infty$ limit.
Moreover, extensive Monte Carlo simulations of the two-dimensional
mixed isovector/isotensor model for $N=3$,
along with preliminary simulations of the two-dimensional $RP^{N-1}$ models
for $N=4,6,8$, showed no sign of any first-order transition
\cite{CEPS_rpn_static}.

One plausible scenario is the following:
All transitions (whether first-order or second-order)
within the ``normal phase'' $|\gamma| < 1/d$
are artifacts of the $N=\infty$ limit,
but transitions (whether first-order or second-order)
into the ``condensate phase'' $|\gamma| = 1/d$
do survive to finite $N$.
In this scenario, the transition curve in the limit $N\to\infty$
would be given by the upper envelope of the transitions shown in
Figures~\ref{phase_diag_2<d<dstar}--\ref{phase_diag_d=3}:
namely, by the line $\betatilde_{T,c'}(\bvt)$ defined by \reff{nematic_lines}
for $2 < d \le d_*$,
and by $\max[\betatilde_{T,c}(\bvt), \, \betatilde_{T,c'}(\bvt)]$
for $d > d_*$.
Most likely, this curve corresponds to the part TF--F1
of the paramagnetic-to-ferromagnetic transition curve.
In particular, point TF would lie at a $\beta_V$ value that is $o(N)$
[e.g.\ most likely of order 1]
and a $\beta_T$ value that is close to $N \betatilde_{T,c'}$
[resp.\ $N \betatilde_{T,c''}$]
for $2 < d \le d_*$ [resp.\ $d > d_*$],
so that as $N\to\infty$ the whole curve N--TF is compressed
to the point $\bvt=0$, $\btt=\betatilde_{T,c'}$
[resp.\ $\btt=\betatilde_{T,c''}$].
Moreover, the whole nematic region (bounded by AF2--TA--N--TF--F2)
is compressed at $N=\infty$ to the line
$\bvt=0$, $\btt \ge \betatilde_{T,c'}$ [resp.\ $\btt \ge \betatilde_{T,c''}$];
this compression is what produces the spurious first-order transition
as this line is crossed.
It would be useful to test these predictions by Monte Carlo simulation
of the 3-dimensional mixed isovector/isotensor model
for various values of $N$.

In dimensions $d \le 2$ we expect that there are no phase transitions
of any kind in the finite $(\beta_V,\beta_T)$-plane.
However, for $1 < d \le 2$ there is an Ising-like phase transition
at $\beta_T = +\infty$, $\beta_V = \beta_{c,Ising}$;
and it is at present unclear whether ``new'' continuum limits
--- that is, continuum theories whose isotensor correlation functions
differ from those of the continuum limit of the pure $N$-vector model ---
can be obtained by approaching this point
\cite{CEPS_PRL,Hasenbusch_96,Niedermayer_96,Catterall_98}
(see also \cite{Cucchieri_97,Seiler_97,Hasenbusch_99}
 for the analogous problem in $d=1$).

\appendix

\section{The Lattice Propagator and Free Energy}  \label{appendix_b}

\subsection{Formulae}     \label{appendix_b.1}

In this appendix we record some needed formulae concerning the
lattice propagator and free energy
\begin{eqnarray}
   f_d(\gamma)   & = &
      \int {d^d p  \over (2\pi)^d}  \,
      \left[ 1 \,-\, \gamma \sum\limits_{i=1}^d \cos p_i \right] ^{-1}
   \label{def1_fd} \\[2mm]
   g_d(\gamma)   & = &
      \int {d^d p  \over (2\pi)^d}  \,
      \log\!\left[ 1 \,-\, \gamma \sum\limits_{i=1}^d \cos p_i \right]
   \label{def1_gd}
\end{eqnarray}
which are well-defined for $|\gamma| \le 1/d$
[with $f_d(\pm 1/d) = \infty$ for $d \le 2$].
Inserting the integral representations
\begin{eqnarray}
   {1 \over z}   & = &   \int\limits_0^\infty  e^{-tz} \, dt         \\[2mm]
   \log z        & = &
      \int\limits_0^\infty  {e^{-t} - e^{-tz}  \over  t}  \, dt
\end{eqnarray}
and performing the momentum-space integrals, we obtain
\begin{eqnarray}
   f_d(\gamma)   & = &
      \int\limits_0^\infty  e^{-t} \, I_0(\gamma t)^d \, dt
   \label{def_int_fd} \\[2mm]
   g_d(\gamma)   & = &
      \int\limits_0^\infty
      {e^{-t} \over t} \, \left[ 1 - I_0(\gamma t)^d \right] \, dt
   \label{def_int_gd}
\end{eqnarray}
where $I_0$ is a modified Bessel function.
This defines an analytic continuation to non-integer $d$.

Note also the relations
\begin{eqnarray}
   {df_d(\gamma) \over d\gamma}  & = &
   {1 \over \gamma} \int\limits_0^\infty  (t-1) e^{-t} \, I_0(\gamma t)^d \, dt
     \label{deriv_fd}  \\[2mm]
   {d g_d(\gamma)  \over  d\gamma}   & = &
      -\, {f_d(\gamma) - 1   \over   \gamma}
   \label{deriv_gd}
\end{eqnarray}
which can be obtained by writing
\be
   {d \over d\gamma} I_0(\gamma t)^d   \;=\;
   {t \over \gamma} \, {d \over dt} I_0(\gamma t)^d
\ee
and integrating by parts.
In particular, we have
\begin{eqnarray}
   \gamma f'_d(\gamma) + f_d(\gamma)   & = &
       \int\limits_0^\infty  t e^{-t} \, I_0(\gamma t)^d \, dt
   \nonumber \\[1mm]
   & = & \int\limits_0^\infty  dt_1 \int\limits_0^\infty dt_2 \,
            e^{-t_1-t_2} \, I_0\Bigl(\gamma (t_1+t_2) \Bigr)^d
   \nonumber \\[1mm]
   & \ge &  \int\limits_0^\infty  dt_1 \int\limits_0^\infty dt_2 \,
            e^{-t_1-t_2} \, I_0(\gamma t_1)^d \, I_0(\gamma t_2)^d
   \nonumber \\[1mm]
   & = &  f_d(\gamma)^2
      \label{eqnB.ineq_fprime}
\end{eqnarray}
for all $d \ge 0$ (with strict inequality for $d>0$),
since $\log I_0$ is a convex function
and satisfies $\log I_0(0)=0$.\footnote{
   The convexity of $\log I_0$ follows easily from H\"older's inequality 
   applied to the integral representation
   $$ I_0(t) \;=\; \int\limits_{-\pi}^\pi {dp \over 2\pi} \, e^{t \cos p}$$
   (see e.g.\ \cite[Theorem I.2.3]{Israel_79}).
   Note also that for {\em integer}\/ $d \ge 1$,
   the inequality \reff{eqnB.ineq_fprime} can be proven directly
   from \reff{def1_fd}, using the Schwarz inequality.
}
It follows that $(f_d - 1)/\gamma f_d$ is an increasing function of
$\gamma$.

The cases $d=1,2$ can be expressed in terms of standard functions:
\be
   f_d(\gamma)   \;=\;
   \cases{ (1 - \gamma^2)^{-1/2}        & for $d=1$   \cr
           \noalign{\vskip 4pt}
           {2 \over \pi} K(2\gamma)     & for $d=2$   \cr
         }
   \label{exact_fd}
\ee
where $K$ is a complete elliptic integral of the first kind.
The case $d=3$ can also be expressed in terms of the elliptic integral $K$,
but the result is a god-awful mess \cite{Joyce_72,Glasser_76,Glasser_77}.
Similarly we have\footnote{
   The explicit expression for $g_2(\gamma )$ is obtained by
   taking the limit $g_2(\gamma ) = \lim_{\epsilon \rightarrow 0}\left[
   \Gamma (\epsilon ) - R(\epsilon ,\gamma )\right]$, where $R(\epsilon ,\gamma )
   = \int_0^{\infty}t^{\epsilon - 1}e^{-t}I_0^2(\gamma t) \, dt =
     \Gamma (\epsilon )
    {{}_4 \! F_3} ({1\over 2},{3\over 2},{\epsilon \over 2},{\epsilon +1\over 2};1,1,1;4\gamma^2) = \Gamma (\epsilon )F_4 ({\epsilon \over 2},{\epsilon +1\over 2};
   1,1;\gamma^2 ;\gamma^2) = \sum_{k=0}^{\infty} {\Gamma (\epsilon +2k )\over k!\Gamma (k+1)} {{}_2 \! F_1} (-k,-k;1;1)(\gamma /2)^{2k}
   $  \cite{Prudnikov}. Here $F_4(a,b;c,c';z,\xi )$ is an Appell's function.
   Since ${{}_2 \! F_1} (-k,-k;1,1)= \Gamma (2k+1)/\Gamma^2(k+1)$,
   we obtain
   $g_2(\gamma )=-\sum_{k=1}^{\infty}{\Gamma (2k)\Gamma (2k+1)\over (k!)^4}(\gamma /2)^{2k} = - {1\over 8}\int_0^{4\gamma^2}
    {{}_3 \! F_2} (1,{3\over 2},{3\over2};2,2;x)dx
   =  -{\gamma^2 \over 2}
               {{}_4 \! F_3} (1,{3\over 2},{3\over 2},1;2,2,2;4\gamma^2)$.
   An alternate expression in terms of ${{}_3 \! F_2}$
   can be deduced from \cite[equation 18]{Glasser_76};
   we thank Professor Glasser for drawing our attention to this.
}
\be
   g_d(\gamma)   \;=\;
   \cases{   \log\!\left( {1 + \sqrt{1 - \gamma^2}  \over  2} \right)
                                                   & for $d=1$   \cr
           \noalign{\vskip 4pt}
           -{\gamma^2 \over 2}
            {{}_4 \! F_3} (1,{3\over 2},{3\over 2},1;2,2,2;4\gamma^2)
                                                   & for $d=2$   \cr
         } 
   \label{exact_gd}
\ee

In general dimension we have the expansions for small $\gamma$:
\begin{eqnarray}
 \hspace*{-1cm}
   f_d(\gamma)   & = &
     1  \,+\,  {d \over 2} \gamma^2  \,+\,
       {3d (2d-1) \over 8} \gamma^4  \,+\,
       {5d (6d^2 - 9d + 4)  \over 16}  \gamma^6 \nonumber \\  & & \,+\,
{35d (24d^3 - 72d^2 + 82d - 33)\over 128}  \gamma^8  \,+\, 
{63d (120d^4 - 600d^3 + 1250d^2 - 1225d + 456)\over 256} \gamma^{10} 
\nonumber \\ & & \,+\, {231d(720d^5 - 5400d^4
 + 17700d^3 - 30600d^2 + 27041d -9460)\over 1024}
\gamma^{12} \nonumber \\ & & \,+\, {429d(5040d^6 - 
52920d^5 + 249900d^4 - 661500d^3 + 1011017d^2 -
 826336d + 274800)\over 2048}
\gamma^{14} \,+\,
\ldots
   \nonumber \\ \label{expansion_fd}  \\[2mm]
 \hspace*{-1cm}
   g_d(\gamma)   & = &
     \hphantom{1}  \,-\, {d \over 4} \gamma^2  \,-\,
       {3d (2d-1) \over 32} \gamma^4  \,-\,
       {5d (6d^2 - 9d + 4)  \over 96}  \gamma^6 
\nonumber \\  & & \,-\,
{35d (24d^3 - 72d^2 + 82d - 33) \over 1024}  \gamma^8 
 \,-\, 
{63d (120d^4 - 600d^3 + 1250d^2 - 1225d + 456)\over 2560} \gamma^{10} 
\nonumber \\ & & \,-\, {77d(720d^5 - 5400d^4
 + 17700d^3 - 30600d^2 + 27041d -9460)\over 4096}
\gamma^{12} \nonumber \\ & & \,-\, {429d(5040d^6 - 
52920d^5 + 249900d^4 - 661500d^3 + 1011017d^2 -
 826336d + 274800)\over 28672}
\gamma^{14} 
 \,-\, \ldots
   \nonumber \\ \label{expansion_gd}
\end{eqnarray}
We remark that for {\em integer}\/ $d \ge 1$,
the Taylor coefficients of $f_d$ and $-g_d$ are all positive,
as can be seen by expanding \reff{def1_fd} and \reff{def1_gd}.
But our numerical computations of the expansions
\reff{expansion_fd}/\reff{expansion_gd}
up to order $\gamma^{22}$
suggest that this property holds {\em only}\/ for integer $d$
(that is, for every noninteger $d$, at least one of the Taylor coefficients
 is negative).

As $\gamma \uparrow 1/d$, we have the asymptotic expansions
\be
   f_d(\gamma)   \;=\;
   \cases{ (d/2\pi)^{d/2} \Gamma(1 - {d \over 2})
               (1-d\gamma)^{(d-2)/2} + O(1)   & for $0 < d < 2$  \cr
           \noalign{\vskip 2mm}
           -{1 \over \pi} \log(1-2\gamma) \,+\, {3 \over \pi} \log 2 \,+\,
               O\Bigl( (1-2\gamma) \log(1-2\gamma) \Bigr)
                                              & for $d=2$        \cr
           \noalign{\vskip 2mm}
           f_* \,+\, (d/2\pi)^{d/2} \Gamma(1 - {d \over 2})
               (1-d\gamma)^{(d-2)/2} + O(1-d\gamma)   & for $2 < d < 4$  \cr
         }
\label{f_d_asymptotic}
\ee

The limiting value $f_*(d) \equiv f_d(1/d)$ is infinite for $d \le 2$
and finite for $d > 2$.
It has the known exact value for $d=3$
\cite{Watson_39,Glasser_77}\footnote{
   The formula in terms of the elliptic integral
   is due to Watson \cite{Watson_39};
   the formula in terms of four gamma functions
   is due to Glasser and Zucker \cite{Glasser_77};
   the reduction to a formula involving only two gamma functions
   is due to Borwein and Zucker \cite{Borwein_92}.
   See also \cite{Glasser_00} for recent extensions.
   (N.B.: The original paper \cite{Glasser_77} has a typographical error
    in the numerical prefactor, as does \cite[p.~21]{Itzykson-Drouffe}.
   The correct prefactor, given here in \reff{fstar3_b},
   has been reported in \cite{Glasser_00}, \cite[p.~126]{Doyle_84}
   and \cite[erratum]{Hara-Slade-Sokal}.)
}
\begin{subeqnarray}
   f_*(3)   &=&  {12 \over \pi^2} \,
                 (18 + 12\sqrt{2} - 10\sqrt{3} - 7\sqrt{6} ) \,
                 K^2 [ ( 2 - \sqrt{3} ) ( \sqrt{3} - \sqrt{2} ) ]   \qquad 
 \\[2mm]
            &=&  {\sqrt{6} \over 32 \pi^3} \,
     \Gamma(\smfrac{1}{24}) \, \Gamma(\smfrac{5}{24}) \,
     \Gamma(\smfrac{7}{24}) \, \Gamma(\smfrac{11}{24})
   \slabel{fstar3_b} \\[2mm]
            &=&  {\sqrt{3} - 1 \over 32 \pi^3} \,
     \Gamma^2(\smfrac{1}{24}) \, \Gamma^2(\smfrac{11}{24})
   \\[2mm]
   & \approx &    1.51638\,60591\,51978
\label{f_*(3)}
\end{subeqnarray}
and the following approximate values for higher dimensions:
\begin{eqnarray}
   f_*(4)   &\approx&  1.23946\,71218      \\
   f_*(5)   &\approx&  1.15630\,81248      \\
   f_*(6)   &\approx&  1.11696\,33732
\end{eqnarray}
As $d \to \infty$ we have the asymptotic expansion
\be
   f_*(d)   \;=\;
      1 \,+\,  {1 \over 2d}  \,+\,
               {3 \over 4d^2}  \,+\,
               {3 \over 2d^3}  \,+\,
               {15 \over 4d^4}  \,+\,
               {355 \over 32d^5} \,+\,
               {595 \over 16d^6} \,+\, \ldots   \;.
\label{f_d_infinity}
\ee

The limiting slope
$f_*'(d) \equiv \lim\limits_{\gamma \uparrow 1/d} df_d(\gamma )/d\gamma$
is infinite for $d \le 4$ and finite for $d>4$.
Its asymptotic expansion as $d \to \infty$ is
\be
   f_*'(d)   \;=\;
      1 \,+\,  {3 \over d}  \,+\,
               {39 \over 4d^2}  \,+\,
               {285 \over 8d^3}  \,+\,
               {2325 \over 16d^4}  \,+\, \ldots   \;.
\ee
The limiting second derivative 
$f_*''(d) \equiv \lim\limits_{\gamma \uparrow 1/d} d^2 f_d(\gamma )/d\gamma^2$
is infinite for $d \le 6$ and finite for $d>6$.
Its asymptotic expansion as $d \to \infty$ is
\be
   f_*''(d)   \;=\;
      d\,+\, 9  \,+\,
               {207 \over 4d}  \,+\,
               {2265 \over 8d^2}  \,+\, \ldots   \;.
\ee

The limiting value $g_*(d) \equiv g_d(1/d)$
is finite in all dimensions $d>0$.
It has the exact values
\begin{eqnarray}
   g_*(1)   &=&  -\log 2 \;\approx\; -0.693147   \\[2mm]
   g_*(2)   &=&  {4G \over \pi} - 2 \log 2 \;\approx\; -0.220051
\end{eqnarray}
where $G \approx 0.915966$ is Catalan's constant,
and the following approximate values for higher dimensions:
\begin{eqnarray}
   g_*(3)   &\approx&  -0.11837\,01593     \\
   g_*(4)   &\approx&  -0.07973\,38971      \\
   g_*(5)   &\approx&  -0.06009\,70332      \\
   g_*(6)   &\approx&  -0.04827\,96878
\end{eqnarray}
Finally, as $d \to \infty$ we have the asymptotic expansion
\be
   g_*(d)   \;=\;
                     \,-\,  {1 \over 4d}  \,-\,
               {3 \over 16d^2}  \,-\,
               {7 \over 32d^3}  \,-\,
               {45 \over 128d^4}  \,-\,
               {269 \over 384d^5} \,-\,
               {805 \over 512d^6} \,+\, \ldots 
\ee

\subsection{Numerical Calculation of $f_d(\gamma )$ and its Derivatives}

In this section we discuss briefly how to make accurate numerical
computations of $f_d(\gamma)$ and its derivatives.
What is slightly tricky is to maintain accuracy
when $|\gamma|$ is near $1/d$.

The function $f_d(\gamma)$ and its first and second derivatives
can be written in the following convenient way:
\begin{eqnarray}
   f_d(\gamma)  &=& A_0 (\gamma ,d)  \\
\gamma  f_d'(\gamma ) &=& A_1(\gamma ,d) - A_0 (\gamma ,d)  \\
\gamma^2  f_d''(\gamma ) &=& A_2 (\gamma ,d) - 4 A_1 (\gamma ,d)
 + 2 A_0 (\gamma ,d)
\end{eqnarray}
where
\be
A_k (\gamma ,d)  \;=\;
    \int\limits_{0}^{\infty} t^k e^{-t} I_0(\gamma t)^d \, dt \;.
 \label{aos001}
\ee
For large values of $t$, the integrand behaves as $t^k e^{-(1-\gamma d)t}$,
so that the integral converges slowly when $\gamma$ is close to $1/d$.
To improve the convergence, we split the interval of $t$-integration into
two subintervals $[0,a]$ and $[a,\infty)$, where $a$ is chosen in such a way 
that for $z \ge a/d$ the function $I_0(z)^d$ is well approximated
by its large-$z$ asymptotic expansion
\be
I_0(z)^d \;\approx\; {e^{dz}\over (2\pi z)^{d/2}}\left[ 1 + {d\over 8 z} +
 {d(d+8)\over 128 z^2 } + {d(d^2+24 d +200)\over 3072 z^3} +
 O \left( 1/z^4\right) \right]\,.
\ee 
Introducing the subtracted quantities
\begin{eqnarray}
F_0 (z,d)  & = &  I_0(z)^d \,-\,  {e^{dz}\over (2\pi z)^{d/2}}   \\[4mm]
F_1 (z,d)  & = &  I_0(z)^d \,-\,
   {e^{dz}\over (2\pi z)^{d/2}} \left( 1 + {d\over 8 z}\right)   \\[4mm]
F_2 (z,d)  & = &  I_0(z)^d \,-\,
   {e^{dz}\over (2\pi z)^{d/2}} \left( 1 + {d\over 8 z} + {d(d+8)\over 128 z^2}
                                \right)
\end{eqnarray}
we can then rewrite  \reff{aos001} as
\begin{eqnarray}
A_0 (\gamma,d) &=& \int\limits_{0}^{a} e^{-t} I_0(\gamma t)^d \, dt \,+\, 
 \int\limits_a^{\infty} e^{-t} F_0 ( \gamma t, d) \, dt
 \,+\, {(1-\gamma d)^{{d\over 2}-1}\over (2\pi \gamma )^{{d\over 2}}}
 \, \Gamma\!\left( 1 - \smfrac{d}{2}, a(1-\gamma d) \right)
     \nonumber \\ \\[4mm]
A_1 (\gamma,d) &=& \int\limits_0^a t e^{-t} I_0(\gamma t)^d \, dt \,+\, 
 \int\limits_a^{\infty} e^{-t}  t F_1 ( \gamma t, d) \, dt \,+\,
 {(1-\gamma d)^{{d\over 2}-2}\over (2\pi \gamma )^{{d\over 2}}}
 \, \Gamma\!\left( 2 - \smfrac{d}{2}, a(1-\gamma d)\right) \nonumber \\
 & & +\; {d\over 8\gamma} \,
         {(1-\gamma d)^{{d\over 2}-1}\over (2\pi \gamma )^{{d\over 2}}}
 \, \Gamma\!\left( 1 - \smfrac{d}{2}, a(1-\gamma d)\right)
     \\[4mm]
A_2 (\gamma,d) &=& \int\limits_0^a e^{-t} t^2  I_0(\gamma t)^d \, dt \,+\, 
 \int\limits_a^{\infty} e^{-t} t^2 F_2 ( \gamma t, d) \, dt  \,+\,
 { (1-\gamma d)^{{d\over 2}-3}\over (2\pi \gamma )^{{d\over 2}} }
 \, \Gamma\!\left( 3 - \smfrac{d}{2}, a(1-\gamma d) \right) \nonumber \\
& & +\; {d\over 8 \gamma } \,
        { (1-\gamma d)^{{d\over 2}-2}\over (2\pi \gamma )^{{d\over 2}}  }
 \, \Gamma\!\left( 2 - \smfrac{d}{2}, a(1-\gamma d) \right) \nonumber \\
& & +\; {d (d+8)\over 128 \gamma^2 } \,
        { (1-\gamma d)^{{d\over 2}-1}\over (2\pi \gamma )^{{d\over 2}}}
 \, \Gamma\!\left( 1 - \smfrac{d}{2}, a(1-\gamma d) \right)
\end{eqnarray}
where 
\be
\Gamma (a,z) \;=\;\int\limits_z^{\infty} t^{a-1}e^{-t} \, dt
\ee
is the incomplete gamma-function. 
In particular, at $\gamma = 1/d$ we have
\be
f_* (d) \;\equiv\; f_d(1/d) \;=\;
   \int\limits_{0}^{a} e^{-t} I_0(t/d)^d \, dt \,+\,
 \int\limits_a^{\infty} e^{-t} F_0(t/d,d) \, dt
 \,+\, {2 \over d-2} \left( d\over 2\pi \right)^{{d\over 2}} a^{1-{d\over 2}}
\ee
for $d>2$.  Using the above formulae, one can compute 
$f_*(3)\approx 1.51638 60591520$ and compare it with the exact
value  \reff{f_*(3)}.

Similar techniques can be used for $g_d(\gamma)$,
but they are not really needed unless $d$ is very close to zero.

\bigskip

{\bf Remark.}
An alternative method for computing
these and similar quantities to very high precision,
with rigorous error bounds, is given in \cite[Appendix B]{Hara-Slade}.
See also \cite[Appendix A and erratum]{Hara-Slade-Sokal} for a brief summary.

\section{Confluent Hypergeometric Function $\ofo (a; b; b\xi )$ in the
   Limit $b\rightarrow \infty$ }    \label{appendix_c}

\subsection{Asymptotic Expansion in Powers of $1/b$ at Fixed $\xi$}
  \label{appendix_c.1}

\begin{figure}[p]
\begin{center}
\epsfbox{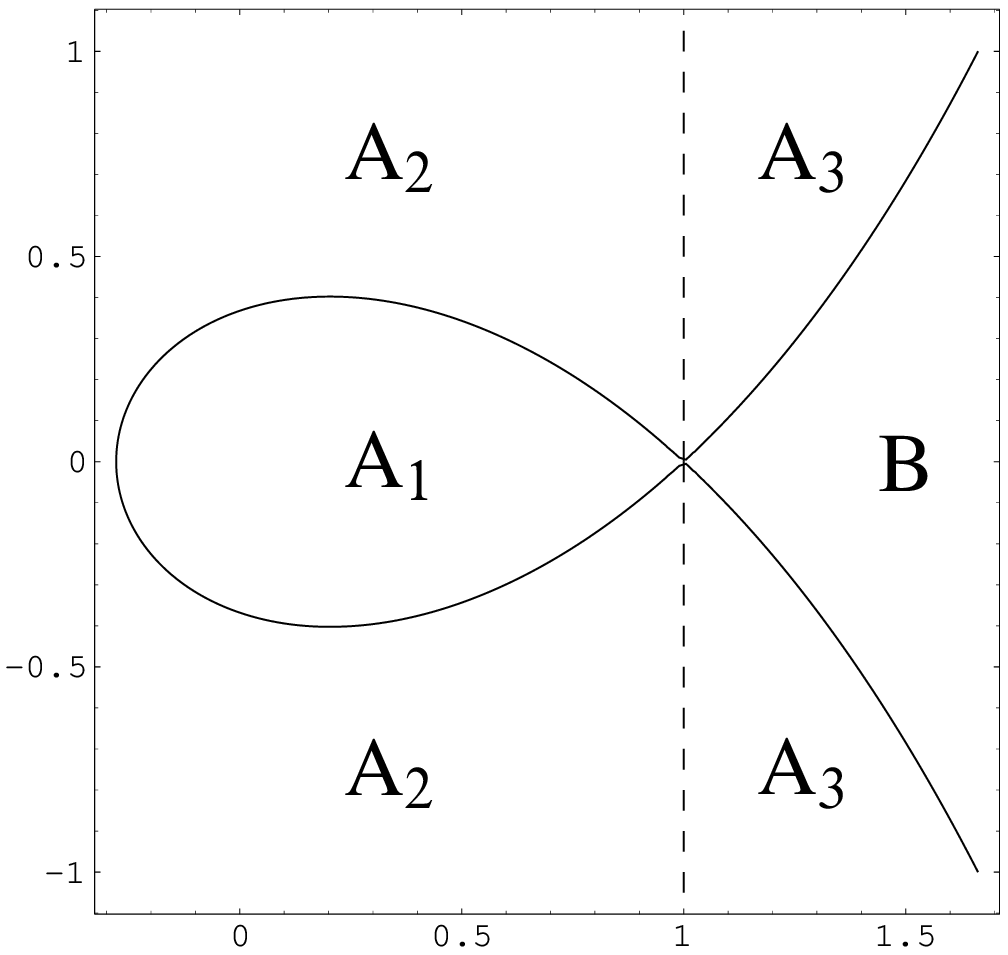}
\end{center}
\caption{
   Szeg\"{o} curve $x^2+y^2=e^{2(x-1)}$ in the complex $\xi = x+iy$ plane:
   we denote by $\scrs_-$ (resp.\ $\scrs_+$)
   the part of this curve at $x < 1$ (resp.\ $x \ge 1$).
   The regions $\scra_1$, $\scra_2$, $\scra_3$ and $\scrb$ are indicated.
}
\label{fig_szego_regions}
\end{figure}

In this appendix we shall develop an asymptotic expansion
for the confluent hypergeometric function $\ofo (a; b; b\xi )$
in the limit $b \to +\infty$,
where $a$ is a fixed positive real number
and $\xi = x +iy$ is a fixed complex number.
Our analysis will be based on the integral representation
\be
   \ofo(a;b;z)  \;=\;
 {\Gamma(b)  \over  \Gamma(a) \Gamma(b-a)}
          \int\limits_0^1  e^{z t} \, t^{a-1} \, (1-t)^{b-a-1} \, dt
     \qquad[\hbox{for }  \real b > \real a > 0]       
  \label{eq.a1}
\ee
combined with Laplace's method for contour integrals
\cite[sections 4.6--4.10]{Olver_74}.
We have
\be
   \ofo(a;b;b\xi)  \;=\;
C(a,b) \int\limits_0^1  e^{bF(t)} \, t^{a-1} \, (1-t)^{-a-1} \, dt      
  \label{eq.a2} \;,
\ee
where
\begin{eqnarray}
C(a,b) & = & {\Gamma(b)  \over  \Gamma(a) \Gamma(b-a)}  \label{eq.a3} \\[2mm]
F(t)   & = &  \xi t + \log (1-t)                        \label{eq.a4}
\end{eqnarray}
The path of integration in \reff{eq.a2} runs initially
along the straight line segment from $t=0$ to $t=1$,
but we are free to deform the contour of integration in the complex $t$-plane.  Writing $t=u+iv$, we have
\begin{subeqnarray}
  {\real F}(u,v)  & = &  xu - yv + \smfrac{1}{2} \log[(1-u)^2 +v^2]
     \\[2mm]
  {\imag F} (u,v) & = &  xv + yu -\arctan\!\left( \frac{v}{1-u} \right)
\end{subeqnarray}
In particular, at the endpoints we have
$F(0) = 0$ and $\real F(1) =-\infty$.
There is a unique saddle point $F'(t_\star) = 0$ at $t_\star = 1- 1/\xi$, i.e.
\begin{subeqnarray}
u_{\star} &=&  1 - x/(x^2+y^2)  \\
v_{\star} &=&  y/(x^2+y^2)
 \label{eq.a_uvstar}
\end{subeqnarray}
The value of $F$ at this saddle point is
\be
    F(t_{\star})  \;=\; \xi -1 -\log \xi  \;,
\ee
and we have in particular
\be
    \real F(t_{\star})  \;=\; x -1 - \smfrac{1}{2} \log (x^2 + y^2) \;.
\ee
For reasons that we will explain shortly,
the curve $\real F(t_{\star}) = 0$
plays a special role in characterizing the asymptotic behavior of
$\ofo (a; b; b\xi )$ in the limit $b \rightarrow \infty$.
Let us therefore denote by $\scrs_-$ (resp.\ $\scrs_+$)
the part of the curve
\be
   x -1 - \smfrac{1}{2} \log (x^2 + y^2)   \;=\;  0
\ee
that lies in the region $x<1$ (resp.\ $x \ge 1$),
and write $\scrs = \scrs_- \cup \scrs_+$
(see Figure \ref{fig_szego_regions}).\footnote{
   This curve was introduced by Szeg\"o \cite{Szego_24}
   in connection with the zeros of the partial sums of
   the Taylor series of $e^z$
   [that is, $s_n(z) = \sum_{k=0}^n z^k/k!$]
   and of the corresponding remainders $e^z - s_n(z)$.
   See also \cite{Saff_76,Saff_78,Pritsker_97}.
}
To the left of $\scrs_+$, i.e.\ in the region
\be
   \scra  \;\equiv\;
   \{ \xi = x+iy \colon\;  x < 1  \hbox{ or }
                           x -1 - \smfrac{1}{2} \log (x^2 + y^2) < 0 \}
\;,
\ee
we shall show that $\ofo (a; b; b\xi )$ has a finite limit as $b \to \infty$.
To the right of $\scrs_+$, i.e.\ in the region
\be
   \scrb  \;\equiv\;
   \{ \xi = x+iy \colon\;  x > 1  \hbox{ and }
                           x -1 - \smfrac{1}{2} \log (x^2 + y^2) > 0 \}
\;,
\ee
we shall show that $\ofo (a; b; b\xi )$ grows exponentially as $b \to \infty$.
More precisely, we shall prove the following asymptotic expansions:

\begin{theorem}
  \label{thmC.1}
\quad\par
\begin{itemize}
   \item[(a)]  For $\xi \in \scra$, we have the asymptotic expansion
\begin{eqnarray}
\hspace*{-2cm}
\ofo (a; b; b\xi ) & = &   
   ( 1-\xi)^{-a}   \Biggl\{  1  - 
       \frac{a(a+1)}{2}  \left( \frac{\xi}{1-\xi}\right)^2  b^{-1}
    \nonumber \\
  & &  \hspace*{-2.5cm} + \;
   \Biggl[ {a(a+1)(a+2)(a+3) \over 8}  \left( \frac{\xi}{1-\xi}\right)^4
        \,+\,  {2a(a+1)(a+2) \over 3}  \left( \frac{\xi}{1-\xi}\right)^3
        \,+\,  {a(a+1) \over 2} \left( \frac{\xi}{1-\xi}\right)^2
        \,-\, 1  \Biggr] b^{-2}
    \nonumber \\
   & & \hspace*{-2.5cm} +\, O(b^{-3})   \Biggr\}
 \label{expansion_region_A}
\end{eqnarray}
uniformly for $\xi$ in compact subsets of $\scra$,
where $(1-\xi)^{-a}$ is defined as the principal branch
(i.e.\ $\arg(1-\xi) = 0$ for $\xi$ real $<1$).
   \item[(b)]  For $\xi \in \scrb$, we have the asymptotic expansion
\begin{eqnarray}
  \ofo (a; b; b\xi )   & = &    e^{b(\xi -1 -\log \xi )} \,
     b^{a - {1 \over 2}} \, {\sqrt{2\pi} \over \Gamma(a)} \,
     \xi (\xi -1)^{a-1} \;\times
        \nonumber \\
  & & \qquad \left[ 1 + \frac{1}{12 b}\left( 1 + {12a (a - 1)\over \xi - 1} +
{6(a-1)(a-2)\over (\xi - 1)^2}\right) + O(1/b^2)\right]\nonumber  \\
 \label{expansion_region_B}
\end{eqnarray}
uniformly for $\xi$ in compact subsets of $\scrb$,
where $\log\xi$ and $(\xi-1)^{a-1}$ are defined as the principal branch
(i.e.\ $\arg \xi = \arg(\xi-1) = 0$ for $\xi$ real $> 1$).
\end{itemize}
\end{theorem}

Let us first consider region $\scra$,
which it is convenient to subdivide as follows:
\begin{subeqnarray}
   \scra_1  & \equiv &
      \{ \xi = x+iy \colon\;  x < 1  \hbox{ and }
                           x -1 - \smfrac{1}{2} \log (x^2 + y^2) \ge 0 \}   \\
   \scra_2  & \equiv &
      \{ \xi = x+iy \colon\;  x \le 1  \hbox{ and }
                           x -1 - \smfrac{1}{2} \log (x^2 + y^2) < 0 \}   \\
   \scra_3  & \equiv &
      \{ \xi = x+iy \colon\;  x > 1  \hbox{ and }
                           x -1 - \smfrac{1}{2} \log (x^2 + y^2) < 0 \}
\end{subeqnarray}
Consider first the region $\real \xi \le 1$.
It is easy to see that $\real F(t) < 0$ for all real $t \in (0,1]$,
so the endpoint $t=0$ is the dominant point
on the line segment of integration in \reff{eq.a2}.
By standard theorems on Laplace-type integrals
\cite[section 4.6]{Olver_74},
we obtain \reff{expansion_region_A}
provided that $\xi \neq 1$.
This proves \reff{expansion_region_A} for
$\xi \in \scra_1 \cup \scra_2$.\footnote{
   We remark that in the smaller domain $|\xi| < 1$,
   \reff{expansion_region_A} can alternatively be proven
   by expanding the series representation
   $$
   \ofo (a; b; b\xi )  \;=\;
   \sum_{k = 0}^{\infty} { (a)_k \over (b)_k} \, {(b\xi )^k \over k!}
   $$
   term-by-term in inverse powers of $b$.
   Slater \cite[p.~66]{Slater} mistakenly asserts that this holds
   for bounded $\xi$, while Tricomi \cite[pp.~111--112]{Tricomi}
   correctly states that it works for $|\xi| < 1$.
}

\begin{figure}[p]
\begin{center}
\vspace*{-2cm} \hspace*{-0cm}
\epsfxsize=0.4\textwidth
\leavevmode\epsffile{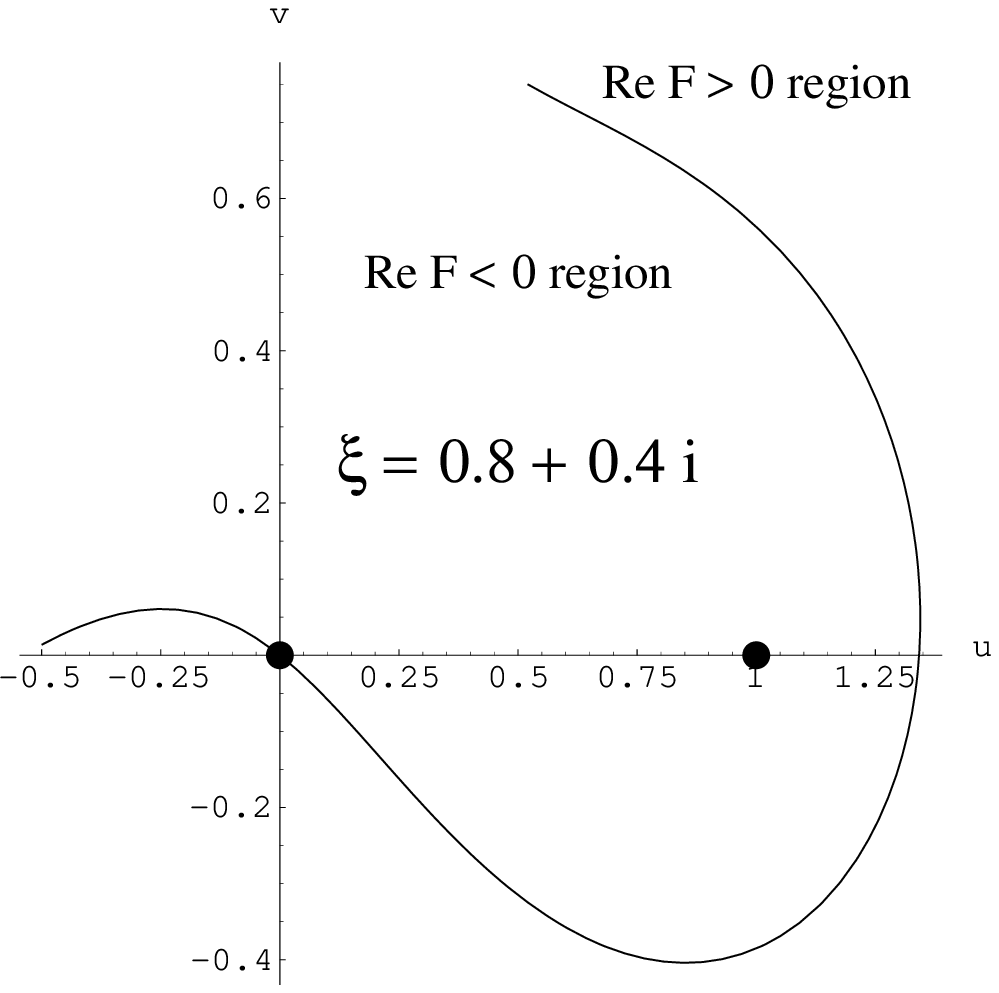}
\hspace{1.5cm}
\epsfxsize=0.4\textwidth
\epsffile{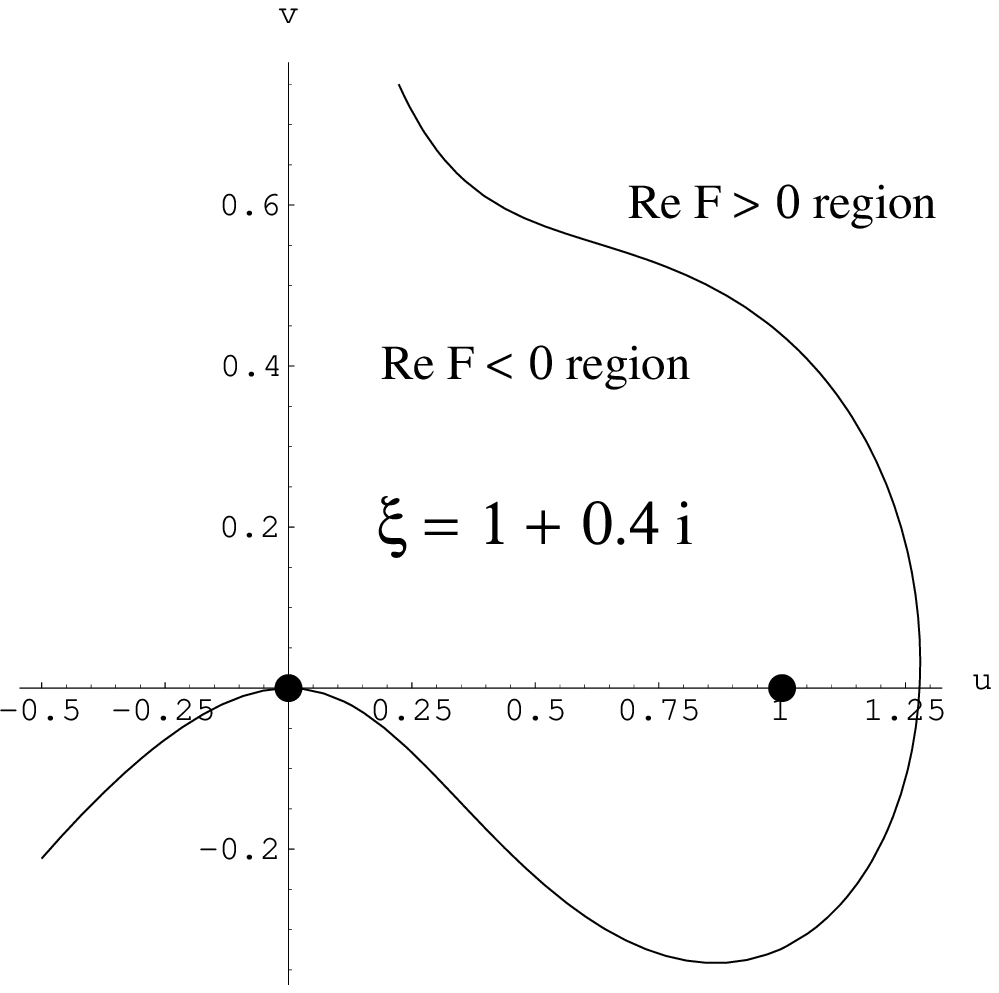}  \\
\vspace*{1cm}
\epsfxsize=0.4\textwidth
\leavevmode\epsffile{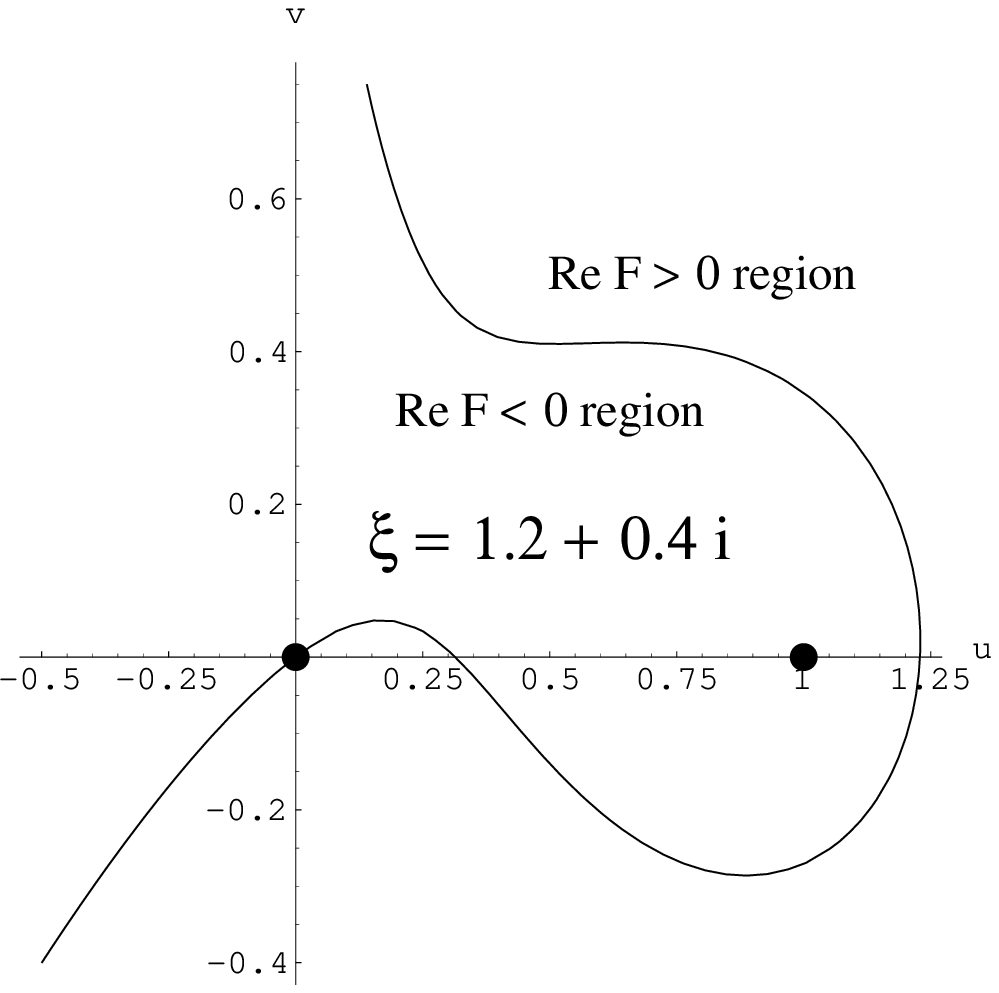}
\hspace{1.5cm}
\epsfxsize=0.4\textwidth
\epsffile{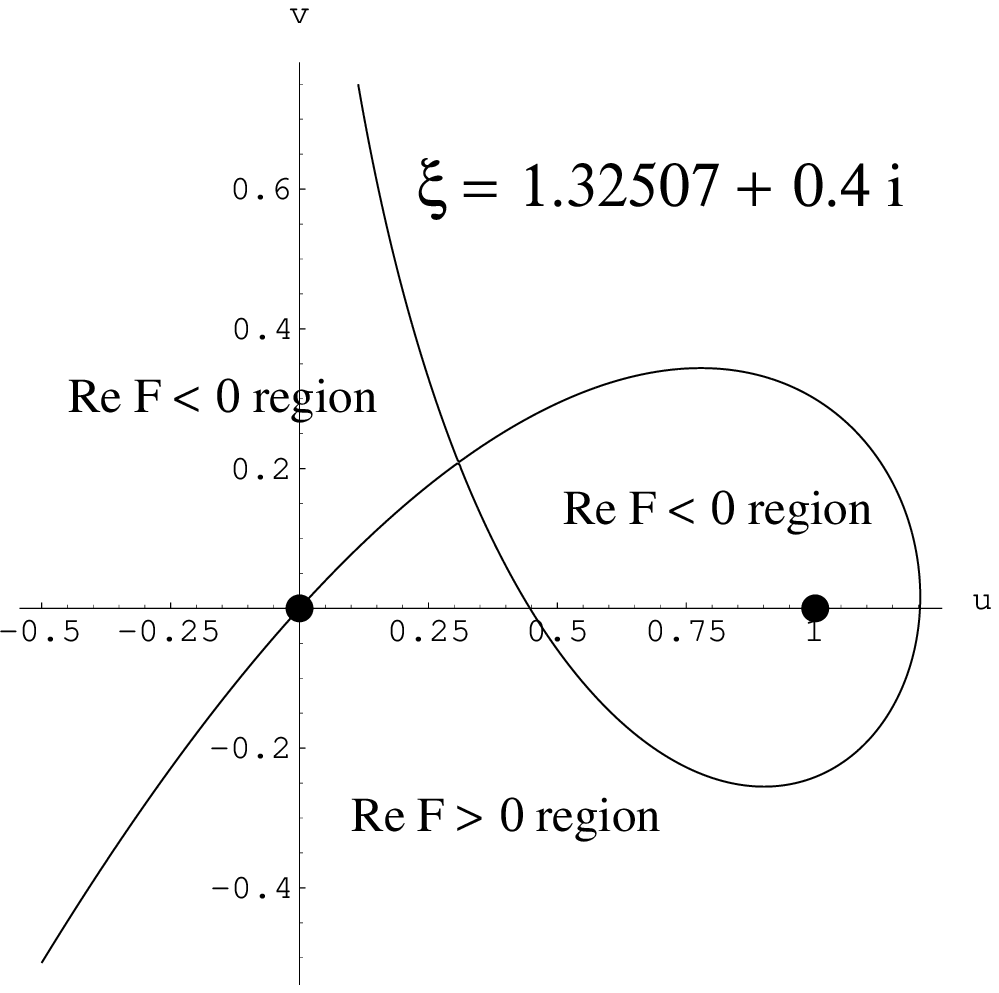}  \\
\vspace*{1cm}
\epsfxsize=0.4\textwidth
\epsffile{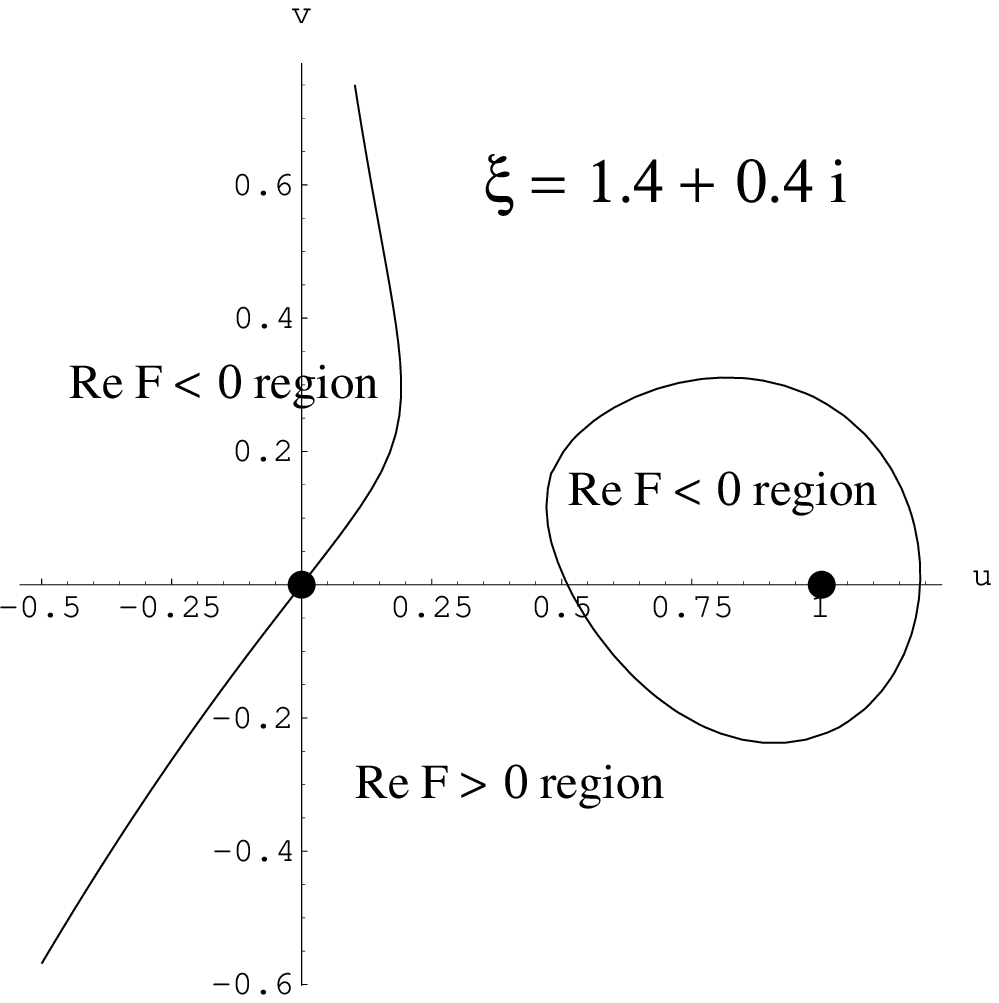} \\
\end{center}
\caption{
   Curve $\real F(t) = 0$ in the complex $t=u+iv$ plane, for
   (a) $\xi \in \scra_2$,
   (b) $\xi$ on the boundary between $\scra_2$ and $\scra_3$,
   (c) $\xi \in \scra_3$,
   (d) $\xi \in \scrs_+$,
   (e) $\xi \in \scrb$.
   The points $t=0$ and $t=1$ can be connected by a path lying in
   the region $\real F < 0$ whenever $t \in \scra$,
   but not otherwise.
}
\label{fig_t-plane_ReF=0}
\end{figure}

Let us now demonstrate that this expansion is valid
in the entire region $\scra$.
If $\real \xi > 1$, the real line segment $0 \le t \le 1$
is no longer a suitable path of integration for \reff{eq.a2},
because we have $\real F(t) > 0$ for small $t>0$,
so that the endpoint $t=0$ is no longer dominant.
What we need to do is to find an alternate contour of integration
${\cal C}$ running from $t=0$ to $t=1$
such that $\real F(t) < 0$ for all $t\in {\cal C}$
except the endpoint $t=0$.
To see that we can indeed connect the endpoints $t=0$ and $t=1$
by such a contour,
let us look at the curve $\real F(u,v) = 0$ in the complex $t$-plane.
Its singular points are given by the
simultaneous solutions of $\real F(u,v) = 0$
and the set of equations
\begin{subeqnarray}
  {\partial\real F(u,v) \over \partial u}  & \equiv &
      x \,+\,  {u-1 \over (u-1)^2+v^2} \;=\; 0    \\[2mm]
  {\partial\real F(u,v) \over \partial v}  & \equiv &
     -y \,+\,  {v   \over (u-1)^2+v^2} \;=\; 0
 \label{eq.a_uv}
\end{subeqnarray}
Now, the unique solution of \reff{eq.a_uv} is given precisely by
the saddle point \reff{eq.a_uvstar};
so singular points of the curve $\real F(u,v) = 0$
can occur only when $\real F(t_\star) = 0$,
i.e.\ when $\xi \in \scrs$
(and in this case the unique singular point is the saddle point of $F$).
Therefore, as we vary $\xi$, the curve $\real F(u,v) = 0$
can change its topological nature only when $\xi$ crosses $\scrs$.
Now, no matter what the value of $\xi$,
the point $t=1$ lies in the region $\real F < 0$,
while the point $t=0$ lies on its boundary.
And we have already shown that when $\xi \in \scra_1 \cup \scra_2$,
the points $t=0$ and $t=1$ can be connected by a path
(namely, the line segment $0 \le t \le 1$)
lying entirely in the region $\real F < 0$
except for the endpoint $t=0$.
And since we can get from $\scra_2$ to anywhere in $\scra_3$
without crossing $\scrs$, it follows that for all $\xi \in \scra_3$
there continues to exist a path ${\cal C}$ from $t=0$ to $t=1$
lying entirely in the region $\real F < 0$
except for the endpoint $t=0$:
see Figure~\ref{fig_t-plane_ReF=0}a--c.
Moreover, the same path ${\cal C}$ will work
for some neighborhood of the initial $\xi$.
Therefore, the endpoint $t=0$ is again dominant,
the standard theorems apply,
and we obtain the same asymptotic expansion \reff{expansion_region_A};
moreover, it is valid uniformly on compact subsets of $\scra$.

\begin{figure}[p]
\begin{center}
\epsfbox{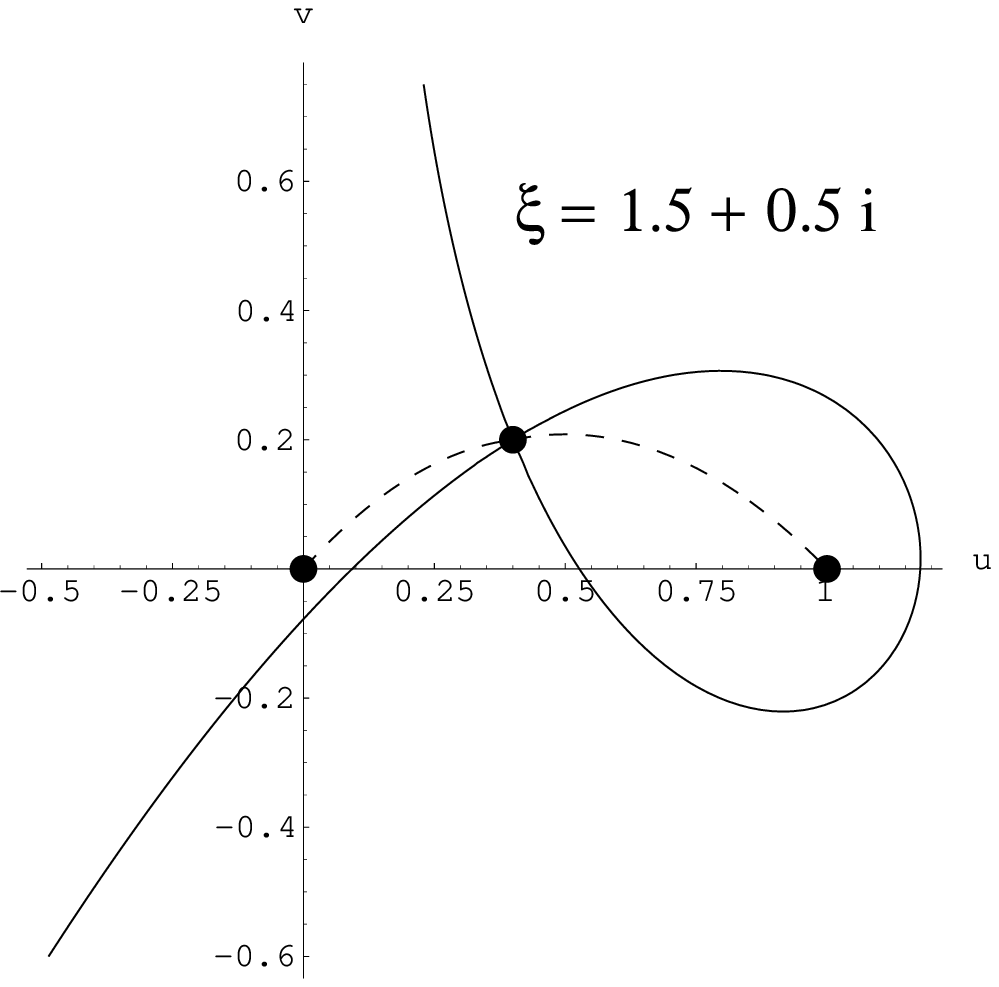}
\end{center}
\caption{
   Curve $\real F(t) = \real F(t_\star)$ in the complex $t=u+iv$ plane,
   for a point $\xi \in \scrb$.
   One possible path $\scrc$ is shown in dashes:
   it runs from $t=0$ to the saddle point $t_\star$ and thence to $t=1$,
   staying entirely in the region $\real F(t) < \real F(t_\star)$
   except at $t=t_\star$.
}
\label{fig_t-plane_ReF=ReFtstar}
\end{figure}

Let us now look at the case when $\xi \in \scrb$.
In this case, $\real F(t_{\star}) > 0$,
so the saddle point dominates over both endpoints
[namely, $\real F(0) = 0$ and $\real F(1) = -\infty$].
It suffices to find a path ${\cal C}$ running from $t=0$ through
the saddle point $t_\star$ and thence to $t=1$,
such that $\real F(t) < \real F(t_\star)$ for all $t\in {\cal C}$
except $t=t_\star$.
Making the change of variables $\widehat{t} = \xi (1-t)$
[valid for $\xi \neq 0$], we find that
\be
   F(t) \,-\, F(t_\star)   \;=\;  -(\widehat{t} - 1 - \log\widehat{t})
   \;,
\ee
so the curve $\real F(t) = \real F(t_\star)$
is precisely a dilated and rotated version of the Szeg\"o curve $\scrs$.
Moreover, the image of $t=0$ (namely, $\widehat{t} = \xi$) lies in $\scrb$,
while the image of $t=1$ (namely, $\widehat{t} = 0$) lies in $\scra_1$.
So each of these two points can be connected to the saddle point $t=t_{\star}$
(whose image is $\widehat{t} = 1$)
by a path whose image lies in $\scrb$ and $\scra_1$, respectively:
see Figure~\ref{fig_t-plane_ReF=ReFtstar}.
This provides the needed path ${\cal C}$.
\qed

\subsection{Complex Zeros of $\ofo (a; b; b\xi )$ as $b\to\infty$}
   \label{appendix_c.2}

We can now use some general results on the limit sets of
complex zeros for certain sequences of analytic functions,
proved recently by one of the authors \cite{Sokal_chromatic_roots},
to deduce from the asymptotic expansions
\reff{expansion_region_A}/\reff{expansion_region_B}
the existence of zeros of $\ofo (a; b; b\xi)$
near the Szeg\"o curve $\scrs_+$ for large $b$.

We begin with a ``real'' form of the theorem:

\begin{theorem}
   \label{thm3.1}
Let $D$ be a domain in $\C$, and let $x_0 \in D \cap \R$.
Let $(f_m)$ be analytic functions on $D$,
and let $(a_m)$ be positive real constants
such that $(|f_m|^{a_m})$ are uniformly bounded on compact subsets of $D$.
Assume further that
\begin{itemize}
   \item[(a)]  For each $x \in D \cap \R$, $f_m(x)$ is real and $> 0$.
   \item[(b)]  For each $x \in D \cap \R$,
     $\lim\limits_{m\to\infty}   a_m \log f_m(x) \equiv g(x)$
     exists and is finite.
   \item[(c)]  $g$ is {\em not}\/ real-analytic at $x_0$.
\end{itemize}
Then, for all $m$ sufficiently large,
there exist zeros $z^*_m$ of $f_m$
such that $\lim\limits_{m\to\infty} z^*_m = x_0$.
\end{theorem}

Theorem~\ref{thm3.1} is in fact well-known to
workers in mathematical statistical mechanics:
it is the contrapositive of an observation going back to
Yang and Lee \cite{Yang-Lee_52} concerning the genesis of phase transitions
(see e.g.\ \cite[Theorem 4.1, p.~51]{Griffiths_72}).
The proof of Theorem~\ref{thm3.1} is, in fact,
a direct transcription of the intuition sketched in Section~\ref{sec6}.
We first need a standard lemma from complex analysis:

\begin{lemma}[log exp Vitali]   \label{lemma3.2}
Let $D$ be a connected open subset of $\C^n$,
and let $\{ g_m \} _{m=1}^\infty$ be a family of analytic functions on $D$
such that
\begin{quote}
\begin{itemize}
  \item[(a)]  For each compact $C \subset D$,
              $\sup\limits_m \sup\limits_{z \in C} \real g_m(z)  <  \infty$
\end{itemize}
\end{quote}
and either
\begin{quote}
\begin{itemize}
  \item[(b1)]  For some determining set $S \subset D$,
     $\lim\limits_{m\to\infty} g_m(z)$ exists for all $z \in S$
  \item[or (b2)]  For some $z_0 \in D$ and all multi-indices $\alpha$,
     $\lim\limits_{m\to\infty} (D^\alpha g_m)(z_0)$ exists.
\end{itemize}
\end{quote}
Then $g_m$ converges, uniformly on compact subsets of $D$,
to an analytic function $g$.
\end{lemma}

\noindent
We recall that $S \subset D$ is called a {\em determining set}\/ for $D$
if $g$ analytic on $D$ and $g \restrict S \equiv 0$ implies $g \equiv 0$.
For example, any real environment
(i.e.\ a set of the form $z_0 + U$, where $U$ is a nonempty open subset of
$\R^n$) is a determining set.
For $n=1$, any set $S$ having an accumulation point in $D$
is a determining set.
A proof of Lemma~\ref{lemma3.2} can be found in \cite[p.\ 343]{Simon_74};
it is a special case of a vastly more powerful result, due to Montel (1912),
on normal families of analytic functions \cite{Montel_27,Schiff_93}.

\bigskip\par\noindent
{\sc Proof of Theorem~\ref{thm3.1}.\ }
Suppose the contrary, i.e.\ suppose that there exists an $\epsilon > 0$
and an infinite sequence $m_1 < m_2 < \ldots$ such that
none of the functions $f_{m_i}$ has a zero in the set
$D_\epsilon = \{ z \in \C \colon\; |z-x_0| < \epsilon \}  \subset D$.
Then, since $D_\epsilon$ is simply connected,
$\log f_{m_i}$ is analytic in $D_\epsilon$
(we take the branch that is real on $D_\epsilon \cap \R$),
and $g_i \equiv (1/a_{m_i}) \log f_{m_i}$ satisfies
$\lim_{i\to\infty} g_i(x) = g(x)$ for $x \in D_\epsilon \cap \R$.
Therefore, by Lemma \ref{lemma3.2} it follows that there exists an
analytic function $\widetilde{g}$ on $D_\epsilon$ that coincides
with $g$ on $D_\epsilon \cap \R$.  But this contradicts hypothesis (c).
\qed

The proof of Theorem \ref{thm6.1} is now an immediate consequence
of \reff{eq6.10} [or \reff{expansion_region_A}/\reff{expansion_region_B}],
once we observe that $|\ofo(a;c;z)| \le e^{|\real z|}$
and that $\ofo(a;c;z) > 0$ for $z$ real.

\bigskip

Let us now state without proof a ``complex'' analogue of \reff{thm3.1}:

\begin{theorem}[Sokal \protect\cite{Sokal_chromatic_roots}]
   \label{thm3.2}
Let $D$ be a domain in $\C$, and let $z_0 \in D$.
Let $(f_m)$ be analytic functions on $D$,
and let $(a_m)$ be positive real constants
such that $(|f_m|^{a_m})$ are uniformly bounded on compact subsets of $D$.
Suppose that there does {\em not}\/ exist a neighborhood $U \ni z_0$
and a function $v$ on $U$ that is either harmonic or else identically $-\infty$
such that $\liminf\limits_{m\to\infty} a_m \log |f_m(z)| \le v(z)
 \le \limsup\limits_{m\to\infty} a_m \log |f_m(z)|$ for all $z \in U$.
Then, for all $m$ sufficiently large,
there exist zeros $z^*_m$ of $f_m$
such that $\lim\limits_{m\to\infty} z^*_m = z_0$.
\end{theorem}

\noindent
The proof, which is not difficult,
can be found in \cite{Sokal_chromatic_roots};
it uses elementary facts from the theory of normal families
\cite{Montel_27,Schiff_93}.

We can use Theorem~\ref{thm3.2} to deduce a vast extension of
Theorem \ref{thm6.1} concerning the zeros of $\ofo (a; b; b\xi)$ for large $b$.
Indeed, from the leading term of the asymptotic expansions
\reff{expansion_region_A}/\reff{expansion_region_B},
we have
\be
   \lim_{b \to +\infty} {1 \over b} \log |\ofo(a;b;b\xi)|
   \;=\;
   \cases{0   & for $\xi \in \scra$   \cr
          \noalign{\vskip 2mm}
          \real(\xi - 1 - \log\xi)   & for $\xi \in \scrb$ \cr
         }
\ee
We can then apply Theorem~\ref{thm3.2} to any point $\xi_0 \in \scrs_+$
(which is the boundary between $\scra$ and $\scrb$) to deduce that:

\begin{theorem}
   \label{thm_1F1_complex_zeros}
Fix $a > 0$ and $\xi_0 \in \scrs_+$.
Then there exist, for all sufficiently large $b > 0$,
complex numbers $z^*_b$ satisfying $\ofo(a;b;z^*_b) = 0$
and $\lim\limits_{b\to\infty} (z^*_b/b) = \xi_0$.
\end{theorem}

To the best of our knowledge, neither Theorem~\ref{thm_1F1_complex_zeros}
nor its real specialization, Theorem~\ref{thm6.1},
can be found in the extensive literature on confluent hypergeometric functions
(see \cite{Buchholz,Tricomi,Slater,Tsvetkoff_41,Kazarinoff_57,Saff_76,Saff_78,%
Temme_78,Pritsker_97,Norfolk_98} among many others).
An old theorem of Tsvetkoff \cite[Theorem 7]{Tsvetkoff_41}
states that for $b > 2a$, every zero of $\ofo(a;b;z)$
lies in the region $\real z > b-2a$
(see also \cite[Proposition 3.2]{Saff_78} for a slight strengthening);
Theorem~\ref{thm6.1} shows that this is asymptotically sharp as $b \to \infty$,
to leading order.

\subsection{Asymptotic Expansion when $\xi = 1 + b^{-1/2} \zeta$}
  \label{appendix_c.3}

In Section \ref{appendix_c.1} we saw that the asymptotic expansion
of $\ofo (a; b; b\xi)$ changes dramatically as $\xi$ crosses
the curve $\scrs_+$,
and in particular on the real axis as it crosses $\xi=1$.
We would now like to look more closely at the crossover behavior
in a neighborhood of the point $\xi=1$.
It turns out that the right ``magnification'' is $b^{1/2}$:
that is, we should look for an expansion of $\ofo (a; b; b\xi)$
when we take $\xi = 1 + b^{-1/2} \zeta$ with $\zeta$ fixed and finite.

\begin{theorem}
  \label{thmC.xxx}
Uniformly for $\zeta$ in compact subsets of the complex plane,
we have the asymptotic expansion
\be
\ofo \left( a; b; b(1 + b^{-1/2} \zeta) \right) 
   \;=\;
  \frac{\Gamma (b)}{\Gamma (b-a)} \, b^{-a/2} \, e^{\zeta^2/4}
  \left[ D_{-a}(-\zeta) + \frac{A_1 (a,\zeta)}{b^{1/2}} +
                         \frac{A_2 (a,\zeta)}{b} + O (b^{-3/2}) \right]
  \;,
\label{B.16}
\ee
where $D_\nu$ is the parabolic cylinder function
\be
   D_\nu(z)  \;=\;  {e^{-z^2/4} \over \Gamma(-\nu)} \,
      \int\limits_0^\infty e^{-zs - s^2/2} s^{-\nu-1} \, ds
\ee
and
\begin{eqnarray}
A_1 (a,\zeta) & = &  a(a+1)D_{-a-1} (-\zeta) -\frac{a(a+1)(a+2)}{3}
                         D_{-a-3}(-\zeta)
   \nonumber \\  \\[4mm]
A_2 (a,\zeta) & = &  \frac{a(a+1)^2 (a+2)}{2} D_{-a-2}(-\zeta)  
  \,-\, \frac{\Gamma (a+4)}{\Gamma (a)}\left( \frac{1}{4}+\frac{a+1}{3}\right)
        D_{-a-4}(-\zeta)
   \nonumber \\
   & & \qquad\qquad\qquad\qquad\qquad\qquad
  \,+\,  \frac{\Gamma (a+6)}{18\Gamma (a)} D_{-a-6}(-\zeta)
\end{eqnarray}
\end{theorem}

\begin{corollary}
  \label{corC.xxx}
For large enough real $b$,
$\ofo (a; b; b\xi )$ has complex zeros
whose asymptotic expansion is given by
\be
\xi_0 = 1 + \frac{\zeta_0}{b^{1/2}} + \frac{\alpha}{b} + \frac{\beta}{b^{3/2}}
+ O (1/b^2 )
\ee
where $\zeta_0$ is any zero of $D_{-a}(-\zeta)$, and
\begin{eqnarray}
\alpha   & = &    a(a+1) { D_{-a-1} (-\zeta_0)  \,-\,
                                \smfrac{1}{3} (a+2) D_{-a-3}(-\zeta_0)
                            \over
                            D_{-a}^{\prime}(-\zeta_0)
                          }
    \label{eqC.alpha} \\[4mm]
\beta & = &  \Biggl\{  \smfrac{1}{2} \alpha^2 D_{-a}'' (-\zeta_0)
      \,+\, \alpha a(a+1) \left[ D_{-a-1}' (-\zeta_0)
              -\smfrac{1}{3}(a+2)D_{-a-3}'(-\zeta_0) \right]  \nonumber  \\
    &  &  \quad+\; 
          \Bigl[ \frac{a(a+1)^2(a+2)}{2}D_{-a-2}(-\zeta_0) \,-\, 
                 \frac{\Gamma(a+4)}{\Gamma(a)}
                    \Bigl( {1\over 4} + {a+1\over 3}\Bigr) D_{-a-4}(-\zeta_0)
                                   \nonumber \\
    & &   \qquad\quad+\; \frac{\Gamma(a+6)}{18\Gamma(a)} D_{-a-6}(-\zeta_0)
          \Bigr] \Biggr\}
  \Bigg/ D_{-a}^{\prime}(-\zeta_0)   \label{eqC.beta}
\end{eqnarray}
\end{corollary}

To prove Theorem~\ref{thmC.xxx},
we start with the integral representation \reff{eq.a2}--\reff{eq.a4}
and set $\xi = 1 + b^{-1/2}\zeta$.  Expanding
\be
   F(t)   \;=\;  \xi t + \log (1-t)
          \;=\;  (\xi-1)t \,-\, \sum\limits_{k=2}^\infty {t^k \over k}
\ee
and making the change of variables $s = b^{1/2}t$,
we can write \reff{eq.a2} in the form
\begin{eqnarray}
  & &
   \ofo(a;b;b(1 + b^{-1/2} \zeta) )  \;=\;
      \nonumber \\[2mm]
  & & \qquad\quad
   C(a,b) \, b^{-a/2} \int\limits_0^{b^{1/2}}
       e^{-s^2/2 + s \zeta}
       \left( 1 - {s\over b^{1/2}}\right)^{-a-1} s^{a-1}
       \exp\!\left[ - \sum\limits_{k=3}^\infty {s^k \over k \, b^{(k-2)/2}}
             \right]
       \, ds    \;.
      \nonumber \\
  \label{eqC.3.integral}
\end{eqnarray}
The blow-up of the term $(1-s/b^{1/2})^{-a-1}$ as $s \uparrow b^{1/2}$
is harmless, as it is more than compensated (for large $b>0$)
by the suppression contained in the $\exp[- \,\cdots]$ factor
[as can be seen by examining \reff{eq.a1} as $t \uparrow 1$].
Therfeore, \reff{eqC.3.integral} becomes, for large $b$,
\begin{subeqnarray}
   \ofo(a;b;b(1 + b^{-1/2} \zeta) )  & = &
   C(a,b) \, b^{-a/2} \int\limits_0^{\infty}
       e^{-s^2/2 + s \zeta} \, s^{a-1} \, ds \,
       \left[1 + O(b^{-1/2})\right]
   \nonumber \\ \\[4mm]
   & = &
   C(a,b) \, b^{-a/2} \, \Gamma(a) \, e^{\zeta^2/4} D_{-a}(-\zeta) \,
       \left[1 + O(b^{-1/2})\right]    \;.
   \nonumber \\
\end{subeqnarray}
The corrections in powers of $b^{-1/2}$ can be easily extracted,
and we arrive at expansion \reff{B.16}.
Corollary~\ref{corC.xxx} then follows using the implicit function theorem.
\qed


\bigskip



\section{Correlation Inequalities}   \label{app.correq}

In this appendix we prove some correlation inequalities for
$N$-component $\sigma$-models with Hamiltonian of the form
\be
  H (\{\bsigma\})   \;=\;
    -\, \sum_{\< xy \>} \scrv_{xy} ( \bsigma_x \cdot \bsigma_y )
  \;,
 \label{ham:general}
\ee
where each spin $\bsigma_x$ is a unit vector in $\R^N$
and each $\scrv_{xy}$ is a real-valued function on $[-1,1]$.

\subsection{Tools \#1: Griffiths Inequalities for the Ising Model}
                                                          \label{app.correq.1}

Consider an Ising model consisting of finitely many spins
$\varepsilon_x = \pm 1$, with Hamiltonian
\be
   H(\{\varepsilon\})   \;=\;
    -\, \sum_{A} J_{A} \varepsilon^A
  \;.
 \label{ham:Ising}
\ee
Here we have used the notation
$\varepsilon^A = \prod\limits_{x} \varepsilon_x^{A_x}$,
where $A = \{A_x\}$ is an arbitrary multi-index
(i.e.\ a vector of non-negative integers).
Of course, for the Ising model it suffices to use $A_x = 0,1$;
the sum in \reff {ham:Ising} runs over all such multi-indices.
Thus, we have allowed quite general multi-spin interactions.
Of course, magnetic fields and pair interactions are included
as special cases.

We then have the following well-known inequalities:

\begin{theorem}[Griffiths' first and second inequalities]  \label{thmA.1}
Assume that $J_{A} \ge 0$ for all $A$.  Then:
\begin{quote}
\begin{description}
     \item[{[G--I]}]  \quad $\;\, \< \varepsilon^A \> \;\ge\; 0\;$ for all $A$.
     \item[{[G--II]}]  \quad
              $\< \varepsilon^A ; \varepsilon^B \>  \,\equiv\,
               \< \varepsilon^A  \varepsilon^B \>  \,-\,
                  \< \varepsilon^A \> \< \varepsilon^B \>  \;\ge\;  0\;$
        for all $A,B$.
\end{description}
\end{quote}
\end{theorem}

\begin{theorem}[Griffiths' comparison inequality]  \label{thmA.2}
Assume that $J_{A} \ge |J'_{A}|$ for all $A$.  Then
\begin{quote}
\begin{description}
     \item[{[G--Comp]}]  \quad
        $\< \varepsilon^A \> _J  \;\ge\;  
         | \< \varepsilon^A \> _{J'} |  \;$  for all $A$.
\end{description}
\end{quote}
\end{theorem}
Elementary proofs of (vastly generalized versions of)
Theorems \ref{thmA.1} and \ref{thmA.2}
can be found in \cite{Sylvester_76} and \cite{Szasz_78}, respectively.

\subsection{Tools \#2: Ginibre Inequalities for the $XY$ Model}
                                                          \label{app.correq.2}

Next consider a classical $XY$ (plane-rotator) model consisting
of finitely many spins $\theta_x \in [0,2\pi)$,
with Hamiltonian
\begin{subeqnarray}
   H(\{\theta\})   & = &
      -\, \sum_{A} \real( J_{A} e^{iA \cdot \theta} )   \\
       & = &   -\, \sum_{A} |J_{A}| \, \cos(A \cdot \theta + \arg J_A )   \;,
 \label{ham:XY}
\end{subeqnarray}
where the sum runs over all vectors $A = \{A_x\}$ of (signed) integers,
the interactions $J_A$ are arbitrary complex numbers,
and $A \cdot \theta \equiv \sum_x A_x \theta_x$.
We then have the following inequalities:

\begin{theorem}[Ginibre's first and second inequalities]  \label{thmA.3}
Assume that $J_{A} \ge 0$ for all $A$.  Then:
\begin{quote}
\begin{description}
     \item[{[Ginibre--I]}]  \quad $\;\, \< \cos A\cdot\theta \> \;\ge\; 0\;$
         for all $A$.
     \item[{[Ginibre--II]}]  \quad
              $\< \cos A\cdot\theta ; \cos B\cdot\theta \>  \,\equiv\,
               \< \cos A\cdot\theta  \cos B\cdot\theta \>  \,-\,
               {\< \cos A\cdot\theta \> \< \cos B\cdot\theta \>  \;\ge\;  0\;}$
        for all $A,B$.
\end{description}
\end{quote}
\end{theorem}

\begin{theorem}[Ginibre comparison inequality]  \label{thmA.4}
Assume that $J_{A} \ge |J'_{A}|$ for all $A$.  Then
\begin{quote}
\begin{description}
     \item[{[Ginibre--Comp]}]  \quad
        $\< \cos A\cdot\theta \> _J  \;\ge\;  
         | \< \cos (A\cdot\theta + \varphi_A) \> _{J'} |  \;$  for all $A$
         and all (real) angles $\varphi_A$.
\end{description}
\end{quote}
\end{theorem}
Proofs of (generalized versions of)
Theorems \ref{thmA.3} and \ref{thmA.4}
can be found \cite{Ginibre_70} and \cite{Messager_78}, respectively.


\subsection{Correlation Inequalities for $S^{N-1}$ $\sigma$-Models}
                                                         \label{app.correq.3}

We now prove some correlation inequalities for
$N$-component $\sigma$-models.
Our method --- which is in essence due to Fr\"ohlich \cite{Frohlich_79} ---
is to embed an Ising or $XY$ model into the $N$-component model,
and then apply the Griffiths or Ginibre comparison inequality
to this embedded model (which has annealed random couplings).\footnote{
   A slightly weaker version of this method was applied earlier by
   Mack and Petkova \cite{Mack_79}.  They were apparently unaware of the
   Griffiths and Ginibre {\em comparison}\/ inequalities, so they relied
   instead on the (somewhat weaker) Griffiths II and Ginibre II inequalities.
   To satisfy the ferromagnetism hypotheses of these latter inequalities,
   they had to impose a ``no-vortex'' (or ``no-monopole'') constraint.
   This constraint is now seen to be optional but {\em not}\/ necessary.
}

We consider $N$-component $\sigma$-models of the type
\be
  H(\{\bsigma\})   \;=\;
    -\, \sum_{\< xy \>} \scrv_{xy}( \bsigma_x \cdot \bsigma_y )
 \label{ham:general_2}
\ee
(and even more general models).
It is convenient to split $\scrv_{xy}$ into its even and odd parts:
\be
   \scrv^\pm_{xy} (s)   \;=\;   { \scrv_{xy} (s)  \pm  \scrv_{xy} (-s)
                                  \over
                                  2
                                }   \;.
\ee
We begin by proving that correlations in the $\sigma$-model
are bounded above by the corresponding correlations in the Ising model:

\begin{theorem}
 \label{thmA.5}
Consider a $\sigma$-model defined by the Boltzmann-Gibbs measure \break
$Z^{-1} e^{-H(\{\bsigma\})}  d\mu(\{\bsigma\})$,
where $H$ is defined by \reff{ham:general_2},
and $d\mu(\{\bsigma\})$ is an arbitrary nonnegative measure
that is invariant under the $Z_2$ gauge transformations
$\bsigma_x \to \eta_x \bsigma_x$ ($\eta_x = \pm 1$).
Then, for any set of sites $x_1,y_1,\ldots,x_n,y_n$, we have
\begin{subeqnarray}
 \left| \< (\bsigma_{x_1} \!\cdot \bsigma_{y_1}) \cdots
           (\bsigma_{x_n} \!\cdot \bsigma_{y_n})    \> _{H,\mu}   \right|
 & \le &
 \< \varepsilon_{x_1} \varepsilon_{y_1} \cdots
    \varepsilon_{x_n} \varepsilon_{y_n}        \> _{Ising,J}
 \,
 \< |\bsigma_{x_1} \!\cdot \bsigma_{y_1}| \cdots
    |\bsigma_{x_n} \!\cdot \bsigma_{y_n}|      \> _{H,\mu}
                                                      \nonumber \\  \\[1mm]
 & \le &
 \< \varepsilon_{x_1} \varepsilon_{y_1} \cdots
    \varepsilon_{x_n} \varepsilon_{y_n}        \> _{Ising,J}
 \,
 \<  (\bsigma_{x_1} \!\cdot \bsigma_{y_1})^2 \cdots
           (\bsigma_{x_n} \!\cdot \bsigma_{y_n})^2   \> _{H,\mu} ^{1/2}
                                                      \nonumber \\  \\[1mm]
 & \le &
 \< \varepsilon_{x_1} \varepsilon_{y_1} \cdots
    \varepsilon_{x_n} \varepsilon_{y_n}        \> _{Ising,J}
 \;,
\end{subeqnarray}
where
\be
   J_{xy}  \;=\;  \| \scrv^-_{xy} \| _\infty
           \,\equiv\;   \sup_{s \in [-1,1]}  |\scrv^-_{xy}(s)|   \;.
\ee
\end{theorem}

\proof
The proof is based on the trivial identity
\be
   \int F(\{\bsigma\}) \, d\mu(\{\bsigma\})
   \;=\;
   \int\!\int F(\{\varepsilon\bsigma\}) \, d\mu(\{\bsigma\})
                                        \, d\nu(\{\varepsilon\})
   \;,
 \label{trivial_ident}
\ee
where $d\nu(\{\varepsilon\})$ is normalized counting measure on
the configurations of Ising spins $\varepsilon_x = \pm 1$,
and $F$ is an arbitrary function.
This identity expresses the $Z_2$ gauge invariance of $d\mu(\{\bsigma\})$.

Without loss of generality we can assume that each $\scrv_{xy}$
is an odd function,
since the even parts of $\scrv_{xy}$ can be absorbed into $\mu$.

Let us apply \reff{trivial_ident} to
$F(\{\bsigma\}) = (\bsigma_{x_1} \!\cdot \bsigma_{y_1}) \cdots
                (\bsigma_{x_n} \!\cdot \bsigma_{y_n}) e^{-H(\{\bsigma\})}$:
\begin{eqnarray}
& &  \int (\bsigma_{x_1} \!\cdot \bsigma_{y_1}) \cdots
          (\bsigma_{x_n} \!\cdot \bsigma_{y_n}) \, e^{-H(\{\bsigma\})}
                                              \, d\mu(\{\bsigma\})
     \nonumber \\[1mm]
& &  \qquad =\;
     \int\!\int  \varepsilon_{x_1} \varepsilon_{y_1} \cdots
                 \varepsilon_{x_n} \varepsilon_{y_n}
                 \, (\bsigma_{x_1} \!\cdot \bsigma_{y_1}) \cdots
                    (\bsigma_{x_n} \!\cdot \bsigma_{y_n})
                 \; e^{-H(\{\varepsilon\bsigma\})}
                 \, d\mu(\{\bsigma\}) \, d\nu(\{\varepsilon\})
     \nonumber \\[1mm]
& &  \qquad =\;
     \int\! d\mu(\{\bsigma\})
            \, (\bsigma_{x_1} \!\cdot \bsigma_{y_1}) \cdots
               (\bsigma_{x_n} \!\cdot \bsigma_{y_n})
     \int\!  \varepsilon_{x_1} \varepsilon_{y_1} \cdots
             \varepsilon_{x_n} \varepsilon_{y_n}
     \, \prod_{\<xy\>}
    e^{\varepsilon_x \varepsilon_y \scrv^-_{xy}(\bsigma_x \cdot \bsigma_y)}
     \, d\nu(\{\varepsilon\})
     \nonumber \\
 \label{eqnA.9}
\end{eqnarray}
We now perform the integration over $\{\varepsilon\}$;
by the Griffiths comparison inequality (Theorem \ref{thmA.2}) we have 
\begin{eqnarray}
& &  \left| \int \varepsilon_{x_1} \varepsilon_{y_1} \cdots
               \varepsilon_{x_n} \varepsilon_{y_n}
     \, \prod_{\<xy\>}
    e^{\varepsilon_x \varepsilon_y \scrv^-_{xy}(\bsigma_x \cdot \bsigma_y)}
     \, d\nu(\{\varepsilon\})
   \right|   
   \nonumber \\[1mm]
& &  \qquad\qquad \le\;
  \< \varepsilon_{x_1} \varepsilon_{y_1} \cdots
     \varepsilon_{x_n} \varepsilon_{y_n}        \> _{Ising,J}
  \; \int \prod_{\<xy\>}
  e^{\varepsilon_x \varepsilon_y \scrv^-_{xy}(\bsigma_x \cdot \bsigma_y)}
     \, d\nu(\{\varepsilon\})
  \;.
\end{eqnarray}
It follows that
\begin{eqnarray}
& &  \left|
     \int (\bsigma_{x_1} \!\cdot \bsigma_{y_1}) \cdots
          (\bsigma_{x_n} \!\cdot \bsigma_{y_n}) \, e^{-H(\{\bsigma\})}
                                              \, d\mu(\{\bsigma\})
     \right|
     \nonumber \\[1mm]
& &  \quad \le\;
     \< \varepsilon_{x_1} \varepsilon_{y_1} \cdots
        \varepsilon_{x_n} \varepsilon_{y_n}        \> _{Ising,J}
     \int\! d\mu(\{\bsigma\})
            \, | (\bsigma_{x_1} \!\cdot \bsigma_{y_1}) \cdots
                 (\bsigma_{x_n} \!\cdot \bsigma_{y_n}) |
     \, \int\! \prod_{\<xy\>}
    e^{\varepsilon_x \varepsilon_y \scrv^-_{xy}(\bsigma_x \cdot \bsigma_y)}
     \, d\nu(\{\varepsilon\})
     \nonumber \\
& &  \quad =\;
     \< \varepsilon_{x_1} \varepsilon_{y_1} \cdots
        \varepsilon_{x_n} \varepsilon_{y_n}        \> _{Ising,J}
     \,
     \int | (\bsigma_{x_1} \!\cdot \bsigma_{y_1}) \cdots
            (\bsigma_{x_n} \!\cdot \bsigma_{y_n}) | \, e^{-H(\{\bsigma\})}
                                              \, d\mu(\{\bsigma\})
     \;.
     \nonumber \\
\end{eqnarray}
The rest follows from the Schwarz inequality together with
the trivial bound $|\bsigma| \le 1$.
\qed

\begin{corollary}
   \label{corA.5a}
For the mixed isovector/isotensor model
\be
   H  \;=\;  - \sum_{\<xy\>}
     \left[ J_{xy} \bsigma_x \cdot \bsigma_y   \,+\,
            {K_{xy} \over 2}  (\bsigma_x \cdot \bsigma_y)^2
     \right]
   \;,
\ee
we have
\be
 \left| \< (\bsigma_{x_1} \!\cdot \bsigma_{y_1}) \cdots
           (\bsigma_{x_n} \!\cdot \bsigma_{y_n})    \> _{J,K}   \right|
 \;\le\;
 \< \varepsilon_{x_1} \varepsilon_{y_1} \cdots
    \varepsilon_{x_n} \varepsilon_{y_n}        \> _{Ising,|J|}
 \;.
\ee
\end{corollary}

{\bf Remarks.}
1.  The bound of Theorem~\ref{thmA.5} and Corollary~\ref{corA.5a}
holds for arbitrary products $\prod (\bsigma_{x_i} \cdot \bsigma_{y_i})$,
but it is only of interest for iso{\em vector}\/ correlation functions.
For iso{\em tensor}\/ correlation functions
(i.e.\ those in which each spin $\bsigma_x$ appears an {\em even}\/ number
 of times),
the corresponding Ising-model correlation is 1, and the bound is trivial.


2.  The only restriction on the measure $\mu$ is that it be
$Z_2$-gauge-invariant.  An interesting family of examples is given by
\be
  d\mu(\{\bsigma\}) \;=\;
  \exp\!\left[ \beta_P \sum\limits_{\scrs}
              \prod\limits_{\<xy\> \in \scrs} \sgn(\bsigma_x \!\cdot \bsigma_y)
        \right]
  \, \prod\limits_x d\Omega(\bsigma_x)
\ee
where the sum runs over plaquettes $\scrs$.
For $\beta_P = 0$, this is the ordinary $\sigma$-model.
For $\beta_P \to +\infty$,
this is the no-vortex $\sigma$-model \cite{Mack_79,Patrascioiu_92b},
in which the constraint
$\prod_{\<xy\> \in \scrs} \bsigma_x \!\cdot \bsigma_y \ge 0$
is imposed for every plaquette $\scrs$.

3.  This proof is based on the embedding
$(\varepsilon,\bsigma) \mapsto \varepsilon\bsigma$.
More general embeddings could be considered, i.e.\ 
$(\varepsilon,\bsigma) \mapsto (I-P) \bsigma + \varepsilon P\bsigma$
where $P$ is any orthogonal projection in $\R^N$.
(If $\rank P = 1$, this is the Wolff embedding \cite{Wolff_89a,Wolff_90};
 if $\rank P = k$, this is the codimension-$k$ embedding discussed in
 \cite{CEPS_swwo4c2};
 if $\rank P = N$, it is the embedding used above.)
We do not know whether embeddings with $\rank P < N$ might be useful
for proving correlation inequalities.
We note, however, that the Wolff embedding has been employed
for analytic purposes (rigorous lower bounds on the correlation function
via a Fortuin-Kasteleyn representation)
by Patrascioiu and Seiler \cite{Patrascioiu_92a} and
Aizenman \cite{Aizenman_94}.

\bigskip

Next we consider bounds relating the $N$-component $\sigma$-model
to an $XY$ model.
For simplicity we consider first an ordinary $N$-vector model
\be
   H(\{\bsigma\})   \;=\;
    -\, \sum_{\< xy \>} J'_{xy} \, \bsigma_x \!\cdot \bsigma_y
  \;,
 \label{ham:N-vector}
\ee
and compare it to an $XY$ model
\be
   H(\{\theta\})   \;=\;
    -\, \sum_{\< xy \>} J_{xy} \, \cos(\theta_x - \theta_y)
  \;.
 \label{ham:XY_pair}
\ee

\begin{theorem}
 \label{thmA.6}
Consider an $N$-vector model defined by the Boltzmann-Gibbs measure\break
$Z^{-1} e^{-H(\{\bsigma\})}  d\mu(\{\bsigma\})$,
where $H$ is defined by \reff{ham:N-vector}
and $d\mu(\{\bsigma\}) = \prod\limits_x d\Omega(\bsigma_x)$.
Consider for comparison an $XY$ model defined by the Boltzmann-Gibbs measure
$Z^{-1} e^{-H(\{\theta\})}  d\nu(\{\theta\})$,
where $H$ is defined by \reff{ham:XY_pair}
and $d\nu(\{\bsigma\}) = \prod\limits_x d\theta_x$.
Assume that $|J'_{xy}| \le J_{xy}$ for all $x,y$.
Then
\be
   \left| \< \bsigma_x \!\cdot \bsigma_y \> _{N,J'} \right|
   \; \le \;
   \< \cos(\theta_x - \theta_y) \> _{XY,J}
 \label{thmA.6.1}
\ee
and
\begin{subeqnarray}
   0  \;\le\;
   \< (\bsigma_x \!\cdot \bsigma_y)^2 \> _{N,J'}
   & \le &
   {1 \over N}  \,+\,
     {(N-1)(N+2)  \over
      N \left[ 2N - (N-2) \< \cos 2(\theta_x - \theta_y) \> _{XY,J} \right]
     }
     \< \cos 2(\theta_x - \theta_y) \> _{XY,J}                 \nonumber \\ \\
   & \le &
   {1 \over N}  \,+\,
     \left( 1 - {1 \over N} \right)  \< \cos 2(\theta_x - \theta_y) \> _{XY,J}
   \;.
 \label{thmA.6.2}
\end{subeqnarray}
\end{theorem}

\proof
Let $R(\theta) \in SO(N)$ be the rotation through an angle $\theta$
in the 1--2 plane.
We will employ an $XY$ embedding
$(\theta_x,\bsigma_x) \mapsto R(\theta_x) \bsigma_x$.
The proof is based on the trivial identity
\be
   \int F(\{\bsigma\}) \, d\mu(\{\bsigma\})
   \;=\;
   \int\!\int F(\{R(\theta) \bsigma\}) \, d\mu(\{\bsigma\})
                                        \, d\nu(\{\theta\})
   \;,
 \label{trivial_ident_XY}
\ee
where $d\mu(\{\bsigma\})$ is uniform measure on the product of spheres,
$d\nu(\{\theta\})$ is uniform measure on the product of circles,
and $F$ is an arbitrary function.

Define $\bsigma_x^\para = (\sigma_x^{(1)}, \sigma_x^{(2)})$
and $\bsigma_x^\perp = (\sigma_x^{(3)}, \ldots, \sigma_x^{(N)})$.
Let us apply \reff{trivial_ident_XY} to
$F(\{\bsigma\}) = (\bsigma_x^\para \!\cdot \bsigma_y^\para)
                  e^{-H(\{\bsigma\})}$:
\begin{eqnarray}
& &  \int (\bsigma_x^\para \!\cdot \bsigma_y^\para)
          \, e^{-H(\{\bsigma\})} \, d\mu(\{\bsigma\})
     \nonumber \\[1mm]
& &  \qquad =\;
     \int\!\int  \cos[ \theta_x - \theta_y +
                       \angle(\bsigma_x^\para , \bsigma_y^\para) ]
                 \, (1 - {\bsigma_x^\perp}^2)^{1/2}
                    (1 - {\bsigma_y^\perp}^2)^{1/2}
                 \; e^{-H(\{ R(\theta) \bsigma\})}
                 \, d\mu(\{\bsigma\}) \, d\nu(\{\theta\})
     \nonumber \\[1mm]
\end{eqnarray}
where $\angle(\bsigma_x^\para , \bsigma_y^\para)$
denotes the angle between $\bsigma_x^\para$ and $\bsigma_y^\para$.
Note that $H(\{ R(\theta) \bsigma\})$,
considered as a function of $\{\theta\}$ for fixed $\{\bsigma\}$,
is an $XY$ model with couplings
$\widetilde{J}_{xy} = J'_{xy} |\bsigma_x^\para| \, |\bsigma_y^\para|
   e^{i \angle(\bsigma_x^\para , \bsigma_y^\para)}$.
Since $|\widetilde{J}_{xy}| \le |J'_{xy}| \le J_{xy}$,
we can perform the integration over $\{\theta\}$,
using the Ginibre comparison inequality (Theorem \ref{thmA.4})
to obtain an upper bound:
\begin{eqnarray}
& &  \left| \int \cos[ \theta_x - \theta_y +
                       \angle(\bsigma_x^\para , \bsigma_y^\para) ]
     \, e^{-H(\{ R(\theta) \bsigma\})}
     \, d\nu(\{\theta\})
   \right|   
   \nonumber \\[1mm]
& &  \qquad\qquad \le\;
  \< \cos(\theta_x - \theta_y) \> _{XY,J}
  \; \int  e^{-H(\{ R(\theta) \bsigma\})}
     \, d\nu(\{\theta\})
  \;.
\end{eqnarray}
It follows that
\begin{eqnarray}
& &  \left| \int (\bsigma_x^\para \!\cdot \bsigma_y^\para)
                 \, e^{-H(\{\bsigma\})} \, d\mu(\{\bsigma\})
     \right|
     \nonumber \\[1mm]
& &  \qquad \le\;
     \< \cos(\theta_x - \theta_y) \> _{XY,J}
     \int\!\int 
             (1 - {\bsigma_x^\perp}^2)^{1/2}
             (1 - {\bsigma_y^\perp}^2)^{1/2}
     \, e^{-H(\{ R(\theta) \bsigma\})}
     \, d\mu(\{\bsigma\})
     \, d\nu(\{\varepsilon\})
     \nonumber \\
& &  \qquad =\;
     \< \cos(\theta_x - \theta_y) \> _{XY,J}
     \,
     \int  (1 - {\bsigma_x^\perp}^2)^{1/2}  (1 - {\bsigma_y^\perp}^2)^{1/2}
           \, e^{-H(\{\bsigma\})}  \, d\mu(\{\bsigma\})
     \;.
     \nonumber \\
\end{eqnarray}
In terms of normalized quantities, this says that
\be
   \left| \< \bsigma_x^\para \!\cdot \bsigma_y^\para \> _{N,J'} \right|
   \;\le\;
   \< \cos(\theta_x - \theta_y) \> _{XY,J}
   \,
   \left\< (1 - {\bsigma_x^\perp}^2)^{1/2}
           (1 - {\bsigma_y^\perp}^2)^{1/2} \right\> _{N,J'}
   \;.
\ee
Now we use the inequality $ab \le \smhalf (a^2 + b^2)$
together with global $O(N)$ invariance to conclude that
\begin{eqnarray}
   \left\< (1 - {\bsigma_x^\perp}^2)^{1/2}
           (1 - {\bsigma_y^\perp}^2)^{1/2} \right\>
   & \le &
   \half \left[ \< 1 - {\bsigma_x^\perp}^2 \>  \,+\,
                  \< 1 - {\bsigma_y^\perp}^2 \>  \right]      \nonumber \\[3mm]
   & = &   {2 \over N}   \;.
\end{eqnarray}
Hence
\be
   \left| \< \bsigma_x^\para \!\cdot \bsigma_y^\para \> _{N,J'} \right|
   \;\le\;
   {2 \over N} \, \< \cos(\theta_x - \theta_y) \> _{XY,J}   \;.
\ee
Using $O(N)$ invariance again in the form
$\< \bsigma_x \!\cdot \bsigma_y \> =
 (N/2) \< \bsigma_x^\para \!\cdot \bsigma_y^\para \>$,
we complete the proof of \reff{thmA.6.1}.

Next let us apply \reff{trivial_ident_XY} to
$F(\{\bsigma\}) = (\bsigma_x^\para \!\cdot \bsigma_y^\para)^2
                  e^{-H(\{\bsigma\})}$.
By the same steps as before, we show that
\be
   \< (\bsigma_x^\para \!\cdot \bsigma_y^\para)^2 \> _{N,J'}
   \;\le\;
   \< \cos^2 (\theta_x - \theta_y) \> _{XY,J}
   \,
   \left\< (1 - {\bsigma_x^\perp}^2)
           (1 - {\bsigma_y^\perp}^2) \right\> _{N,J'}
   \;.
 \label{eqA.23}
\ee
Now we exploit $O(N)$ invariance in the form of identities
\begin{subeqnarray}
   \< T_x^{(\alpha\beta)} \>    & = &    0      \\[2mm]
   \< T_x^{(\alpha\beta)} T_y^{(\kappa\lambda)} \>    & = &
         {1 \over N^2 + N - 2}  \,
         (\delta^{\alpha\kappa} \delta^{\beta\lambda} +
          \delta^{\alpha\lambda} \delta^{\beta\kappa} -
            {2 \over N}
          \delta^{\alpha\beta} \delta^{\kappa\lambda} )   \,
         G_T(x,y) \quad
\end{subeqnarray}
where the isotensor field $T_x^{(\alpha\beta)}$
is defined in \reff{def_isotensor_field}.
After some algebra 
we obtain
\begin{subeqnarray}
 \< (\bsigma_x^\para \!\cdot \bsigma_y^\para)^2 \> _{N,J'}
   & = &   {2 \over N^2}  \,+\,  {6N-4 \over N(N^2+N-2)} G_T(x,y)     \\[4mm]
 \< {\bsigma_x^\perp}^2 \>  \;=\; \< {\bsigma_y^\perp}^2 \>
   & = &   {N-2 \over N}                                              \\[4mm]
 \< {\bsigma_x^\perp}^2 {\bsigma_y^\perp}^2 \>
   & = &   \left( {N-2 \over N} \right) ^{\! 2}   \,+\,
                       {4N-8 \over N(N^2+N-2)} G_T(x,y)
\end{subeqnarray}
so that \reff{eqA.23} can be written as
\be
   {2 \over N^2}  \,+\,  {6N-4 \over N(N^2+N-2)} G_T(x,y)
   \;\le\;
   \smhalf (1 + \gamma_{xy})
   \left[ {4 \over N^2} + {4N-8 \over N(N^2+N-2)} G_T(x,y) \right]
\ee
where $\gamma_{xy} \equiv \< \cos 2(\theta_x - \theta_y) \> _{XY,J}$.
The bounds \reff{thmA.6.2} follow by simple algebra.
\qed

{\bf Remarks.}
1.  Note that the additive term $1/N$ in \reff{thmA.6.2}
is just what is needed to give the (properly truncated)
isotensor correlation function, i.e.
\be
   G_T(x,y)_{N,J'}   \;\le\;
     \left( 1 - {1 \over N} \right)  \< \cos 2(\theta_x - \theta_y) \> _{XY,J}
   \;.
\ee

2.  A partially alternate proof of Theorem \ref{thmA.6} can be based
on the parametrization
$\bsigma_x = \Bigl( (1 - {\bsigma_x^\perp}^2)^{1/2} \cos \theta_x,
                    (1 - {\bsigma_x^\perp}^2)^{1/2} \sin \theta_x,
                    \bsigma_x^\perp
             \Bigr)$.
Thus, $\theta_x$ is here the polar coordinate of $\bsigma_x^\para$
(whereas it was previously an {\em additional}\/ rotation applied to
$\bsigma_x^\para$).
Under the identification
$\bsigma_x \leftrightarrow (\theta_x,\bsigma_x^\perp)$,
the uniform measure $d\mu(\{\bsigma\})$ becomes
$d\nu(\{\theta\}) \, d\widetilde{\mu}(\{\bsigma^\perp\})$
for a suitable measure $\widetilde{\mu}$
(whose form is of no interest to us).
Now the Hamiltonian $H(\{\bsigma\})$,
considered as a function of $\{\theta\}$ with $\{\bsigma^\perp\}$ fixed,
has couplings
$\widehat{J}_{xy} =
 J'_{xy} (1 - {\bsigma_x^\perp}^2)^{1/2} (1 - {\bsigma_x^\perp}^2)^{1/2}$,
and so is {\em ferromagnetic}\/ if the original model is
(i.e.\ if $J'_{xy} \ge 0$).
The remainder of the proof is as before, but in the ferromagnetic case
we can now use the
Ginibre II inequality rather than the stronger Ginibre comparison inequality.
However, the first proof has the advantage that it generalizes readily
to {\em any}\/ $\sigma$-model whose invariance group contains a
$U(1)$ subgroup, e.g.\ to principal chiral models.

2. One could consider different embeddings of $XY$ spins into
$N$-vector spins, e.g.\ different one-parameter subgroups
$R(\theta) \in SO(N)$.
We do not know whether such embeddings might be useful
for proving correlation inequalities.

\bigskip
\medskip

Finally, let us consider the $N$-component mixed isovector/isotensor model
\be
   H(\{\bsigma\})   \;=\;
    -\, \sum_{\< xy \>} \left[ J'_{xy} \, \bsigma_x \!\cdot \bsigma_y
           \,+\, \smhalf K'_{xy} \, (\bsigma_x \!\cdot \bsigma_y)^2 \right]
  \;,
 \label{A.mixed.N}
\ee
and compare it to a generalized $XY$ model
\be
   H(\{\theta\})   \;=\;
    -\, \sum_{\< xy \>} \left[ J_{xy} \, \cos(\theta_x - \theta_y)
           \,+\, K_{xy} \, \cos 2(\theta_x - \theta_y)  \right]
  \;.
 \label{A.mixed.XY}
\ee

\begin{theorem}
 \label{thmA.7}
Consider an $N$-component mixed isovector/isotensor model
defined by the Boltzmann-Gibbs measure
$Z^{-1} e^{-H(\{\bsigma\})}  d\mu(\{\bsigma\})$,
where $H$ is defined by \reff{A.mixed.N},
and $d\mu(\{\bsigma\}) = \prod\limits_x d\Omega(\bsigma_x)$.
Consider for comparison a generalized $XY$ model
defined by the Boltzmann-Gibbs measure
$Z^{-1} e^{-H(\{\theta\})}  d\nu(\{\theta\})$,
where $H$ is defined by \reff{A.mixed.XY},
and $d\nu(\{\bsigma\}) = \prod\limits_x d\theta_x$.
Assume that 
\be
   \left\{  \begin{array}{ll}
               |J'_{xy}|     & \hbox{if } |K'_{xy}| \le |J'_{xy}|   \\[2mm]
               {1 \over 4} |K'_{xy}| \left( 1 + {|J'_{xy}| \over |K'_{xy}|}
                                     \right) ^{\! 2}
                             & \hbox{if } |K'_{xy}|  >  |J'_{xy}|
            \end{array}
   \right\}
   \;\le\;   J_{xy}
 \label{thmA.7.hypI}
\ee
and
\be
   \smfrac{1}{4}  |K'_{xy}|   \;\le\;   K_{xy}
\ee
for all $x,y$.
Then inequalities \reff{thmA.6.1} and \reff{thmA.6.2}
hold for $\< \,\cdot\, \> _{N,J',K'}$ and
$\< \,\cdot\, \> _{XY,J,K}$.
\end{theorem}

{\bf Remark.}
The left-hand side of \reff{thmA.7.hypI}
is bounded above by $|J'_{xy}| + {1 \over 4} |K'_{xy}|$.
So one can get a simpler (but weaker) version of the theorem
by using this formula in the hypothesis.

\proof
The proof is identical to that of Theorem \ref{thmA.6}.
We find that $H(\{ R(\theta) \bsigma\})$,
considered as a function of $\{\theta\}$ for fixed $\{\bsigma\}$,
is a generalized $XY$ model with couplings
\begin{subeqnarray}
  \widetilde{J}_{xy}   & = &
      (J'_{xy} + K'_{xy} \bsigma_x^\perp \!\cdot\! \bsigma_y^\perp)
      |\bsigma_x^\para| \, |\bsigma_y^\para|
      e^{i \angle(\bsigma_x^\para , \bsigma_y^\para)}
   \\[2mm]
  \widetilde{K}_{xy}   & = &
      {1 \over 4} K'_{xy} 
      |\bsigma_x^\para|^2 \, |\bsigma_y^\para|^2
      e^{2i \angle(\bsigma_x^\para , \bsigma_y^\para)}
\end{subeqnarray}
By straightforward calculus we see that
$|\widetilde{J}_{xy}| \le$
the left-hand side of \reff{thmA.7.hypI}.
So we can use Ginibre comparison inequality (Theorem \ref{thmA.4})
as before.
The rest of the proof concerns the treatment of the various
{\em observables}\/ (as opposed to {\em Hamiltonians}\/),
and so is unchanged.
\qed

\section*{Acknowledgments}

We wish to thank Paolo Butera, Sergio Caracciolo, Herv\'e Kunz,
Andrea Pelissetto and Paolo Rossi for many helpful discussions,
and Larry Glasser for correspondence.
In particular, it was Paolo Butera who pointed out to one of us (A.D.S.),
nearly a decade ago, that the
$N=\infty$ phase transition in $RP^{N-1}$ models occurs also in
dimension $d=1$ --- a comment that directly inspired the present work.
This research was supported in part by
U.S.\ National Science Foundation grants PHY-9520978 and PHY-9900769.
It was also aided by the gracious hospitality of the Scuola Normale Superiore.

\renewcommand{\baselinestretch}{1}
\large\normalsize
%
%
%
%

\clearpage


%
%

\clearpage

\clearpage

\clearpage

\clearpage

\clearpage

\clearpage

\clearpage

\clearpage

\clearpage

\clearpage

\clearpage

\clearpage

\clearpage

\clearpage

\clearpage

\clearpage

\clearpage

\clearpage

\clearpage

\clearpage

\clearpage

\clearpage

\clearpage

\begin{figure}[p]
\begin{center}
\epsfbox{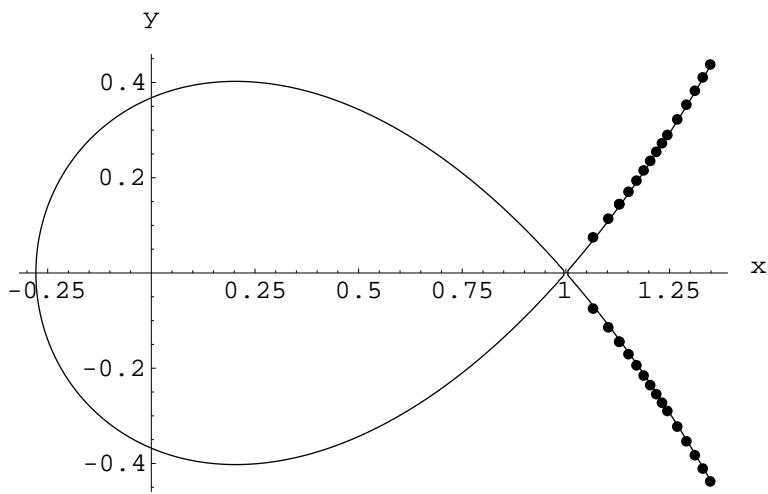}
\end{center}
\caption{
   Szeg\"{o} curve $x^2+y^2=e^{2(x-1)}$
   together with some zeros of $\ofo (a; b; b(x + iy))$
   for $a=1/2$, $b=1000$.
   Note the accumulation of zeros densely on $\scrs_+$.
}
\end{figure}

\clearpage

\end{document}